\documentclass[12pt,letterpaper,english,aps,tightenlinesletterpaper,groupaddress,secnumarabic,nofootinbib]{revtex4}
\usepackage{mathptmx}

\usepackage[T1]{fontenc}
\usepackage[latin9]{inputenc}
\usepackage{babel}
\usepackage{float}
\usepackage{amsmath}
\usepackage{amssymb}
\usepackage{graphicx}
\usepackage[unicode=true,pdfusetitle,
 bookmarks=true,bookmarksnumbered=false,bookmarksopen=false,
 breaklinks=false,pdfborder={0 0 1},backref=section,colorlinks=false]
 {hyperref}
\hypersetup{
 citecolor=blue}

\makeatletter

\pdfpageheight\paperheight
\pdfpagewidth\paperwidth

\@ifundefined{textcolor}{}
{%
 \definecolor{BLACK}{gray}{0}
 \definecolor{WHITE}{gray}{1}
 \definecolor{RED}{rgb}{1,0,0}
 \definecolor{GREEN}{rgb}{0,1,0}
 \definecolor{BLUE}{rgb}{0,0,1}
 \definecolor{CYAN}{cmyk}{1,0,0,0}
 \definecolor{MAGENTA}{cmyk}{0,1,0,0}
 \definecolor{YELLOW}{cmyk}{0,0,1,0}
}

\usepackage{babel}
\usepackage{tensind}

\makeatother

\begin{document}
\title{An Analytic Treatment of Underdamped Axionic Blue Isocurvature Perturbations}
\author{Daniel J. H. Chung}
\email{danielchung@wisc.edu}

\affiliation{Department of Physics, University of Wisconsin-Madison, Madison, WI
53706, USA}
\author{Sai Chaitanya Tadepalli}
\email{stadepalli@wisc.edu}

\affiliation{Department of Physics, University of Wisconsin-Madison, Madison, WI
53706, USA}
\begin{abstract}
Previous computations of strongly blue tilted axionic isocurvature
spectra were computed in the parametric region in which the lightest
time-dependent mass is smaller than the Hubble expansion rate during
inflation, leading to an overdamped time evolution. Here we present
the strongly blue tilted axionic isocurvature spectrum in an underdamped
time evolution parametric regime. Somewhat surprisingly, there exist
parametric regions with a strong resonant spectral behavior that leads
to a rich isocurvature spectral shape. We focus on computing this
resonant spectrum analytically in a large parametric region amenable
to such computations. Because the spectrum is sensitive to nonperturbative
classical field dynamics, a wide variety of analytic techniques are
used including a time-space effective potential obtained by integrating
out high frequency fluctuations.

\tableofcontents{} 
\end{abstract}
\maketitle

\section{Introduction}

Axions are well motivated from the perspective as being a solution
to the strong CP problem \citep{Kim:2008hd,DiLuzio:2020wdo,Peccei:2010ed,Peccei:2006as}
where the experimental bounds (see e.g. \citep{Agrawal:2017cmd})
have pushed the Peccei-Quinn (PQ) symmetry breaking scale $f_{\mathrm{PQ}}$
to large values such that the axions are extremely weakly interacting
with the Standard Model (SM). The largeness of $f_{\mathrm{PQ}}$
at the same time presents an opportunity for axions to be the dominant
component of the cosmological dark matter both from the perspective
of its interaction strengths and the cosmological energy density \citep{Kawasaki:2013ae,Kolb:1990vq,Marsh:2015xka,Sikivie:2006ni,Preskill:1982cy,Abbott:1982af,Dine:1982ah, Berezhiani:1992rk,Sakharov:1994id,Khlopov:1999tm}.
Since axions are important from both particle physics as well as cosmological
perspectives, several experiments have been devoted to its search
\citep{B_hre_2013,2017,Ouellet_2019,Braine_2020,Gramolin_2020,Kwon_2021,Brubaker_2017,mcallister2017organ,alesini2017klash,Alesini_2021,geraci2017progress,Budker_2014,Brun_2019,Armengaud_2014,Armengaud2019}.
A few reviews on direct detection can be found here \citep{ringwald2012exploring,Redondo_2011,Ringwald_2014,Stadnik_2017,carosi2013probing,Graham_2013}
and one can also refer to recent reports \citep{agrawal2021feeblyinteracting,Irastorza_2018}
for a list of various experimental searches and methods.

In most popular axion scenarios where a SM singlet field $\vec{\phi}$
obtains a large vacuum expectation value (VEV) to fix $\langle|\vec{\phi}|\rangle=f_{\mathrm{PQ}}$,
the potential for the singlet has a quartic term, which makes the
$|\vec{\phi}|$ fast roll to its minimum during inflation if $f_{\mathrm{PQ}}\gg H$
where $H$ is the expansion rate during inflation. In such cases,
inflation driven by a different field than $\vec{\phi}$ completes
well after $\langle|\vec{\phi}|\rangle$ settles to the minimum of
the potential. In these situations where the axions are spectator
fields during inflation, the isocurvature spectrum associated with
the axion field is nearly scale invariant. The weakness of the axion
interactions with the SM fields allow the axion isocurvature perturbations
to survive thermalization to be observable through cosmological observables
such as the CMB and galaxy surveys. Production of spectator axion
isocurvature perturbations, its model dependences and associated observational
constraints have been widely studied \citep{Kasuya2009,Kasuya1997,Kawasaki1995,Nakayama2015,Harigaya2015,Kadota2014,Kitajima2014,Kawasaki2014,Higaki2014,Jeong2013,Kobayashi2013,Kasuya2009a,Hamann2009,Hertzberg2008,Beltran2006,Fox2004,Estevez2016,Kearney2016,Nomura2015,Kadota2015,Hikage2012,Langlois2003,Mollerach1990,Axenides1983,Jo2020,Iso2021,Bae2018,Visinelli2017,Takeuchi2014,Bucher:2000hy}.

The work of \citep{Kasuya:2009up} has pointed out that if the PQ
charged SM singlet $\vec{\phi}$ moves along a flat direction lifted
only by masses of $O(H)$ that is typical in supersymmetric embeddings
of the SM, then because the amplitude of the isocurvature perturbations
are proportional to $1/|\vec{\phi}|$ the isocurvature fluctuations
of the spectator axion fields can have a strongly blue tilt. Such
situations allow the isocurvature to be negligible on large scales
probed by the cosmic microwave background (CMB) yet become large on
short length scales. Unlike the compensated isocurvature perturbations
\citep{Grin_2011,Mu_oz_2016,Gordon_2009,He_2015} which hide the total
matter gravitational effects at linear order, the strongly blue isocurvature
perturbations can give large gravitational effects at linear order
on short length scales. Also, unlike the phenomena explored in works
such as \citep{Dimastrogiovanni_2016} where the $O(H)$ mass field
mixing effects with the curvature perturbations lead to observables,
here we are exploring situations where the $O(H)$ mass field is stable
similar to the ideas of \citep{Linde_1997} and can be observed gravitationally
in standard probes such as CMB and large scale structure. Besides
being important for the completion of QCD axion phenomenology, a discovery
of a strongly blue tilted isocurvature spectra will generically indicate
the existence of a dynamical degree of freedom during inflation which
has a time dependent mass, quite model-independently \citep{Chung:2015tha}.
The transition region from the strongly blue tilted region to the
flat region of the isocurvature spectra within the supersymmetric
axion model of \citep{Kasuya:2009up} was investigated by \citep{Chung:2016wvv}.
All of the previous computations of the spectrum focused on the overdamped
scenarios in which the mass of $\vec{\phi}$ flat direction is smaller
than $3H/2$. Even in the fits to data that were done in \citep{Planck:2018jri,Chung:2017uzc},
the parameters were restricted to overdamped scenarios because the
spectrum was naively expected to be negligible in the underdamped
scenarios.

In this paper, we compute the strongly blue axionic isocurvature spectrum
in the underdamped case where the mass of $\vec{\phi}$ flat direction
is larger than $3H/2$, focusing on a parametric region where both
the spectral shape is interesting and analytic computation is possible.
Somewhat surprisingly, the isocurvature spectrum can exhibit a set
of rich spectral resonant shapes with a large enhancement in the amplitude
that crucially depends on the underdamped nature of the $\vec{\phi}$
dynamics. Because the spectrum in this resonant parametric region
depends on non-perturbative classical dynamics of $\vec{\phi}$, a
set of nonperturbative mathematical methods is employed to obtain
the analytic spectrum. These include piecewise polynomial solutions
to differential equations in a couple of time regions and piecewise
effective time-space potential (ETSP) modeling after integrating out
fast oscillations. This allowed us to compute a transfer matrix solution
to the isocurvature mode equations. Because of this lack of perturbativity,
the derivation of analytic expressions as well as the results are
quite lengthy.\footnote{A Mathematica package to evaluate the spectrum using the analytic
methods is given in ``https://pages.physics.wisc.edu/\textasciitilde stadepalli/Blue-Axion-IsoCurvSpec-Underdamped.nb''.} The readers interested in just the main results can refer to Eq.~(\ref{eq:maineq})
where the quantity that is most cumbersome to evaluate is $T_{c}$
as explained there.

Intuitively, the isocurvature spectral range that we can give our
analytic results is for the wave vector $k$ range where the heavy
modes can be decoupled in the axionic model of our interest with multiple
degrees of freedom. For the interesting oscillatory part of the spectrum
arising from a resonance of background field dynamics, we focus on
the parametric region where the velocity of the $\vec{\phi}$ flat
direction field is below a particular critical amount to avoid heavy
mode mixing and the background field dynamics becoming chaotic. Due
to the already extreme length of the present paper, we defer the discussion
of chaotic dynamics and numerical fitting functions that may be useful
for data applications to a separate paper. The intuition behind why
there is an interesting resonance in the underdamped scenarios while
such resonances do not occur in overdamped scenarios is because the
overdamped scenarios have field dynamics characterized by exponentials
of the form $\exp(-\sqrt{9/4-c_{+}}T)$ (where $T$ is a time parameter
and $c_{+}$ is a mass squared parameter for the $\vec{\phi}$ flat
direction during the initial period) which turns into resonant oscillation
producing $\exp(-i\sqrt{c_{+}-9/4}T)\ni\cos(\sqrt{c_{+}-9/4}T)$ for
the underdamped case. This $\cos$ factor which has a zero will allow
the field to reach a dynamically interesting small field region while
the kinetic energy is enhanced by an expansion parameter $(f_{\mathrm{PQ}}/H)^{2}\gg1$,
which translates to a factor of at least an $O(10)$ large enhancement
in the spectral amplitude over a range of $k$ values when compared
to the overdamped scenario. Additionally, the resonance condition
has the initial condition dependent coincidence requirement of interaction
induced mixing between two dynamical degrees of freedom being efficient
as will be explained.

Although several of our plots are given with reference to a parameter
$\omega_{a}^{2}$ related to the QCD axion, it is written to divide
out the effects of the QCD phase transition. Hence, all of our results
can easily be used with the axion field interpreted as a general axion-like-particle.

The order of presentation will be as follows. After a brief review
of the underlying axion model in Sec.~\ref{sec:A-brief-review},
we explain in Sec.~\ref{sec:Decoupling} the decoupling of the heavy
modes that can be viewed as the main characterization of the analytic
formula presented in this work. Sec.~\ref{sec:Behavior-of-} explains
one of the technically difficult part of this work: analytically computing
the time $T_{c}$ when the resonant transition occurs. Sec.~\ref{sec:Numerically-Motivated-model}
presents the parameterization of the ETSP that results from integrating
out the fast oscillations of the classical background fields (which
still exists after decoupling heavy quantum modes). Sec.~\ref{sec:Analytic-spectrum}
maps the parameters of the previous section to the underlying axion
model parameter space spanned by \{$c_{+},c_{-},F$$\}$ and explains
the derivation of the isocurvature spectrum without making assumptions
about how many large dips there are in the ETSP. Sec.~\ref{sec:features-of-isocurvspectrum}
presents a closed form analytic expressions for the isocurvature power
spectrum in a certain restricted region of the underlying model space
supporting a single large dip in the ETSP. Sec.~\ref{sec:Parametric-dependences-of}
explains how the isocurvature spectrum changes as the axion model
parameters $\{c_{+},c_{-},F\}$ are varied. Sec.~\ref{sec:Conclusions}
summarizes this work. An extensive set of appendix sections contain
some of the details omitted in the main text.

Appendix~\ref{sec:Taylor-expansion-consistency} contains an alternative
method of computing a critical time $T_{c}$ required for the isocurvature
computations. It serves as an independent check of the computation
of $T_{c}$ presented in Sec.~\ref{sec:Behavior-of-}. Appendix~\ref{sec:Small}
describes how the less striking non-resonant situations can be computed
within this paper's framework. Appendix~\ref{sec:Adiabatic-approximation-for}
describes a method from \citep{UVoscillations} of integrating out
the fast oscillations to obtain an effective differential equation
containing smaller frequencies. Appendix~\ref{sec:Flat-deviation}
discusses the dynamics of a composite field object that will be useful
in integrating out fast oscillations in the axion model of interest
in this paper. Appendix~\ref{sec:UV-and-IR-phi-fields} applies the
results of Appendix~\ref{sec:Adiabatic-approximation-for} and \ref{sec:Flat-deviation}
to integrate out the fast oscillations in the axion model. Appendix~\ref{sec:Lightest-eigenvector}
describes the crucial dynamics associated with the lightest eigenvector
rotation that will be helpful in constructing the ETSP of Sec.~\ref{sec:Numerically-Motivated-model}
as well as the maps to the $\{c_{+},c_{-},F\}$ space in Sec.~\ref{sec:Analytic-spectrum}.
Appendix~\ref{sec:Lighter-mass-eigenvalue} describes the time dependence
of the lightest mass eigenvalue which will be useful in constructing
the ETSP as well as the parametric map in Sec.~\ref{sec:Analytic-spectrum}.
Appendix~\ref{sec:Justification-for-the} describes the form of the
ETSP parameterization used in Sec.~\ref{sec:Numerically-Motivated-model}.
Appendix~\ref{sec:mB2} discusses the slowly varying part of the
lightest mass squared eigenvalue function that governs the physics
of one of the parameters of Sec.~\ref{sec:Numerically-Motivated-model}.
Appendix~\ref{sec:Decoupling-of-heavy} explains the details of the
effects coming from the heavy modes considered in Sec.~\ref{sec:Decoupling}.

\section{\label{sec:A-brief-review}A brief review of blue axionic isocurvature
perturbations}

In \citep{Kasuya:2009up}, a supersymmetric axion model is studied
with the following well known renormalizable superpotential 
\begin{equation}
W=h(\Phi_{+}\Phi_{-}-F_{a}^{2})\Phi_{0}
\end{equation}
where the subscripts on $\Phi$ indicate $U(1)_{PQ}$ global Peccei-Quinn
(PQ) charges. Note that this is also the most general renormalizable
superpotential transforming under a $U(1)_{R}$ as 
\begin{equation}
\Phi_{0}\rightarrow e^{ir}\Phi_{0}
\end{equation}
\begin{equation}
\Phi_{+}\Phi_{-}\rightarrow\Phi_{+}\Phi_{-}
\end{equation}
\begin{equation}
W\rightarrow e^{ir}W.
\end{equation}
The F-term potential is 
\begin{equation}
V_{F}=h^{2}|\Phi_{+}\Phi_{-}-F_{a}^{2}|^{2}+h^{2}(|\Phi_{+}|^{2}+|\Phi_{-}|^{2})|\Phi_{0}|^{2}.
\end{equation}
A special property of this class of potentials is the existence of
flat directions: i.e. in this particular model, it is 
\begin{equation}
\Phi_{+}\Phi_{-}=F_{a}^{2}\,\,\,\,\,\,\,\,\,\,\,\,\Phi_{0}=0.\label{eq:flatdirection}
\end{equation}
The existence of this flat direction is important because this is
the reason why the effective PQ parameters will be rolling with a
mass of order $H$ during inflation (instead of being much heavier
and having already settled down), taking advantage of the inflationary
$\eta$-problem: i.e. the Kaehler potential induced scalar potential
is 
\begin{equation}
V_{K}=c_{+}H^{2}|\Phi_{+}|^{2}+c_{-}H^{2}|\Phi_{-}|^{2}+c_{0}H^{2}|\Phi_{0}|^{2}
\end{equation}
where $c_{+,-,0}$ are positive $O(1)$ constants. The parameter $c_{+}$
dominantly controls the blue spectral index. This setup implicitly
assumes that the inflaton sector can be arranged to have $H\ll F_{a}$
such that the flat directions are only lifted by the quadratic terms
at the renormalizable level.

Looking along the flat direction of Eq.~(\ref{eq:flatdirection}),
we set $\Phi_{0}=0$. The resulting relevant effective potential during
inflation is 
\begin{equation}
V\approx h^{2}|\Phi_{+}\Phi_{-}-F_{a}^{2}|^{2}+c_{+}H^{2}|\Phi_{+}|^{2}+c_{-}H^{2}|\Phi_{-}|^{2}.\label{effpotential}
\end{equation}
During inflation, the minimum of $V$ lies at 
\begin{align}
|\Phi_{\pm}^{\mbox{min}}| & =\sqrt{\frac{\sqrt{c_{\mp}}}{\sqrt{c_{\pm}}}F_{a}^{2}-\frac{c_{\mp}}{h^{2}}H^{2}}\label{eq:fieldmin}\\
 & \approx\left(\frac{c_{\mp}}{c_{\pm}}\right)^{1/4}F_{a}.
\end{align}
The key initial condition is that $\Phi_{\pm}$ starts out away from
the minimum with a magnitude much larger than $O(F_{a})$ and rolls
towards the minimum during inflation. This implies the $U(1)_{PQ}$
symmetry is broken during inflation. Hence, there will be a linear
combination of the phases of $\Phi_{\pm}$ which will be the Nambu-Goldstone
boson associated with the broken $U(1)_{PQ}$. In particular, with
the parameterization

\begin{equation}
\Phi_{\pm}\equiv\frac{\varphi_{\pm}}{\sqrt{2}}\exp\left(i\frac{a_{\pm}}{\sqrt{2}\varphi_{\pm}}\right)\label{eq:angularparam}
\end{equation}
where $\varphi_{\pm}$ and $a_{\pm}$ are real, the axion is 
\begin{equation}
a=\frac{\varphi_{+}}{\sqrt{\varphi_{+}^{2}+\varphi_{-}^{2}}}a_{+}-\frac{\varphi_{-}}{\sqrt{\varphi_{+}^{2}+\varphi_{-}^{2}}}a_{-}\label{eq:axion}
\end{equation}
while the heavier combination 
\begin{equation}
b=\frac{\varphi_{-}}{\sqrt{\varphi_{+}^{2}+\varphi_{-}^{2}}}a_{+}+\frac{\varphi_{+}}{\sqrt{\varphi_{+}^{2}+\varphi_{-}^{2}}}a_{-}\label{eq:biszero-1}
\end{equation}
is governed by the potential 
\begin{equation}
V_{b}=-h^{2}F_{a}^{2}\varphi_{+}\varphi_{-}\cos\left(\frac{\sqrt{\varphi_{+}^{2}+\varphi_{-}^{2}}}{\varphi_{+}\varphi_{-}}b\right).\label{eq:bheavy}
\end{equation}
Since the $b$ field is heavy (\emph{i.e.} $(\varphi_{+}^{2}+\varphi_{-}^{2})F_{a}^{2}/(\varphi_{+}\varphi_{-})\gg H^{2}$),
it is not dynamically important. Hence, one can gain some intuition
for how the axion composition time evolves by setting $b=0$. When
$\varphi_{+}$ is large, the axion is dominantly $a_{+}$ and later
when $\varphi_{+}$ becomes comparable to $\varphi_{-}$, the axion
is a mixture of $a_{-}$ and $a_{+}$.

According to model \citep{Kasuya:2009up}, the background equations
are as follows.

\begin{align}
\ddot{\Phi}_{+}(t)+3H\dot{\Phi}_{+}(t)+c_{+}H^{2}\Phi_{+}+h^{2}(\Phi_{+}\Phi_{-}-F_{a}^{2})\Phi_{-} & =0\\
\ddot{\Phi}_{-}(t)+3H\dot{\Phi}_{-}(t)+c_{-}H^{2}\Phi_{-}+h^{2}(\Phi_{+}\Phi_{-}-F_{a}^{2})\Phi_{+} & =0
\end{align}
where $\Phi_{\pm}$ has been phase rotated to be real (which is referred
to as $\tilde{\Phi}$ in \citep{Chung:2016wvv}).

The background system can be rescaled as follows 
\begin{align}
\ddot{\phi}_{+}(T)+3\dot{\phi}_{+}(T)+c_{+}\phi_{+}+\xi(\phi_{+},\phi_{-})\phi_{-} & =0\label{eq:backgroundeom0}\\
\ddot{\phi}_{-}(T)+3\dot{\phi}_{-}(T)+c_{-}\phi_{-}+\xi(\phi_{+},\phi_{-})\phi_{+} & =0\label{eq:backgroundeom}
\end{align}
where 
\begin{equation}
\phi_{\pm}\equiv\Phi_{\pm}\frac{h}{H}
\end{equation}
\begin{equation}
F=hF_{a}/H\label{eq:Fscale}
\end{equation}
\begin{equation}
\xi(\phi_{+},\phi_{-})\equiv\phi_{+}\phi_{-}-F^{2}\label{eq:xidef}
\end{equation}
and 
\begin{equation}
T\equiv tH.
\end{equation}

The mode equations can be written in these coordinates as \citep{Chung:2016wvv}
\begin{equation}
\left(\partial_{T}^{2}+3\partial_{T}\right)I+\left(\frac{Ka(0)}{a(T)}\right)^{2}I+\tilde{M}^{2}I=0\label{eq:modeeq}
\end{equation}
where 
\begin{equation}
K\equiv\frac{k}{a(0)H}\label{eq:KDEF}
\end{equation}
\begin{equation}
a(T)=a(0)\exp(T)
\end{equation}
where $I=(I_{+},I_{-})$ and the mass matrix can be rewritten as 
\begin{equation}
\tilde{M}^{2}(T)\equiv\left(\begin{array}{cc}
c_{+} & F^{2}\\
F^{2} & c_{-}
\end{array}\right)+\left(\begin{array}{cc}
\phi_{-}^{2}(T) & 0\\
0 & \phi_{+}^{2}(T)
\end{array}\right).\label{eq:massmat}
\end{equation}
Note that we are neglecting the slow roll effects since the $\epsilon$
in models where this scenario is of greatest interest is negligibly
small during most of inflation. Note also that as explained in \citep{Chung:2015pga},
Eq.~(\ref{eq:modeeq}) represents the non-sourced part of the isocurvature
modes: i.e. the isocurvature modes. The full $\delta\Phi_{\pm}(x)$
field contains gravitational infall inhomogeneities sourced by the
adiabatic inflaton inhomogeneities.

The expression for the isocurvature can be written as 
\begin{equation}
\Delta_{S}^{2}(t,\vec{k})\approx4\omega_{a}^{2}\frac{k^{3}}{2\pi^{2}}I^{\dagger}\left(\begin{array}{cc}
r_{+}^{2} & 0\\
0 & r_{-}^{2}
\end{array}\right)I\label{eqspec}
\end{equation}
\begin{equation}
r_{\pm}\equiv\sqrt{2}\sqrt{\frac{\phi_{\pm}^{2}(t)}{(\phi_{+}^{2}(t)+\phi_{-}^{2}(t))^{2}\theta_{+}^{2}(t_{i})}}
\end{equation}
\begin{eqnarray}
\omega_{a} & \equiv & \frac{\Omega_{a}}{\Omega_{{\rm cdm}}}\label{eq:darkmatterfraction}\\
 & = & W_{a}\theta_{+}^{2}(t_{i})\left(\frac{\sqrt{2}\left(\tilde{\Phi}_{+}^{2}(t_{f})+\tilde{\Phi}_{-}^{2}(t_{f})\right)^{1/2}}{10^{12}{\rm GeV}}\right)^{n_{{\rm PT}}}
\end{eqnarray}
where $W_{a}\approx1.5$ and $n_{{\rm PT}}\approx1.19$ and $t_{f}$
is the time just before QCD phase transition\footnote{The fields $\Phi_{\pm}$ have settled down long before this.}.

The background field equations Eq.~(\ref{eq:backgroundeom}) control
the behavior of isocurvature modes of Eq.~(\ref{eq:modeeq}). Hence,
to understand the isocurvature modes, we need to understand the solution
space of Eq.~(\ref{eq:backgroundeom}) in addition to solving Eq.~(\ref{eq:modeeq}).
In the parametric region of $c_{+}<9/4$, the background solutions
only have a single bump deviation from the time behavior of the lightest
mass squared eigenvalue rising with a constant log slope connecting
to a plateau region in $T$ space.

$\phi_{+}$ starts from a near Planckian value (but restricted to
sub-Planckian to have a good chance of the EFT being valid) and moves
towards $F$ in the approximate solution 
\begin{equation}
\phi_{+}(T)=\phi_{+}(0)e^{-3T/2}\left[\cos(\omega T)+\frac{\epsilon_{0}+3/2}{\omega}\sin(\omega T)\right]\label{eq:zerothphiplus}
\end{equation}
where we have labeled the initial time as $T=0$ and 
\begin{equation}
\omega\equiv\sqrt{c_{+}-9/4}\label{eq:omega}
\end{equation}
\begin{equation}
\epsilon_{0}\equiv\frac{\dot{\phi}_{+}(0)}{\phi_{+}(0)}\label{eq:eps0}
\end{equation}
while $\phi_{-}$ stays near $F^{2}/\phi_{+}$ which is the approximate
minimum of the potential. Hence, during the initial time period, the
background fields $\phi_{\pm}$ (whose nonzero VEV breaks PQ symmetry)
reduce to a single radial degree of freedom. The potentially interesting
and nontrivial aspect of this background system's time evolution occurs
in two cases: a) when $\phi_{+}(T)$ reaches $O(F)$ during the time
when $\phi_{-}\ll F$; b) $T_{c}$ when the energy transfer from $\phi_{-}$
to $\phi_{+}$ becomes significant (this will be quantified in Sec.~\ref{subsec:Resonant-scenarios}).
Both of these time periods are dynamically potentially interesting
because the mass matrix undergoes transitions such that the mass eigenvalues
and the eigenvectors have time variations that are nonadiabatic (change
fast compared to time scale of $H^{-1}$). As we will explain, in
most cases, only event b) leaves significant imprints on the isocurvature
spectrum $\Delta_{S}^{2}(k)$.

\section{\label{sec:Decoupling}Decoupling}

The dimensionality of the mass matrix indicates that there are two
different mass modes. The key dS physics is that at late times, the
massive eigenmodes decay away while the lighter mode is important.
This means that we do not care about the full equations but only the
projected equation onto the lightest eigenvector. Let 
\begin{equation}
I=\sum_{n=1}^{2}y_{n}(k,T)e_{n}(T)
\end{equation}
where $e_{n}(T)$ are real normalized eigenvectors of $\tilde{M}^{2}$
with the the $n=1$ modes being the lighter eigenvalue mode.\footnote{For example, when $\phi_{\pm}(T)$ have reached the values corresponding
to Eq.~(\ref{eq:fieldmin}), the lightest eigenvector is 
\begin{equation}
e_{1}=\frac{\left(-\sqrt{c_{-}},\sqrt{c_{+}}\right)}{\sqrt{c_{-}+c_{+}}}.
\end{equation}
} We will call this the instantaneous normalized eigenvector basis.
The mode Eq.~(\ref{eq:modeeq}) becomes 
\begin{equation}
\mathcal{O}_{1}y_{1}=S_{12}y_{2}\label{eq:lightmode}
\end{equation}
\begin{equation}
\mathcal{O}_{2}y_{2}=S_{21}y_{1}\label{eq:heavymode}
\end{equation}
\begin{equation}
\mathcal{O}_{n}\equiv\left(\partial_{T}^{2}+3\partial_{T}\right)+\left[-\partial_{T}e_{n}\cdot\partial_{T}e_{n}+\frac{k^{2}}{\left(a(T)H\right)^{2}}+m_{n}^{2}(T)\right]\label{eq:kineticoperator}
\end{equation}
\begin{equation}
S_{ns}(T)=-e_{n}\cdot\partial_{T}^{2}e_{s}-3\partial_{T}e_{s}\cdot e_{n}-2e_{n}\cdot\partial_{T}e_{s}\partial_{T}\label{eq:Sns}
\end{equation}
where $m_{n}^{2}(T)$ are the time dependent eigenvalues of $\tilde{M}^{2}$.
One can solve Eq.~(\ref{eq:heavymode}) formally using the Green's
function satisfying 
\begin{equation}
\mathcal{O}_{n}G_{n}(T,T')=\delta(T-T').
\end{equation}
This gives 
\begin{equation}
y_{2}(T)=y_{2}^{h}(T)+\int dT'G_{2}(T,T')S_{21}y_{1}(T')
\end{equation}
where $y_{2}^{h}$ is the solution to $\mathcal{O}_{2}y_{2}^{h}=0$.
Putting this into Eq.~(\ref{eq:lightmode}) gives 
\begin{equation}
\mathcal{O}_{1}y_{1}=S_{12}y_{2}^{h}(T)+S_{12}\int dT'G_{2}(T,T')S_{21}(T')y_{1}(T').\label{eq:integral-equation}
\end{equation}
This is the integro-differential equation that needs to be solved
with Bunch-Davies (BD) boundary conditions to compute the isocurvature
perturbations.

There are two independent solutions to Eq~(\ref{eq:integral-equation}),
both of which are excited to some extent by the quantization with
BD boundary conditions. However, the heavy mode is not excited appreciably
for the BD boundary conditions as has been checked explicitly. Hence,
we focus on the mode with the boundary condition with an initial magnitude
of $y_{2}(T_{i})\ll y_{1}(T_{i})$ which means 
\begin{equation}
y_{2}^{h}(T)=0.
\end{equation}
In this case, we see that the right hand side (RHS) of Eq~(\ref{eq:integral-equation})
can be neglected for the evolution of $y_{1}$ as long as 
\begin{equation}
\left|\frac{\left(e_{2}\cdot\partial_{T}^{2}e_{1}\right)\left(e_{1}\cdot\partial_{T}^{2}e_{2}\right)}{m_{2}^{2}\left(m_{1}^{2}+H^{2}\right)}\right|\ll1.\label{eq:neglectheavy2}
\end{equation}
Before the two fields transition at $T_{c}$ when the mass eigenvalues
change as a function of time nonadiabatically \footnote{The transition is defined in Sec.~\ref{subsec:Beyond-perturbation-theory}.},
we can estimate $m_{2}^{2}\sim h^{2}\phi_{+}^{2}$ and 
\begin{equation}
\left(e_{2}\cdot\partial_{T}^{2}e_{1}\right)\left(e_{1}\cdot\partial_{T}^{2}e_{2}\right)\sim\left[c_{+}F_{a}^{2}H^{2}/\phi_{+}^{2}\right]^{2}
\end{equation}
which means that Eq.~(\ref{eq:neglectheavy2}) is satisfied and $y_{2}$
can be neglected. On the other hand, at $T=T_{c}$, the RHS of Eq~(\ref{eq:integral-equation})
may be important since at that time there is only one scale of $F$
in the system. During this transition time, the time width of the
transition is fixed by 
\begin{equation}
\Delta T_{{\rm c}}\sim\frac{1}{F}.
\end{equation}
The heavy mixing effect is then quantified in the vicinity of $T_{c}$
in terms of a new parameter $\chi_{{\rm HM}}$ defined in Sec.~\ref{sec:Decoupling-of-heavy}
as 
\begin{equation}
\chi_{{\rm HM}}\left(l_{1}^{2},l_{2}^{2}\right)\approx\frac{1}{l_{2}^{2}-l_{1}^{2}}\left(1+2\sqrt{-l_{1}^{2}}+\frac{8/3}{\sqrt{-l_{1}^{2}}}\left(-1+e^{-\sqrt{-l_{1}^{2}}}\cos\left[l_{2}\right]\right)\right)
\end{equation}
where $l_{i}^{2}=(m_{i}^{2}-\dot{e}_{i}^{2})/\dot{e}_{i}^{2}$ and
HM stands for Heavy-Mixing.

One can then show that as long as (Refer to Sec.~(\ref{sec:Decoupling-of-heavy}))
\begin{equation}
\max\left(\chi_{{\rm HM}}\right)\lesssim O\left(r_{a}\right)\label{eq:neglectheavyatTc}
\end{equation}
the effect of heavy mode mixing and the associated RHS of Eq.~(\ref{eq:integral-equation})
can be neglected. Close to transition, as $\dot{e}_{i}^{2}$ tends
to $O(F^{2})$, $m_{1}^{2}$ becomes negative due to nonperturbative
effects of $O(F^{2})$ while the heavier mass eigenvalue $m_{2}^{2}\sim O(\sqrt{c_{+}/c_{-}}F^{2})$.
The details of the physics and the derivation are discussed in Appendix
\ref{sec:Decoupling-of-heavy}. Therefore, we shall work with only
those cases that satisfy the condition in Eq.~(\ref{eq:neglectheavyatTc}).
Later we will express these cases more explicitly in terms of the
Lagrangian parameters.

\section{\label{sec:Behavior-of-}Behavior of $\phi_{\pm}$ near the first
crossing of $\phi_{\pm}$}

For analytically solvable cases, the details of $\phi_{\pm}(T)$ near
the time when 
\begin{equation}
\phi_{+}(T_{1})=\phi_{-}(T_{1})\label{eq:T1def}
\end{equation}
for the first time will be important. Hence, in this section, we provide
an analytic approximation of this time behavior.

\subsection{\label{subsec:Perturbative-solution}Perturbative solution}

For $T\ll T_{1},$ the system can be solved by making the following
expansion: 
\begin{equation}
\phi_{+}(T)=\frac{1}{\lambda}\phi_{+}^{(0)}+O(\lambda^{0})+O(\lambda^{1})+O(\lambda^{2})+\lambda^{3}\phi_{+}^{(1)}\label{eq:phiplamexp}
\end{equation}
\begin{equation}
\phi_{-}(T)=\lambda\phi_{-}^{(0)}+O(\lambda^{2})+\lambda^{3}\phi_{-}^{(1)}+O(\lambda^{4})+\lambda^{5}\phi_{-}^{(2)}\label{eq:phimlamexp}
\end{equation}
where the near Planck scale initial conditions for $\phi_{+}^{(0)}$
gives rise to the prominence of $\phi_{+}^{(0)}$ justifying $\lambda^{-1}$,
the near flat direction solution that we seek fixes the $\lambda$
power for $\phi_{-}^{(0)}$, and the rest of the $\lambda$ powers
are simply increasing powers where we omit some of them (such as $O(\lambda^{0})$
in $\phi_{+}(T)$ expansion) because they will not contribute (as
one can check by introducing them). In other words, one can consider
the expansion in $\lambda$ defined here to be that of smallness of
\begin{equation}
\lambda\leftrightarrow O\left(\sqrt{\frac{\phi_{-}}{\phi_{+}}}\right)\label{eq:lambdadef}
\end{equation}
which is valid over a finite time interval before $T_{1}$.

Putting Eqs.~(\ref{eq:phiplamexp}) and (\ref{eq:phimlamexp}) into
Eqs.~(\ref{eq:backgroundeom0}) and (\ref{eq:backgroundeom}) and
collecting powers of $\lambda$, we find the following: 
\begin{equation}
\lambda^{-1}:\hspace{1em}\partial_{T}^{2}\phi_{+}^{(0)}+3\partial_{T}\phi_{+}^{(0)}+c_{+}\phi_{+}^{(0)}=0\hspace{1em}\hspace{1em}\phi_{+}^{(0)}\phi_{-}^{(0)}-F^{2}=0
\end{equation}
\begin{align}
\lambda^{1}: & \hspace{1em}\partial_{T}^{2}\phi_{-}^{(0)}+3\partial_{T}\phi_{-}^{(0)}+\left(\phi_{+}^{(0)}\right)^{2}\phi_{-}^{(1)}+c_{-}\phi_{-}^{(0)}=0
\end{align}
\begin{align}
\lambda^{3} & :\hspace{1em}\partial_{T}^{2}\phi_{+}^{(1)}+3\partial_{T}\phi_{+}^{(1)}+c_{+}\phi_{+}^{(1)}+F^{2}\phi_{-}^{(1)}=0\hspace{1em},\\
 & :\partial_{T}^{2}\phi_{-}^{(1)}+3\partial_{T}\phi_{-}^{(1)}+F^{2}\phi_{+}^{(1)}+\phi_{-}^{(2)}\phi_{+}^{(0)2}+c_{-}\phi_{-}^{(1)}=0.
\end{align}
The $\lambda^{-1}$ order has a simple solution $\phi_{+}$ identical
to Eq.~(\ref{eq:zerothphiplus}) which can also be rewritten as 
\begin{align}
\phi_{+}(T)\approx\phi_{+}^{(0)}(T) & =\phi_{+}(0)e^{-\frac{3}{2}T}\sec(\varphi)\cos(\omega T-\varphi).\label{eq:approxsol}
\end{align}
where 
\begin{equation}
\tan\varphi\equiv\frac{3/2+\epsilon_{0}}{\omega}.\label{eq:tanphi}
\end{equation}
The matching order $\phi_{-}(T)$ solution is 
\begin{equation}
\phi_{-}(T)\approx\phi_{-}^{(0)}=\frac{F^{2}}{\phi_{+}^{(0)}}.\label{eq:phimapprox}
\end{equation}
Note that when $\phi_{+}$ initially does not have much kinetic energy
(i.e.~$\epsilon_{0}\ll1$), $\varphi$ takes on values that monotonically
decrease from $\pi/2$ to order unity as $c_{+}$ increases from $9/4$
to $10$. The $\lambda^{1}$ order also has a simple, local solution:
\begin{equation}
\phi_{-}^{(1)}=-\frac{1}{\phi_{+}^{(0)2}}\left[\partial_{T}^{2}\phi_{-}^{(0)}+3\partial_{T}\phi_{-}^{(0)}+c_{-}\phi_{-}^{(0)}\right].\label{eq:perturbative}
\end{equation}
The $\lambda^{3}$ order has a nonlocal solution: 
\begin{equation}
[\partial_{T}^{2}\phi_{+}^{(1)}+3\partial_{T}\phi_{+}^{(1)}+c_{+}]G_{+}(T,T')=\delta(T-T')
\end{equation}
\begin{equation}
\phi_{+}^{(1)}=-F^{2}\int dT'G_{+}(T,T')\phi_{-}^{(1)}(T')
\end{equation}
\begin{equation}
\phi_{-}^{(2)}=\frac{-1}{\phi_{+}^{(0)2}}\left[\partial_{T}^{2}\phi_{-}^{(1)}+3\partial_{T}\phi_{-}^{(1)}+F^{2}\phi_{+}^{(1)}+c_{-}\phi_{-}^{(1)}\right].
\end{equation}
Nonetheless, this perturbative expansion by design breaks down near
$T_{1}$ since the $\phi_{-}/\phi_{+}$ hierarchy represented by $\lambda$
is lost.

Interestingly enough, the correction to 
\begin{equation}
\phi_{+}\approx\frac{1}{\lambda}\phi_{+}^{(0)}\label{eq:lambdainv}
\end{equation}
is $O(\lambda^{3})$ which means that the ratio of the next to leading
order to the leading order is $O(\lambda^{4})$. In contrast, the
next to leading order to leading order ratio for $\phi_{-}$ is $O(\lambda^{2})$.
To understand this, note that unlike in the equation of motion for
$\phi_{-},$ $\phi_{+}^{(0)}$ is the \textbf{exact }solution to Eq.~(\ref{eq:backgroundeom0})
if $\phi_{-}=\phi_{-}^{(0)}$. In contrast, $\phi_{-}^{(0)}$ is not
the exact solution to Eq.~(\ref{eq:backgroundeom}) with $\phi_{+}=\phi_{+}^{(0)}$.
This means that even though the perturbative expansion of Eqs.~(\ref{eq:phiplamexp})
and (\ref{eq:phimlamexp}) for both $\phi_{\pm}$ break down at $T_{1}$,
the approximation for $\phi_{-}$ breaks down faster in the region
\begin{equation}
\sqrt{\frac{\phi_{-}}{\phi_{+}}}\approx\frac{1}{2}
\end{equation}
corresponding to an error of the leading order approximation in this
region being 
\begin{equation}
\frac{\Delta\phi_{+}}{\phi_{+}^{(0)}}\sim O\left(\frac{1}{16}\right)\hspace{1em}\hspace{1em}\frac{\Delta\phi_{-}}{\phi_{-}^{(0)}}\sim O\left(\frac{1}{4}\right).
\end{equation}

\begin{figure}
\begin{centering}
\includegraphics[scale=0.8]{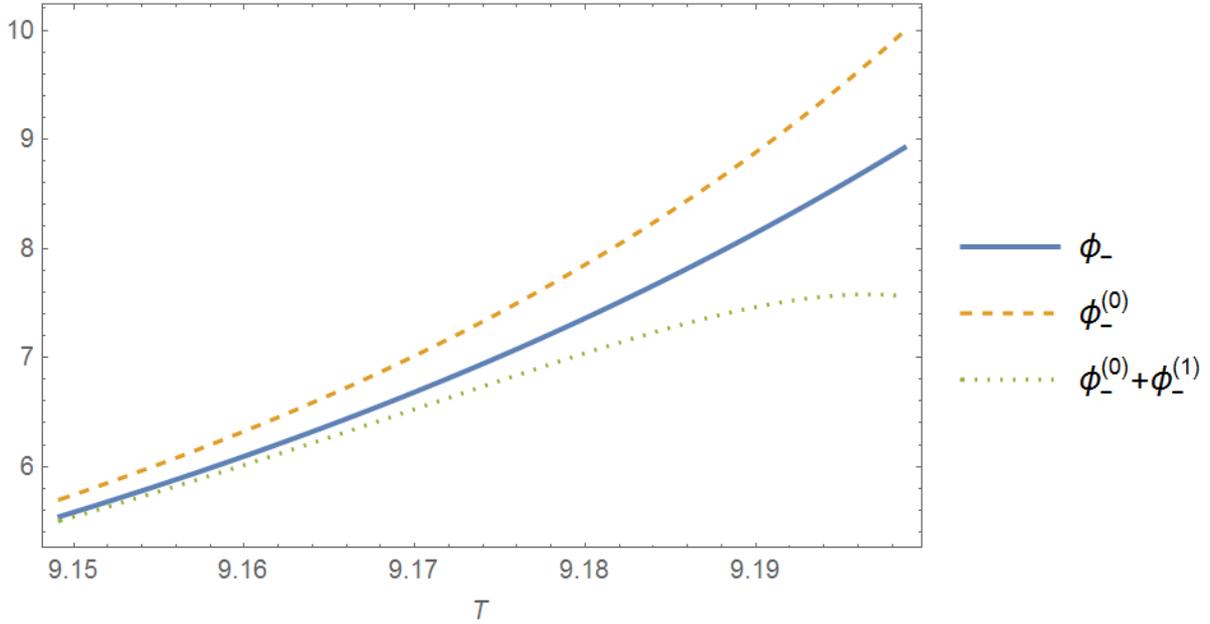} 
\par\end{centering}
\caption{\label{fig:Numerical-background-solution}Numerical background solution
compared to the perturbative solution near $T_{1}=9.248$ and $\{c_{+}=2.35,c_{-}=0.5,F=20.2,\epsilon_{0}=0,\phi_{+}(0)=3.32\times10^{8}\}$.
For $T\lesssim T_{1}-1/F$, the perturbative solution corrected by
$\phi_{-}^{(1)}$ does much better than the leading order solution
$\phi_{-}^{(0)}$ for $T\lesssim9.18$. The increasing deviation at
$T_{1}$ is expected as explained in the text.}
\end{figure}

As a preliminary check on the perturbative $\phi_{\pm}$ solution,
we can compare the numerical solution to the perturbative solution
for $\phi_{-}$ for the case of $\{c_{+}=2.35,c_{-}=0.5,F=20.2,\epsilon_{0}=0,\phi_{+}(0)=3.32\times10^{8}\}$
as shown in Fig.~\ref{fig:Numerical-background-solution}. The improvement
from $\phi_{-}^{(1)}$ is manifest before the expected breakdown of
small $\lambda=O(\sqrt{\phi_{-}/\phi_{+}}$) expansion at $T_{1}$
defined by Eq.~(\ref{eq:T1def}).

\subsection{\label{subsec:Beyond-perturbation-theory}Beyond perturbation theory}

Because the background solutions are sensitive to the details of $\phi_{-}(T_{1})$
and the perturbation theory breaks down when $\phi_{+}(T_{1})=\phi_{-}(T_{1})$
(see Eq.~(\ref{eq:lambdadef})), we need a method to solve for the
background fields more accurately at $T_{1}$. As we will justify
later, because most interesting isocurvature spectral behavior comes
from the models in which $\phi_{+}(T_{1})$ is near a zero crossing
(i.e.~$\phi_{+}^{(0)}(T_{1}+O(1/F))=0$), this section will mainly
focus on such cases. We will mainly use the method of interpolation
using a cubic order polynomial between the time when the perturbation
in $\lambda$ starts to break down and $T_{1}$. We will also check
this method in Appendix \ref{sec:Taylor-expansion-consistency} using
a Taylor expansion approach.

The interpolation polynomial is parameterized as
\begin{align}
\phi_{+}(T) & =p_{0}+p_{1}(T-T_{s})+p_{2}(T-T_{s})^{2}+p_{3}(T-T_{s})^{3}\label{eq:phippoly}\\
\phi_{-}(T) & =q_{0}+q_{1}(T-T_{s})+q_{2}(T-T_{s})^{2}+q_{3}(T-T_{s})^{3}\label{eq:phimpoly}
\end{align}
where we choose $T_{s}$ to be the time when $\phi_{-}(T)$ begins
to deviate significantly (to be defined) from $\phi_{-}^{(0)}(T)$.
We will then choose the interpolation point $T_{I}$ taken to be the
midpoint between $T_{s}$ and $T_{1}$ based on the idea that such
a choice approximately minimizes set of competing errors.\footnote{Since this midpoint choice is an ansatz, we will actually choose the
midpoint between $T_{s}$ and $T_{z}$ where $T_{z}$ defined below
is close to $T_{1}$.} The coefficients $\{p_{n},q_{m}\}$ will be constrained at $T_{I}$
through the original differential equations.

In choosing the time $T_{s}$ to be where the perturbative solution
starts to break down, we expect the deviation to come from the neglect
of the $\ddot{\phi}_{-}^{(0)}$ in the zeroth order perturbative solution.
Hence, we set $T_{s}$ to be the time when 
\begin{equation}
\ddot{\phi}_{-}^{(0)}\sim\left(\frac{1}{n}\right)\phi_{+}^{(0)2}\phi_{-}^{(0)}\label{eq:Tseq}
\end{equation}
where $n$ parameterizes the $(1/n)$ accuracy we want to achieve
in the approximation. For concreteness, we will take $n=10$ in the
analysis below. To solve this equation analytically in a closed form,
it is useful to obtain a polynomial form. Hence we expand $\phi_{+}^{(0)}(T)$
about $\phi_{+}^{(0)}(T_{z})$ where 
\begin{equation}
\phi_{+}^{(0)}(T_{z})=0
\end{equation}
or equivalently 
\begin{equation}
T_{z}=\frac{1}{\omega}\left(\frac{\pi}{2}+\varphi\right)\label{eq:Tz}
\end{equation}
where $\varphi$ is defined by Eq.~(\ref{eq:tanphi}) and assume
that $\phi_{+}^{(0)}(T)$ near $T_{s}$ is well described by a quadratic
expansion of $\phi_{+}^{(0)}(T)$ about $T_{z}$. We will justify
this through self-consistency after the analysis.

To simplify the parametric dependence, define a new dimensionless
parameter $\alpha$ describing the slope of the zero crossing: 
\begin{align}
\alpha & \equiv\frac{\left|\partial_{T}\phi_{+}^{(0)}(T_{z})\right|}{F^{2}}\label{eq:alphadefinition}\\
 & =\omega\frac{\phi_{+}(0)}{F^{2}}\,\sec\varphi\,e^{-3/2T_{z}}.\label{eq:alpha-formula}
\end{align}
In terms of initial conditions $\phi_{+}(0)$ and $\epsilon_{0}$,
the $\alpha$ parameter can be expressed as
\begin{equation}
\alpha=\frac{\phi_{+}(0)}{F^{2}}\sqrt{\omega^{2}+(3/2+\epsilon_{0})^{2}}e^{-3/2[\frac{1}{\omega}\left(\frac{\pi}{2}+\arctan\frac{3/2+\epsilon_{0}}{\omega}\right)]}.\label{eq:alphacplus}
\end{equation}
Putting Eqs.~(\ref{eq:approxsol}) and (\ref{eq:phimapprox}) into
Eq.~(\ref{eq:Tseq}), we obtain 
\begin{align}
T_{s} & =T_{z}-\frac{\left(2n\right)^{\frac{1}{4}}}{\sqrt{\alpha}F}+O\left(\frac{27n}{16\alpha^{2}F^{4}}\right)\label{eq:Tssolution}
\end{align}
where the above expansion is valid for $F\gg1$. We now choose the
interpolation point $T_{I}$ to be the midpoint 
\begin{equation}
T_{I}=\frac{T_{s}+T_{z}}{2}
\end{equation}
which in most cases will be a point that lies in the interval $[T_{s},T_{1}]$.\footnote{In situations where $T_{I}$ coincides with $T_{1}$, one can increase
the value of $n$ to achieve the desired interpolation.}

We then obtain 8 equations to solve for the 8 coefficients of Eqs.~(\ref{eq:phippoly})
and (\ref{eq:phimpoly}) using the background differential Eqs.~(\ref{eq:backgroundeom0}),
(\ref{eq:backgroundeom}), and the values and the derivatives of the
perturbative solutions Eqs\@.(\ref{eq:approxsol}) and (\ref{eq:phimapprox}).
Solving for these coefficients, we obtain 
\begin{align}
p_{0} & =\phi_{+}^{(0)}(T_{s})\label{eq:p0sol}\\
q_{0} & =\phi_{-}^{(0)}(T_{s})\\
p_{1} & =\dot{\phi}_{+}^{(0)}(T_{s})\\
q_{1} & =\dot{\phi}_{-}^{(0)}(T_{s})\label{eq:q1sol}\\
p_{2} & =-\frac{3}{2}p_{1}-c_{+}p_{0}\\
q_{2} & =-\frac{3}{2}q_{1}-c_{-}q_{0}
\end{align}
\begin{align}
p_{3} & \approx\frac{\left(6+9\epsilon+c_{-}\epsilon^{2}+\left(\phi_{+(2)}(T_{I})\right)^{2}\epsilon^{2}\right)\left(3p_{1}+p_{2}\left(2+6\epsilon\right)+\phi_{-(2)}(T_{I})\xi_{(2)}(T_{I})+c_{+}\phi_{+(2)}(T_{I})\right)}{\epsilon D}\\
 & -\frac{\left(-F^{2}+2\phi_{+(2)}(T_{I})\phi_{-(2)}(T_{I})\right)\left(3q_{1}+q_{2}\left(2+6\epsilon\right)+\phi_{+(2)}(T_{I})\xi_{(2)}(T_{I})+c_{-}\phi_{-(2)}(T_{I})\right)}{D/\epsilon}
\end{align}

\begin{align}
q_{3} & \approx p_{3}\left(+\longleftrightarrow-,p\longleftrightarrow q\right)\label{eq:cubiccoeff}
\end{align}
where 
\begin{align}
\xi_{(2)} & =\xi(\phi_{+(2)},\phi_{-(2)})\\
D= & \epsilon^{4}\left(-F^{2}+2\phi_{+(2)}(T_{I})\phi_{-(2)}(T_{I})\right)^{2}-\nonumber \\
 & \left(6+9\epsilon+c_{-}\epsilon^{2}+\left(\phi_{+(2)}(T_{I})\right)^{2}\epsilon^{2}\right)\left(6+9\epsilon+c_{+}\epsilon^{2}+\left(\phi_{-(2)}(T_{I})\right)^{2}\epsilon^{2}\right)
\end{align}
\begin{equation}
\epsilon\equiv T_{I}-T_{s}\label{eq:epsilondef}
\end{equation}
and $\phi_{\pm(2)}$ are defined by Eqs.~(\ref{eq:phippoly}) and
(\ref{eq:phimpoly}) with the cubic terms dropped: e.g. $\phi_{+(2)}(T)\equiv p_{0}+p_{1}(T-T_{s})+p_{2}(T-T_{s})^{2}$.
Note that since we are using the perturbative solutions $\phi_{\pm}^{(0)}$
(i.e.~Eqs.~(\ref{eq:approxsol}) and (\ref{eq:phimapprox})) for
Eqs.~(\ref{eq:p0sol}) through (\ref{eq:q1sol}), we cannot make
$n$ too small (otherwise, the perturbative solutions will be unjustified).
To address this and as a general check, we also solve the background
system using a Taylor expansion method in Appendix \ref{sec:Taylor-expansion-consistency}.
We find reasonable agreement with the current method if we take $n\approx10$.

\begin{figure}
\begin{centering}
\includegraphics[scale=0.7]{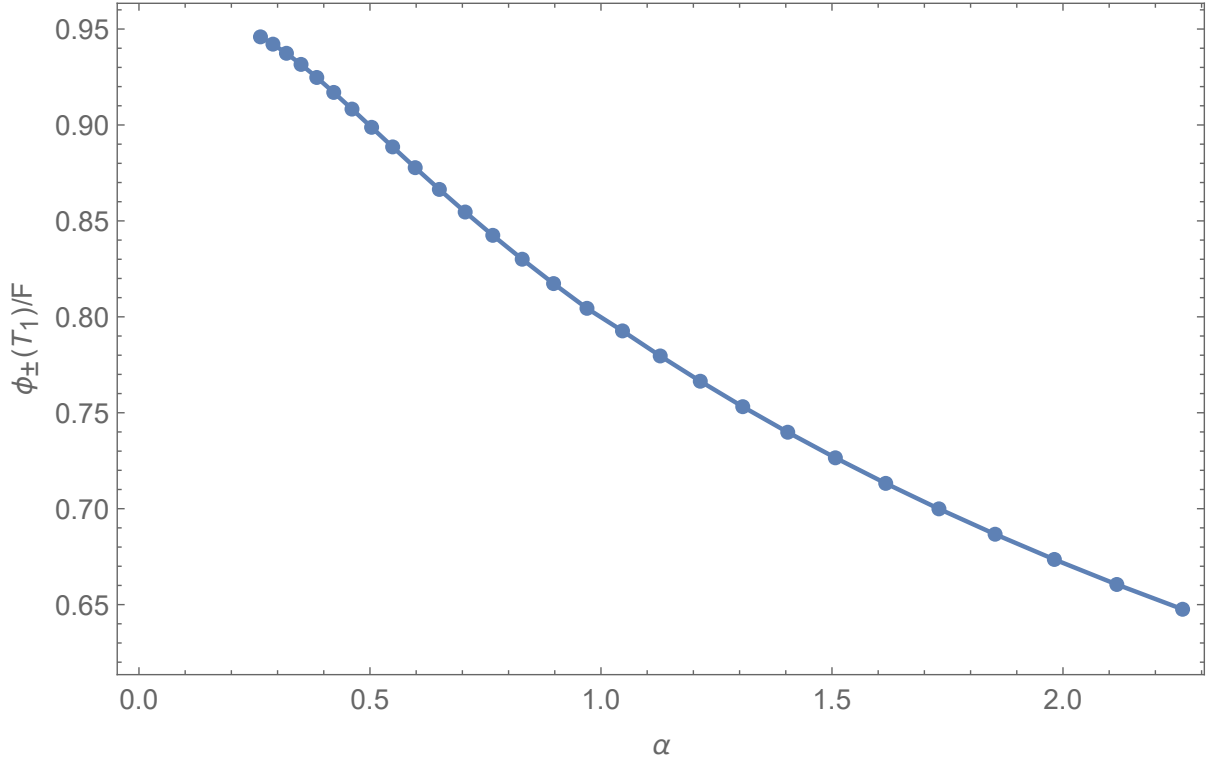} 
\par\end{centering}
\caption{\label{fig:phipT1closetoF}This plot shows that when $\phi_{\pm}(T)$
crosses each other for the first time at $T_{1}$ (evaluated by using
the highly nontrivial Eqs.~(\ref{eq:phippoly}) and (\ref{eq:phimpoly})),
its value is close to $F$ for the parametric region $\alpha\lesssim1$
discussed in the text. This curve is insensitive to the choice of
$\{c_{\pm},F,\phi_{+}(0),\epsilon_{0}\}$ except through $\alpha$
given by Eq.~(\ref{eq:alphadefinition}). We will use this feature
to find an analytic approximation to the isocurvature spectrum.}
\end{figure}

Now that the background solution is approximately fixed, we can use
Eqs.~(\ref{eq:phippoly}) and (\ref{eq:phimpoly}) to solve for the
crossing time $T_{1}$ and the field value there: i.e.~solve for
$\phi_{+}(T_{1})=\phi_{-}(T_{1})$. A plot of $\phi_{+}(T_{1})/F$
is given in Fig.~\ref{fig:phipT1closetoF}, showing that for $\alpha\lesssim1$,
the crossing occurs when $\phi_{+}(T_{1})\approx F$. In terms of
$\alpha$ and $F$, we obtain the following equations for $T_{1}$
and $\phi_{\pm}\left(T_{1}\right)$ by fitting to examples obtained
from the analytic expressions:

\begin{align}
T_{1} & \approx T_{z}-\frac{0.7}{\alpha F}+O\left(\frac{1}{F^{2}}\right),\label{eq:Tcsolution}\\
\phi_{\pm}\left(T_{1}\right) & \approx F\left(1-0.2\alpha\right)+O\left(\frac{1}{F}\right).
\end{align}
By the self-consistency of the solution and the method of construction
involving Eq.~(\ref{eq:Tssolution}), the time $T_{1}$ itself is
$O\left(\left(2n\right)^{\frac{1}{4}}/(\sqrt{\alpha}F)\right)$ away
from $T_{z}$. Note that as $\alpha$ becomes small, $\epsilon=\left(2n\right)^{\frac{1}{4}}/(\sqrt{\alpha}F)$
increases such that the cubic polynomial expansion of the background
fields in Eqs.~(\ref{eq:phippoly}) and (\ref{eq:phimpoly}) is insufficient
and higher order terms become significant. Hence the cubic expansion
nonperturbative method utilized here is valid when 
\begin{equation}
\frac{p_{3}\left(T_{I}-T_{s}\right)^{3}}{\phi_{+(2)}(T_{I})}\lesssim O\left(r_{a}\right)\qquad\frac{q_{3}\left(T_{I}-T_{s}\right)^{3}}{\phi_{-(2)}(T_{I})}\lesssim O\left(r_{a}\right)
\end{equation}
where $T_{I}-T_{s}=\epsilon$. Expanding around $\alpha=0$ yields:
\begin{equation}
\frac{3(48+n)}{432+5n+96\alpha^{2}\left(48+n\right)}\lesssim O\left(r_{a}\right)
\end{equation}
which gives us the following lower bound on $\alpha$ 
\begin{equation}
\alpha>\alpha_{{\rm L}}\equiv\frac{\sqrt{144+3n-432r_{a}-5nr_{a}}}{4\sqrt{6nr_{a}+288r_{a}}}.\label{eq:alpha1}
\end{equation}
For $r_{a}\sim0.2$ and $n=10$ we obtain $\alpha_{{\rm L}}\sim0.25$
serving as a reasonable cutoff for a 20\% accurate computation.

In Sec.~\ref{sec:Decoupling}, we remarked that as long as $\chi_{{\rm HM}}\lesssim O(r_{a})$,
the decoupling of the lighter and heavier modes is justified. Fig.~\ref{fig:chiHMplot}
in Appendix \ref{sec:Decoupling-of-heavy} suggests that this is true
when $\alpha$ is less than an upper bound given as $\alpha_{{\rm U}}$.
Using $r_{a}\sim0.2$ we infer that the decoupling is satisfied at
the first crossing of the background fields for 
\begin{equation}
\alpha\lesssim\alpha_{{\rm U}}\equiv1.\label{eq:alphaU}
\end{equation}
Later in Appendix \ref{sec:UV-and-IR-phi-fields}, we will show that
this upper bound is consistent with another analytic procedure where
we integrate out the high frequency UV modes. Thus, the nonperturbative
methods and the analytic techniques utilized in this paper are applicable
within a specific parameteric region defined by the parameter $\alpha$.
Henceforth in this paper, we will limit ourselves to the study of
underdamped axionic isocurvature power spectrum applicable to those
cases where 
\begin{align}
0.25 & \lesssim\alpha\lesssim1.\label{eq:alphaapplicable}
\end{align}
Interestingly, both the lower and upper bounds are nearly $F$-independent
for $F\gg1$, implying that the parameter $\alpha$ is a suitable
parameterization for studying resonant underdamped isocurvature modes.
Because we will be interested in resonant cases (to be defined below),
$\alpha$ will never be very small in the cases of our main interest.
For completeness, the small $\alpha$ cases ($\alpha<\alpha_{{\rm L}}$)
are discussed in Appendix \ref{sec:Small}.

Additionally, using the Eqs.~(\ref{eq:phippoly}) and (\ref{eq:phimpoly})
we note that the flat-deviation $\left|\xi(\phi_{+},\phi_{-})\right|\sim O(F^{2})$
at the crossing $T_{1}$. This is a unique feature of the underdamped
scenario where the flat-deviation can tend to $O(F^{2})$ if the background
fields cross close to the $\phi_{+}^{(0)}$ zero-crossing ($T_{1}\sim T_{z}-O(1/F)$).
Post $T_{1}$, the flat-deviation oscillates rapidly with a frequency
of $O(F)$ and an $O(F^{2}$) amplitude that decays in time with the
Hubble friction. These rapid oscillations are identified as resonance.
Accordingly, the axion mode function is now characterized by the $F$
scale dynamics till the flat-deviation decays or becomes insignificant.
This is unlike the overdamped or non-resonant scenarios where the
flat-deviation is negligible and the mode amplitude dynamics is defined
primarily by the $H$ scale throughout.

\subsection{\label{subsec:Resonant-scenarios}Resonant scenarios}

In this work, we focus on initial conditions where the $\phi_{+}$
and $\phi_{-}$ initially follow the flat direction of the potential.
This corresponds to the initial trajectories approximated by $\phi_{\pm}^{(0)}$
of Eq.~(\ref{eq:approxsol}) for which the flat deviation $\xi(\phi_{+}^{(0)},\phi_{-}^{(0)})=0$.
For certain parametric cases, there is a significant force on $\phi_{+}$
by $\phi_{-}$ through $\xi\phi_{-}$ when the two fields meet. Such
forces cause displacements of $\phi_{+}$ towards the ``steep''
direction in the potential where $\xi$ is significant. This in turn
causes strong oscillatory behavior of both $\phi_{+}$ and the order
unity coupled $\phi_{-}$.\footnote{The term ``strong oscillatory'' here refers to the frequency being
much larger than that of $\omega$.} We now present a quantitative condition for this class of scenarios
which we call \emph{resonant} scenarios.

During each $T_{{\rm cross}}$ when $\phi_{+}(T_{{\rm cross}})=\phi_{-}(T_{{\rm cross}})$,
the effective coupling force $f_{+}$ on $\phi_{+}$ can be expressed
as 
\begin{equation}
f_{+}(T_{{\rm cross}})=-\xi\phi_{-}|_{T_{{\rm cross}}}
\end{equation}
whose magnitude measures deviation of $\phi_{+}$ from the flat direction
trajectory. This deviation is a sufficient condition for the force
in the ``steep'' direction to be significant. Hence, we define resonant
scenarios to be the cases in which 
\begin{enumerate}
\item $\left.\xi\phi_{-}\right|_{T=T_{c}}\gtrsim O(0.1)\ddot{\phi}_{+}(T_{c})$ 
\item $|\dot{\phi}_{+}(T_{c})|\gtrsim R_{c}F^{2}$ 
\end{enumerate}
where $T_{c}$ is the first $T_{{\rm cross}}$ that satisfies these
conditions. The first of the conditions ensures sufficient coupling
force $f_{+}$ so that $\phi_{+}$ deviates significantly from the
perturbed solution $\phi_{+}^{(0)}$ while the second condition here
is required for $\phi_{+}$ to oscillate with an amplitude whose significance
is determined by the choice of $R_{c}$. For specificity, we will
choose $R_{c}=\alpha_{{\rm L}}$.

In summary, we can define $T_{c}$ to be the time at which 
\begin{equation}
\phi_{+}(T_{c})=\phi_{-}(T_{c})\label{eq:equality}
\end{equation}
for which $\phi_{+}$ has a large kinetic energy and a large deviation
from the flat direction. In this paper, we restrict ourselves to only
those cases where the fields transition at the first crossing. Therefore
$T_{c}=T_{1}$ and hence forth we drop the notation $T_{1}$ for crossing/transition.
While this choice may seem very restrictive, in principle the model
presented in this paper is still applicable to other cases where $T_{1}$
does not correspond to $T_{c}$ (under certain conditions). Such cases
will be studied in a separate paper \citep{futurepaper}.

\section{\label{sec:Numerically-Motivated-model}Numerically Motivated model}

After the transition time $T_{c}$ defined in Eq.~(\ref{eq:equality}),
the $\phi_{+}$ field takes a large dip towards the negative $\phi_{+}$
direction. During a $O(1/F)$ time period surrounding $T_{c}$, the
effective mass squared eigenvalues in the instantaneously diagonal
mass matrix basis has a large dip of $O(F^{2})$. We can capture this
behavior in terms of an approximate step function when solving the
mode equation in the background of this $\phi_{\pm}$ system.
\begin{equation}
\partial_{T}^{2}y_{1}(k,T)+3\partial_{T}y_{1}(k,T)+\left(K^{2}e^{-2T}+\left[-\dot{e}_{1}\cdot\dot{e}_{1}+m_{1}^{2}(T)\right]\right)y_{1}(k,T)=0.\label{eq:schematic}
\end{equation}

\begin{figure}
\begin{centering}
\includegraphics[scale=0.5]{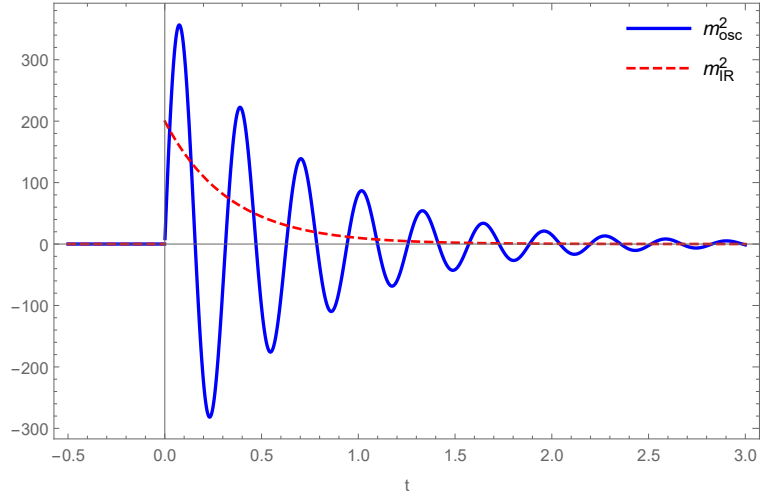} 
\par\end{centering}
\caption{\label{fig:Sample-plots-highlighting} Shown is an effective IR mass
squared (dashed) obtained from integrating out high frequency oscillations
here illustrated with $m_{{\rm osc}}^{2}\equiv Ae^{-\frac{3}{2}t}\sin[ft]$
(solid) where $t$ is the dependent variable of this toy function.
The effective IR mass squared contribution $m_{{\rm IR}}^{2}\equiv\frac{1}{2}\left(A/f\right)^{2}e^{-3t}$
obtained through methods of Appendix \ref{sec:Adiabatic-approximation-for}
is exponentially decaying with an additional factor of $A/\left(2f^{2}\right)$
coming from the UV propagator. Note that the IR ETSP contribution
from the UV modes is positive, consistent with the fact that the UV
oscillations are of the decoupling type (i.e. they are not destabilizing).}
\end{figure}

To model this, define the effective mass squared in instantaneous
normalized eigenvector basis shown in Eq.~(\ref{eq:schematic}) as
\begin{equation}
m_{y_{1}}^{2}\equiv-\dot{e}_{1}\cdot\dot{e}_{1}+m_{1}^{2}\label{eq:my1sq}
\end{equation}
where $m_{1}^{2}$ is the lightest eigenvalue of Eq.~(\ref{eq:massmat})
\begin{equation}
m_{1}^{2}=\frac{\tilde{M}_{11}^{2}+\tilde{M}_{22}^{2}}{2}\left(1-\sqrt{1+4\frac{F^{4}-\tilde{M}_{11}^{2}\tilde{M}_{22}^{2}}{(\tilde{M}_{22}^{2}+\tilde{M}_{11}^{2})^{2}}}\right)\label{eq:lighteigval}
\end{equation}
where $\tilde{M}_{ij}^{2}$ are the elements of $\tilde{M}^{2}$.
Because we will be only interested in the long time behavior of the
zero mode here, we do not need to solve the mode equation with high
time resolution. Hence, we use a double perturbative expansion in
amplitude and frequency as explained in Appendix \ref{sec:Adiabatic-approximation-for}
to separate out the low resolution behavior we are interested in.
After integrating out the UV modes, we find an effective IR mode mass
squared $m^{2}(T)$ that has only a small number of features. This
is illustrated schematically in Fig.~\ref{fig:Sample-plots-highlighting}.
In the context of Eq.~(\ref{eq:lighteigval}), combined with $m_{1}^{2}(T<T_{c})$
being $c_{+}$, we see that integrating out the UV modes has generated
an effective jump in $m^{2}(T)$. As one can see in the figure, the
effective $m^{2}(T)$ is significantly simpler than the original $m_{y_{1}}^{2}(T)$.

\begin{figure}[t]
\begin{centering}
\includegraphics[scale=0.9]{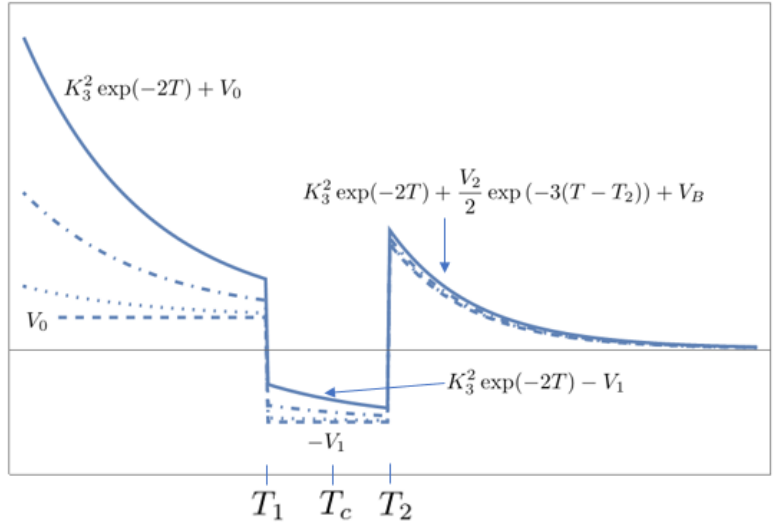} 
\par\end{centering}
\caption{\label{fig:Schematic-massmodel}Schematic diagram of the mass-model
highlighting key features for a single dip case. The dashed curve
represents $m^{2}(T)$. The dotted and dot-dashed curves have the
addition of $K_{1}^{2}\exp(-2T)$ and $K_{2}^{2}\exp(-2T)$, respectively,
to $m^{2}(T)$. The $K_{n}$ values have the hierarchy of $K_{3}>K_{2}>K_{1}$.
The constant $V_{B}$ applicable for the region $T>T_{2}$ is typically
small as suggested implicitly in this schematic figure. Given that
this figure is schematic, the $K_{n}$ here should not be confused
with objects such as $K_{2}$ in Eq.~(\ref{eq:K2cutoff}).}
\end{figure}

In the step function approximation this can be modeled as
\begin{equation}
m^{2}\approx\begin{cases}
V_{0} & T_{0}<T<T_{1}\\
-V_{1} & T_{1}<T<T_{2}\\
V_{B}{\rm sqw}(T,T_{2},T_{B}-T_{2})+\frac{V_{2}}{2}e^{-3(T-T_{2})}-\sum_{i=3}^{n}V_{i}\,{\rm sqw}(T,T_{i},\Delta_{i}) & T_{2}<T
\end{cases}\label{eq:model}
\end{equation}
where 
\begin{equation}
{\rm sqw}(T,T_{i},\Delta_{i})=\begin{cases}
1 & T_{i}\leq T\leq T_{i}+\Delta_{i}\\
0 & otherwise
\end{cases},
\end{equation}
and 
\begin{equation}
{\rm Pset}\equiv\left\{ V_{i},V_{B},T_{B},T_{i},\Delta_{i}\right\} 
\end{equation}
are the model parameters for this steplike approximation. In particular,
$T_{1}$ is defined as the time when IR averaged $m_{y_{1}}^{2}$
makes a negative jump, and $T_{2}$ is defined as the time after $T_{1}$
when IR averaged $m_{y_{1}}^{2}$ makes a positive jump. Later on,
we will see that $V_{B}$ here represents the step approximation of
a smooth decaying nonoscillatory nonequilibrium time-dependent axion
mass function whose extinction point corresponds to the PQ symmetry
breaking vacuum, where the Goldstone theorem condition is satisfied.
The rest of the square well bumps are supposed to be approximations
to oscillatory nonequilibrium time-dependent axion mass function.
This in turn means that 
\[
T_{B}-T_{2}\gg\Delta_{i}
\]
in this parameterization. Fig.~\ref{fig:Schematic-massmodel} shows
a schematic depiction of the mass-model highlighting its key features
for a single dip case.

Although this model can in principle be parameterized with an arbitrary
number of steplike features controlled by $\{V_{i},T_{i},\Delta_{i}\}$,
we will in practice consider at most two such features (i.e.~$T_{i}$
will have at most $i\in\{0,1,2,3\}$ in this paper where the two dips
occur in the intervals $[T_{1},T_{2}]$ and $[T_{3},T_{3}+\Delta_{3}]$
with $T_{2}=T_{1}+\Delta_{1}$). The second dip$\left(V_{3}\right)$
occurs when $\alpha\gtrsim\alpha_{2}$ where $\alpha_{2}$ is defined
in Eq.~(\ref{eq:alphacutoff}). It corresponds to the situation where
the background fields $\phi_{\pm}$ cross each other again after $T_{c}$.
Since we choose $c_{+}>c_{-}$ throughout this paper, there will always
be an even number of crossings between the two background fields after
$T_{c}$. Every such crossing corresponds to a dip in the ETSP within
the framework of our mass-model. By limiting the current analysis
to two dips, we consider only those cases where a third dip is less
than $O(1)$ in magnitude. This correponds to all cases where $\alpha\lesssim\alpha_{3}$
with $\alpha_{3}$ defined by Eq.~(\ref{eq:alpha3conditionalexpr}).
Other cases can be treated by including additional steplike features
as elucidated previously. With an underlying theory such as Eq.~(\ref{effpotential}),
the parameters ${\rm Pset}$ can be computed in terms of $\{c_{+},c_{-},F\}$.
However, here we will first solve this system analytically, then later
express the parameters in terms of $\{c_{+},c_{-},F$\}.

\subsection{\label{subsec:Piecewise-solution-(Scattering}Piecewise solution
(scattering matrix approach)}

In this subsection, we would like to derive an expression for $(y_{1},\dot{y}_{1})$
at some final time $T_{N}$ given its value at some initial time $T_{0}$,
assuming that we know the approximate forms of the solution in $N$
discrete time regions.

Consider a time region $R_{j}$ with boundaries $[T_{j},T_{j+1}]$.
As per this convention, the first region lying between $T_{0}$ and
$T_{1}$ is termed $R_{0}$. The $y_{1}$-mode function within any
region can be expressed through superposition of linearly independent
basis functions $\psi_{1,2}^{(R_{j})}$:
\begin{equation}
y_{1}(K,T)=\begin{cases}
c_{1}^{(R_{0})}\psi_{1}^{(R_{0})}+c_{2}^{(R_{0})}\psi_{2}^{(R_{0})} & T\in[T_{0},T_{1}]\\
c_{1}^{(R_{1})}\psi_{1}^{(R_{1})}+c_{2}^{(R_{1})}\psi_{2}^{(R_{1})} & T\in[T_{1},T_{2}]\\
... & ...
\end{cases}
\end{equation}
with the Wronskian $W^{(R_{j})}(T)=\dot{\psi}_{2}^{(R_{j})}\psi_{1}^{(R_{j})}-\psi_{2}^{(R_{j})}\dot{\psi_{1}}^{(R_{j})}$.
We will take different approximate forms of $\psi_{1,2}^{(R_{n})}$
in each of the regions $R_{n}$ and match the value and its derivatives
at the boundaries to construct $y_{1}$ in the entire domain $\bigcup_{n}R_{n}$
as will be described below.

Let us define $Y$, $\Psi$, and $C$ matrices by rewriting 
\[
\left[\begin{array}{c}
y_{1}\\
\dot{y}_{1}
\end{array}\right]=\left[\begin{array}{cc}
\psi_{1}^{(R_{j})} & \psi_{2}^{(R_{j})}\\
\dot{\psi}_{1}^{(R_{j})} & \dot{\psi}_{2}^{(R_{j})}
\end{array}\right]\left[\begin{array}{c}
c_{1}^{(R_{j})}\\
c_{2}^{(R_{j})}
\end{array}\right]
\]
as 
\begin{equation}
Y=\Psi^{(R_{j})}\,C^{(R_{j})}.
\end{equation}
where 
\begin{equation}
\Psi^{(R_{j})}\equiv\left[\begin{array}{cc}
\psi_{1}^{(R_{j})} & \psi_{2}^{(R_{j})}\\
\dot{\psi}_{1}^{(R_{j})} & \dot{\psi}_{2}^{(R_{j})}
\end{array}\right]
\end{equation}
and $Y\equiv(y_{1},\dot{y}_{1})$. The coefficients $C^{(R_{j})}$
within the region $R_{j}$ are given by the expression 
\begin{equation}
C^{(R_{j})}=\Psi^{(R_{j})-1}(T_{j}^{+})Y(T_{j}^{-})
\end{equation}
Here $T_{j}^{-}$ indicates the incoming $y_{1}$ mode function from
the left-hand side. The solution $Y$ at $T_{j+1}^{-}$ as the mode
exits region $R_{j}$ is $\Psi^{(R_{j})}(T_{j+1}^{-})\,C^{(R_{j})}$.
The mode function at $T=T_{j+1}^{-}$ can be constructed as,

\begin{align*}
Y(T_{j+1}^{-}) & =\Psi^{(R_{j})}(T_{j+1}^{-})\,C^{(R_{j})}\\
 & =\Psi^{(R_{j})}(T_{j+1}^{-})\,\Psi^{(R_{j})-1}(T_{j}^{+})\,Y(T_{j}^{-})\\
 & \equiv S(T_{j+1},T_{j})Y(T_{j}^{-})
\end{align*}
where the matrix $S(T_{j+1},T_{j})$ acts as a scattering-propagator
for the $y_{1}$ mode function from time $T_{j}$ to $T_{j+1}$ through
the time slice $R_{j}$. The mode function at time $T_{N}$ after
passing through $N$ piecewise continuous regions is given as:

\begin{equation}
Y(T_{N}^{-})=\prod_{j=0}^{N-1}S(T_{j+1},T_{j})Y(T_{0}).\label{eq:final-S}
\end{equation}

\subsection{\label{subsec:Independent-analytic-functions}Independent analytic
functions in each piecewise-region}

We shall now give the general linearly independent basis functions
$\psi_{1,2}$ for the form of functions that appear with in our model.
Let us consider the following second order ODE as a generic case for
$y_{1}$ mode equation,
\begin{equation}
\partial_{T}^{2}y_{1}(K,T)+3\partial_{T}y_{1}(K,T)+\left(K^{2}e^{-2T}+m^{2}(T)\right)y_{1}(K,T)=0.\label{eq:modeeq-1}
\end{equation}
In each piecewise region ($[T_{j},T_{j+1}]$ in Eq.~(\ref{eq:model}))
of our $m^{2}(T)$ model, its behavior is either a constant or an
exponentially decaying function. Let's consider these two situations
case by case. 
\begin{enumerate}
\item $\text{constant}\equiv c$ where $c$ can either be positive or negative.
For this case, the above ODE has following linearly independent solutions
\begin{align}
\psi_{1} & =e^{-\frac{3}{2}T}J_{\sqrt{9/4-c}}\left(Ke^{-T}\right),\label{eq:constpot1}\\
\psi_{2} & =e^{-\frac{3}{2}T}Y_{\sqrt{9/4-c}}\left(Ke^{-T}\right).\label{eq:constpot2}
\end{align}
\item $\text{exponentially decaying}\equiv Ve^{-3T}$ for some arbitrary
$V>0$. The effective frequency squared 
\begin{equation}
K^{2}e^{-2T}+Ve^{-3T}\label{eq:effectivefreq}
\end{equation}
now has two different order of decaying exponentials. 
\end{enumerate}
A fundamental intuition for these mode function time evolution is
that whenever $K^{2}e^{-2T}+m^{2}(T)>0$, $|y_{1}(T)|$ has a tendency
to decay while for the opposite sign, $|y_{1}(T)|$ has a tendency
to increase. This can be viewed as the result of the equation with
$K^{2}e^{-2T}+m^{2}(T)=0$ having a constant solution (similar to
the usual inflationary adiabatic mode) which makes $K^{2}e^{-2T}+m^{2}(T)=0$
a ``point'' of criticality in a family of differential equations
represented by Eq.~(\ref{eq:modeeq-1}). This means that the sign
of $m^{2}(T)$ is fundamental to understanding the mode amplitude
evolution as a function of time. Moreover, this behavior of mode functions
is a fundamental element of quantum fields in curved spacetime.

To solve Eq.~(\ref{eq:modeeq-1}) analytically, we define an approximate
frequency squared $U(T)$ as explained below and further divide the
region of interest into sub-regions such that the ODE can be approximated
as

\begin{equation}
\ddot{y}_{1}+3\dot{y}_{1}+U(T)y_{1}(T)\approx0.
\end{equation}
The idea for the approximation is that competing terms of the form
\begin{equation}
A_{1}e^{-2T}+A_{2}e^{-3T}\label{eq:twoterm}
\end{equation}
have only one term dominating except for at most a brief period when
the two terms become comparable. During this ``comparable'' time
period, the Taylor expansion of the time dependence is 
\begin{align}
A_{1}e^{-2T}+A_{2}e^{-3T} & \approx2A_{1}e^{-2T_{X}}\left(1-\frac{5}{2}(T-T_{X})\right)\\
 & \approx2A_{1}e^{-2T_{X}}\exp\left(-\frac{5}{2}(T-T_{X})\right)\label{eq:approx}
\end{align}
where we have linearly expanded about the equality time $T_{X}$ when
the $A_{1}$ and $A_{2}$ terms are equal. Note that the first of
Eq.~(\ref{eq:twoterm}) will dominate over the second term in a time
period of $\Delta T\sim O(1)$. During this time period about $T_{X}$,
the fractional error between Eq.~(\ref{eq:approx}) and the exact
Eq.~(\ref{eq:twoterm}) is 
\begin{equation}
\frac{{\rm exact}-{\rm approx}}{{\rm exact}}=1-\frac{1}{\cosh\left(\frac{T-T_{X}}{2}\right)}\label{eq:error}
\end{equation}
which is about 0.2 for the maximum value of $T-T_{X}=2\ln2$ that
we take below. This lack of sensitivity is an accidental property
of the $\cosh(x)$ which has a flat region at $x=0$.

Now, let's discuss in detail how this approximation is implemented
in the model of Eq.~(\ref{eq:model}). At time $T_{2}$, there is
a jump in the $m^{2}$ of Eq.~(\ref{eq:model}) due to the term $V_{2}$.
We will denote the jump amplitude in the effective frequency squared
$U(T)$ as $V$ in this generically parameterized analysis here. Because
the $V$ term decays faster than the $K^{2}$ term, $U(T)$ will need
to take into account the $K^{2}$ term. We define $T_{V}$ as the
time when $V$ term is equal to the full $U(T)$ that includes the
$K^{2}$ term. Subsequently, $U(T)$ decays according to the approximate
expression of Eq.~(\ref{eq:approx}). Eventually, the $V$ term in
$U(T)$ will be negligible, and only the $K^{2}$ term will need to
be kept. Since the $K^{2}$ term decays slower than the approximate
$U(T)$ in Eq.~(\ref{eq:approx}), the expression for $U(T)$ will
need to be changed to keeping just the $K^{2}$ term when $U(T)$
term equals the $K^{2}$ term at $T_{K}$.

The previous paragraph can be explicitly expressed in terms of the
effective frequency equation as 
\begin{equation}
U(T)-c=\begin{cases}
Ve^{-3T_{X}}e^{-3\left(T-T_{X}\right)} & T_{2}<T<T_{V}\\
\left(K^{2}e^{-2T_{X}}+Ve^{-3T_{X}}\right)e^{-\frac{5}{2}\left(T-T_{X}\right)} & T_{V}<T<T_{K}\\
K^{2}e^{-2T_{X}}e^{-2\left(T-T_{X}\right)} & T_{K}<T<T_{\infty}
\end{cases}\label{eq:UT1}
\end{equation}
Thus, the time interval $\left[T_{2},T_{\infty}\right]$ is sub-divided
into three piecewise regions where each region is characterized by
a distinct exponential decay rate such that the system of differential
equation in Eq.~(\ref{eq:modeeq-1}) is now analytically solvable
in each sub-region. We define $T_{X}$ as when $K^{2}e^{-2T}\text{ and }Ve^{-3T}$
are equal 
\begin{equation}
T_{X}=\ln\left(\frac{V}{K^{2}}\right)\label{eq:TX}
\end{equation}
while $T_{V}$ and $T_{K}$ are defined as the time boundaries that
connect the piecewise regions continuously:
\begin{align}
T_{V} & =T_{X}-2\ln\left(1+\frac{K^{2}e^{-2T_{X}}}{Ve^{-3T_{X}}}\right)=T_{X}-2\ln2=\ln\left(\frac{V}{4K^{2}}\right)
\end{align}
\begin{align}
T_{K} & =T_{X}+2\ln2=\ln\left(\frac{4V}{K^{2}}\right).
\end{align}
To further improve the accuracy of the above piecewise technique,
the amplitude of the exponentials in each sub-region is evaluated
as an integrated average of $K^{2}e^{-2T}+Ve^{-3T}$ as follows: 
\begin{equation}
\frac{\int dT\left(K^{2}e^{-2T}+Ve^{-3T}\right)}{\int dTe^{-nT}}
\end{equation}
where $n\in\{3,5/2,2\}$ in each sub-region. Using the definition
of $T_{X}$ and the amplitude defined above, the expression for $U(T)$
in Eq\@.~(\ref{eq:UT1}) simplifies to
\begin{equation}
U(T)-c=\begin{cases}
\left(V+K^{2}\frac{3/2\left(e^{T_{2}}+e^{T_{V}}\right)}{1+2\cosh\left(T_{2}-T_{V}\right)}\right)e^{-3T} & T_{2}<T<T_{V}\\
\left(\frac{3125}{1364}K\sqrt{V}\right)e^{-\frac{5}{2}T} & T_{V}<T<T_{K}\\
\left(K^{2}+V\frac{1+2\cosh\left(T_{K}-T_{\infty}\right)}{3/2\left(e^{T_{K}}+e^{T_{\infty}}\right)}\right)e^{-2T} & T_{K}<T<T_{\infty}
\end{cases}.
\end{equation}
In each sub-region now the ODE has a Bessel solution of order $\left(\sqrt{9/4-c}\right)/n$
for an ETSP of the form $U(T)=A^{2}e^{-nT}+c$: 
\begin{align}
\psi_{1} & =e^{-\frac{3}{2}T}J_{\frac{\sqrt{9-4c}}{n}}\left(\frac{2}{n}Ae^{-\frac{n}{2}T}\right)\label{eq:expopot1}\\
\psi_{2} & =e^{-\frac{3}{2}T}Y_{\frac{\sqrt{9-4c}}{n}}\left(\frac{2}{n}Ae^{-\frac{n}{2}T}\right)\label{eq:expopot2}
\end{align}
such that the general solution is a superposition of $\psi_{1,2}$.

\section{\label{sec:Analytic-spectrum}Isocurvature spectrum relation to model
parameters}

In this section, we give analytic expressions for the numerically
motivated model parameters and provide isocurvature power spectrum
results in certain regions of the underlying model space \{$c_{+},c_{-},F$$\}$.
The parameter region is most efficiently divided by $\alpha$ introduced
in Eq.~(\ref{eq:alphadefinition}). Small $\alpha$ resonance corresponds
to the dynamics of the background field with a $\dot{\phi}_{+}(T_{c}$)
(where $T_{c}$ defined in Sec.~\ref{subsec:Resonant-scenarios})
that is neither too small (in which case the dynamics is not resonant)
or large (in which case, the dynamics becomes difficult to predict
due to the large series of nonlinear interactions involved). More
precisely, we define this set of resonant cases by Eq.~(\ref{eq:alphaapplicable}):
\begin{equation}
0.25\lesssim\alpha\lesssim1.
\end{equation}
We present below the analytic formula for the isocurvature spectrum
in this corner of the parameter space.

\subsection{\label{subsec:General-map-of}General map of analytic model parameters
to $\{c_{+},c_{-},F\}$}

As defined previously, the mass-model has following set of parameters:
\begin{equation}
{\rm Pset}\equiv\left\{ V_{i},V_{B},T_{B},T_{i},\Delta_{i}\right\} .
\end{equation}
The final $y_{1}$ mode amplitude is evaluated in terms of these model
parameters. Below we will give a map of these model parameters in
terms of $\{c_{+},c_{-},F\}$ and then provide analytic expressions
for their evaluations. We limit ourselves to $i=3$ that cover up
to double-dip cases in Eq.~(\ref{eq:model}). With $\alpha\lesssim\min\left(\alpha_{3},\alpha_{{\rm U}}\right)$
the general map is

\begin{align}
V_{0} & \approx c_{+}\\
V_{1} & \approx\left|\min\left(m_{1}^{2}-\dot{e}_{1}\cdot\dot{e}_{1}\right)\right|\label{eq:e1prsqdip}\\
V_{2} & \approx A^{2}\left\langle \beta^{2}\right\rangle \\
T_{0} & =0\\
T_{1} & \approx T_{c}-\left(\frac{3.11-1.05\alpha}{2F}\right)\\
T_{2} & \approx T_{c}+\left(\frac{3.11-1.05\alpha}{2F}\right)
\end{align}
with the additional second dip for $\alpha\gtrsim\alpha_{2}$ given
by the following expressions
\begin{align}
V_{3} & \approx\begin{cases}
\left(\dot{e}_{1}^{2}\right)_{\max}e^{-3(T_{3}-T_{c})} & \frac{Ae^{-3/2(T_{3}-T_{c})}}{2F^{2}}>0.15\\
\left(\dot{e}_{1}^{2}\right)_{\max}\left(\frac{\dot{g}_{s}\left(T_{3}\right)}{\dot{g}_{s}\left(T_{c}\right)}\right)^{2} & \frac{Ae^{-3/2(T_{3}-T_{c})}}{2F^{2}}<0.15
\end{cases}\\
\phi_{-s}\left(T_{3}\right) & \approx F\qquad\text{for \ensuremath{T_{3}>T_{c}}}\\
\Delta_{3} & \approx\frac{0.72}{\sqrt{V_{3}}}.
\end{align}
Further, the background mass parameter $V_{B}$ for single and double
dip cases is defined as follows 
\begin{align}
V_{B} & \approx\begin{cases}
c_{-}+\frac{1}{\left(T_{L}-T_{2}\right)}\left(\frac{1063}{3072}+\frac{106793c_{-}}{393216c_{+}}\right) & \alpha<\alpha_{\text{2}}\\
V_{B2}(T) & \alpha_{2}<\alpha
\end{cases}\\
T_{B} & \approx\begin{cases}
T_{L} & \alpha<\alpha_{\text{2}}\\
T_{3}+\Delta_{3}+\frac{2}{\Lambda} & \alpha_{\text{2}}<\alpha
\end{cases}
\end{align}
\begin{equation}
V_{B2}(T)\equiv\begin{cases}
c_{-} & T_{2}<T<T_{3}+\Delta_{3}\\
\frac{\Lambda}{2}\int_{0}^{\infty}\frac{c_{-}c_{+}\left(c_{+}n^{4}-c_{-}\right)\left(-1+n^{4}\right)}{\left(c_{+}n^{4}+c_{-}\right)^{2}}dT & T_{3}+\Delta_{3}<T<T_{B}\\
0 & {\rm otherwise}
\end{cases}
\end{equation}
where 
\begin{align}
T_{c} & \approx T_{z}-\frac{0.7}{\alpha F}\\
-A & \equiv\min(\xi)\\
\left\langle \beta^{2}\right\rangle  & \approx F^{-2}\left(0.138+\frac{.14}{1.1+\exp\left(11\left(\alpha-0.72\right)\right)}\right).
\end{align}

\begin{align}
g_{s}(T\gtrsim T_{2}) & \approx\phi_{-s}^{2}-\frac{F^{4}}{\phi_{-s}^{2}}\\
\bar{\Omega} & \approx2.05F+\frac{0.133F+0.045F^{2}}{1+\exp\left(7.86\left(\alpha-0.744+0.0008F\right)\right)}
\end{align}
\begin{align}
n\equiv n(T) & \approx1-\frac{1}{3}\exp\left(-\Lambda T\right)\\
n_{1} & =\sqrt{1-4c_{-}/9}\\
n_{2} & =\sqrt{1-8c_{-}/9}\\
T_{L} & \approx T_{2}-\left(\frac{3}{c_{-}}\right)\ln\left(\frac{2\sin\left(\pi n_{1}\right)2^{2-n_{1}}\Gamma\left(1-n_{1}\right)\phi_{-\min}\,x^{n_{1}}}{\pi\left(3\phi_{-s}(T_{2})x\,\partial_{x}J{}_{n_{1}}\left(x\right)+\left(3\phi_{-s}(T_{2})+2\dot{\phi}_{-s}(T_{2})\right)J_{n_{1}}\left(x\right)\right)}\right)_{x=\frac{A\sqrt{2}}{3\bar{\Omega}}}\quad\text{for \ensuremath{c_{-}\ll1}}\\
\Lambda & \approx\frac{3}{2}-\sqrt{\frac{9}{4}-\frac{4c_{-}c_{+}}{c_{-}+c_{+}}}\\
\frac{3}{F} & \approx\left.\frac{J_{1}\left(\sqrt{2}A/(3\bar{\Omega})\right)}{\sqrt{2}A/(3\bar{\Omega})}\right|_{\alpha=\alpha_{2}}\label{eq:alphacutoff}
\end{align}
where $\phi_{-s}(T>T_{2})$ is given in Eq.~(\ref{eq:phims_sol}),
and $\phi_{-s}(T_{2})$ and $\dot{\phi}_{-s}(T_{2})$ are given in
Eq.~(\ref{eq:phims_atT2}) while $\phi_{-\min}=\phi_{-}\left(T_{\infty}\right)$.
Quite noticeably, the analysis turns very arduous by the addition
of a second dip. Precisely for this reason, in Sec.~\ref{sec:features-of-isocurvspectrum}
we will give closed form analytic expressions for the axion isocurvature
spectrum corresponding to single dip cases only. Although we will
sketch the motivation and the details of the derivation in the Appendices
\ref{sec:Lightest-eigenvector}, \ref{sec:Lighter-mass-eigenvalue},
\ref{sec:Justification-for-the}, and \ref{sec:mB2}, here we describe
the intuition behind this map of the approximation parameters to the
underlying model.

The parameter $V_{0}$ represents the effective axion mass before
$T_{1}$. During this time the mass is nearly constant because the
background fields are following a flat direction such that the potential
does not change as the fields change. The $V_{1}$ dip at $T_{1}\approx T_{c}$
is a type of frame dependent eigenvalue rotation mass effect as seen
in Eq.~(\ref{eq:e1prsqdip}). It is characterized by the superposition
of the $-(\dot{e}_{1})^{2}$ and $m_{1}^{2}$ dips close to $T_{c}$
with a phase separation $\mu$ between the lighter eigenvalue and
the corresponding eigenvector rotation gradient effects. The phase
separation is $\alpha$ dependent. Fields with small $\alpha$ tend
to have an almost coincident superpositioning of the dips and thus
correspond to a small $\mu$. This $\alpha$-dependence can be understood
by referring to the location of the two dips and their subsequent
superposition. From Eq.~(\ref{eq:e1primemax}) and Appendix \ref{sec:Lightest-eigenvector},
we infer that the location of the first $-(\dot{e}_{1})^{2}$ dip
corresponds to the time when the $\phi_{+}$ field tends to $F$.
Meanwhile Fig.~\ref{fig:phipT1closetoF} suggests that fields with
small $\alpha$ transition close to $F$ and the $m_{1}^{2}$ dip
reaches a minimum soon after transition. Therefore, as $\alpha$ increases,
the background fields transition farther from $F$ such that the separation
between the two dips widens resulting in an increased phase separation
$\mu$. For smaller $\alpha$ , the two dips are almost coincident
resulting in a smaller phase separation.

After the $V_{1}$ dip, at time $T_{2}$, there is a jump in the effective
mass squared due to the strong nonlinear interactions through $\xi\phi_{\pm}$
Eq.~(\ref{eq:xidef}). The jump amplitude is approximately $V_{2}$,
and after the jump, there is an exponential decay (see Fig.~\ref{fig:Sample-plots-highlighting})
which captures the results of the UV modes that have been integrated
out. This UV mode averaging has the effect of multiplying the ETSP
$A\sim O(F^{2})$ by the amplitude of the UV mode $A$ divided by
the propagator $1/k^{2}\sim1/F^{2}$ resulting in 
\begin{equation}
\frac{A^{2}}{F^{2}}\sim O(F^{2}).
\end{equation}
$V_{3}$ represents the next $(\dot{e}_{1}\cdot\dot{e}_{1})$ dip
whose physics is similar to the first $(\dot{e}_{1}\cdot\dot{e}_{1})$
dip modified by the Hubble friction. Meanwhile, the parameter $V_{B}$
for $\alpha<\alpha_{2}$ is a constant of $O(c_{-})$ that represents
the average mass squared function $m_{B}^{2}$ over the time interval
from $\left[T_{2},T_{B}\right]$ as detailed in Appendix \ref{sec:mB2}.
The dynamics of the $m_{B}^{2}$ function is controlled by the slow-varying
IR components of the background fields. For $\alpha\lesssim\alpha_{\text{2}}$
it is effectively positive and leads to the decay of the mode amplitude
whereas for $\alpha>\alpha_{2}$ it is negative and results in mode
amplification.

The width of each dip can be evaluated analytically by taking the
ratio of the net area under the peak to its maximum amplitude. In
principle, the width of a $(\dot{e}_{1}\cdot\dot{e}_{1})$ dip increases
for larger $i$ as the amplitude of each dip decreases. This can be
qualitatively understood by the fact that the velocity of the fields
governing the dip widths are proportional to the amplitude of the
fields which is proportional to the square-root of 
\begin{equation}
\dot{e}_{1}\cdot\dot{e}_{1}\sim\sum_{nm}d_{nm}(\phi_{+},\phi_{-})\frac{\dot{\phi}_{n}\dot{\phi}_{m}}{F^{2}}.
\end{equation}
Hence, as shown in Appendix \ref{sec:Lightest-eigenvector}, the dip
width can be parameterized as 
\begin{equation}
\Delta_{(\dot{e}_{1})^{2}}\approx\frac{0.72}{\sqrt{\dot{e}_{1}\cdot\dot{e}_{1}}}.
\end{equation}
The fact that the $0.72$ appears in the above expression approximately
independently of other parameters is due to the fact that we are focusing
on the parametric region where the maximum field excursion parametric
dependences are canceled (See Appendix \ref{sec:Lightest-eigenvector}).
Using the analytically obtained polynomial fit for $(\dot{e}_{1}\cdot\dot{e}_{1})_{\max}$
from Appendix \ref{sec:Lightest-eigenvector}, the width $\Delta$
of the first $(\dot{e}_{1}\cdot\dot{e}_{1})$ dip can be expressed
in terms of $\alpha$ as 
\begin{equation}
\Delta_{(\dot{e}_{1})_{\max}^{2}}\approx\frac{\left(2.93-1.86\alpha\right)}{F}\qquad\forall\;0.25\lesssim\alpha\lesssim1
\end{equation}
which highlights that the dip width reduces with an increasing $\alpha$
or with the incoming velocity of the $\phi_{+}$ field.

A similar expression for the width $T_{2}-T_{1}$ of the $V_{1}$
dip is given below: 
\begin{equation}
T_{2}-T_{1}\approx\frac{\left(3.11-1.05\alpha\right)}{F}\qquad\forall\;0.25\lesssim\alpha\lesssim1.
\end{equation}
By rewriting the width $T_{2}-T_{1}$ in terms of $\Delta_{(\dot{e}_{1})_{\max}^{2}}$,
we obtain the following relation 
\begin{equation}
T_{2}-T_{1}\approx\Delta_{(\dot{e}_{1})_{\max}^{2}}+\frac{\alpha}{F}.
\end{equation}
As expected the width of the $V_{1}$ dip is broader than the width
of the first $(\dot{e}_{1}\cdot\dot{e}_{1})$ dip. For small $\alpha\sim O(0.1)$
scenarios, the width of the $V_{1}$ dip is nearly equivalent to that
of the $(\dot{e}_{1}\cdot\dot{e}_{1})$. This situation corresponds
to a small phase separation $\mu$ such that the $-(\dot{e}_{1}\cdot\dot{e}_{1})$
and $m_{1}^{2}$ dips almost coincide.

Next within our model, the logarithmic functional dependence of $T_{3}-T_{1}$
comes from the exponentially decaying frequency of $\dot{e}_{1}\cdot\dot{e}_{1}$
oscillations as explained in Appendix \ref{sec:Lightest-eigenvector}.

From above parameter assignments, we shall now give expressions for
$(\dot{e}_{1}\cdot\dot{e}_{1})_{{\rm max}}$ and $A$. These are defined
as follows:
\begin{align}
(\dot{e}_{1}\cdot\dot{e}_{1})_{{\rm max}} & \approx\left.\left(\frac{\dot{g}}{5F^{2}}\right)^{2}\left(1+\frac{4}{5}\left(\frac{g+F^{2}}{F^{2}}\right)+\frac{2}{25}\left(\frac{g+F^{2}}{F^{2}}\right)^{2}-\frac{28}{125}\left(\frac{g+F^{2}}{F^{2}}\right)^{3}\right)\right|_{\phi_{+}\rightarrow F}\label{eq:e1primemax}
\end{align}
\begin{equation}
g=\tilde{M}_{11}^{2}-\tilde{M}_{22}^{2}
\end{equation}
\begin{equation}
\tilde{M}_{11}^{2}\equiv c_{+}+\phi_{-}^{2}
\end{equation}
\begin{equation}
\tilde{M}_{22}^{2}\equiv c_{-}+\phi_{+}^{2}
\end{equation}
\begin{equation}
\dot{g}=2\dot{\phi}_{-}\phi_{-}-2\dot{\phi}_{+}\phi_{+}
\end{equation}
where the $\phi_{\pm}$ fields are as defined in Eqs.~(\ref{eq:phippoly})
and (\ref{eq:phimpoly}). By $\phi_{+}\rightarrow F$, we are denoting
functions such as $\phi_{-}(T)$ are to be evaluated at the specific
time $T_{F}$ when $\left|\phi_{+}(T_{F})\right|=F$.

\begin{figure}
\begin{centering}
\includegraphics[scale=0.75]{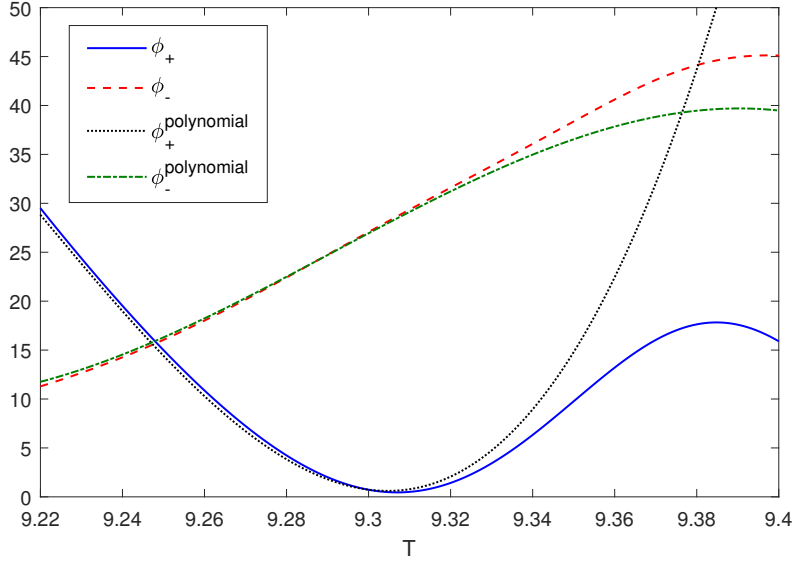} 
\par\end{centering}
\caption{\label{fig:Comparison-of-Taylor}Shown are the polynomial background
$\phi_{\pm}(T)$ solutions given in Eqs.~(\ref{eq:phippoly}) and
(\ref{eq:phimpoly}) with parameters changed to those of Eq.~(\ref{eq:posttransparam}).
These solutions accurately track the numerical solutions over the
interval $[T_{c},T_{m}]$ where $T_{m}$ is the time at which the
minimum of the $\phi_{+}(T)$ occurs. Shown in the figure for comparison
is the numerical solution with $c_{+}=2.35$ and a standard fiducial
set $P_{A}$ of parameters that we will use throughout this paper:
$P_{A}\equiv\{F=20.2,\,c_{-}=0.5,\,\epsilon_{0}=0,\,\phi_{+}(0)=0.1M_{p}/H\}$.}
\end{figure}

\begin{figure}
\begin{centering}
\includegraphics[scale=0.35]{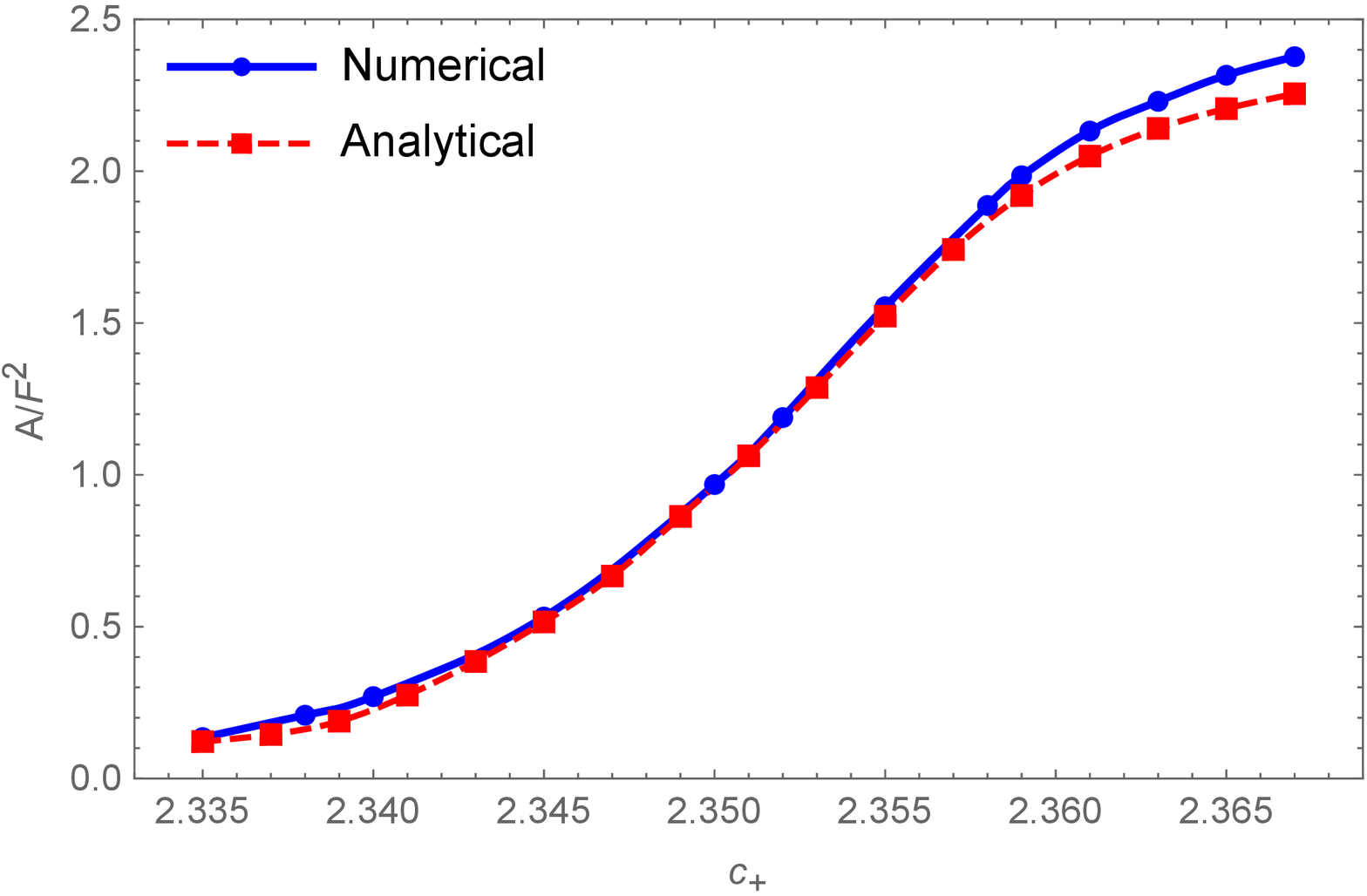}\hspace{0.1in}\includegraphics[scale=0.36]{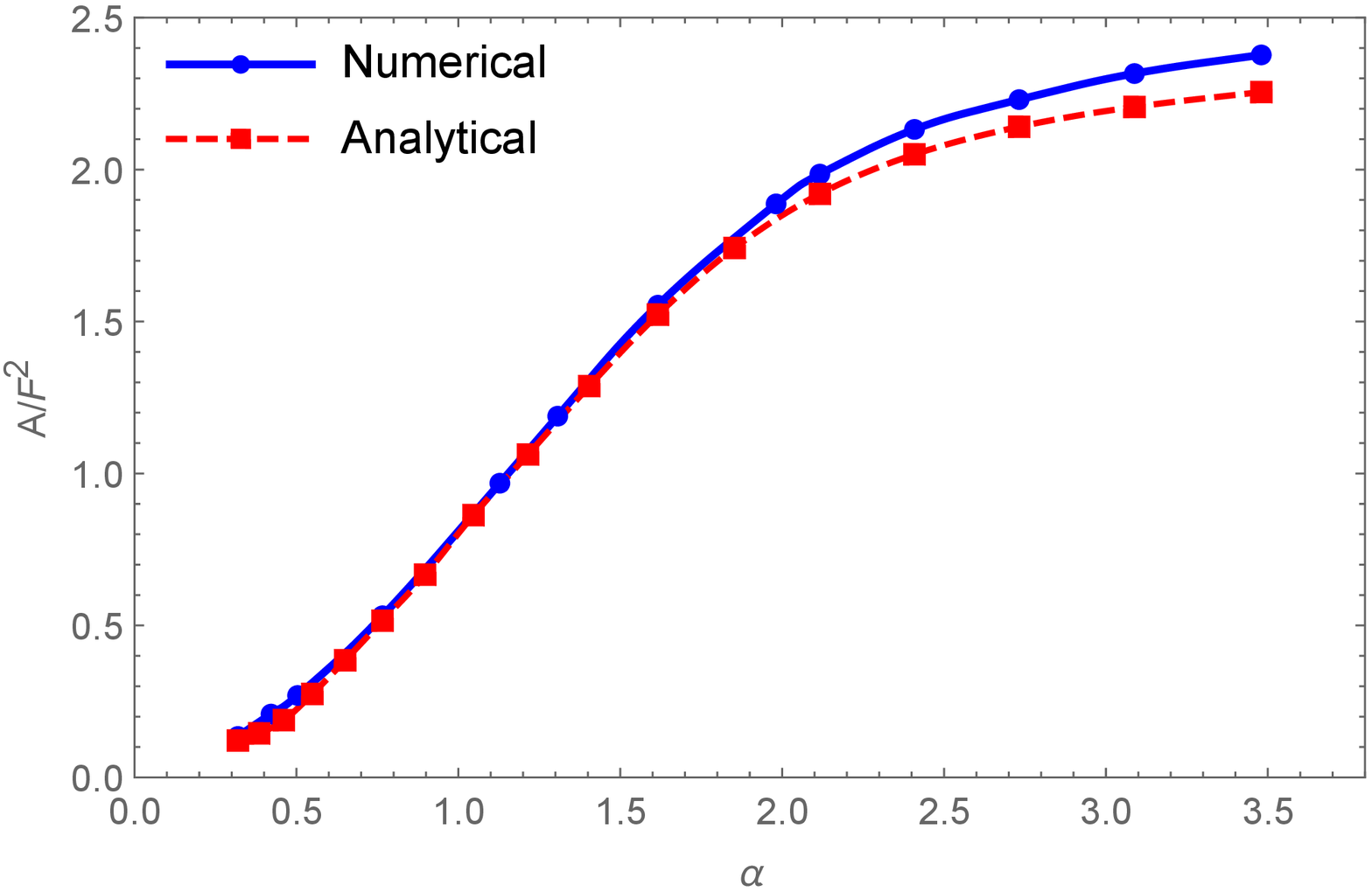} 
\par\end{centering}
\caption{\label{fig:Plots-of-Avscp}Eq.~(\ref{eq:Aanalytic}) using analytic
cubic order polynomial expansion is compared with the value of $A$
obtained by putting the numerically solved $\phi_{\pm}$ into Eq.~(\ref{eq:exact}).
Fiducial parameter set $P_{A}$ of Fig.~\ref{fig:Comparison-of-Taylor}
is used.}
\end{figure}

Next we estimate the amplitude $A$ of the flat-deviation $\xi\left(\phi_{+},\phi_{-}\right)$.
As shown in Appendix \ref{sec:Flat-deviation}, $\xi$ can be approximately
represented via sinusoidal oscillations that drive the resonant exchange
of energy between the $\phi_{+}$ and $\phi_{-}$ fields. For $T>T_{c}$,
we can express $\xi$ as 
\begin{equation}
\xi\left(\phi_{+},\phi_{-}\right)\approx-Ae^{-\frac{3}{2}\left(T-T_{m}\right)}\cos\left(\int_{T_{m}}^{T}\Omega(t)dt\right)
\end{equation}
where $T_{m}\approx T_{c}+O\left(1/F\right)$ and $\Omega\sim O(F)$
is an approximate frequency of oscillations. In order to determine
$A$, we solve for the $\phi$ fields post-transition using another
set of cubic polynomials with primed coefficients 
\begin{equation}
\mbox{parameters for }T>T_{c}:\,p'_{i},q'_{i}\label{eq:posttransparam}
\end{equation}
where the primed coefficients are used to distinguish between non-primed
ones in Eqs.~($\ref{eq:phippoly},\ref{eq:phimpoly}$). The 8 coefficients
are evaluated using similar expressions as in Sec.~\ref{subsec:Beyond-perturbation-theory}
where the initial conditions must now be evaluated at the resonant
transition time $T_{1}=T_{c}$ (instead of $T_{s}$) and choose instead
the interpolation point $T_{I}=T_{c}+\epsilon$ (where $\epsilon$
is defined in Eq.~(\ref{eq:epsilondef})). Finally we evaluate $A$
as 
\begin{align}
-A & \equiv\min(\xi)\\
 & =\min(\phi_{+}(T)\phi_{-}(T)-F^{2}).\label{eq:exact}
\end{align}
For $\alpha\lesssim1$ cases where the minima of $\xi$ roughly corresponds
to the first minima of $\phi_{+}$, we can estimate $A$ by evaluating
the location $T_{m}$ where $\dot{\phi}_{+}(T_{m})=0$.
\begin{align}
T_{m} & =T_{c}+\delta T_{m}\\
\delta T_{m} & \equiv\frac{-p'_{2}+\sqrt{p'_{2}\,^{2}-3p'_{1}p'_{3}}}{3p'_{3}}
\end{align}
which gives us,
\begin{align}
-A\approx\xi(T_{m}) & =-\left.\frac{\ddot{\phi}_{+}}{\phi_{-}}\right|_{T=T_{m}}\\
 & \approx-\frac{2p'_{2}+6p'_{3}\delta T_{m}}{q'_{0}+q'_{1}\delta T_{m}+q'_{2}\delta T_{m}^{2}+q'_{3}\delta T_{m}^{3}}.\label{eq:Aanalytic}
\end{align}
The coefficients turn out to be 
\begin{equation}
q_{0}'\sim p_{0}'\sim O(F)
\end{equation}
\begin{equation}
q_{1}'\sim p_{1}'\sim q_{2}'\sim p_{2}'\sim O(F^{2})
\end{equation}
\begin{equation}
q_{3}'\sim p_{3}'\sim O(F^{3})\label{eq:p3q3p}
\end{equation}
and the actual parametric dependence of $q_{3}'$ and $p_{3}'$ with
$c_{\pm}$ is extremely complicated as can be seen in Sec.~\ref{subsec:Beyond-perturbation-theory}.
We show comparison with the numerical results for $c_{+}=2.35$ in
Figs.~\ref{fig:Comparison-of-Taylor} and \ref{fig:Plots-of-Avscp}.

Remarkably, despite the complicated parametric dependence of Eq.~(\ref{eq:p3q3p})
in terms of $c_{+}$, the parametric dependence of $A$ in terms of
$\alpha$ is very simple as can be seen in Fig.~\ref{fig:Plots-of-Avscp}.
For $\alpha\lesssim2$, we can give a third order polynomial fit for
the amplitude $A$ 
\begin{equation}
A(\alpha,c_{-}=0.5)\approx F^{2}\left(-0.089+0.479\alpha+0.599\alpha^{2}-0.170\alpha^{3}\right)\quad\hspace{1em}\hspace{1em}0.25\leq\alpha\lesssim2\label{eq:Aapprox}
\end{equation}
where all the $c_{+}$ dependence is contained in $\alpha(c_{+})$
through Eq.~(\ref{eq:alphacplus}). In this expression, $\alpha$
is not bounded from above by $\alpha_{{\rm U}}\sim1$ because the
determination of $A$ or $(\dot{e}_{1})_{\max}^{2}$ is independent
of mode decoupling or ETSP evaluation. Instead the above evaluation
is valid as long as the $\phi$ fields can be successfully expressed
via a cubic expansion. As $\alpha$ gets close to 2, the $|\phi_{+}|$
field becomes much larger than $F$ after crossing zero. Correspondingly
the $\phi_{-}$ field undergoes rapid oscillations due to the heavy
mass coming from $\left|\phi_{+}\right|>F$. These rapid $\phi_{-}$
oscillations cannot be captured by the cubic polynomial in $T$ and
the analytic method described above breaks down. Consequently, the
analytic estimation of the minima of $\xi$ soon after $T_{c}$ is
lower than the one obtained numerically as shown in Fig.~(\ref{fig:Plots-of-Avscp}).

Note that $\alpha$ depends only on $c_{+}$ in Eq.~(\ref{eq:alphacplus})
because by definition we are neglecting the back reaction from $\phi_{-}$
in considering the initial velocity condition of $\dot{\phi}_{+}$.
This type of parameterization is natural since $\dot{\phi}_{+}$ is
very large ($O(F^{2}))$ in the resonant scenario where we are giving
our analytic results. Nonetheless, given that the analytic results
are formulaically (as opposed to numerically) fitting the actual background
field solutions to polynomials before and after $T_{c}$ that contain
$c_{-}$, our analytic results fully capture the $c_{-}$ dependence
in the resonant cases considered here.

We remark that the presence of the $V_{3}$ dip is implicitly dependent
upon the strength of the flat-deviation amplitude $A$. The $V_{3}$
dip is associated with a second $\dot{e}_{1}.\dot{e}_{1}$ dip which
occurs when the two background fields cross each other again after
transition at $T_{c}$. This second crossing after $T_{c}$ is controlled
by the condition given in Eq.~(\ref{eq:alpha2conditionalexpr}) that
is dependent upon $A$ and $F$. Including the expression for the
amplitude $A$ from above, we find that the second dip occurs for
$\alpha\gtrsim\alpha_{2}$.

Although the interpolation method used in this analytic result seems
somewhat ad hoc, the results are consistent with a more systematic
expansion around $T_{c}$ as described in Appendix \ref{sec:Taylor-expansion-consistency}.

\subsection{\label{subsec:The-isocurvature-spectrum}The isocurvature spectrum}

The axion isocurvature spectrum is given in Eq.$\,$\ref{eqspec}.
We note that 
\begin{equation}
\lim_{T\rightarrow T_{\infty}}I_{-}=-rI_{+}
\end{equation}
where we define $r=\sqrt{c_{+}/c_{-}}$ and simplify Eq.$\,$\ref{eqspec}
further in terms of $y_{1}$ mode function,
\begin{equation}
\Delta_{s}^{2}(k)\approx\omega_{a}^{2}\frac{4K^{3}}{\pi^{2}}\left|y_{1}(K,T_{\infty})\right|^{2}\frac{r\left(1+r^{4}\right)}{\left(1+r^{2}\right)^{3}}\frac{1}{\theta_{+}^{2}F^{2}}.\label{eq:isocintermsofy1}
\end{equation}
The $y_{1}(K,T_{\infty})$ is solved using the model of Eq.~(\ref{eq:model}),
the approximations of Sec.~\ref{subsec:Independent-analytic-functions},
and its associated model parameters of Sec.~\ref{subsec:General-map-of}.
Let's now sketch the steps involved in a bit more detail.

First, we set up the approximate BD equivalent leading adiabatic order
boundary condition for the $y_{1}(K,T)$ mode equation at time $T_{0}$
for $K$ modes that satisfy $K^{2}\tau_{0}^{2}\gg c_{+}-2$:
\begin{equation}
y_{1}(K,T_{0})=\frac{1}{a(T_{0})\sqrt{2K}}e^{-iK\tau_{0}}\hspace{1em}\hspace{1em}\partial_{T}y_{1}(K,T_{0})=\frac{\left(iK-a(T_{0})\right)}{\left(a(T_{0})\right)^{2}\sqrt{2K}}e^{-iK\tau_{0}}
\end{equation}
where 
\begin{equation}
\tau\equiv-\frac{1}{a(T)H}\label{eq:conformaltime}
\end{equation}
is the conformal time with the scale factor $a(T)=e^{T}$ and evaluate
the first scattering matrix $S(T_{1},T_{0})$ of Eq.~(\ref{eq:final-S})
using the solutions of Sec.~\ref{subsec:Independent-analytic-functions}
evaluated with Eqs.~($\ref{eq:constpot1}$) and (\ref{eq:constpot2}).
Using the solution from $S(T_{1},T_{0})$ for the region $[T_{0},T_{1}]$
we obtain the initial conditions for $y_{1}(K,T)$ at $T=T_{1}$,
when the nonadiabatic rotation of the mass eigenvector becomes strong.
Next until $T_{2}$ when the mass squared jumps, the solutions used
to evaluate $S(T_{2},T_{1})$ are again Eqs.~($\ref{eq:constpot1}$)
and (\ref{eq:constpot2}), but with a tachyonic constant mass squared.
Afterwards, until time $T_{3}$ when the rotation of the mass eigenvector
becomes strong again, the effective frequency-squared is 
\begin{equation}
V_{2}e^{-3\left(T-T_{2}\right)}/2\label{eq:jumppotential}
\end{equation}
(this is what we will call the jump ETSP which is obtained after integrating
out the UV modes). The solution in this region to be used in $S(T_{3},T_{2})$
of Eq.~(\ref{eq:final-S}) is governed by the superposition of Eqs.~($\ref{eq:expopot1}$)
and (\ref{eq:expopot2}) via the approximations of the $U(T)$ in
Eq.~(\ref{eq:UT1}). The initial conditions at $T_{2}$ should be
modified due to the UV integration (highlighted in Appendix \ref{sec:Adiabatic-approximation-for}).
This is primarily done by scaling the $y_{1}\left(T_{2}\right)$ and
its derivative by a $Q$ matrix as follows: 
\begin{equation}
Q=\left(\begin{array}{cc}
1 & 0\\
-\sqrt{V_{2}} & 1
\end{array}\right).
\end{equation}
The region $(T_{2},T_{3})$ is sub-divided into time intervals whose
boundaries are dependent on $K$ and the amplitude of the jump ETSP
through the Eq.~(\ref{eq:TX}). The gentle time dependent changes
in $U(T)$ exponents are most critical for the intermediate modes
that satisfy the condition $Ke^{-T}<1$ slightly after the transition
time $T_{1}$ while the jump ETSP is still significant. Starting at
time $T_{3}$, the cycle repeats with the eigenvector rotation becoming
strong, although with a smaller magnitude than at $T_{1}$. The final
$y_{1}(K,T_{\infty})$ is obtained via Eq.~(\ref{eq:final-S}) where
we select $T_{N}=T_{\infty}$. In our calculations we set $T_{0}=0$
and $T_{\infty}=35$ after which the background fields oscillations
are negligible.

Next we remark that this model has been constructed using only the
lightest mass eigenmodes to keep it analytically tractable. Even then,
we see that the analytic results are complicated and borders on ``intractable''.
Hence, this model is applicable up to a maximum $K$ mode before the
coupling from the heavier $y_{2}$ mode becomes significant: i.~e.
$K\lesssim K_{2}$\textbf{ 
\begin{equation}
K_{2}\equiv m_{2}\frac{a(T_{c})}{a(0)}\label{eq:K2cutoff}
\end{equation}
} (where $m_{2}\sim O(F)$). However, if $K$ becomes sufficiently
large far beyond this heavy mode coupling values, the dynamics eventually
becomes identical to the usual massless axion dynamics. This usual
plateau isocurvature spectrum exists for $K\gtrsim K_{P}$ where
\begin{equation}
K_{P}\approx\begin{cases}
\frac{a(T_{2})}{a(0)}\exp\left(T_{L}-T_{2}\right)\left(\frac{1}{3r_{a}}\right)^{\frac{1}{\Lambda}} & \alpha<\alpha_{2}\\
\frac{a(T_{2})}{a(0)}\exp\left(2\left(T_{4}-T_{2}\right)\right)\left(\frac{1}{3r_{a}}\right)^{\frac{1}{\Lambda}} & \alpha_{2}<\alpha<\alpha_{{\rm 3}}
\end{cases}\label{eq:KPdef}
\end{equation}

\begin{align}
\Lambda & \approx\frac{3}{2}-\sqrt{9/4-\frac{4c_{-}c_{+}}{c_{-}+c_{+}}}
\end{align}
corresponding to the wave vector modes that leave the horizon after
the background fields have settled to a $\left(1-r_{a}\right)$ fraction
of their respective minima. Therefore, the only part of the spectrum
that we do not have a prediction for (in this small $\alpha$ case)
is the $K$ range $[K_{2},K_{P}]$.

\begin{figure}
\begin{centering}
\includegraphics[scale=0.3]{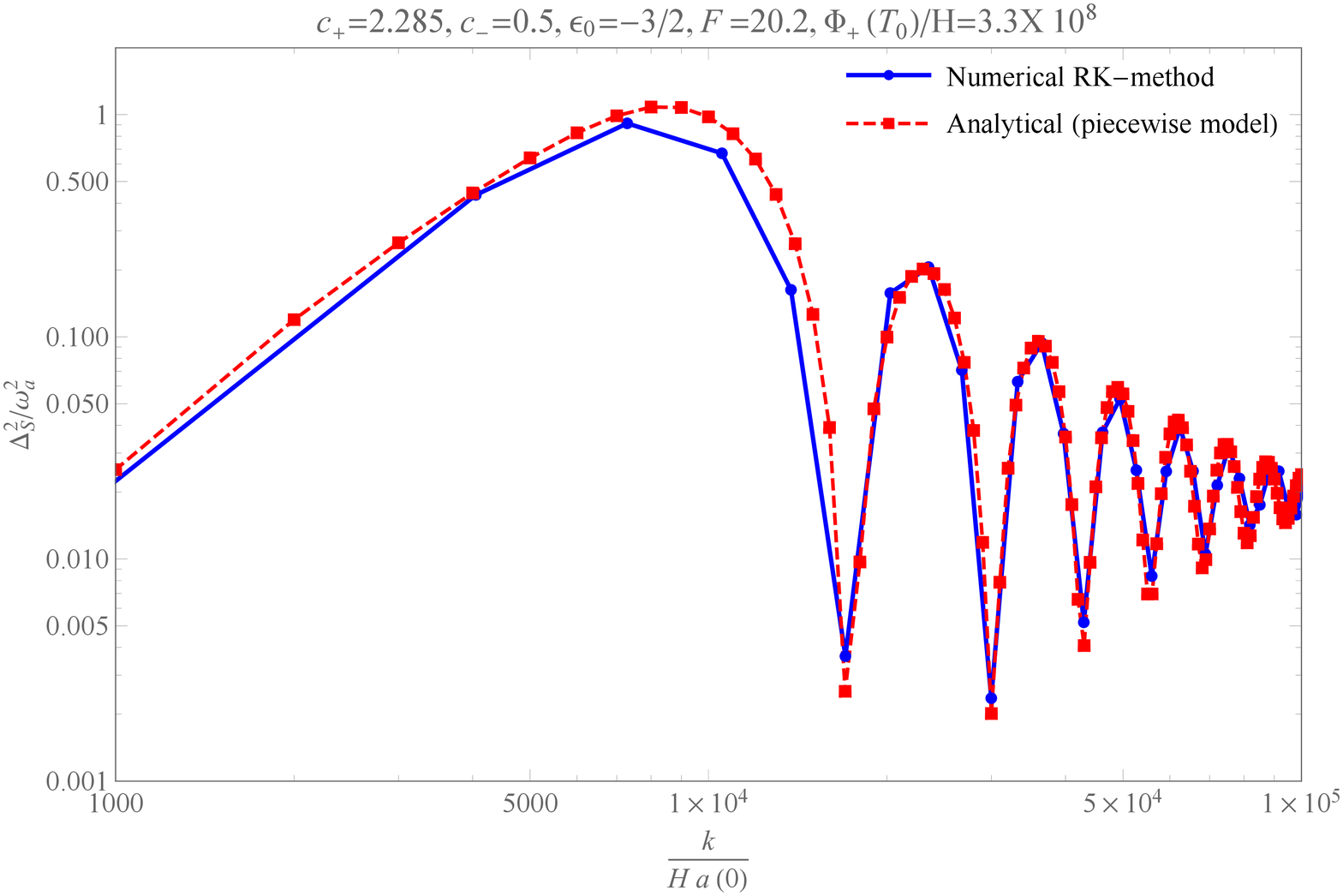} 
\par\end{centering}
\begin{centering}
\hspace{-0.4cm}\includegraphics[scale=0.35]{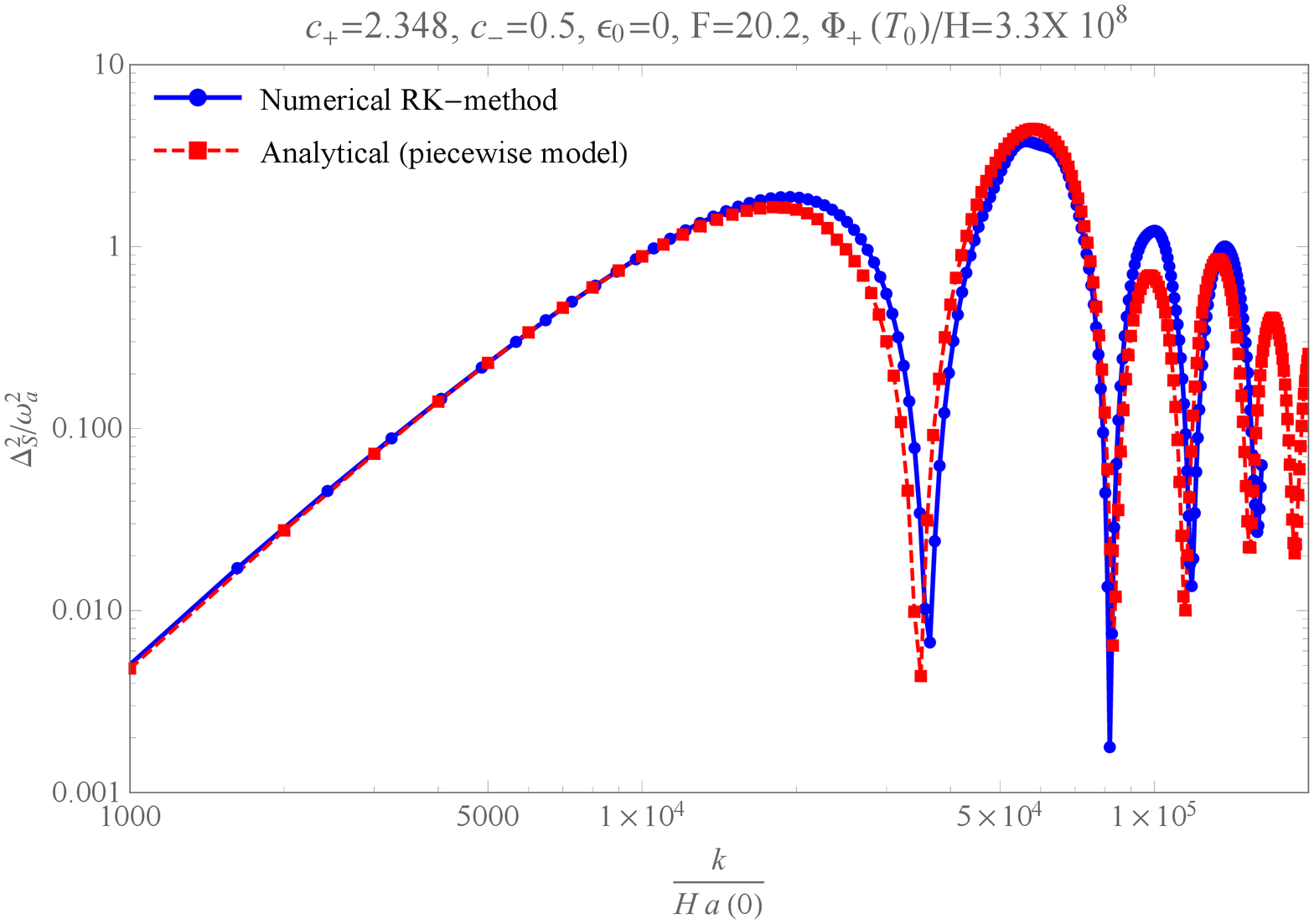} 
\par\end{centering}
\begin{centering}
\hspace{-0.1cm}\includegraphics[scale=0.35]{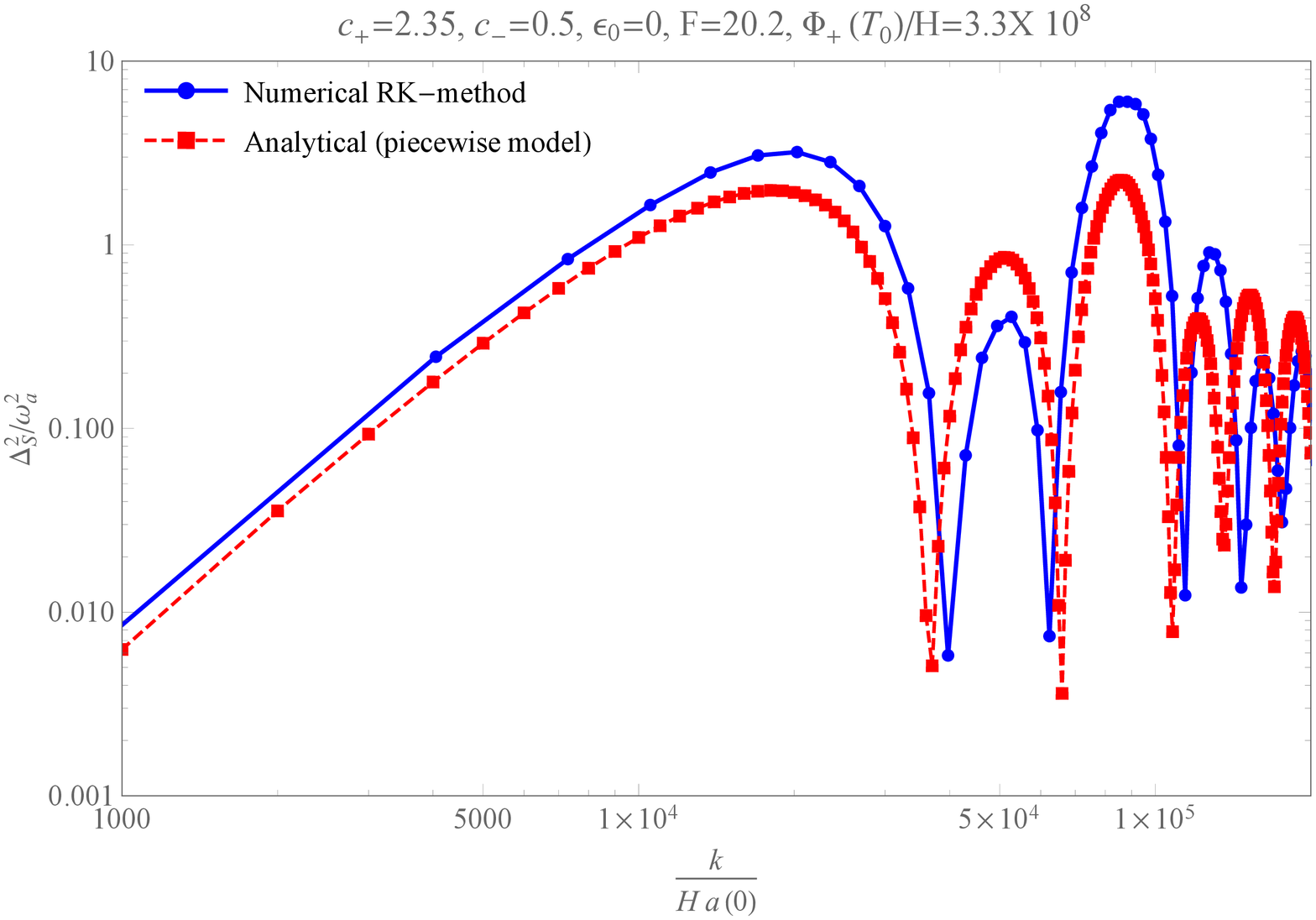} 
\par\end{centering}
\caption{\label{fig:Chart-showing-comparison} These plots illustrate the analytic
spectrum computed using Eq.~(\ref{eq:final-S}) for the parameter
set $P_{A}$ used in Fig.~\ref{fig:Comparison-of-Taylor} except
with $\epsilon_{0}$ also varied (recall $\epsilon_{0}=-3/2$ corresponds
to a dynamically reasonable initial velocity situation of $\dot{\phi}_{+}(0)=-3\phi_{+}(0)/2$)
as denoted in the title of each plot. They are compared with the Runge-Kutta
solution to the mode Eq.~(\ref{eq:modeeq}). The interesting feature
of the second and the third peaks being higher than the first peak
will be explained in Sec.~\ref{sec:features-of-isocurvspectrum}.}
\end{figure}

In Fig.~\ref{fig:Chart-showing-comparison}, we give plots of the
axion isocurvature spectrum generated by the above mass-model for
$c_{+}$ values $2.285$, 2.348, and $2.35$ with distinct initial
conditions corresponding to increasing values of $\alpha$ of $0.52$,
0.97, and $1.13$ respectively.\footnote{The $c_{+}=2.35$ case has a corresponding $\alpha=1.128$ which is
slightly larger than $\alpha_{{\rm U}}$ defined in Eq.~(\ref{eq:alphaU}),
but since this is only at the cusp of the approximations breaking
down, the agreement with the numerical results are reasonable.} For comparison, numerically obtained spectrum is also included. From
the plots we infer that the mass-model is successful in generating
the isocurvature spectrum within the parametric region of applicability.
The model generates a blue power spectrum for small $K$ modes with
an approximate spectral index $n_{I}-1\approx3$. The spectrum peaks
at the first bump and subsequently undergoes oscillations that quickly
die away. As we shall see later the location of the first bump and
the frequency of subsequent spectral oscillations (bumps) is related
to the transition time $T_{c}$. The discrepancies between the numerical
and analytic computations are noticeable for $K$ modes beyond the
first bump because during the time these modes exit the horizon, the
axion mass is oscillating with a large amplitude in the resonant scenarios.
As an example, consider in Fig.~\ref{fig:Chart-showing-comparison}
the case of $c_{+}=2.35$ for which the discrepancy is the largest
for the $K$-region $[5\times10^{4},10^{5}]$.

These discrepancies can be explained through the limitations of $\xi$
modeling and the integrating out approximation. More specifically,
we noted in Sec.~\ref{subsec:Independent-analytic-functions} that
the analytic approximation of the ETSP $U(T)$ in Eq.~(\ref{eq:UT1})
to solve the $y_{1}$ mode function is most critical for these intermediate
K-modes as long as the modes leave the horizon while the jump ETSP
is still significant. The jump ETSP $V_{2}e^{-3\left(T-T_{2}\right)}/2$
is obtained after the UV modes have been integrated out as shown previously
in Fig.~\ref{fig:Sample-plots-highlighting} and detailed in Appendix
\ref{sec:Justification-for-the}. In the UV integration procedure,
we have made the assumption that the flat-deviation $\xi$ is purely
sinusoidal with a constant amplitude and a slow-varying time-dependent
frequency $O(F)$. Post UV integration, we averaged out the remaining
slow-varying prefactors to obtain a constant amplitude jump ETSP $V_{2}$
(See Appendix \ref{sec:Justification-for-the} for details). These
assumptions along with a pure harmonic approximation can be insufficient
to accurately map the sensitivity of the spectrum to subtle variations
of the mass eigenvalue for the intermediate modes that leave the horizon
around the same time.

One then concludes that background fields with weaker resonance $O(\xi)<\frac{F^{2}}{2}$
(since that situation will be less sensitive to the limitations of
$\xi$ modeling in general) will show far smaller discrepancy between
numerical and analytical spectra for these intermediate modes as observed
for $c_{+}=2.285$. Nonetheless, note that the mass-model successfully
generates the distinct feature of a larger second bump and a larger
third bump than the first for $c_{+}=2.348$ and $c_{+}=2.35$, respectively.
We will discuss this more in the next section.

\section{\label{sec:features-of-isocurvspectrum}Explanation of the features
of isocurvature spectrum}

We will now give analytic expressions for $y_{1}(K,T)$ mode functions
for a specific class of simplified mass-model. This will allow us
to explain the parametric dependence of the isocurvature spectrum
at different scales and provide closed form analytic expressions for
the isocurvature power spectrum in a certain restricted region of
the underlying model space.

For the following discussion, we will restrict ourselves to single
dip cases corresponding to $\alpha<\alpha_{2}$. Under this condition,
our time-dependent piecewise mass-model in Eq.~(\ref{eq:model})
simplifies to the following form
\begin{equation}
m^{2}\approx\begin{cases}
V_{0} & T_{0}<T<T_{1}\\
-V_{1} & T_{1}<T<T_{2}\\
V_{B}\,{\rm sqw}(T,T_{2},T_{B}-T_{2})+\frac{V_{2}}{2}e^{-3\left(T-T_{2}\right)} & T_{2}<T<T_{\infty}
\end{cases}
\end{equation}
where the mass-model is now limited to a single dip $-V_{1}$, an
exponentially decaying jump ETSP $V_{2}$, and an $O(c_{-})$ mass
squared term $V_{B}$. As explained in Sec.~(\ref{subsec:Independent-analytic-functions}),
the above mass-model is used within the corresponding $y_{1}$ differential
equation of Eq.~(\ref{eq:modeeq-1}).

Furthermore, as discussed in Sec.(\ref{subsec:Independent-analytic-functions}),
the above differential equation is analytically intractable for certain
$K$ modes where the following two terms are of similar order of magnitude
at $T_{2}$ 
\[
K^{2}e^{-2T_{2}}e^{-2\left(T-T_{2}\right)},\quad\frac{V_{2}}{2}e^{-3\left(T-T_{2}\right)}\,.
\]
To solve this system analytically, we sub-divided the time region
$\left[T_{2},T_{\infty}\right]$ into regions where either one of
the aforementioned two terms is dominant over the other.

Hence, in order to obtain a simplified closed form analytic expression
for the isocurvature spectrum for $\alpha<\alpha_{2}$ cases, we will
utilize the following approach. As a first step, we will evaluate
the isocurvature power spectrum with the $V_{2}$ term neglected.
This is clearly applicable to all resonance cases where $V_{2}<O(c_{-})$.
This assumption immensely simplifies our model and consequently allows
us to obtain tractable analytic expressions. Subsequently, the effect
of the $V_{2}$ jump ETSP is added in the form of a correction factor
$f_{\mathrel{\mathrm{correction}}}$. Through this two-step procedure,
we give an approximate analytic expression for the isocurvature power
spectrum which will allow us to discover some important generic features.
If one is only interested in the results, we refer the reader to Eqs.~(\ref{eq:maineq})
and (\ref{eq:amplitude}). We will now give details regarding the
aforementioned approach.

\subsection{Step 1: $V_{2}<O(c_{-})$}

When $V_{2}<O(c_{-})$, the mass-model simplifies to 
\begin{equation}
m^{2}\approx\begin{cases}
V_{0} & T_{0}<T<T_{1}\\
-V_{1} & T_{1}<T<T_{2}\\
V_{B}\,{\rm sqw}(T,T_{2},T_{B}-T_{2}) & T_{2}<T<T_{\infty}
\end{cases}.\label{eq:reduced-model-singledip}
\end{equation}
Defining $u$ as
\begin{align}
y_{1} & =e^{-\frac{3}{2}T}u
\end{align}
Eq.~(\ref{eq:modeeq-1}) becomes
\begin{equation}
\partial_{T}^{2}u+\left(K^{2}e^{-2T}+m^{2}-9/4\right)u=0
\end{equation}
that has an incoming BD normalized solution
\begin{equation}
u(K,T<T_{1})\approx\frac{\sqrt{\pi}}{2\sqrt{2}}e^{i(\frac{i\omega\pi}{2}+\frac{\pi}{4})}H_{i\omega}^{1}\left(Ke^{-T}\right)\quad\forall Ke^{-T_{0}}\gg|\omega|\label{eq:BD-u-solution}
\end{equation}
with $\omega=\sqrt{c_{+}-9/4}$ for $V_{0}=c_{+}$.

For the rest of our discussion, we will use the following asymptotic
forms of the Hankel function $H_{i\omega}^{1}(z)$ for a real argument
$z$:
\begin{equation}
H_{i\omega}^{1}(z)\approx\begin{cases}
\frac{1+i\cot\left(i\omega\pi\right)}{\Gamma\left(i\omega+1\right)}\left(\frac{z}{2}\right)^{i\omega}-i\frac{\Gamma\left(i\omega\right)}{\pi}\left(\frac{z}{2}\right)^{-i\omega} & 0<z\ll\sqrt{1+i\omega}\\
\sqrt{\frac{2}{\pi z}}e^{i\left(z-\frac{i\omega\pi}{2}-\frac{\pi}{4}\right)} & z\gg\left|-\omega^{2}-\frac{1}{4}\right|
\end{cases}.\label{eq:Hankelapproximations}
\end{equation}
Next we solve within $T_{1}<T<T_{2}$ for a single dip $-V_{1}$ and
express the solution at $T=T_{2}$ as
\begin{equation}
\left[\begin{array}{c}
u\\
\partial_{T}u
\end{array}\right]_{T=T_{2}}=\cosh[-b\,\Delta T]\left[\begin{array}{cc}
1 & -\frac{\tanh[-b\,\Delta T]}{b}\\
-b\tanh[-b\,\Delta T] & 1
\end{array}\right]\left[\begin{array}{c}
u\\
\partial_{T}u
\end{array}\right]_{T=T_{1}}\label{eq:matrix1}
\end{equation}
with $\Delta T=T_{2}-T_{1}$ and 
\begin{equation}
b^{2}\equiv V_{1}+9/4-K^{2}e^{-(T_{2}+T_{1})}.\label{eq:bsq}
\end{equation}
Note that Eq.~(\ref{eq:matrix1}) is the simplified result coming
from the average value for $K^{2}e^{-2T}$ term within the short interval
$\Delta T$, while $V_{1}$ containing the initial kinetic energy
information generically satisfies 
\begin{equation}
V_{1}\sim O(F^{2})\gg O(10).\label{eq:V1largeassume}
\end{equation}
Next we solve within $\left[T_{2},T_{\infty}\right]$. To further
simplify our evaluations, we consider that for $V_{B}\ll9/4$, the
effect of the model parameter $V_{B}$ can be factored in through
the exponential decay of the final mode amplitude. Equivalently, consider
the differential equation 
\begin{equation}
\partial_{T}^{2}u+\left(K^{2}e^{-2T}+V_{B}-9/4\right)u=0\label{eq:VB-u-eqn}
\end{equation}
then the mode function $u(T)$ has the following asymptotic solution
\begin{equation}
\lim_{T\rightarrow T_{\infty}}u(T,V_{B}\neq0)\approx\lim_{T\rightarrow T_{\infty}}e^{\left(-\frac{3}{2}+\sqrt{\frac{9}{4}-V_{B}}\right)\left(T-\tilde{T}\right)}u(T,V_{B}=0)
\end{equation}
where $u(T,V_{B}\neq0)$ is the solution of Eq.~(\ref{eq:VB-u-eqn})
with $V_{B}\neq0$ while $u(T,V_{B}=0)$ is the solution with the
$V_{B}$ term equal to zero. Hence, if we neglect the $V_{B}$ term
and solve for $u$ in terms of Bessel functions of order $3/2$ (similar
to a massless axion), we obtain the final mode function as follows:
\begin{align}
\lim_{T_{3}\rightarrow T_{\infty}}\left[\begin{array}{c}
y_{1}\\
\partial_{T}y_{1}
\end{array}\right]_{T=T_{3}} & =\mathfrak{D}\frac{\cosh[-b\,\Delta T]}{K^{\frac{3}{2}}\sqrt{-K\tau_{2}}}\times\nonumber \\
 & \left[\begin{array}{cc}
\frac{3\cos[-K\tau_{2}]}{2}-\left(\frac{3/2}{-K\tau_{2}}+K\tau_{2}\right)\sin[-K\tau_{2}] & -\cos[-K\tau_{2}]+\frac{\sin[-K\tau_{2}]}{-K\tau_{2}}\\
0 & 0
\end{array}\right]\times\nonumber \\
 & \left[\begin{array}{cc}
1 & -\frac{\tanh[-b\,\Delta T]}{b}\\
-b\tanh[-b\,\Delta T] & 1
\end{array}\right]\left[\begin{array}{c}
u\\
\partial_{T}u
\end{array}\right]_{T=T_{1}}\label{eq:k-mode-eqn-amplitude-simple-model}
\end{align}
where $\tau_{2}$ is the conformal time corresponding to the time
$T_{2}$ (see Eq.~(\ref{eq:conformaltime})) and the factor 
\begin{equation}
\mathfrak{D}\equiv e^{\left(-\frac{3}{2}+\sqrt{\frac{9}{4}-V_{B}}\right)\left(T_{\infty}-\tilde{T}\right)}\label{eq:Dfirst}
\end{equation}
accounts for the mode amplitude decay/amplification through a positive/negative
$V_{B}$ parameter as explained previously. Using the simplified expression
in Eq.~(\ref{eq:k-mode-eqn-amplitude-simple-model}), we examine
the axion isocurvature spectrum for our model for different $K$ ranges.

\subsubsection{\textbf{Modes that leave the horizon early: $-K\tau_{2}\ll1$}}

Starting with the above equation, we simplify in terms of \textbf{$-K\tau_{2}\ll1$,}
\begin{align}
\lim_{T_{3}\rightarrow T_{\infty}}\left[\begin{array}{c}
y_{1}\\
\partial_{T}y_{1}
\end{array}\right]_{T=T_{3}} & \approx\mathfrak{D}\frac{\cosh[-b\,\Delta T]}{K^{\frac{3}{2}}\sqrt{Ke^{-T_{2}}}}\left[\begin{array}{cc}
\frac{\left(Ke^{-T_{2}}\right)^{2}}{2} & \frac{\left(Ke^{-T_{2}}\right)^{2}}{3}\\
0 & 0
\end{array}\right]\times\nonumber \\
 & \left[\begin{array}{cc}
1 & -\frac{\tanh[-b\,\Delta T]}{b}\\
-b\tanh[-b\,\Delta T] & 1
\end{array}\right]\left[\begin{array}{c}
u\\
\partial_{T}u
\end{array}\right]_{T=T_{1}}\nonumber \\
 & \approx\mathfrak{D}\cosh[-b\,\Delta T]e^{-\frac{3}{2}T_{2}}\left[\begin{array}{cc}
\frac{1}{2} & \frac{1}{3}\\
0 & 0
\end{array}\right]\left[\begin{array}{cc}
1 & -\frac{\tanh[-b\,\Delta T]}{b}\\
-b\tanh[-b\,\Delta T] & 1
\end{array}\right]\left[\begin{array}{c}
u\\
\partial_{T}u
\end{array}\right]_{T=T_{1}}
\end{align}

\begin{align}
\lim_{T\rightarrow T_{\infty}}y_{1}(K,T) & \approx e^{-\frac{3}{2}T_{2}}\mathfrak{D}\cosh\left[b\Delta T\right]\frac{\sqrt{\pi}}{2\sqrt{2}}e^{i(\frac{i\omega\pi}{2}+\frac{\pi}{4})}\left[\left(\frac{1}{2}-\frac{b}{3}\tanh\left[-b\Delta T\right]\right)H_{i\omega}^{1}\left(Ke^{-T_{1}}\right)\right.\nonumber \\
 & \left.+\left(\frac{1}{3}-\frac{\tanh\left[-b\Delta T\right]}{2b}\right)\partial_{T}H_{i\omega}^{1}\left(Ke^{-T_{1}}\right)\right].
\end{align}
Since $b^{2}\sim V_{1}\gg1$ (see Eq.~(\ref{eq:V1largeassume})),
the mode amplitude is dominated by $H_{i\omega}^{1}\left(Ke^{-T_{1}}\right)$
rather than its derivative.
\begin{align}
\lim_{T\rightarrow T_{\infty}}y_{1}(K,T) & \approx e^{-\frac{3}{2}T_{2}}\mathfrak{D}\cosh\left[b\Delta T\right]\frac{\sqrt{\pi}}{2\sqrt{2}}e^{i(\frac{i\omega\pi}{2}+\frac{\pi}{4})}\left[\left(\frac{1}{2}-\frac{b}{3}\tanh\left[-b\Delta T\right]\right)H_{i\omega}^{1}\left(Ke^{-T_{1}}\right)\right]
\end{align}
where $\Delta T=T_{2}-T_{1}$. Observe that the mode amplitude is
dependent upon the freezeout Hankel function at $T=T_{1}$ such that
$Ke^{-T_{1}}\ll1$. At the outset, the isocurvature spectrum appears
to have a blue spectral index $n_{I}\approx4$ for these scales. However
for $\omega<1$, the Hankel function has $K$-dependence as given
in Eq.~(\ref{eq:Hankelapproximations}). We apply this to the Hankel
function in our case where $Ke^{-T_{1}}\ll1\leq\sqrt{1+i\omega}$
to yield the following $K$-dependence for the power spectrum:
\begin{align}
\Delta_{s}^{2}(K) & \propto K^{3}\left|1-i\frac{\Gamma\left(i\omega\right)\Gamma\left(i\omega+1\right)}{\pi\left(1+i\cot\left(i\omega\pi\right)\right)}e^{-2i\omega\ln\left(\frac{Ke^{-T_{1}}}{2}\right)}\right|^{2}.
\end{align}
For $\omega<1$, the term $\frac{\Gamma\left(i\omega\right)\Gamma\left(i\omega+1\right)}{\pi\left(1+i\cot\left(i\omega\pi\right)\right)}\sim O(1)$
such that the power spectrum has oscillations in the long wavelength
region with a log-$K$ dependence. Therefore, one observes deviation
of the spectral index from $4$ which is sinusoidal in log-$K$. These
deviations decay as $e^{-\pi\sqrt{c_{+}-9/4}}$ and become insignificant
for large $c_{+}$ fields or when $\omega>1$.

\subsubsection{\textbf{Scales near the first bump ($-K\tau_{1}\rightarrow1$)}}

Next we consider $K$ modes that approach $-1/\tau_{2}$ such that
the modes leave the horizon after the axion effective frequency squared
undergoes oscillations. We start with the Eq.~(\ref{eq:k-mode-eqn-amplitude-simple-model})
giving us the mode amplitude $y_{1}\left(K,T\right)$:
\begin{align}
\lim_{T\rightarrow T_{\infty}}y_{1}(K,T) & \approx\mathfrak{D}\frac{\cosh[-b\,\Delta T]}{K^{\frac{3}{2}}\sqrt{Ke^{-T_{2}}}}\nonumber \\
 & \left[\left(\frac{3}{2}\cos[-K\tau_{2}]-\left(\frac{3/2}{Ke^{-T_{2}}}-Ke^{-T_{2}}\right)\sin[-K\tau_{2}]\right)\left(u-\frac{\tanh[-b\,\Delta T]}{b}\partial_{T}u\right)+\right.\nonumber \\
 & \left.\left(-\cos[-K\tau_{2}]+\frac{1}{Ke^{-T_{2}}}\sin[-K\tau_{2}]\right)\left(-b\tanh[-b\,\Delta T]u+\partial_{T}u\right)\right].
\end{align}
Since $b^{2}\sim V_{1}\gg1$ (Eq.~(\ref{eq:V1largeassume})), the
mode amplitude simplifies as
\begin{align}
\lim_{T\rightarrow T_{\infty}}y_{1}(K,T) & \approx\mathfrak{D}\frac{\cosh\left[b\Delta T\right]}{K^{\frac{3}{2}}\sqrt{Ke^{-T_{2}}}}\frac{\sqrt{\pi}}{2\sqrt{2}}e^{i(\frac{i\omega\pi}{2}+\frac{\pi}{4})}\left[\left(\frac{3}{2}+b\tanh\left[-b\Delta T\right]\right)\cos[-K\tau_{2}]\right.\nonumber \\
 & \left.-\frac{1}{Ke^{-T_{2}}}\left(\frac{3}{2}+b\tanh\left[-b\Delta T\right]-\left(Ke^{-T_{2}}\right)^{2}\right)\sin[-K\tau_{2}]\right]H_{i\omega}^{1}\left(Ke^{-T_{1}}\right).
\end{align}
Putting this into Eq.~(\ref{eq:isocintermsofy1}), we find the isocurvature
amplitude to be proportional to 
\begin{align}
\sqrt{\Delta_{s}^{2}(K)} & \propto\frac{\left|e^{i(\frac{i\omega\pi}{2}+\frac{\pi}{4})}H_{i\omega}^{1}\left(Ke^{-T_{1}}\right)\right|}{\sqrt{Ke^{-T_{2}}}}\left[\left(\frac{3}{2}+b\tanh\left[-b\Delta T\right]\right)\cos[-K\tau_{2}]\right.\nonumber \\
 & \left.-\frac{1}{Ke^{-T_{2}}}\left(\frac{3}{2}+b\tanh\left[-b\Delta T\right]-\left(Ke^{-T_{2}}\right)^{2}\right)\sin[-K\tau_{2}]\right].
\end{align}
The location of the first bump $K_{\text{first-bump}}$is determined
by solving 
\begin{equation}
\frac{d}{dK}\sqrt{\Delta_{s}^{2}(K_{\text{first-bump}})}=0
\end{equation}
since one can show that there are no small oscillatory features in
the rising part of the spectrum in the region $-K\tau_{1}<1$. To
solve for the peak $K_{\text{first-bump}}$, we approximate $T_{2}\approx T_{1}$
since $T_{2}-T_{1}=\Delta T\sim O\left(1/F\right)\ll1$ to obtain
the following transcendental equation for the small $c_{+}-2$ limiting
case $Ke^{-T_{1}}\sim1\gg\left|\omega^{2}-\frac{1}{4}\right|$:\textbf{
\begin{equation}
\left(1-\frac{B}{z^{2}}\right)\cot\left(z\right)=\frac{B\left(z^{2}-2\right)}{2z^{3}}\label{eq:transcend1}
\end{equation}
\begin{equation}
\cot z=\frac{2-z^{2}}{2z}
\end{equation}
}where $B=3/2+b\tanh\left[-b\Delta T\right]$ and $z=Ke^{-T_{2}}$.
Since $B\gg1$, the solution to the above expression is nearly independent
of $B$ or more explicitly the properties such as the amplitude and
velocity of the field oscillations after $T_{c}$. We obtain the solution
of 
\begin{equation}
K_{\text{first-bump}}e^{-T_{2}}=\frac{\pi+\sqrt{\pi^{2}-8}}{2}+O\left(B^{-1}\right)\approx2
\end{equation}
analytically by expanding in the limit $B\gg1$.

Calculations in large $c_{+}-2$ limiting case $Ke^{-T_{1}}\sim1\ll\left|\sqrt{1+iw}\right|$
yield similar results by solving an analog of Eq\@.~(\ref{eq:transcend1}):\textbf{
\begin{align}
\cot\left(z\right) & \approx\left(\frac{1}{z}-\frac{2z}{3}\right)+O\left(\frac{1}{B}\right)
\end{align}
}leading to 
\begin{equation}
K_{\text{first-bump}}e^{-T_{2}}=\frac{1}{16}\left(9\pi-6+\sqrt{81\pi^{2}-108\pi-156}\right)\approx2.48.
\end{equation}
Thus, we have shown that the isocurvature power has the first large
bump at approximately 
\begin{equation}
K_{\text{first-bump}}=\frac{k}{a(0)H}\approx2e^{T_{2}}\label{eq:Kbump}
\end{equation}
with 25\% accuracy. Eq.~(\ref{eq:Kbump}) shows that the location
of the first bump is the scale that leaves the horizon at time $T_{2}$
near the resonant transition time $T_{c}$. Since the transition $T_{c}$
is dependent upon the mass $c_{+}$ and the initial conditions $\phi_{+}(0)$
and $\epsilon_{0}$ (see Eq.~(\ref{eq:alphacplus})), background
fields with smaller $\alpha$ tend to transition later such that they
have a larger $T_{c}$. Under these circumstances the corresponding
first-bump location $K_{\text{first-bump}}$ will be pushed to even
smaller scales and become unobservable due to limitations in the experimental
sensitivity of short length scales. This is qualitatively the same
as the situation in which $c_{+}<9/4$, which was the subject of previous
works on this topic \citep{Chung:2016wvv}.

\subsubsection{\textbf{Scales that lie within oscillating spectrum: $3\lesssim-K\tau_{2}<K_{2}$}}

Next we consider $K$ modes satisfying $3\lesssim-K\tau_{2}<K_{2}$
that leave the horizon after the axion effective frequency squared
undergoes oscillations. The upper limit for these $K$-modes is set
by the heavy mode coupling as elucidated in Eq.~(\ref{eq:K2cutoff}).
Meanwhile, Eq.~(\ref{eq:k-mode-eqn-amplitude-simple-model}) gives
the mode amplitude $y_{1}\left(K,T\right)$ in the limit $Ke^{-T_{1}}\gg\left|\omega^{2}-1/4\right|$
to be
\begin{align}
y_{1}(K,T_{\infty}) & \approx\mathfrak{D}\frac{e^{-iK\tau_{1}}}{2K^{3/2}}\times\nonumber \\
 & \left[\cosh\left[-b\Delta T\right]\left(\left(-ie^{\Delta T/2}-\frac{3e^{-\Delta T/2}}{-2K\tau_{2}}\right)\cos\left[-K\tau_{2}\right]+\left(-e^{-\Delta T/2}+\frac{ie^{\Delta T/2}}{-K\tau_{2}}\right)\sin\left[-K\tau_{2}\right]\right)+\right.\nonumber \\
 & \left.\left.\sinh\left[-b\Delta T\right]\left(\left(\frac{i3e^{\Delta T/2}}{2b}+\frac{be^{-\Delta T/2}}{-K\tau_{2}}\right)\cos\left[-K\tau_{2}\right]+\right.\left(\frac{-ie^{\Delta T/2}K\tau_{2}}{b}-\frac{be^{-\Delta T/2}}{K^{2}\tau_{2}^{2}}\right)\sin\left[-K\tau_{2}\right]\right)\right].\label{eq:oscillybeg}
\end{align}
Considering that $\Delta T\sim O\left(1/F\right)\ll1$, we can rewrite
this expression as
\begin{align}
\lim_{T\rightarrow T_{\infty}}y_{1}(K,T) & \approx\mathfrak{D}\frac{e^{-iK\tau_{1}}}{2K^{3/2}}\times\nonumber \\
 & \left[\cosh\left[-b\Delta T\right]\left(-ie^{iK\tau_{2}}\right)+\sinh\left[-b\Delta T\right]\left(\left(\frac{b}{-K\tau_{2}}\right)\cos\left[-K\tau_{2}\right]+\left(i\frac{-K\tau_{2}}{b}\right)\sin\left[-K\tau_{2}\right]\right)\right].\label{eq:simposcill}
\end{align}

We notice from the above expression that the spectrum oscillates via
the $\cos\left[-K\tau_{2}\right]$ term for the intermediate K-modes
(scales) about a background with an initial decay envelope proportional
to $1/K^{2}$. The amplitude of these oscillations is controlled by
the dip amplitude $V_{1}$ operating for a short time interval $\Delta T$
(see Eq.~(\ref{eq:e1prsqdip})), which is largely controlled by the
mass squared eigenvalue and eigenvector rotation. Meanwhile, the $K$-spacing
of these oscillations is approximately 
\begin{align}
\Delta k & \equiv\Delta Ka(0)H\approx\pi a(T_{2})H\\
 & \approx K_{\text{first-bump}}\,a(0)H.\label{eq:firstbunp}
\end{align}
Therefore, the location of the first bump and the frequency of the
first few spectral oscillations are directly related to the transition
time $T_{c}$ of the background fields. This can be understood by
the following discussion.

For our simplified model, as the background fields transition, the
mass squared $m^{2}$ dips to a negative ETSP $-V_{1}$ for a time
period $\Delta T=T_{2}-T_{1}$. For all $K$-modes that are still
sub-horizon at transition, the incoming mode amplitude picks up a
phase that is dependent upon the momentum $K$ of the mode sampled
at the transition where $T_{c}\approx(T_{2}+T_{1})/2$. Later when
these modes exit the horizon the resulting mode amplitude is oscillatory
in $K$-space with a $K$-spacing that is dependent upon the transition
time $T_{c}$. As a result, the power spectrum for these scales oscillates
while the imaginary part of the phase controls the amplitude of these
oscillations.
\begin{figure}
\begin{centering}
\includegraphics[scale=0.42]{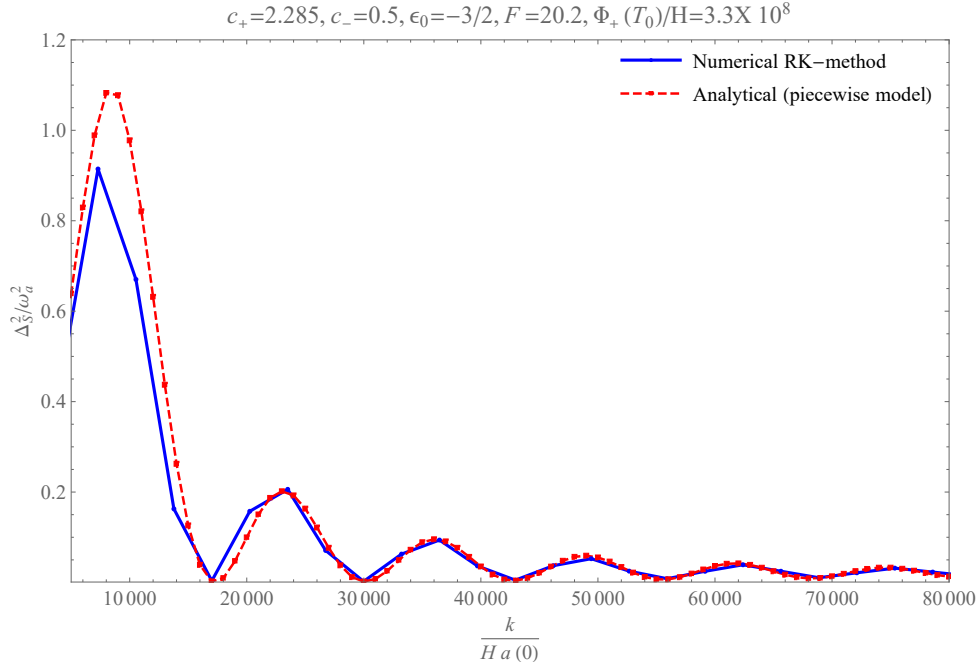} 
\par\end{centering}
\caption{Plot of the spectra made using Eq.~(\ref{eq:final-S}) highlighting
the analytic form of the spectral oscillations on a linear amplitude
scale for $k$-modes for $k/(a(T_{c})H)>2$ (i.e.~modes which become
superhorizon after the nonadiabatic transition) where $a(T_{c})/a(0)=O(10^{4})$.
The first peak/bump occurs around $k\sim2a(T_{2})H$ where $a(T_{2})\approx a(T_{c})$.
The spectrum oscillates with a $k$-period $\pi e^{T_{2}}a(0)H$ and
the initial decay behavior of the envelope is approximately proportional
to $k^{-2}$. For $k\gtrsim e^{T_{2}}a(0)H\sqrt{V_{1}}$ the envelope
decay then transitions to $k^{-1}$ behavior, oscillating about a
background spectral amplitude of $O(F^{-2})$. Shown are the results
with the same parameters set as Fig.~\ref{fig:Comparison-of-Taylor},
except with $\epsilon_{0}=-3/2$ (sizable initial velocity). Note
that the height of the first bump is $\sim O(10)$ larger than the
height of the final massless axion plateau which on a linear scale
is negligible.}
\end{figure}

\subsubsection{Scales leave the horizon late: $K>K_{P}$}

As remarked near Eq.~(\ref{eq:KPdef}), for the plateau part of the
spectrum ($K>K_{P}$), the isocurvature perturbation modes return
to the usual massless form. In this case, it is better to work in
the final massless axion basis. Because the axion field $a$ in Eq.~(\ref{eq:axion})
is not normalized canonically, the canonically normalized axion is
$A=a/\sqrt{2}$. This allows us to write the plateau part of the spectrum
as

\begin{align}
\frac{\Delta_{S}^{2}(K)}{\omega_{a}^{2}} & =4\left(\frac{H(t_{K})}{2\pi A(t_{K})}\right)^{2}\\
 & =\frac{2}{\left(\frac{c_{-}}{c_{+}}\right)^{1/2}+\left(\frac{c_{+}}{c_{-}}\right)^{1/2}}\left(\frac{h}{2\pi\theta_{+}(0)F}\right)^{2}\label{eq:hfactor}
\end{align}
where $K>K_{P}$ and we have approximated $H(t_{K})$ to be constant
(neglecting corrections of slow roll parameters $O(\epsilon)$ which
are typically negligible in the physical scenarios of interest in
this scenario). The appearance of $h$ in Eq.~(\ref{eq:hfactor})
is merely dividing out the scaling of $F_{a}$ in Eq.~(\ref{eq:Fscale})
This flat part of the spectrum has also been numerically confirmed.

\subsection{\label{subsec:V2correction}Step 2: Adding correction for the $V_{2}$
term}

Eq.~(\ref{eq:k-mode-eqn-amplitude-simple-model}) gives the final
mode amplitude for axionic isocurvature perturbation for $\alpha<\alpha_{2}$
cases consisting of a $-V_{1}$ single dip and a negligible $V_{2}$
jump ETSP where the effect of the $V_{B}$ parameter (average of $m_{B}^{2}$)
is included through the exponential decay factor $\mathfrak{D}^{2}$
in Eq.~(\ref{eq:Dfirst}). Using the expression for $V_{2}$ under
Eq.~(\ref{eq:e1prsqdip}) and the expression for $A$ from Eq.~(\ref{eq:Aapprox}),
we infer that for $F$ scales greater than $O(20)$, the amplitude
of the resonant UV oscillations is larger than $O(c_{-})$ such that
the $V_{2}/2$ term cannot be neglected. Below we will briefly discuss
the corrections coming from $V_{2}$.

Since $V_{2}$ is positive, its inclusion in Eq.~(\ref{eq:model})
for the $y_{1}$ mode function leads to a decay of all the modes that
are super-horizon at $T_{2}$ (see the discussion below Eq.~(\ref{eq:effectivefreq})).
Consequently, a significant $V_{2}$ leads to a reduction in the isocurvature
power spectrum for a range of small $K$ modes. On the other hand,
for all modes that are sub-horizon at $T_{2}$, the effect is significantly
diminished since the ETSP decays exponentially with a decay factor
of $3$. Another way to understand this correction is to note that
a large resonant UV oscillations of the background fields imply a
significant interaction energy compared to the mass energy at transition.
This increases the effective mass for the perturbation modes thereby
reducing their amplitudes.

\begin{figure}
\begin{centering}
\includegraphics[scale=0.4]{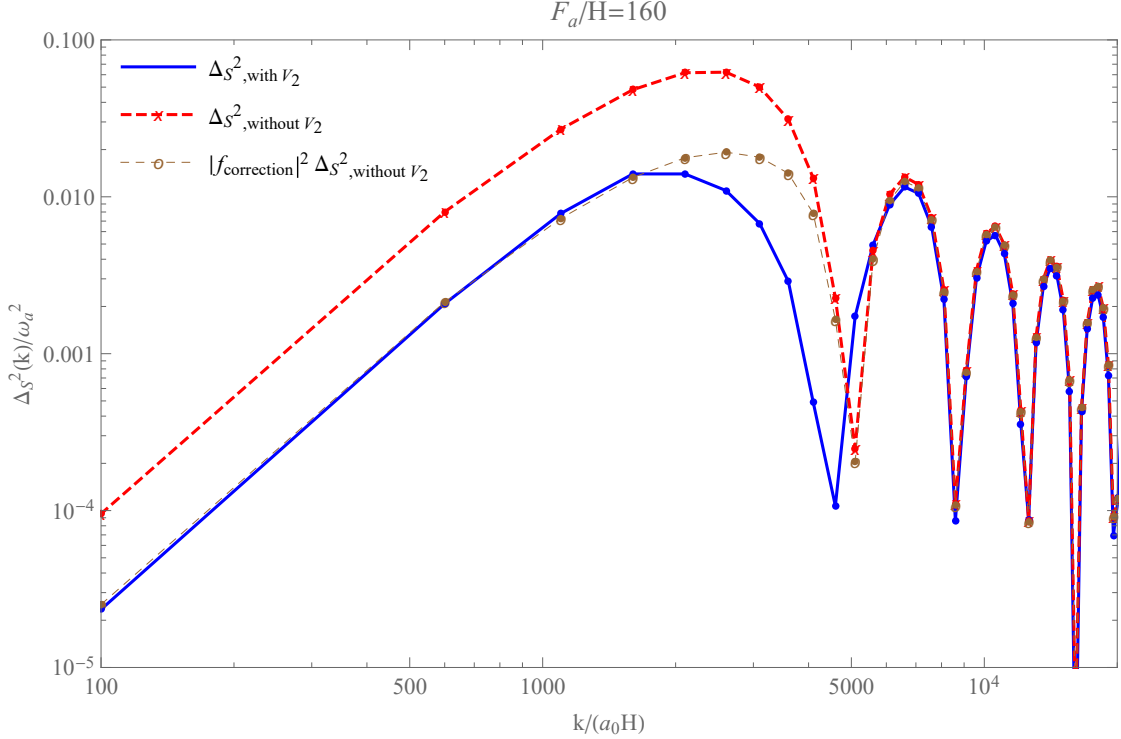} 
\par\end{centering}
\caption{\label{fig:effect of jump ETSPs} This figure highlights the effect
of jump ETSP $V_{2}$ on isocurvature power spectrum derived using
the scattering matrices of Eq.~(\ref{eq:final-S}). The thick (blue)
curve corresponds to the actual power spectrum where modification
due to $V_{2}$ parameter has been accounted for. The thick dashed
(red) curve neglects the effect due to the $V_{2}$ jump ETSP. The
thin dashed (brown) curve adds the correction factor in Eq\@.(\ref{eq:fcorr})
to the spectrum without the $V_{2}$ jump ETSP. A positive jump ETSP
$V_{2}$ leads to a decay of all modes super-horizon at $T_{2}$.
This can result in significant attenuation of spectral power for these
modes as shown in the plots. The above plots are constructed using
the following parameter set $\{F=161.6,\,c_{+}=2.415,\,c_{-}=0.5,\,\epsilon_{0}=0,\,\phi_{+}(0)=0.1M_{p}/H\}$.}
\end{figure}

We now give approximate analytic expressions to include the effect
of the jump ETSP $V_{2}$ to the previously derived isocurvature perturbation
mode amplitude. The effect of $V_{2}$ ETSP is included as

\begin{equation}
f_{\mathrm{correction}}\left(K\right)\approx\frac{1+l(K)}{2}+\frac{1-l(K)}{2}\left(\frac{\left(Ke^{-T_{2}}-\sqrt{\frac{V_{2}}{2}}\right)}{\sqrt{1+\left(Ke^{-T_{2}}-\sqrt{\frac{V_{2}}{2}}\right)^{2}}}\right)\label{eq:fcorr}
\end{equation}
where 
\begin{equation}
l(K)=\left[\left(a_{1}+\frac{a_{2}}{3}+c_{-}\frac{3a_{1}+2a_{2}}{27}\right)\frac{2\sin\left(\pi n_{1}\right)\Gamma\left(1-n_{1}\right)V_{2}^{n_{1}/2}2^{-n_{1}/2}3^{-n_{1}}}{\pi\left(a_{1}\frac{\sqrt{2V_{2}}}{3}\left.\partial_{x}J{}_{n_{1}}\left(x\right)\right|_{x=\frac{\sqrt{2V_{2}}}{3}}+\left(a_{1}+2a_{3}/3\right)J_{n_{1}}\left(\frac{\sqrt{2V_{2}}}{3}\right)\right)}\right]^{-1}\label{eq:lk}
\end{equation}
with $a_{1,2,3}$ defined as 
\begin{align}
a_{1} & \approx y_{1}(K,T_{2})\\
a_{2} & \approx\partial_{T}y_{1}(K,T_{2})\\
a_{3} & \approx-a_{1}\sqrt{V_{2}}+a_{2}.
\end{align}
Note that the expression $l(K)$ is primarily derived for all super-horizon
modes $Ke^{-T_{2}}\ll\sqrt{\frac{V_{2}}{2}}$. Since $l(K)$ must
tend to $1$ for modes that satisfy $Ke^{-T_{2}}\gg\sqrt{\frac{V_{2}}{2}}$,
we have constructed $f_{\mathrm{correction}}\left(K\right)$ as a
smooth function connecting these two asymptotic values of $l(K)$.
Hence, the above correction factor is an interpolated approximation
for the intermediate modes lying between the two asymptotic scales.
An important consequence of this interpolation is that it does not
show a gradual shift in the location of the first bump towards smaller
$K$ values due to an increasing $V_{2}$ jump ETSP as can be observed
in Fig\@.~\ref{fig:effect of jump ETSPs}. To accurately model this
gradual shift of the first bump, one needs to evaluate an improved
correction factor for the intermediate modes by solving the scattering
matrices of Sec.~(\ref{subsec:Piecewise-solution-(Scattering}) with
the $V_{2}$ jump ETSP included explicitly. Furthermore, we remark
that $l(K)$ is nearly a constant since Eq.~(\ref{eq:matrix1}) and
Eq.~(\ref{eq:bsq}) suggest that 
\begin{equation}
\frac{\partial_{T}y_{1}(K,T_{2})}{y_{1}(K,T_{2})}\approx-b\tanh\left[-b\left(T_{2}-T_{1}\right)\right],
\end{equation}
which is independent of $K$ for all modes with $Ke^{-T_{2}}\ll\sqrt{V_{1}}$
at $T_{2}$.

\subsection{Isocurvature power spectrum}

In summary, the isocurvature power spectrum for background fields
with $\alpha_{{\rm L}}\lesssim\alpha\lesssim\alpha_{2}$ (where $\alpha_{{\rm L}}$
and $\alpha_{2}$ are given in Eqs.~(\ref{eq:alpha1}) and (\ref{eq:alphacutoff})
respectively) corresponding to a single $-V_{1}$ dip can be expressed
(where $\alpha$ is defined in Eq.~(\ref{eq:alphacplus})) as follows:

\begin{align}
\Delta_{S}^{2}(K) & \approx\left|f_{\mathrm{correction}}(K)\right|^{2}\times\begin{cases}
C_{1}K^{3}\left|1-i\frac{\Gamma\left(i\omega\right)\Gamma\left(i\omega+1\right)}{\pi\left(1+i\cot\left(i\omega\pi\right)\right)}e^{-2i\omega\ln\left(\frac{-K\tau_{1}}{2}\right)}\right|^{2} & -K\tau_{c}\ll1\\
C_{2}\mathfrak{D}^{2}\left|H_{i\omega}^{1}\left(-K\tau_{1}\right)\right|^{2}\left(-K\tau_{2}\right)\left(\sin\left(-K\tau_{2}\right)\phantom{\frac{\cos[-K\tau_{2}]}{-K\tau_{2}}}\right.\\
\left.+\left(3/2+b\tanh\left[-b\Delta T\right]\right)\left(\frac{\cos[-K\tau_{2}]}{-K\tau_{2}}-\frac{\sin[-K\tau_{2}]}{\left(-K\tau_{2}\right)^{2}}\right)\right)^{2} & 0.5\lesssim-K\tau_{c}<3\\
C_{3}\mathfrak{D}^{2}\cosh^{2}\left[b\Delta T\right]\times\\
\left|\left(-ie^{iK\tau_{2}}\right)+\tanh\left[-b\Delta T\right]\times\right.\\
\left.\left(\left(\frac{b}{-K\tau_{2}}\right)\cos\left[-K\tau_{2}\right]+\left(i\frac{-K\tau_{2}}{b}\right)\sin\left[-K\tau_{2}\right]\right)\right|^{2} & 3\lesssim-K\tau_{c}<K_{2}\\
C_{4}\times1 & K>K_{P}
\end{cases}\label{eq:maineq}
\end{align}
with coefficients $C_{1,2,3,4}$ given as
\begin{align}
C_{1} & \approx C\mathfrak{D}^{2}\frac{\pi}{8}e^{-\omega\pi}\cosh^{2}\left[b\Delta T\right]\frac{e^{-3T_{2}}}{3}\left(\frac{3}{2}-b\tanh\left[-b\Delta T\right]\right)^{2}\left|\frac{1+i\cot\left(i\omega\pi\right)}{\Gamma\left(i\omega+1\right)}\right|^{2}\nonumber \\
C_{2} & \approx C\frac{\pi}{8}e^{-\omega\pi}\cosh^{2}\left[b\Delta T\right]\label{eq:amplitude}\\
C_{3} & \approx C\frac{1}{4}\nonumber \\
C_{4} & \approx\omega_{a}^{2}\frac{h^{2}}{2\pi^{2}\theta_{+}^{2}F^{2}}\left(\frac{r}{1+r^{2}}\right)
\end{align}
where 
\begin{equation}
C=\omega_{a}^{2}\frac{4}{\pi^{2}}\frac{r\left(1+r^{4}\right)}{\left(1+r^{2}\right)^{3}}\frac{h^{2}}{\theta_{+}^{2}F^{2}}\label{eq:Cterm}
\end{equation}
for $r=\sqrt{c_{+}/c_{-}}$, and 
\begin{align}
\mathfrak{D} & \approx\exp\left(\left(-\frac{3}{2}+\sqrt{\frac{9}{4}-V_{B}}\right)\left(T_{B}-\tilde{T}\right)\right)\label{eq:Dterm}
\end{align}
which accounts for the mode amplitude decay through the $V_{B}$ parameter:
\begin{align*}
V_{B} & \approx c_{-}+\frac{1}{\left(T_{L}-T_{2}\right)}\left(\frac{1063}{3072}+\frac{106793c_{-}}{393216c_{+}}\right)
\end{align*}
for parameters $\tilde{T}$

\begin{equation}
\tilde{T}=\max\{T_{2},\ln\left(2K/3\right)\}
\end{equation}
and $T_{B}=T_{L}$ given in Eq.~(\ref{eq:TLeqn}) which for $c_{-}\ll1$
reduces to

\begin{equation}
T_{L}\approx T_{2}-\left(\frac{3}{c_{-}}\right)\ln\left(\frac{2\sin\left(\pi n_{1}\right)2^{2-n_{1}}\Gamma\left(1-n_{1}\right)\phi_{-\min}\left(\frac{A\sqrt{2}}{3\bar{\Omega}}\right)^{n_{1}}}{\pi\left(3\phi_{-s}(T_{2})\frac{A\sqrt{2}}{3\bar{\Omega}}\left.\partial_{x}J{}_{n_{1}}\left(x\right)\right|_{x=\frac{A\sqrt{2}}{3\bar{\Omega}}}+\left(3\phi_{-s}(T_{2})+2\dot{\phi}_{-s}(T_{2})\right)J_{n_{1}}\left(\frac{A\sqrt{2}}{3\bar{\Omega}}\right)\right)}\right)
\end{equation}
where $\phi_{-s}$ and its derivative in Eq.~(\ref{eq:TLeqn}) for
$T_{L}$ are given in Eq.~(\ref{eq:phims_atT2}), $\bar{\Omega}$
is given in Eq.~(\ref{eq:avgomega}) and $n_{1}=\sqrt{1-4c_{-}/9}$
. As noted near Eq.~(\ref{eq:KPdef}), there is a gap in the analytic
spectrum in the region $[K_{2},K_{P}]$. The correction factor $f_{\mathrm{correction}}(K)$
is defined in Eq.~(\ref{eq:fcorr}).

The coefficients $C_{n}$ have been defined such as to be approximately
scale-independent. Since $V_{1}\gg1$, the term $b=\sqrt{V_{1}+9/4-K^{2}e^{2T_{c}}}$
in $C_{1,2}$ is approximately $K$-independent. Similarly, $\mathfrak{D}^{2}$
is independent of $K$ for long wavelengths and hence it is absorbed
into $C_{1}$. Meanwhile $-K\tau_{c}\sim2$ gives us the approximate
location of the first bump. If $V_{2}$ is neglected, the $\mathfrak{D}^{2}$
term has the following approximate form for $A\sqrt{2}/(3\Omega)\lesssim1$,
\begin{align}
\mathfrak{D}^{2} & \approx e^{\frac{2}{3}\left(c_{-}\left(\tilde{T}-T_{2}\right)-\frac{1063}{3072}\right)}\frac{1}{F^{2}}16\sqrt{\frac{c_{+}}{c_{-}}}\left(1-4c_{-}/9\right)\left(\frac{1}{\alpha/2+1/5+4/F}\right)^{2}\qquad K\lesssim K_{2},\label{eq:Dsquare}
\end{align}
where we remark that $\mathfrak{D}^{2}$ eventually tends to $1$
for extremely small scales that correspond to the massless axion.
To evaluate the expressions given in Eqs.~(\ref{eq:maineq}) and
(\ref{eq:amplitude}) into a numerical amplitude, the parameters $\{T_{1},T_{2},T_{c},V_{1},\Delta T=T_{2}-T_{1}\}$
can be computed through 
\begin{align}
T_{1} & \approx T_{c}-\left(\frac{3.11-1.05\alpha}{2F}\right)\label{eq:deltat}\\
T_{2} & \approx T_{c}+\left(\frac{3.11-1.05\alpha}{2F}\right)
\end{align}
\begin{equation}
T_{c}\approx T_{z}-\frac{0.7}{\alpha F}
\end{equation}
\begin{equation}
V_{1}=\left|\min\left(m_{1}^{2}-\dot{e}_{1}\cdot\dot{e}_{1}\right)\right|\label{eq:V1}
\end{equation}
obtained from Sec.~\ref{subsec:General-map-of}. The definition of
$T_{z}$ can be found in Eq.~(\ref{eq:Tz}), and the variables $\tau_{1,2,c}$
are the conformal times corresponding to $T_{1,2,c}$ obtained through
Eq.~(\ref{eq:conformaltime}). In turn, to evaluate $\alpha$ in
Eq.~(\ref{eq:deltat}), use Eqs.~(\ref{eq:alpha-formula}) and (\ref{eq:Tssolution}).
To evaluate $V_{1}$ of Eq.~(\ref{eq:V1}), put Eqs.~(\ref{eq:phippoly})
and (\ref{eq:phimpoly}) into Eqs.~(\ref{eq:lightesteig}) and (\ref{e1primesq})
and minimize by varying time $T$.

For cases with $\alpha>\alpha_{2}$ where a second dip is also significant,
the shape of the isocurvature spectrum is modified as the parameter
$C_{3}$ becomes a $K$-dependent function. The modes now carry additional
phases that are dependent upon the dynamics of the $V_{3}$ dip. This
situation is similar to the explanation provided in Sec.~\ref{subsec:The-isocurvature-spectrum}.
These cases are solved using the scattering matrices of Eq.~(\ref{eq:final-S})
with the full set of model parameters. Eq.~(\ref{eq:maineq}) and
the more general computational procedure presented in Sec.~\ref{sec:Analytic-spectrum}
are the main results of this paper.

\subsection{Discussion\label{subsec:Discussion}}

By substituting Eqs.~(\ref{eq:amplitude}), and (\ref{eq:Dsquare})
into Eq.~(\ref{eq:maineq}) we obtain an approximate order of magnitude
estimate for the amplitude of the first bump corresponding to $-K_{\text{first-bump}}\tau_{2}\sim2$
for $c_{-}\ll1$:

\begin{equation}
\frac{\Delta_{S}^{2(c_{+}>9/4)}(K_{\mathrm{first-bump}})}{C}\sim O(1)\pi\left(e^{-\omega\pi}\left|H_{i\omega}^{1}\left(2\right)\right|^{2}\right)\left(1-4c_{-}/9\right)\sqrt{\frac{c_{+}}{c_{-}}}\frac{V_{1}}{F^{2}}\left(\frac{.5}{1/F+.12}\right)^{2}
\end{equation}
where 
\begin{equation}
e^{-\omega\pi}\left|H_{i\omega}^{1}\left(2\right)\right|^{2}\approx0.31+O(\omega^{2})
\end{equation}
near $\omega\rightarrow0^{+}$ and varies slowly (fractional power
of $\omega$) for $\omega\sim O(1)$. The above expression is an approximation
and higher order corrections such as the presence of additional dips
(for instance $-V_{3}$ dip due to the second crossing after $T_{c}$)
can lead to further increase in the amplitude.

Next, we compare the above expression to a corresponding one for the
overdamped scenario. In a previous work \citep{Chung2017}, numerical
fitting functions were developed to estimate the isocurvature power
spectrum for overdamped cases. It was found that the bump amplitude
was maximally approximately a factor of 3 compared to the massless
axion plateau. However, those fitting functions were evaluated in
a corner of parameteric region with $0.5<c_{-}<1$ such that they
were largely independent of $c_{-}$ up to the required accuracy.
In order to include the $c_{-}$ parametric dependence, we note that
within the framework of our mass-model, the overdamped scenario can
be studied by considering a single dip followed by the $V_{B}$ parameter
in Eq.~(\ref{eq:reduced-model-singledip}). Using the cubic-polynomial
expressions in Eqs.~(\ref{eq:phippoly}) and (\ref{eq:phimpoly}),
it is easy to show that for the overdamped scenario $V_{1}\approx1.5\sim O(1)$
and $\Delta T\sim O(1)$. Using these estimates, we evaluate the amplitude
at first bump as 
\begin{equation}
\frac{\Delta_{S}^{2(c_{+}<9/4)}(K_{\mathrm{first-bump}})}{C}\sim O(0.08)\pi\left(\frac{c_{+}}{c_{-}}\right)^{0.5}
\end{equation}
where it is worth noting that the residual $\left(c_{+}/c_{-}\right)^{0.5}$
dependence is obtained by fitting the mode amplification due to the
slow-varying $m_{B}^{2}$ function. To understand this, note that
for the overdamped scenario, the background fields settle to their
respective minima along a trajectory where $\phi_{-s}<\phi_{\min}$.
Hence, the associated $m_{B}^{2}$ function is negative and can lead
to amplification of the mode function (see Sec.~(\ref{sec:mB2})
for further details). Also, note that unlike the underdamped scenario,
there is no large kinetic energy at transition and thus the absence
of any $O(F)$ amplitude enhancing corrections for the overdamped
case.

We now compare the isocurvature spectral amplitude between the two
cases and obtain 
\begin{equation}
\frac{\Delta_{s}^{2(c_{+}>9/4)}(K_{\mathrm{first-bump}})}{\Delta_{s}^{2(c_{+}<9/4)}(K_{\mathrm{first-bump}})}\sim O(10)\left(e^{-\omega\pi}\left|H_{i\omega}^{1}\left(2\right)\right|^{2}\right)\left(1-4c_{-}/9\right)\left(\frac{.5}{1/F+.12}\right)^{2}\frac{V_{1}}{F^{2}}.
\end{equation}
For $V_{1}>O(10)$, the $\alpha$ dependence is 
\begin{equation}
\frac{V_{1}}{F^{2}}\sim0.67\alpha-0.05+\frac{1}{F}\qquad\forall\alpha\gtrsim0.1
\end{equation}
which is an approximate numerical value ($O(10\%)$ accurate) resulting
from evaluating $V_{1}$ as explained below Eq.~(\ref{eq:V1}). Thus,
the ratio of the isocurvature power spectrum between overdamped ($c_{+}<9/4)$
and underdamped ($c_{+}>9/4)$ scenarios is approximately
\begin{equation}
\left.\frac{\Delta_{s}^{2(c_{+}>9/4)}(K)}{\Delta_{s}^{2(c_{+}<9/4)}(K)}\right|_{K_{\mathrm{first-bump}}}\sim O(10)e^{-\omega\pi}\left|H_{i\omega}^{1}\left(2\right)\right|^{2}\left(\frac{0.5}{1/F+0.12}\right)^{2}\left(0.67\alpha-0.05+\frac{1}{F}\right)\quad\forall\alpha\gtrsim0.1.\label{eq:ratio}
\end{equation}
At $\alpha\sim0.1$, the above ratio is approximately $\gtrsim1$
and hence as $c_{+}\rightarrow9/4$ and $V_{1}\rightarrow O(1)$,
Eq.~(\ref{eq:ratio}) tends to unity, giving us a check of the formulas
based on the consistency with the results of \citep{Chung2017}. Also,
Eq.~(\ref{eq:ratio}) becomes $F$ independent for large $F$.

Furthermore, the ratio of the amplitude of first bump to the massless
axion plateau is
\begin{equation}
\frac{\Delta_{S}^{2(c_{+}>9/4)}(K_{\mathrm{first-bump}})}{\Delta_{S}^{2}(K>K_{\mathrm{P}})}\approx8\frac{r\left(1+r^{4}\right)}{\left(1+r^{2}\right)^{2}}O(1)\pi e^{-\omega\pi}\left|H_{i\omega}^{1}\left(2\right)\right|^{2}\frac{V_{1}}{F^{2}}\left(\frac{1}{2/F+0.24}\right)^{2}.\label{eq:ratio-1}
\end{equation}
Note that as $c_{+}\gg O(10)$, $\left(e^{-\omega\pi}\left|H_{i\omega}^{1}\left(2\right)\right|^{2}\right)\sim1/\omega\sim1/\sqrt{c_{+}}$
canceling the $r$ enhancement factor. Therefore the ratio of the
bump amplitudes saturates to a constant for large $c_{+}$ values.
For $c_{+}\sim O(1)$ in the resonant case of our interest, Eq.~(\ref{eq:V1})
can be approximated as 
\begin{equation}
\frac{\Delta_{S}^{2(c_{+}>9/4)}(K_{\mathrm{first-bump}})}{\Delta_{S}^{2}(K>K_{\mathrm{P}})}\approx O(30)\alpha\left(\frac{4/3}{8/F+1}\right)^{2}\label{eq:ratio-2}
\end{equation}
which shows how the underdamped scenarios enhance the bump amplitude.
This large $O(30)$ number ultimately can be traced to the combination
of two coincident effects: a) enhancement of the kinetic energy due
to a time phase accident in the context of oscillatory background
solutions which exists only in the underdamped cases b) $\xi$-involving
interaction energy dominating over the mass energy. For instance,
consider the ratio of the kinetic energy $(KE)$ to the net potential
energy (mass + interaction energy $ME+IE$) at transition for the
two cases. Using Eq.~(\ref{eq:Tcsolution}), we can approximate this
ratio for $\alpha>\alpha_{{\rm L}}$ for the resonant underdamped
case as follows 
\begin{align}
\left(\frac{KE}{ME+IE}\right)_{\mathrm{c_{+}>9/4}} & =\frac{\dot{\phi}_{+}^{2}+\dot{\phi}_{-}^{2}}{\xi^{2}+c_{+}\phi_{+}^{2}+c_{-}\phi_{-}^{2}}\\
 & \sim\frac{\alpha^{2}F^{4}}{F^{4}\left(\left(1-0.2\alpha\right)^{2}-1\right)^{2}+O(F^{2})}\\
 & \sim O(8)\label{eq:order8}
\end{align}
where we note that for underdamped resonant cases, the interaction
energy $\xi^{2}$ is $O(F^{4})$. Remarkably, the parametric dependences
have canceled out in Eq.~(\ref{eq:order8}).

A similar evaluation for the overdamped situation where the interaction
energy $\xi^{2}$ is $O(1)$ yields 
\begin{align}
\left(\frac{KE}{IE+ME}\right)_{\mathrm{c_{+}\lesssim9/4}} & =\frac{\dot{\phi}_{+}^{2}+\dot{\phi}_{-}^{2}}{\xi^{2}+c_{+}\phi_{+}^{2}+c_{-}\phi_{-}^{2}}\\
 & \sim\frac{F^{2}}{9/4F^{2}}\\
 & \sim O(0.5)
\end{align}
where we note $\xi$ being insignificant in both the numerator and
the denominator. Thus, we observe an approximate $O(10)$ enhancement
in the spectral power for the underdamped cases compared to those
of the overdamped. Since, the first bump in overdamped cases is maximally
approximately a factor of 3 compared to the massless axion plateau,
we obtain an effective enhancement factor of $O(30)$ for the underdamped
scenario as observed in Eq.~(\ref{eq:ratio-2}).

As highlighted previously in Sub-sec.~(\ref{subsec:V2correction}),
a large $V_{2}$ jump ETSP leads to an attenuation of spectral power
for all modes super-horizon at $T_{2}$ while having a decreasingly
small effect on the sub-horizon modes. An interesting consequence
of this is that for large enough $V_{2}$, the amplitude of a subsequent
bump (second or higher) can appear much greater than the first bump.
This can be understood as follows. From Eq.~(\ref{eq:maineq}), we
infer that the amplitude of the spectral bumps (oscillations) in the
absence of the $V_{2}$ jump ETSP can be expressed as
\begin{equation}
\Delta_{S,V_{2}=0}^{2}(K)\approx\mathcal{A}_{0}\left(\frac{K_{{\rm first-bump}}}{K}\right)^{2}\quad\forall K_{{\rm first-bump}}\leq K\lesssim\tau_{c}^{-1}\sqrt{V_{1}}
\end{equation}
where $\mathcal{A}_{0}$ is the amplitude of the first-bump and is
independent of $V_{2}$.

With the inclusion of $V_{2}$, the spectral power for all super-horizon
modes at $T_{2}$ is attenuated due to a $V_{2}$ dependent correction
factor $\left|f_{{\rm correction}}\right|<1$ as shown in Sec.~\ref{subsec:V2correction}.
For $V_{2}\gg1$, the corrected spectral amplitude of the first bump
can be written as
\begin{equation}
\Delta_{S,V_{2}\gg1}^{2}(K_{{\rm first-bump}})\approx\left|f_{{\rm correction}}\left(K_{{\rm first-bump}}\right)\right|^{2}\mathcal{A}_{0}\label{eq:fb-V2corr}
\end{equation}
where the $K$ dependence of $f_{{\rm correction}}$ is approximately
a constant for all super-horizon modes. On the other hand, the spectral
power in Eq.~(\ref{eq:maineq}) for the sub-horizon modes have the
property $\left|f_{{\rm correction}}(K)\right|\sim1$:

\begin{equation}
\Delta_{S,V_{2}\gg1}^{2}(K)\approx\left|f_{{\rm correction}}(K)\right|^{2}\mathcal{A}_{0}\left(\frac{K_{{\rm first-bump}}}{K}\right)^{2}\quad\forall K_{{\rm V}}\lesssim K\lesssim\tau_{c}^{-1}\sqrt{V_{1}}\label{eq:sb-V2corr}
\end{equation}
where $K_{{\rm V}}\equiv\tau_{c}^{-1}\sqrt{V_{2}}$ and defines mode
when the $\left|f_{{\rm correction}}(K)\right|\sim1$. The $\left|f_{{\rm correction}}(K)\right|$
smoothly interpolates between $1$ and $\left|f_{{\rm correction}}(K_{{\rm first-bump}})\right|$
in the spectral region {[}$K_{{\rm first-bump}},K_{V}]$.

Comparing Eqs\@.~(\ref{eq:fb-V2corr}) and (\ref{eq:sb-V2corr}),
we conclude that it is possible for certain high $K$ modes to have
a larger spectral power than the first-bump. Quantitatively, this
is true for the following approximate range of $K$ modes
\begin{align}
K_{{\rm V}} & \lesssim K\lesssim\left|\frac{K_{{\rm first-bump}}}{f_{{\rm correction}}\left(K_{{\rm first-bump}}\right)}\right|\equiv K_{{\rm f}}.
\end{align}
This is generically observed as a larger second or third bump than
the first (see Fig.~\ref{eq:maineq}). The amount of this relative
enhancement can be evaluated as:
\begin{align}
\frac{\Delta_{S}^{2}(K)}{\Delta_{S}^{2}(K_{{\rm first-bump}})} & \sim O\left(V_{2}^{1/2}\right)\left(\frac{K_{{\rm V}}}{K}\right)^{2}\quad\forall K_{{\rm V}}\lesssim K\lesssim K_{{\rm f}}\label{eq:highKbumpenhancement}
\end{align}
where the factor of $V_{2}^{-3/2}$ comes from $\left|f_{\mathrm{correction}}(K)\right|^{2}\approx\left|l(K)\right|^{2}$
in Eq\@.(\ref{eq:lk}) for $V_{2}\gg1$. Hence, the enhancement is
approximately proportional to $\sqrt{V_{2}}$ and increases with the
$F$ scale. Thus, we remark that large $O(F^{2}$) resonant effects
together with any additional dips ($-V_{i}\text{ for}\;i\geq3$) corresponding
to higher order corrections, can result in spectral power enhancement
by a factor greater than the $O(30)$ as derived previously for the
resonant underdamped cases. Unlike the $O(30)$ factor whose origin
was discussed in Eq.~(\ref{eq:order8}), this high $K$ mode enhancement
is dependent on the parameter $F$.

\section{\label{sec:Parametric-dependences-of}Parametric dependences of the
isocurvature spectrum}

One qualitative predictability difference between the overdamped and
underdamped axionic scenarios where the PQ symmetry is broken before
the end of inflation stems from the fact that there is an attractor
solution for the background fields as well as for the linear perturbations
in the overdamped scenarios. This means that given a Lagrangian in
overdamped scenarios, the cosmological predictions have less dependence
on the initial conditions. As an analogy, in the case of slow roll
single field inflation, one only needs to specify the initial field
value and not its time derivative to specify a prediction for the
observables. For the usual cosmological axion scenarios, the radial
field associated with PQ symmetry breaking is considered to be sitting
at the minimum of its potential. This means that the only initial
condition dependence of the axion isocurvature during inflation is
$\theta_{i}f_{a}$. In the case of the current underdamped scenario
of interest, there is an additional phase space dependence of the
PQ symmetry breaking radial field directions as well. Because of the
non-attractor behavior for the underdamped dynamics along the flat
direction, the dominant additional phase space degree of freedom is
the initial field $\phi_{+}(0)$ value and the kinetic energy of the
radial field along the flat direction parameterized by $\epsilon_{0}$.

In this section we will study the dependence of the axion isocurvature
spectrum on the model parameters $\{c_{+},c_{-},F\}$ and the initial
conditions $\{\epsilon_{0},\phi_{+}(0)\}$. The effect of each parameter
variation is discussed keeping all of the others fixed.

\subsection{$c_{+}$}

The numerical model presented in this paper and the associated axion
isocurvature spectrum have been derived for background fields within
the parametric region given by $\alpha_{{\rm L}}\lesssim\alpha\lesssim\min\left(\alpha_{3},\alpha_{{\rm U}}\right)$.
Above the upper bound $\alpha_{{\rm U}}$, the analytic methods utilized
in this paper break down due to the significant heavy-mode mixing
from Eq.~(\ref{eq:neglectheavyatTc}) at $T_{c}$, and at around
the same upper bound, the adiabatic approximation technique also breaks
down.\footnote{Beyond $\alpha>\alpha_{3}$, the background fields will cross at least
twice after transition. For large $F$, these crossings can occur
close enough such that effective heavy mixing from the superposition
of each crossing becomes significant. Moreover, as $\alpha\rightarrow1$,
the background field dynamics turns highly chaotic after transition
and a closed form prediction of the mode amplitude in terms of $\alpha$
is not feasible. Field configurations with large $c_{+}$ tend to
fall under this category. These configurations associated with $\alpha\gtrsim1$
and the accompanying isocurvature spectrum are a subject of interest
and will be explored in a separate companion paper \citep{futurepaper}.} For $\alpha$ less than the lower bound, the cubic polynomial expansion
of the background fields is insufficient and higher order terms become
significant. The parameter $\alpha$ can be computed as a function
of underlying Lagrangian model parameters and the initial conditions
from Eq.~(\ref{eq:alpha-formula}).

\begin{figure}
\begin{centering}
\includegraphics[scale=0.9]{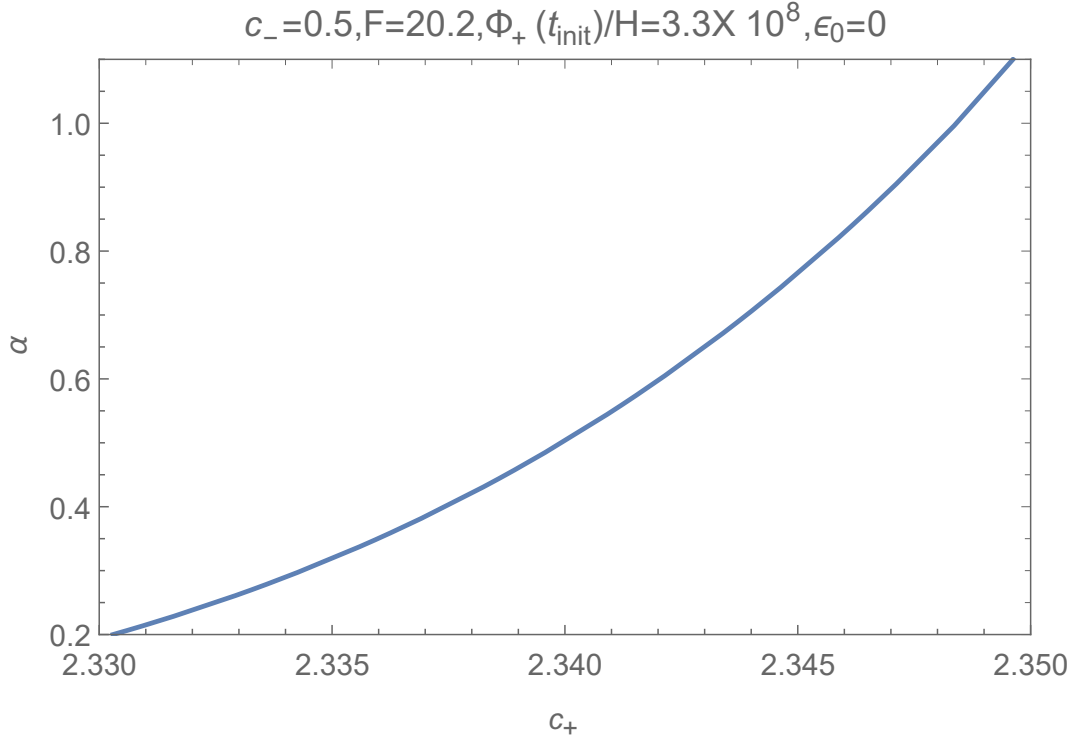} 
\par\end{centering}
\caption{\label{fig:alpha_vs_cp}Plot Eq.~(\ref{eq:alphacplus}) showing regions
of $c_{+}$ with corresponding value of $\alpha$ within the range
of Eq.~(\ref{eq:alphaapplicable}). The other parameters are set
to the fiducial set $P_{A}$ of Fig.~\ref{fig:Comparison-of-Taylor}.}
\end{figure}

In Fig.~\ref{fig:alpha_vs_cp} we plot the $c_{+}$ dependence of
$\alpha$ within the range of Eq.~(\ref{eq:alphaapplicable}) close
to $c_{+}=9/4$ corresponding to the range of $c_{+}$ values where
$T_{c}$ is close to (but before) the first zero-crossing of the $\phi_{+}^{(0)}$
field.\footnote{Recall $\phi_{+}(T)$ is approximately equal to $\phi_{+}^{(0)}(T)$
before $T_{c}$.} The monotonic increase in $\alpha$ is captured through the following
expression: 
\begin{equation}
\alpha\propto\exp\left(-\frac{3}{2}\left[\frac{\pi}{\sqrt{c_{+}-9/4}}\right]\right).\label{eq:alphadependence}
\end{equation}

\begin{figure}
\begin{centering}
\includegraphics[scale=0.5]{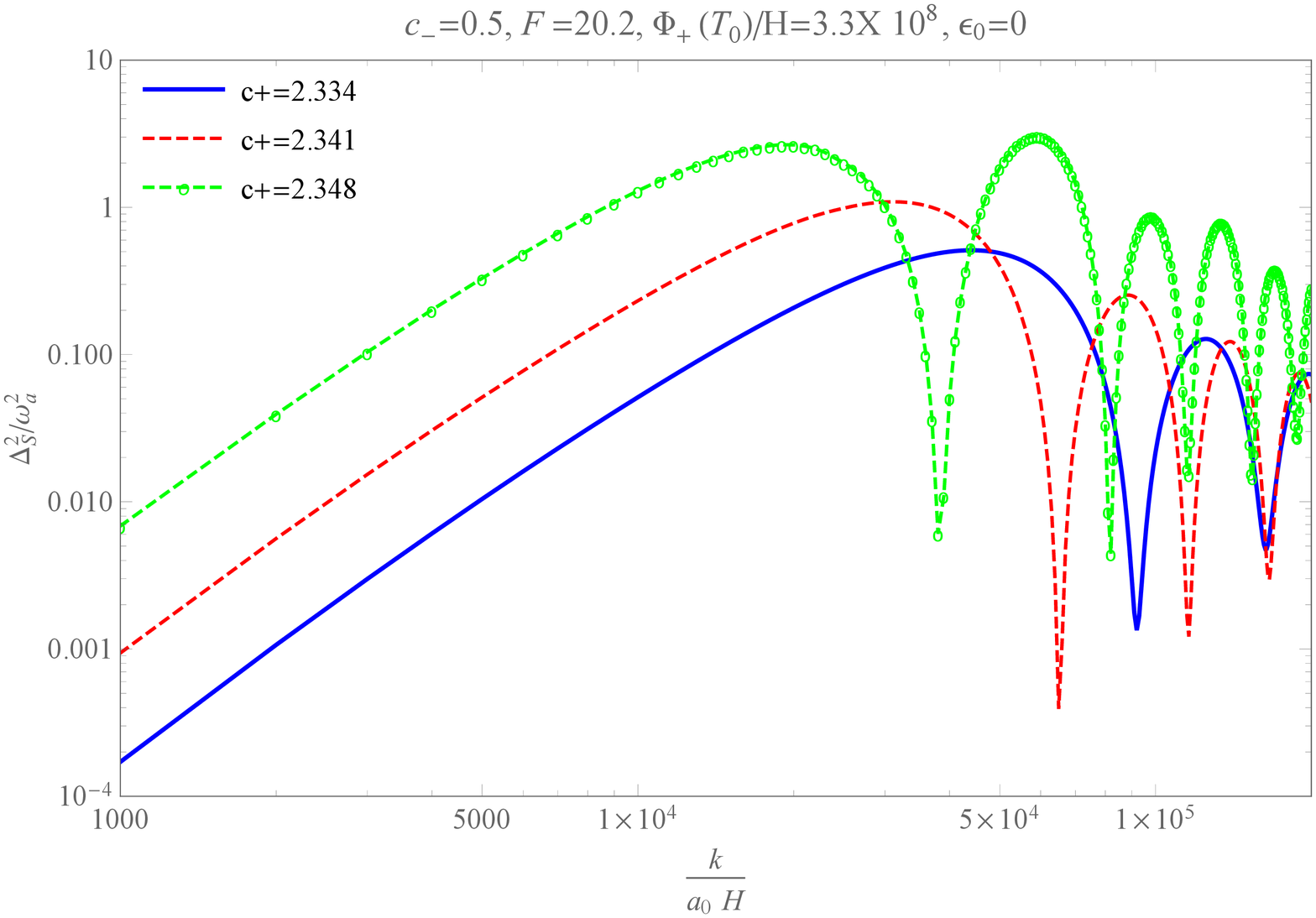} 
\par\end{centering}
\caption{\label{fig:alpha-dep}Plot of spectra made using Eq.~(\ref{eq:final-S})
for increasing values of $c_{+}$ at transition. Note that other parameters
are set at the fiducial parameter set $P_{A}$ used in Fig.~\ref{fig:Comparison-of-Taylor}.}
\end{figure}

Next, in Fig.~\ref{fig:alpha-dep}, we present isocurvature spectra
for three different $c_{+}$ values. We remark that as $c_{+}$ is
increased, the $\phi_{+}^{(0)}$ field rolls down faster owing to
the frequency $\omega=\sqrt{c_{+}-9/4}$.\footnote{The exact dependence of the velocity on $\omega$ is dependent on
whether the system is in the resonant, non-resonant, overdamped, underdamped
situations. Here, we will here be focusing only on the resonant cases,
which is the main focus of our work.} Consequently, the first zero-crossing of the $\phi_{+}^{(0)}$ field
occurs earlier with an increasing $c_{+}$ (See Eq.~(\ref{eq:Tz})).
This in turn increases the $\exp\left(-3T_{c}/2\right)$ factor controlling
$\dot{\phi}_{+}$ leading to a larger kinetic energy at $T_{c}$.
The larger kinetic energy is later converted into the interaction
energy at transition whose larger value is responsible for the growth
of the isocurvature amplitude through the resonant effects. Hence,
increasing $c_{+}$\footnote{See below regarding how much one can increase $c_{+}$ .}
results in an amplification of first bump in the isocurvature spectrum.
Moreover, an increasing height of the isocurvature spectrum is accompanied
by a receding location of the first bump $K_{\text{first-bump}}$
towards lower values. This is again explained by decreasing $T_{c}$
as $c_{+}$ is increased because Eq.~(\ref{eq:Kbump}) implies 
\begin{equation}
K_{\text{first-bump}}\approx O(1)e^{T_{c}}.
\end{equation}

When one compares the first bump amplitude to the plateau amplitude
in the $c_{+}=2.348$ case of Fig.~\ref{fig:alpha-dep}, one sees
that the ratio can be about 30. This is already explained in Eq.~(\ref{eq:ratio-2})
of Sec.~\ref{subsec:Discussion} where we evaluated an enhancement
factor of 
\begin{equation}
O(30)\alpha\left(\frac{4/3}{8/F+1}\right)^{2}
\end{equation}
for modes lying within the range $0.5\lesssim-K\tau_{c}<3$ compared
to the modes in the massless axion plateau region of the spectrum
($K>K_{\mathrm{P}}$).

To understand this qualitatively, one first notes that there is the
possibility in the underdamped scenarios of a large kinetic energy
in the falling $\phi_{+}$ when the $\phi_{-}$ interaction with $\phi_{+}$
becomes strong. This large kinetic energy leads to nonadiabatic effects
post $T_{c}$ such that the axion mode amplitude obtains an $O(10)$
enhancement compared to an overdamped scenario.

Note in Fig.~\ref{fig:alpha-dep} that the frequency of the $K$-space
oscillations also increase with $c_{+}$. This has already been explained
quantitatively in Eq.~(\ref{eq:firstbunp}). To understand this another
way qualitatively, note that the temporarily negative lightest mass
squared eigenvalue and the nonadiabatic rotation of the lightest eigenvector
pump the mode amplitude (see Eq.~(\ref{eq:massmat})). Therefore,
the $\text{\ensuremath{k}}-$space oscillation frequency is reflective
of the nonadiabaticity producing mode dynamics at time $\sim T_{c}$
where the modes have the characteristic phase $\exp\left(-ik\exp(-T_{c})\right)$,
giving a $k$-space oscillation period of $O\left(\exp(T_{c})\right)$
as can be seen explicitly in Eq.~(\ref{eq:firstbunp}).

The examples presented in this paper are limited to a range of $c_{+}$
values where the transition occurs close to the first zero-crossing.
One might then worry that the mass model is no longer applicable for
higher $c_{+}$ values because according to Fig.~\ref{fig:alpha_vs_cp},
the $\alpha$ value naively seems to increase to violate the approximation
methods used. However, as we will discuss in a separate paper \citep{futurepaper},
$\alpha$ is a discontinuous function of $c_{+}$. As we will show
there, the present mass model is still applicable for a range of larger
$c_{+}$ values (although the $c_{+}$ regions where the model is
applicable are not continuously connected).

\subsection{$F$}

The $F$ dependence of the isocurvature spectrum is multi-faceted.
The $C$ term in Eq.~(\ref{eq:Cterm}), which is mostly about the
normalization of the axion field, suggests a $1/F^{2}$ proportionality
of the power spectrum. This is an expected result since the time-dependent
massive axion isocurvature spectrum has a $1/\phi_{+}^{2}$ dependence
within the long wavelength region (as well as in both the plateau
regions). Thus, the variation of the power spectrum in the massless
plateau region has an $1/F^{2}$ proportionality similar to the overdamped
scenario.

\begin{figure}
\begin{centering}
\includegraphics[scale=0.5]{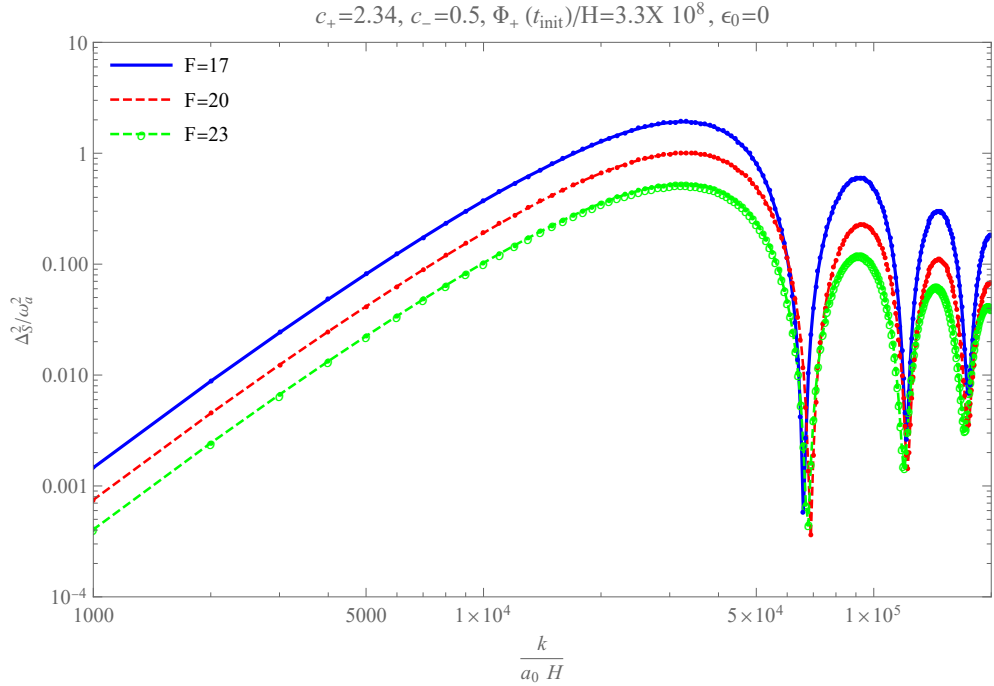} 
\par\end{centering}
\caption{\label{F_dependence_1}Illustrated is how the spectra (made using
Eq.~(\ref{eq:final-S})) varies as $F$ increases. The other parameters
are fixed at approximately the set $P_{A}$ as in Fig.~\ref{fig:Comparison-of-Taylor}.
The value of $\alpha$ in this plot varies as $\{0.38,0.51,0.71\}$
as $F$ is reduced.}
\end{figure}

For scales that lie within the oscillating part of the spectrum, additional
$F$ dependences arise from the $\mathfrak{D^{2}}$ and $b^{2}$ terms
of Eq.~(\ref{eq:maineq}). As shown in Sec.~\ref{subsec:Discussion},
the spectrum in this region has the following proportionality 
\begin{align}
\frac{\Delta_{S}^{2}(K)}{C} & \propto\frac{V_{1}}{F^{2}}\left(\frac{.5}{1/F+.12}\right)^{2}.
\end{align}
Using the analytic expressions derived previously, the ratio $V_{1}/F^{2}$
has the following polynomial form within the parametric region given
by Eq.~(\ref{eq:alphaapplicable}),
\begin{align}
\frac{V_{1}}{F^{2}} & \approx c_{1}+c_{2}\alpha\label{eq:V1parametrization}
\end{align}
where $c_{1}\sim O(0.1)$ and $c_{2}\sim O(1)$. With all other parameters
fixed, we have the relationship $\alpha\propto1/F^{2}$. Including
the $C$ term, the power spectrum has the following effective $F$
dependence in the oscillating region
\begin{equation}
\frac{\Delta_{S}^{2}(K)}{\omega_{a}^{2}}\sim\left(\frac{c_{1}}{F^{2}}+\frac{c_{3}}{F^{4}}\right)\left(\frac{.5F}{1+.12F}\right)^{2}
\end{equation}
where
\begin{equation}
c_{3}=\omega\phi_{+}(0)\,\sec\varphi\,e^{-3/2T_{z}}\gg c_{1}.
\end{equation}
As $F$ becomes large, the power spectrum tends to the expected $1/F^{2}$
proportionality. One can understand this by noting that for all other
parameters fixed, an increase in $F$ results in a rapid reduction
in $\alpha$ such that $c_{2}\alpha$ can subsequently become smaller
than $c_{1}$. This is an interesting behavior which can be explained
more clearly by noting that for resonant underdamped fields, $T_{c}$
occurs close to the $\phi_{+}^{(0)}$ zero-crossing: 
\begin{equation}
\dot{\phi}_{+}(T_{c})\approx\dot{\phi}_{+}^{(0)}(T_{z})-\int_{T_{s}}^{T_{c}}dT\xi\phi_{-}.
\end{equation}
At $T_{c}$, the $\phi_{-}$ field is $O(F)$ and the strong coupling
force $\xi\phi_{-}<0$ and $\dot{\phi}_{+}(T_{z})<0$ resulting in
$\left|\dot{\phi}_{+}(T_{c})\right|<\left|\dot{\phi}_{+}(T_{z})\right|$.
Using Eq.~(\ref{eq:alphadefinition}) and $V_{1}\sim\left|\dot{\phi}_{+}(T_{c})\right|$,
we deduce that the $c_{2}$ parameter in Eq.~(\ref{eq:V1parametrization})
is associated with $\left|\dot{\phi}_{+}^{(0)}(T_{z})\right|=\alpha F^{2}$
whereas $c_{1}$ is related to the integral of the coupling term $\xi\phi_{-}$.
Fig.~\ref{F_dependence_1} shows plots of power spectra for a fixed
$c_{+}$ value with different $F$ scales highlighting a $\sim1/F^{n}$
reduction of the power spectrum for $n\sim3-4$.

\subsection{$c_{-}$}

Next we consider the $c_{-}$ dependence of the power spectrum. We
will consider two situations here. In the first, we discuss fields
with $\alpha\lesssim\alpha_{\text{2}}$ such that the fields evolve
after the transition without any crossings. This situation gives rise
to a slowly-varying $m_{B}^{2}>0$. The second case is where the fields
have large resonant amplitude such that the fields cross each other
at least once after $T_{c}$. As explained in the Appendix \ref{sec:mB2},
such a crossing results in a situation where the $\phi_{-}$ field
settles to its minimum from below ($\phi_{-}<\phi_{-\min}$). Due
to this unique alignment, the $m_{B}^{2}$ function now becomes negative
as the fields settle to their minima asymptotically. In both cases
we shall see that the spectrum increases for smaller $c_{-}$ values
with the essential dynamics dictated by the settling of $m_{B}^{2}$
during different temporal phases. The discussion is limited to $c_{-}<9/4$.

Let us now consider the first case. The $m_{B}^{2}$ function results
in an approximate exponential decay of the mode amplitude through
the 
\begin{equation}
y_{1}(T_{\infty})\propto\exp\left(-\frac{1}{3}\int_{\tilde{T}}^{T_{\infty}}m_{B}^{2}dT\right)\label{eq:controllingexponential}
\end{equation}
factor where the integral is evaluated through the Eq.~(\ref{eq:mb2integral}).
The integral can be divided into two temporal phases. During the first
phase, the dominant $\phi_{-}$ field rolls down from its peak amplitude
with a decay constant equal to $c_{-}/3$. Later as the fields get
closer to their minima during the second phase, the decay constant
increases by nearly a factor of four to $\approx4c_{-}c_{+}/\left(c_{-}+c_{+}\right)$.
Since $m_{B}^{2}$ starts out close to $c_{-}$ and eventually settles
to zero, the integral is dominated by the first phase. In this first
phase, $m_{B}^{2}$ behaves as the flat direction mass which decreases
with $c_{-}$ decreasing, resulting in $y_{1}(T_{\infty})\propto\left(c_{+}/c_{-}\right)^{1/4}$:
i.e.~a smaller $c_{-}$ results in a larger effective mode amplitude.

\begin{figure}
\centering{}\includegraphics[scale=0.5]{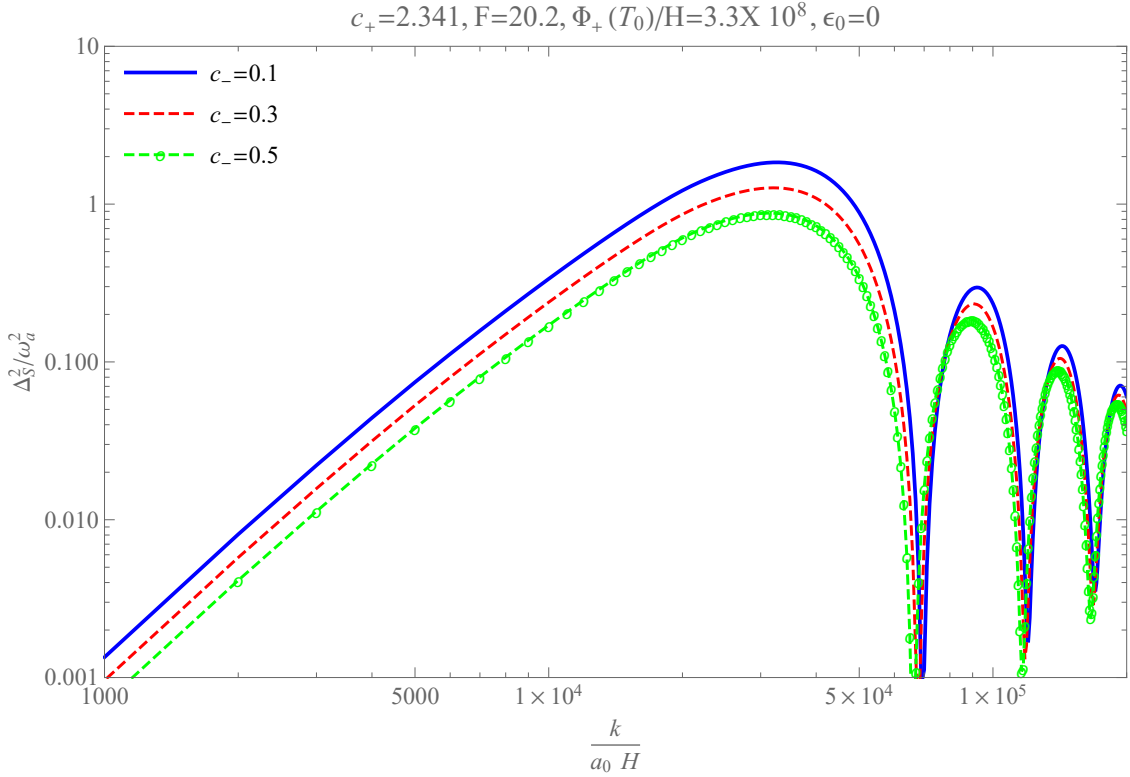}\caption{\label{fig:-dependence-of}This plot shows the $c_{-}$ dependence
of the spectra. The other parameters are fixed at approximately the
set $P_{A}$ as in Fig.~\ref{fig:Comparison-of-Taylor}.}
\end{figure}

In the second case, the $\phi_{-}$ field approaches its minimum from
below. Unlike the previous case, the first temporal phase is insignificant
due to the oscillating IR fields and a dominant $V_{2}$ ETSP. Hence,
the effective mode amplification due to the negative $m_{B}^{2}$
is brought about during the second phase through Eq.~(\ref{eq:controllingexponential}).
As $c_{-}$ decreases, $m_{B}^{2}$ decreases much slower due to the
$\sim4c_{-}/3$ exponential decay rate of $\phi_{-}$. Moreover, a
smaller $c_{-}$ results in a larger value of $|m_{B}^{2}|$. As a
result of these two effects, we see from Eq.~(\ref{eq:controllingexponential})
that the mode function undergoes larger amplification. This is shown
in Fig.~\ref{fig:-dependence-of}.

\subsection{$\epsilon_{0}$ and $\phi_{+}(0)$}

\begin{figure}
\begin{centering}
\includegraphics[scale=0.5]{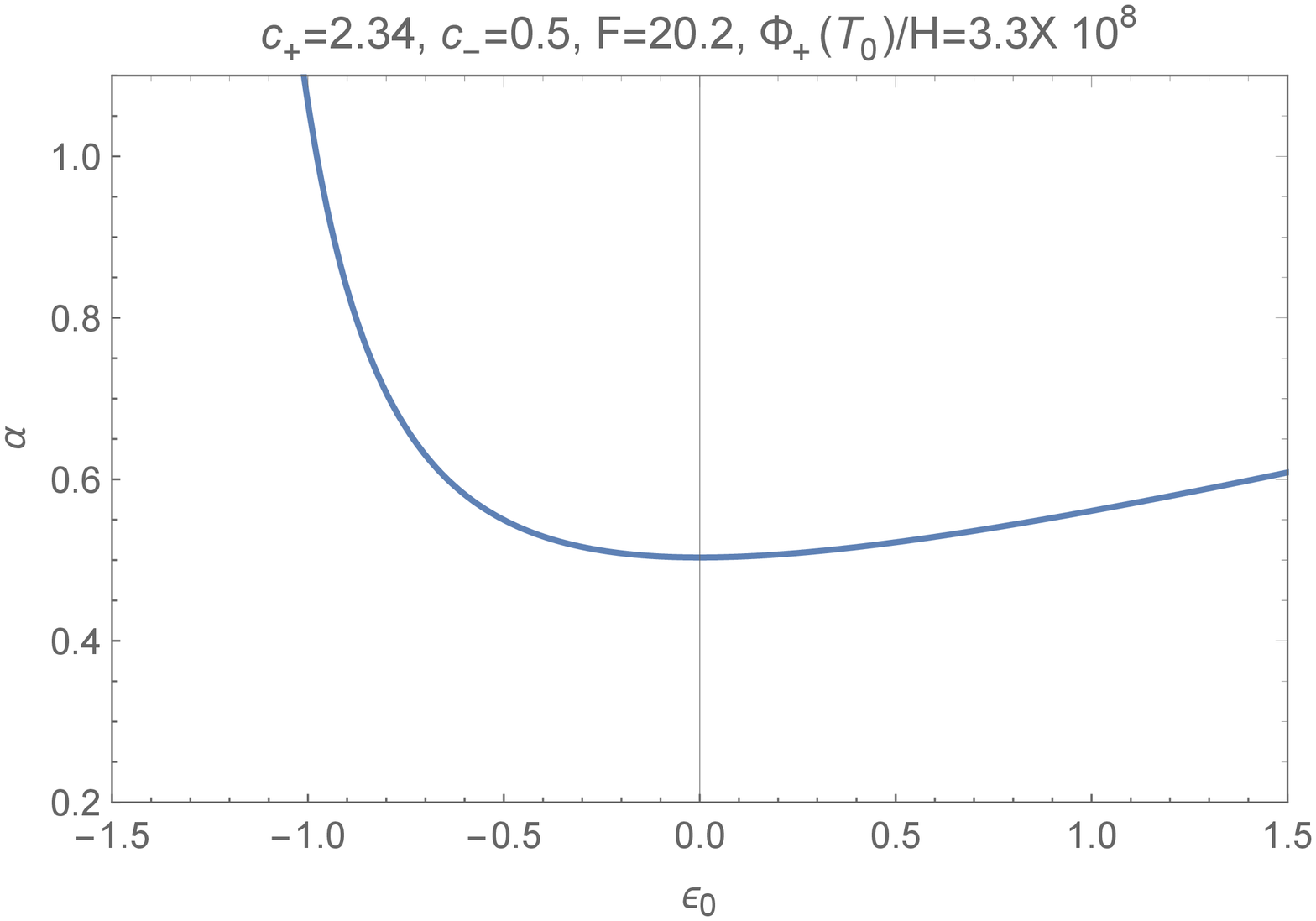} 
\par\end{centering}
\begin{centering}
\includegraphics[scale=0.5]{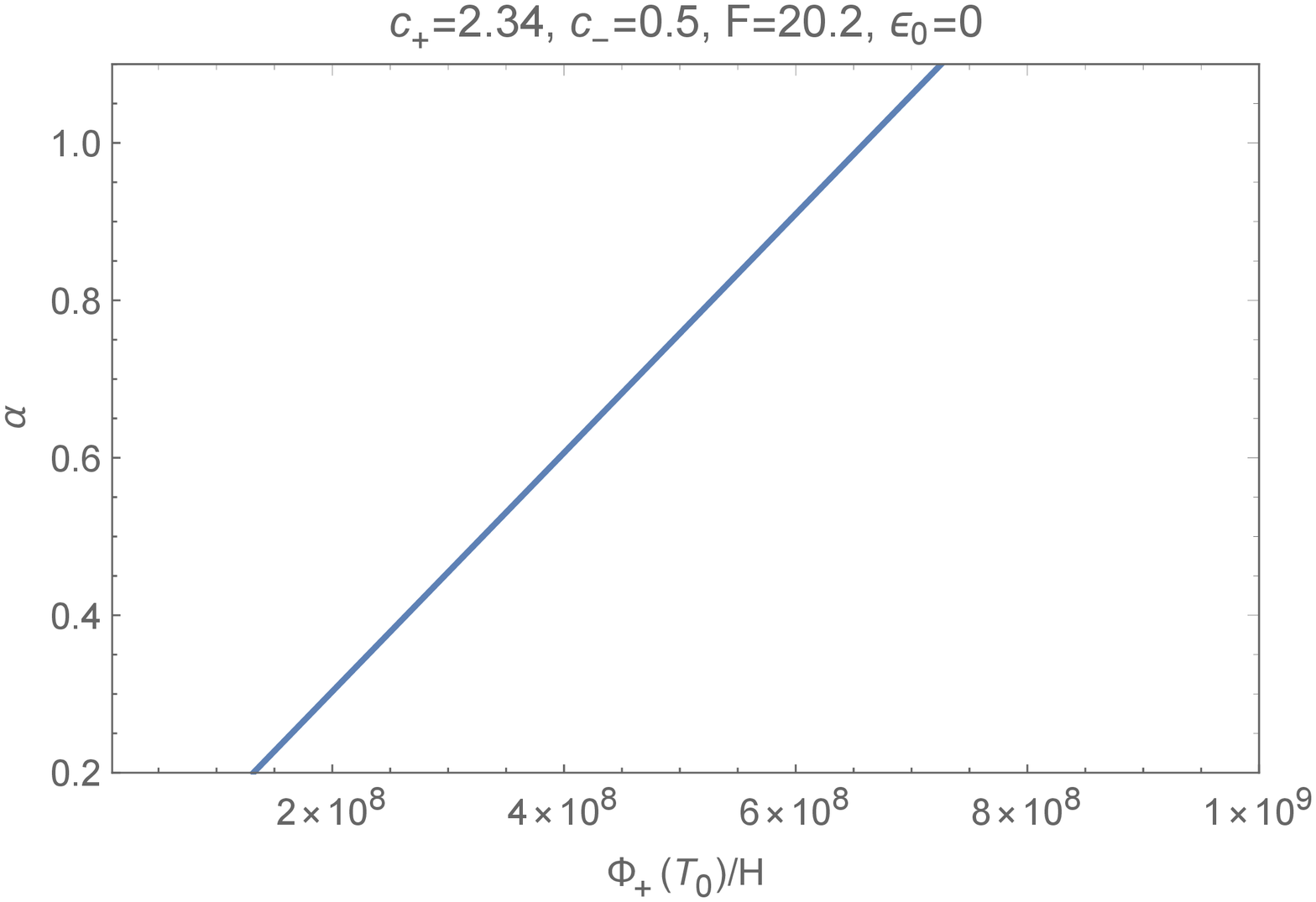} 
\par\end{centering}
\caption{\label{fig:spectrum- epsilon and phip0 dependence}Plots highlighting
the dependence of $\alpha$ on the initial conditions (see Eq.~(\ref{eq:alphacplus})).
Keeping all other parameters fixed, a larger initial energy density
results in a larger isocurvature amplitude of the initial bump.}
\end{figure}

Here we discuss the parameters $\epsilon_{0}$ and $\phi_{+}(0)$
that define the initial conditions for the underdamped rolling fields.
Varying these initial conditions directly alter $T_{c}$ and the value
of $\alpha$. Hence, the effect of these two parameters is best understood
by studying the $\alpha$ expression from Eq.~(\ref{eq:alphacplus}).
We consider the minimal case with $\epsilon_{0}=0$ and expand $\alpha$
to quadratic order in $\epsilon_{0}$ in Eq.~(\ref{eq:alphacplus})
\begin{equation}
\alpha\approx\frac{\phi_{+}(0)}{F^{2}}\sqrt{c_{+}}e^{-\frac{3}{2}\left[\frac{1}{\sqrt{c_{+}-9/4}}\left(\frac{\pi}{2}+\arctan\frac{3/2}{\sqrt{c_{+}-9/4}}\right)\right]}\left(1+\frac{\epsilon_{0}^{2}}{2c_{+}}\right)+O\left(\epsilon_{0}^{3}\right).\label{eq:alpha-function}
\end{equation}
We discover that $\alpha$ has a local minimum at $\epsilon_{0}=0$
evinced by the absence of the linear term in $\epsilon_{0}$. This
is expected since an increase in the initial kinetic energy leads
to a larger $\dot{\phi}_{+}$ at the zero-crossing. As seen in Fig.~\ref{fig:spectrum- epsilon and phip0 dependence},
the dependence of $\alpha$ on $\epsilon_{0}$ in Eq.~(\ref{eq:alphacplus})
is nonlinear beyond the quadratic nonlinearities in Eq.~(\ref{eq:alpha-function}).
As seen in Eq.~(\ref{eq:alpha-function}), the parameter $\phi_{+}(0)$
leads to a monotonic increase in the value of $\alpha$ within the
range of Eq.~(\ref{eq:alphaapplicable}) with the intuition that
the initial energy is increased as in the intuition for the $\epsilon_{0}$
increase. A representative set of $\alpha$ values as a function of
$\phi_{+}(0)$ is shown in Fig.~\ref{fig:spectrum- epsilon and phip0 dependence}.

As $|\epsilon_{0}|$ is continually increased, $T_{c}$ becomes nearer
to the next higher $\phi_{+}^{(0)}(T)$ zero-crossing time, thereby
reducing the overall spectral power. Similarly, if $\phi_{+}(0)$
is continually increased then the transition time $T_{c}$ moves to
the next higher zero-crossing time. The subsequent height of the power
spectrum can be analyzed by evaluating the value of $\alpha$ at the
new transition. Due to the length of this paper and the natural parametric
scope of small $c_{+}$ in this paper, we discuss the isocurvature
spectra reflecting higher zero-crossing of $\phi_{+}^{(0)}$ in a
separate companion paper \citep{futurepaper}.

\section{\label{sec:Conclusions}Summary}

In this paper, we provided an analytic expression for the blue axionic
isocurvature spectrum in the underdamped nonequilibrium axion scenarios
for a particular parametric region corresponding to a mild amplitude
($\alpha\in[\alpha_{L},\alpha_{2}]$) of resonantly oscillating PQ
symmetry-breaking radial fields. The main expression given by Eq.~(\ref{eq:maineq})
exhibits an amplitude of Eq.~(\ref{eq:amplitude}) and an oscillating
spectrum whose $k$-space oscillation period is of the order the value
of the first $k$-break location $K_{{\rm first-bump}}$ in the spectrum
(Eq.~(\ref{eq:Kbump})). Remarkably, the first bump amplitude of
the spectrum is enhanced by a factor of $O(30)$ compared to the plateau
amplitude of the spectrum associated with the massless axions as explained
near Eq.~(\ref{eq:order8}). Furthermore, in some cases with large
$O(F^{2}$) resonant effects, the spectral power can receive larger
than the $O(30)$ enhancement of the first peak as explained near
Eq.~(\ref{eq:highKbumpenhancement}).\footnote{Here we have scaled the PQ symmetry breaking parameter $F_{a}$ as
$F\equiv hF_{a}/H$ where $h$ is a quartic coupling and $H$ is the
Hubble expansion rate during inflation.} In contrast, for the overdamped nonequilibrium axion scenarios (see
\citep{Chung:2016wvv}), the relative amplitude ratio is only a factor
of few (maximally approximately a factor of 3). The $\text{\ensuremath{k}}$-oscillation
spacing is reflective of the mode amplitude-setting dynamics at time
$\sim T_{c}$ whose phase is $\exp\left(-ik\exp(-T_{c})\right)$,
giving a $k$-space oscillation period of $O\left(\exp(T_{c})\right)$.

Technically, the computation was carried out using a combination of
a parametric restriction where the heavy modes are decoupled (Sec.~\ref{sec:Decoupling-of-heavy}),
perturbation theory (Sec.~\ref{subsec:Perturbative-solution}), analytic
fitting to polynomials in the nonperturbative region (Sec.~\ref{subsec:Beyond-perturbation-theory}),
piecewise ETSP modeling (Eq.~(\ref{eq:model})), and the technique
of integrating out fast oscillations (Appendix \ref{sec:Adiabatic-approximation-for}).
This allowed us to compute a transfer matrix solution to the isocurvature
mode equations (Eq.~(\ref{eq:final-S})). Overall, based on comparing
to sample numerical calculations, the accuracy of the computation
is about 20\%-50\% with $r_{a}$ set to 0.2. Most of the uncertainty
is coming from the technique of integrating out the fast oscillations
(for example, estimation of $V_{2}$) and the approximations made
regarding the IR components of the $\phi_{\pm}$ fields after the
transition at $T_{c}$. A Mathematica package to evaluate the spectrum
using the analytic methods can be accessed from ``https://pages.physics.wisc.edu/\textasciitilde stadepalli/Blue-Axion-IsoCurvSpec-Underdamped.nb''.
Even though the analytic formula is complicated, compared to the pure
numerical solver, the speedup factor is about $O(100)$.

In this paper, we have focused on presenting analytic spectral results
in the resonant oscillatory $k$-range for moderate $\alpha$ values,
where $\alpha$ characterizes the velocity of the falling $\phi_{+}$
field near the transition time $T_{c}$ (see Eq.~(\ref{eq:alphadefinition})).
This would be helpful even if a purely numerical approach to the problem
were to be used in data fitting since this will serve as a solvable
check on the system. In a companion paper \citep{futurepaper}, we
will present results for larger $\alpha$ situations in the high $c_{+}$
limit to complete the understanding of the possibly observable blue
isocurvature spectra. In such cases, there are no purely analytic
results, but there will appear a novel stochastic model that parameterizes
the small $k$ range amplitude variation with $\alpha$. In that paper,
we will also exploit various symmetries to obtain relationships between
parameters which is useful more as a transferable technique than quantitative
predictions. Finally, we will also defer to that paper a discussion
of general fitting functions for this underdamped blue axionic isocurvature
class of models that do not refer to underlying Lagrangian parameters
$\{c_{\pm},F\}$.

There are many interesting possible followup topics to be investigated.
Recent Planck results \citep{Planck:2018jri} show that axion-like
curvaton models with uncorrelated blue-tilted spectrum is the most
favored of the isocurvature models. The fits also indicate a possibility
of measuring a spectral index of $1.55<n_{I}<3.67$ at 95\% CL consistently
with the recent findings of \citep{Chung:2017uzc}. It would be interesting
to study whether there are hints of resonant isocurvature spectra
presented in this paper in the existing and future data. Another possible
usage of this enhanced oscillatory peaks presented in this paper would
be to investigate the formation of primordial black holes similar
to the investigation of the curvaton models in \citep{Inomata2021,Ando2018,Kawasaki2013,Kawasaki-PBH2013,Bugaev-PBH2013,BUGAEV_2013,Young_2013,Ando:2018nge,Chen_2019,Passaglia:2021jla}.
Similarly it is equally appealing to study second order gravitational
waves and non-gaussianities through either a QCD axion or a curvaton
isocurvature mode as in \citep{Kawasaki2021,Kawasaki:2008pa,Ando2018,BUGAEV_2013,Hikage2012,Inomata2021,Inomata2019,Langlois_2011,Beltr_n_2008,Linde_1997}. 
\begin{acknowledgments}
This work was supported in part by the Ray MacDonald Fund at UW-Madison.
We also thank Amol Upadhye for preliminary investigations into this
problem. 
\end{acknowledgments}

\appendix

\section{\label{sec:Taylor-expansion-consistency}Taylor expansion consistency
check}

In Sec.~\ref{subsec:Beyond-perturbation-theory}, we gave a solution
to $T_{1}$ using an analytic fit to cubic polynomial, utilizing the
information from the differential equation and perturbative solution.
In this section, as a check, we give an alternate derivation of $T_{1}$
assuming $T_{1}=T_{c}$ where $T_{c}$ satisfies 
\begin{equation}
\phi_{+}(T_{c})=\phi_{-}(T_{c})=rF\label{eq:tccond}
\end{equation}
where 
\begin{equation}
0.1\ll r<1.\label{eq:rrangedef}
\end{equation}
We restrict ourselves to solving the case of the resonant case (see
Sec.~\ref{subsec:Resonant-scenarios}) in which 
\begin{equation}
\cos\left[\omega T_{c}-\varphi\right]\ll1
\end{equation}
(see Eq.~(\ref{eq:tanphi}) for definition of $\varphi$ and Eq.~(\ref{eq:zerothphiplus})
for the initial conditions on $\phi_{\pm}$). The reason why we dwell
on the accuracy of the value of $T_{c}$ is because the numerical
results in the resonant situations are very sensitive to the value
of $T_{c}$ due to the large $\dot{\phi}_{+}(T_{c})\sim O(F^{2})\gg H\phi_{+}(T_{c})\sim HF$
and the fact that the complex mode amplitude is sensitive to the time
phase of real background field $\phi_{+}(T)$. The solution method
presented here involves a combination of perturbation theory, Taylor
expansion, and successive linearization approximations. The most difficult
aspect of the computation is in estimating the errors associated with
the approximation, and it is this feature that the present section's
approach is an advantage from that of Sec.~\ref{subsec:Beyond-perturbation-theory}.
However, the formalism here is cumbersome compared to that of Sec.~\ref{subsec:Beyond-perturbation-theory},
and the differential equation solution for $\xi=(\phi_{+}\phi_{-}-F^{2})$
used below has limited parametric applicability. The results nonetheless
serve as a check on Sec.~\ref{subsec:Beyond-perturbation-theory}
and provides an error estimate.

Since the computation is long, we first give the results:

\begin{equation}
T_{c}=T_{z}-\Upsilon F^{-1}\label{eq:upstc-1}
\end{equation}
\begin{equation}
\Upsilon\approx\Upsilon_{1}+\Upsilon_{2}\lesssim1\label{eq:upsilon-1}
\end{equation}
\begin{equation}
\Upsilon_{1}\equiv-\frac{j_{1}\times\left(3\frac{F^{4}}{\phi_{+}(0)^{2}}e^{3\text{\ensuremath{T_{z}}}}\left(8F^{2}+8Fj_{1}+3j_{1}^{2}\right)\cos^{2}(\varphi)+j_{1}^{4}\omega^{2}\left(4\omega^{2}-9\right)e^{\frac{3j_{1}}{F}}\right)}{j_{1}^{2}\omega^{2}e^{\frac{3j_{1}}{F}}\left(-8F^{2}+12Fj_{1}+j_{1}^{2}\left(4\omega^{2}-9\right)\right)-\frac{1}{\phi_{+}(0)^{2}}\Upsilon_{212}}
\end{equation}
\begin{equation}
\Upsilon_{2}\equiv\frac{\Upsilon_{21}-\frac{\Upsilon_{221}\left(\Upsilon_{2221}+\Upsilon_{2222}+\Upsilon_{2223}\right)}{\Upsilon_{223}}}{\frac{\Upsilon_{231}\left(\Upsilon_{2321}+\Upsilon_{2322}+\Upsilon_{2323}\right)}{\Upsilon_{233}}+\Upsilon_{24}}\label{eq:ups2def}
\end{equation}

\begin{equation}
\Upsilon_{21}\equiv-\frac{3e^{\frac{3\text{\ensuremath{j_{1}}}}{F}}F\text{\ensuremath{j_{1}}}\omega\Upsilon_{211}\phi_{+}(0)\cos(\varphi)}{4\left(\Upsilon_{212}+e^{\frac{3\text{\ensuremath{j_{1}}}}{F}}\text{\ensuremath{j_{1}}}^{2}\phi_{+}^{2}(0)\omega^{2}\left(8F^{2}-12\text{\ensuremath{j_{1}}}F+\text{\ensuremath{j_{1}}}^{2}\left(9-4\omega^{2}\right)\right)\right)}
\end{equation}
\begin{equation}
\Upsilon_{211}\equiv\left(32F^{4}-16\text{\ensuremath{j_{1}}}F^{3}+24\text{\ensuremath{j_{1}}}^{2}\left(3-2\omega^{2}\right)F^{2}-8\text{\ensuremath{j_{1}}}^{3}\left(5\omega^{2}-9\right)F+3\text{\ensuremath{j_{1}}}^{4}\left(9-4\omega^{2}\right)\right)
\end{equation}
\begin{equation}
\Upsilon_{212}\equiv3e^{3T_{z}}\left(16F^{2}+12\text{\ensuremath{j_{1}}}F+3\text{\ensuremath{j_{1}}}^{2}\right)\cos^{2}(\varphi)F^{4}
\end{equation}
\begin{equation}
\Upsilon_{221}\equiv\cos(\varphi)\left(\Upsilon_{2211}+e^{\frac{3\text{\ensuremath{j_{1}}}}{F}}\text{\ensuremath{j_{1}}}^{2}\phi_{+}^{2}(0)\omega^{2}\left(8F^{2}-12\text{\ensuremath{j_{1}}}F+\text{\ensuremath{j_{1}}}^{2}\left(9-4\omega^{2}\right)\right)\right)
\end{equation}
\begin{equation}
\Upsilon_{2211}\equiv3e^{3T_{z}}\left(16F^{2}+12\text{\ensuremath{j_{1}}}F+3\text{\ensuremath{j_{1}}}^{2}\right)\cos^{2}(\varphi)F^{4}
\end{equation}
\begin{equation}
\Upsilon_{2221}\equiv9e^{6T_{z}}\left(8F^{2}+8\text{\ensuremath{j_{1}}}F+3\text{\ensuremath{j_{1}}}^{2}\right)^{2}\left(38F^{2}+30\text{\ensuremath{j_{1}}}F+9\text{\ensuremath{j_{1}}}^{2}\right)\cos^{4}(\varphi)F^{6}
\end{equation}
\begin{align}
\Upsilon_{2222} & \equiv6e^{3\left(\frac{\text{\ensuremath{j_{1}}}}{F}+T_{z}\right)}\text{\ensuremath{j_{1}}}^{2}\phi_{+}^{2}(0)\omega^{2}\cos^{2}(\varphi)F^{3}\left(960F^{5}+384\text{\ensuremath{j_{1}}}F^{4}-16\text{\ensuremath{j_{1}}}^{2}\left(14\omega^{2}+45\right)F^{3}\right.\nonumber \\
 & \left.-8\text{\ensuremath{j_{1}}}^{3}\left(40\omega^{2}+117\right)F^{2}-24\text{\ensuremath{j_{1}}}^{4}\left(7\omega^{2}+18\right)F-9\text{\ensuremath{j_{1}}}^{5}\left(4\omega^{2}+9\right)\right)
\end{align}
\begin{align}
\Upsilon_{2223} & \equiv2e^{\frac{6\text{\ensuremath{j_{1}}}}{F}}\text{\ensuremath{j_{1}}}^{4}\phi_{+}^{4}(0)\omega^{4}\left(192F^{4}-384\text{\ensuremath{j_{1}}}F^{3}-48\text{\ensuremath{j_{1}}}^{2}\left(2\omega^{2}-3\right)F^{2}\right.\nonumber \\
 & \left.+96\text{\ensuremath{j_{1}}}^{3}\omega^{2}F+\text{\ensuremath{j_{1}}}^{4}\left(16\omega^{4}+81\right)\right)
\end{align}
\begin{align}
\Upsilon_{223} & \equiv16\text{\ensuremath{j_{1}}}\omega\phi_{+}(0)\left(-3e^{3T_{z}}\left(12F^{2}+10\text{\ensuremath{j_{1}}}F+3\text{\ensuremath{j_{1}}}^{2}\right)\cos^{2}(\varphi)F^{3}\right.\nonumber \\
 & \left.-2e^{\frac{3\text{\ensuremath{j_{1}}}}{F}}\text{\ensuremath{j_{1}}}^{2}(2F-3\text{\ensuremath{j_{1}}})\phi_{+}^{2}(0)\omega^{2}\right)^{3}
\end{align}
\begin{equation}
\Upsilon_{231}\equiv\cos(\varphi)\left(\Upsilon_{212}+e^{\frac{3\text{\ensuremath{j_{1}}}}{F}}\text{\ensuremath{j_{1}}}^{2}\phi_{+}^{2}(0)\omega^{2}\left(8F^{2}-12\text{\ensuremath{j_{1}}}F+\text{\ensuremath{j_{1}}}^{2}\left(9-4\omega^{2}\right)\right)\right)^{2}
\end{equation}
\begin{align}
\Upsilon_{2321} & \equiv9e^{6T_{z}}\left(4224F^{6}+10688\text{\ensuremath{j_{1}}}F^{5}+12272\text{\ensuremath{j_{1}}}^{2}F^{4}+8112\text{\ensuremath{j_{1}}}^{3}F^{3}+\right.\nonumber \\
 & \left.3258\text{\ensuremath{j_{1}}}^{4}F^{2}+756\text{\ensuremath{j_{1}}}^{5}F+81\text{\ensuremath{j_{1}}}^{6}\right)\cos^{4}(\varphi)F^{6}
\end{align}
\begin{align}
\Upsilon_{2322} & \equiv12e^{3\left(\frac{\text{\ensuremath{j_{1}}}}{F}+T_{z}\right)}\text{\ensuremath{j_{1}}}^{2}\phi_{+}^{2}(0)\omega^{2}\cos^{2}(\varphi)F^{3}\left(896F^{5}+128\text{\ensuremath{j_{1}}}F^{4}-48\text{\ensuremath{j_{1}}}^{2}\left(6\omega^{2}+11\right)F^{3}\right.\nonumber \\
 & \left.-8\text{\ensuremath{j_{1}}}^{3}\left(46\omega^{2}+45\right)F^{2}-9\text{\ensuremath{j_{1}}}^{4}\left(20\omega^{2}+9\right)F-36\text{\ensuremath{j_{1}}}^{5}\omega^{2}\right)
\end{align}
\begin{align}
\Upsilon_{2323} & \equiv2e^{\frac{6\text{\ensuremath{j_{1}}}}{F}}\text{\ensuremath{j_{1}}}^{4}\phi_{+}^{4}(0)\omega^{4}\left(384F^{4}-864\text{\ensuremath{j_{1}}}F^{3}-8\text{\ensuremath{j_{1}}}^{2}\left(32\omega^{2}-81\right)F^{2}\right.\nonumber \\
 & \left.+72\text{\ensuremath{j_{1}}}^{3}\left(4\omega^{2}-3\right)F+3\text{\ensuremath{j_{1}}}^{4}\left(16\omega^{4}-24\omega^{2}+27\right)\right)
\end{align}
\begin{align}
\Upsilon_{233} & \equiv32F\text{\ensuremath{j_{1}}}^{2}\omega\phi_{+}(0)\left(3e^{3T_{z}}\left(12F^{2}+10\text{\ensuremath{j_{1}}}F+3\text{\ensuremath{j_{1}}}^{2}\right)\cos^{2}(\varphi)F^{3}\right.\nonumber \\
 & \left.+2e^{\frac{3\text{\ensuremath{j_{1}}}}{F}}\text{\ensuremath{j_{1}}}^{2}(2F-3\text{\ensuremath{j_{1}}})\phi_{+}^{2}(0)\omega^{2}\right)^{4}
\end{align}
\begin{equation}
\Upsilon_{24}\equiv\frac{e^{\frac{3\text{\ensuremath{j_{1}}}}{F}-3T_{z}}\phi_{+}(0)\omega\left(8F^{2}-12\text{\ensuremath{j_{1}}}F+\text{\ensuremath{j_{1}}}^{2}\left(9-4\omega^{2}\right)\right)\sec(\varphi)}{8F^{3}}\label{eq:ups24def}
\end{equation}
where $T_{z}$ is given by Eq.~(\ref{eq:Tz}) and 
\begin{equation}
j_{1}\approx2.
\end{equation}
This $j_{1}$ is very insensitive to the parametric details because
of the $1/6$ power in 
\begin{equation}
j_{1}\approx\left(\frac{4\Upsilon^{3}R_{1}}{\left|\frac{4(3\Upsilon-2)F}{\Upsilon(\Upsilon+2)^{3}}+\frac{18(\Upsilon-2)}{(\Upsilon+2)^{3}}\right|}\right)^{1/6}
\end{equation}
where 
\begin{equation}
R_{1}\equiv\frac{\mathcal{A}_{1}F}{2}+\frac{3}{2}\mathcal{A}_{2}+\frac{\text{1}}{4F}\mathcal{A}_{3}+\frac{81\left(4c_{-}+27\right)j_{1}^{2}}{32F^{3}}+\frac{9j_{1}\left(3\left(4c_{-}+45\right)j_{1}\tau(j_{1})+4c_{-}-4\omega^{2}+27\right)}{8F^{2}}
\end{equation}
and $\mathcal{A}_{n}$ are functions of $j_{1}$ themselves given
in Eqs.~(\ref{eq:a1}), (\ref{eq:a2}), and (\ref{eq:a3}). With
the fiducial value of $j_{1}=2$ without solving for $j_{1}$ self-consistently,
the estimated error on $j_{1}$ is around 30\% for an $O(2)$ variation
in $c_{+}$ around $2.35$. Note that $j_{1}$ here is the analog
of the $\left(2n\right)^{1/4}/\sqrt{\alpha}$ in Eq.~(\ref{eq:Tssolution}),
and $j_{1}=2$ for $c_{+}=2.35$ is consistent with taking $n=10$.
This is one of the main consistency checks of this appendix on Sec.~\ref{sec:Behavior-of-}.
The $c_{+}$ parametric dependences of $T_{c}$ values of this appendix
section for $c_{+}$ near 2.35 agree with the presentation of Sec.~\ref{sec:Behavior-of-}
providing another independent consistency check. If one wants better
accuracy, it is straight forward (but tedious) to iterate using Eqs.~(\ref{eq:upsilon-1}),
(\ref{eq:minimize}), (\ref{eq:eps12}), and (\ref{eq:E2}). One of
the most interesting aspects of this is Eq.~(\ref{eq:cancellation})
that shows that $j_{1}\approx$2 and $0.5\lesssim\Upsilon\lesssim1$
to be a generic prediction in the resonant case. These results can
also be viewed as an alternate method of computing $T_{c}$ that can
be combined with Eq.~(\ref{eq:maineq}) to evaluate the isocurvature
spectrum.

The improvement in the $\phi_{-}(T_{c})$ solution can be seen by
comparing the solid line and long dashed in line in Fig.~\ref{fig:Combined-error-for}.
In the parametric case of 
\begin{equation}
\{c_{+}=2.35,c_{-}=0.5,F=20.2,\epsilon_{0}=0,\phi_{+}(0)=3.32\times10^{8}\}\label{eq:fiduci}
\end{equation}
the agreement with numerics is about 6\% in $\Upsilon\approx0.6$
(or equivalently about $0.06\times(0.6/F)/T_{c}\approx0.02\%$ in
$T_{c}\approx9.248$ which illustrates a very high precision in $T_{c}$
is required to get the $\phi_{-}(T_{c})$ to be accurate to 6\%).
For the more general case, we estimate an error for $\Upsilon$ of
less than about 35\% assuming that the error in the prediction for
$\phi_{-}$ dominates. The $\Upsilon_{1}/\Upsilon_{2}$ ratio for
this case of Eq.~(\ref{eq:fiduci}) is about 5. The reason why the
$\phi_{-}(T_{c})$ computation is very sensitive to $T_{c}$ is because
small changes $\Delta T_{c}$ leads to large changes $\Delta\phi_{-}(T_{c})$
since Eq.~(\ref{eq:tccond}) implies 
\begin{equation}
\Delta\phi_{-}(T_{c})\approx\left[\partial_{T}\phi_{+}(T_{c})-\partial_{T}\phi_{-}(T_{c})\right]\Delta T_{c}
\end{equation}
where $\partial_{T}\phi_{+}(T_{c})\gg\phi_{-}$.

In the rest of this section, we derive these results. Readers not
interested in the details can skip most of the rest of this section.

\subsection{$\phi_{\pm}$ behavior in resonant scenarios}

Here, we construct the $\phi_{\pm}$ solution in the region near $T_{c}$,
where the perturbative expansion Eqs.~(\ref{eq:phiplamexp}) and
(\ref{eq:phimlamexp}) break down. The tools we will use to construct
this solution are 1) different derivative approximation $\partial_{T}^{n}\phi_{\pm}(T)\approx\partial_{T}^{n}\phi_{\pm}^{(0)}(T)$
for different $n$ break down at different times $T$; 2) an expansion
of a different differential equation of composite operators that restricts
the functional space of $\phi_{\pm}$ about a special point where
$\dot{\phi}_{+}=0$. Note that the Taylor expansion method of 1) is
non-perturbative although very limited in its time-range extension
of analytic computation. For sudden transitions that is being studied
here, even this limited method yields nearly an order of magnitude
improvement in accuracy in the estimate.

\subsubsection{Region $[T_{c}-j_{1}/F,T_{c}]$}

As noted when discussing $\phi_{-}^{(1)}$ in Eq.~(\ref{eq:perturbative}),
the $\phi_{-}^{(0)}$ solution becomes a bad solution exponentially
fast near $T=T_{c}$. Hence, we will define below a time period $[T_{c}-j_{1}/F,T_{c}]$
just before $T_{c}$ to match the known Eq.~(\ref{eq:phimapprox})
to a finite order polynomial in this time region. The reason why the
finite order polynomial will turn out to be a better approximation
than the original perturbative solutions will be due to the fact that
different Taylor expansion derivative approximation $\partial_{T}^{n}\phi_{\pm}(T_{c}-j_{1}/F)\approx\partial_{T}^{n}\phi_{\pm}^{(0)}(T_{c}-j_{1}/F)$
for different $n$ break down at different times $T$. We will choose
$j_{1}$ from the condition that the finite order polynomial and $\partial_{T}^{n}\phi_{\pm}(T_{c}-j_{1}/F)\approx\partial_{T}^{n}\phi_{\pm}^{(0)}(T_{c}-j_{1}/F)$
be a good approximation at the same time.

Start by a quadratic Taylor expansion of $\phi_{-}(T)$ about $T=T_{c}$
\begin{equation}
\phi_{-}(T)=\phi_{-}\left(T_{c}-\frac{j_{1}}{F}\right)+\dot{\phi}_{-}\left(T_{c}-\frac{j_{1}}{F}\right)\left(T-T_{c}+\frac{j_{1}}{F}\right)+\frac{1}{2}\ddot{\phi}_{-}\left(T_{c}-\frac{j_{1}}{F}\right)\left(T-T_{c}+\frac{j_{1}}{F}\right)^{2}+....
\end{equation}
We truncate this at the quadratic order and replace the coefficients
with leading perturbative solution:
\begin{align}
\phi_{-}(T) & \approx\phi_{-}^{(0)}\left(T_{c}-\frac{j_{1}}{F}\right)+\dot{\phi}_{-}^{(0)}\left(T_{c}-\frac{j_{1}}{F}\right)\left(T-T_{c}+\frac{j_{1}}{F}\right)+\frac{1}{2}\ddot{\phi}_{-}^{(0)}\left(T_{c}-\frac{j_{1}}{F}\right)\left(T-T_{c}+\frac{j_{1}}{F}\right)^{2}\nonumber \\
 & +\mathcal{E}_{1}(T)+\mathcal{E}_{2}(T)\label{eq:linear}
\end{align}

\noindent where the error estimate $\mathcal{E}_{1}$ is for the error
incurred in matching the Taylor expansion coefficients to $\phi_{-}^{(0)}$
derivatives and $\mathcal{E}_{2}$ is the error incurred for the quadratic
Taylor expansion truncation to the exact solution. It is important
to keep in mind that the left hand side of Eq.~(\ref{eq:linear})
is not the approximate $\phi_{-}^{(0)}(T)$ but meant to be the \textbf{exact
solution} that is valid even at $T_{c}$. If one forgets that, then
this equation seems like an approximation of $\phi_{-}^{(0)}(T)$
as a quadratic function instead of the exact solution \textbf{in a
small neighborhood.}

How can a Taylor expansion of $\phi_{-}^{(0)}$ do better than keeping
the original $\phi_{-}^{(0)}$ itself? After all, why stop at just
quadratic order in $\phi_{-}^{(0)}$ Taylor expansion if one can get
higher derivatives using $\phi_{-}^{(0)}?$ The answer is that each
successive derivative Taylor expansion coefficient evaluated at $T_{c}-j_{1}/F$
becomes an increasingly poorer approximation of the exact solution's
derivative $\partial_{T}^{n}\phi_{-}\neq\partial_{T}^{n}\phi_{-}^{(0)}$.
We will demonstrate this explicitly.

In matching the exact solution to $\phi_{-}^{(0)}$ at $T=T_{c}-j_{1}/F$,
the error incurred for the zeroth order Taylor expansion can be estimated
using the perturbative solution Eq.~(\ref{eq:perturbative}) since
at $T_{c}-j_{1}/F$, the $\lambda$ perturbation of Eq.~(\ref{eq:perturbative})
is still valid 
\begin{align}
\mathcal{E}_{10} & =\left|\frac{\phi_{-}^{(1)}}{\phi_{-}^{(0)}}\right|_{T=T_{c}-\frac{j_{1}}{F}}\\
 & =\left\{ \frac{\frac{27}{4}+c_{-}+\omega^{2}\sec^{2}\left[\omega T-\varphi\right]+\omega\tan\left[\omega T-\varphi\right]\left(6+\omega\tan\left[\omega T-\varphi\right]\right)}{\left[\phi_{+}^{(0)}(T)\right]^{2}}\right\} _{T=T_{c}-\frac{j_{1}}{F}}
\end{align}

\begin{align}
\mathcal{E}_{11} & =\left|\frac{\partial_{T}\phi_{-}^{(1)}}{\partial_{T}\phi_{-}^{(0)}}\right|_{T=T_{c}-\frac{j_{1}}{F}}\\
 & =\left\{ \frac{81+12c_{-}+4\omega\left(10\omega\sec^{2}\left[\omega T-\varphi\right]+18\tan\left[\omega T-\varphi\right]-3\omega\right)}{4\left[\phi_{+}^{(0)}(T)\right]^{2}}\right\} _{T=T_{c}-\frac{j_{1}}{F}}
\end{align}
\begin{align}
\mathcal{E}_{12} & =\left|\frac{\partial_{T}^{2}\phi_{-}^{(1)}}{\partial_{T}^{2}\phi_{-}^{(0)}}\right|_{T=T_{c}-\frac{j_{1}}{F}}\\
 & =\left\{ \frac{1}{4\left[\phi_{+}^{(0)}(T)\right]^{2}}\left[\frac{112\omega^{4}\sec^{4}(\omega T-\varphi)+4\mathcal{T}_{1}(T)+4\mathcal{T}_{2}(T)}{4\omega^{2}\sec^{2}(\omega T-\varphi)+\left(3+2\omega\tan(\omega T-\varphi)\right)^{2}}\right]\right\} _{T=T_{c}-\frac{j_{1}}{F}}
\end{align}
\begin{equation}
\mathcal{T}_{1}(T)\equiv9[3+2\omega\tan(\omega T-\varphi)]^{2}\left[\frac{27}{4}+c_{-}+\left(6+\omega\tan(\omega T-\varphi)\right)\omega\tan(\omega T-\varphi)\right]
\end{equation}
\begin{equation}
\mathcal{T}_{2}(T)\equiv2\omega^{2}\sec^{2}(\omega T-\varphi)\left[189+6c_{-}+2\omega\tan(\omega T-\varphi)\left(129+44\omega\tan(\omega T-\varphi)\right)\right]
\end{equation}
where 
\begin{equation}
\phi_{-}\left(T_{c}-\frac{j_{1}}{F}\right)=\phi_{-}^{(0)}\left(T_{c}-\frac{j_{1}}{F}\right)\left(1+\mathcal{E}_{10}\right)
\end{equation}
\begin{equation}
\partial_{T}\phi_{-}\left(T_{c}-\frac{j_{1}}{F}\right)=\partial_{T}\phi_{-}^{(0)}\left(T_{c}-\frac{j_{1}}{F}\right)\left(1+\mathcal{E}_{11}\right)
\end{equation}
\begin{equation}
\partial_{T}^{2}\phi_{-}\left(T_{c}-\frac{j_{1}}{F}\right)=\partial_{T}^{2}\phi_{-}^{(0)}\left(T_{c}-\frac{j_{1}}{F}\right)\left(1+\mathcal{E}_{12}\right).
\end{equation}

\begin{figure}
\begin{centering}
\includegraphics[scale=0.8]{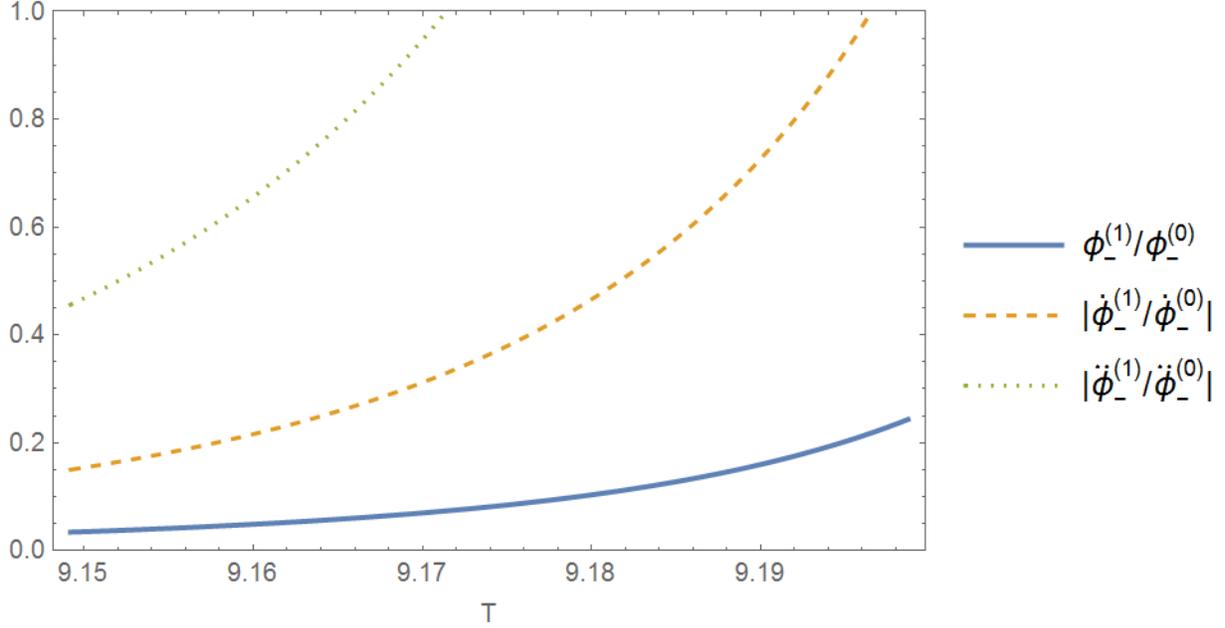} 
\par\end{centering}
\caption{\label{fig:Uncertainty-in-the}Uncertainty in the Taylor expansion
coefficients as a function of matching time $T$ for the same parameters
as in Fig.~\ref{fig:Numerical-background-solution} for which $T_{c}=9.248$
(the rightmost $T$ is about $T_{c}-1/F$ and the leftmost $T$ is
about $T_{c}-2/F$). Note that as expected, the higher derivatives
become non-perturbative faster as $T$ approaches $T_{c}$, which
is consistent with the notion that truncated Taylor expansions are
better approximations over a longer time period as $T_{c}$ is approached.
This shows that we want to match the Taylor expansion coefficients
to the zeroth order perturbative solution at an earlier time as much
as possible.}
\end{figure}

In the resonant cases in which $j_{1}/F$ is not large enough to destroy
the small cosine approximation $\cos(\omega T_{c}-\varphi)\approx\omega\left(T_{z}-T_{c}\right)$,
we can use the following relationship: 
\begin{align}
\cos^{2}(\omega(T_{c}-\frac{j_{1}}{F})-\varphi) & \approx\omega^{2}\left(\frac{j_{1}+\Upsilon}{F}\right)^{2}\ll1
\end{align}
where we have defined 
\begin{equation}
\Upsilon\equiv(T_{z}-T_{c})F.
\end{equation}
This can be used to rewrite these errors as 
\begin{align}
\mathcal{E}_{10} & =\frac{e^{-3j_{1}/F}}{r_{1}^{2}\left(j_{1}\Upsilon^{-1}+1\right)^{2}}\left[\frac{\frac{27}{4}+c_{-}}{F^{2}}+\frac{1}{\left(j_{1}+\Upsilon\right)^{2}}+\frac{\sqrt{1-\omega^{2}\left(\frac{j_{1}+\Upsilon}{F}\right)^{2}}}{j_{1}+\Upsilon}\left(\frac{6}{F}+\frac{\sqrt{1-\omega^{2}\left(\frac{j_{1}+\Upsilon}{F}\right)^{2}}}{j_{1}+\Upsilon}\right)\right]
\end{align}
where 
\begin{equation}
r_{1}F\equiv\phi_{+}(0)e^{-3T_{c}/2}\sec(\varphi)\omega\Upsilon/F\label{eq:r1def}
\end{equation}
has been defined to suggest the appropriate scale to understand this
expression in the resonant case. Note that although it looks dimensionally
wrong, it is actually consistent since we have divided out the $H$
scale here. Our final value of computed $\Upsilon$ will determine
$r_{1}$. Similarly, the error for the higher derivative coefficients
are
\begin{align}
\mathcal{E}_{11} & =\frac{e^{-3j_{1}/F}}{r_{1}^{2}\left(j_{1}\Upsilon^{-1}+1\right)^{2}}\left[\frac{\frac{81}{4}+3c_{-}}{F^{2}}+\frac{10}{\left(j_{1}+\Upsilon\right)^{2}}+\frac{18}{F}\frac{\sqrt{1-\omega^{2}\left(\frac{j_{1}+\Upsilon}{F}\right)^{2}}}{j_{1}+\Upsilon}-\frac{3\omega^{2}}{F^{2}}\right]
\end{align}

\begin{align}
\mathcal{E}_{12} & =\frac{e^{-3j_{1}/F}}{r_{1}^{2}\left(j_{1}\Upsilon^{-1}+1\right)^{2}}\left[\frac{\frac{7}{\left(j_{1}+\Upsilon\right)^{2}}+\left(j_{1}+\Upsilon\right)^{2}\frac{\mathcal{T}_{1}(T_{c}-\frac{j_{1}}{F})+\mathcal{T}_{2}(T_{c}-\frac{j_{1}}{F})}{4F^{4}}}{1+\frac{\left(j_{1}+\Upsilon\right)^{2}}{4F^{2}}\left(3+2F\frac{\sqrt{1-\omega^{2}\left(\frac{j_{1}+\Upsilon}{F}\right)^{2}}}{j_{1}+\Upsilon}\right)^{2}}\right]
\end{align}
\begin{align}
\mathcal{T}_{1}(T_{c}-\frac{j_{1}}{F}) & \approx9\left[3+2\frac{\sqrt{1-\omega^{2}\left(\frac{j_{1}+\Upsilon}{F}\right)^{2}}}{\left(\frac{j_{1}+\Upsilon}{F}\right)}\right]^{2}\times\nonumber \\
 & \left[\frac{27}{4}+c_{-}+\frac{\sqrt{1-\omega^{2}\left(\frac{j_{1}+\Upsilon}{F}\right)^{2}}}{\left(\frac{j_{1}+\Upsilon}{F}\right)}\left(6+\frac{\sqrt{1-\omega^{2}\left(\frac{j_{1}+\Upsilon}{F}\right)^{2}}}{\left(\frac{j_{1}+\Upsilon}{F}\right)}\right)\right]
\end{align}
\begin{equation}
\mathcal{T}_{2}(T_{c}-\frac{j_{1}}{F})\approx2\frac{1}{\left(\frac{j_{1}+\Upsilon}{F}\right)^{2}}\left[189+6c_{-}+2\frac{\sqrt{1-\omega^{2}\left(\frac{j_{1}+\Upsilon}{F}\right)^{2}}}{\left(\frac{j_{1}+\Upsilon}{F}\right)}\left(129+44\frac{\sqrt{1-\omega^{2}\left(\frac{j_{1}+\Upsilon}{F}\right)^{2}}}{\left(\frac{j_{1}+\Upsilon}{F}\right)}\right)\right].
\end{equation}
One can see from Fig.~\ref{fig:Uncertainty-in-the} (in which we
plot $\mathcal{E}_{10},\mathcal{E}_{11},\mathcal{E}_{12}$) how the
higher order Taylor expansion coefficients become more uncertain at
earlier before reaching $T_{c}$. In this numerical example case,
we see that to keep the second order Taylor expansion coefficient
accurate to about 20\%, we need Taylor expand at about $T_{c}-2/F\approx9.15$.
The actual error $\mathcal{E}_{1}$ at $T_{c}$ can be better than
20\% depending on which Taylor expansion term contributes the most.

Hence, we conclude based on coefficient errors alone 
\begin{equation}
\mathcal{E}_{1}(T_{c})\approx\left|\phi_{-}^{(1)}\left(T_{c}-\frac{j_{1}}{F}\right)+\dot{\phi}_{-}^{(1)}\left(T_{c}-\frac{j_{1}}{F}\right)\frac{j_{1}}{F}+\frac{1}{2}\ddot{\phi}_{-}^{(1)}\left(T_{c}-\frac{j_{1}}{F}\right)\left(\frac{j_{1}}{F}\right)^{2}\right|
\end{equation}
\begin{align}
\phi_{-}^{(1)}\left(T_{c}-\frac{j_{1}}{F}\right) & =\frac{e^{\frac{9}{2}T_{c}}F^{2}\cos^{3}\varphi\sec^{3}(\omega T-\varphi)}{\phi_{+}^{3}(0)}\times\nonumber \\
 & \left[\frac{27}{4}+c_{-}+\omega^{2}\sec^{2}\left[\omega T-\varphi\right]+\omega\tan\left[\omega T-\varphi\right]\left(6+\omega\tan\left[\omega T-\varphi\right]\right)\right]_{T_{c}-\frac{j_{1}}{F}}
\end{align}
\begin{align}
\dot{\phi}_{-}^{(1)}\left(T_{c}-\frac{j_{1}}{F}\right) & =\frac{e^{\frac{9}{2}T_{c}}F^{2}\cos^{3}\varphi\sec^{3}(\omega T-\varphi)}{\phi_{+}^{3}(0)}\left(\frac{3}{2}+\omega\tan\left[\omega T-\varphi\right]\right)\times\nonumber \\
 & \left[\frac{81}{4}+3c_{-}+\omega\left(10\omega\sec^{2}\left[\omega T-\varphi\right]+18\tan\left[\omega T-\varphi\right]-3\omega\right)\right]_{T_{c}-\frac{j_{1}}{F}}
\end{align}
\begin{equation}
\ddot{\phi}_{-}^{(1)}\left(T_{c}-\frac{j_{1}}{F}\right)=\frac{e^{\frac{9}{2}T_{c}}F^{2}\cos^{3}\varphi\sec^{3}(\omega T-\varphi)}{\phi_{+}^{3}(0)}\left[7\omega^{4}\sec^{4}(\omega T-\varphi)+\tilde{\mathcal{T}}_{1}+\tilde{\mathcal{T}_{2}}\right]_{T_{c}-\frac{j_{1}}{F}}
\end{equation}
\begin{equation}
\tilde{\mathcal{T}}_{1}=\frac{\omega^{2}}{2}\sec^{2}\left(\omega T-\varphi\right)\left[189+6c_{-}+2\omega\tan(\omega T-\varphi)\left(129+44\omega\tan(\omega T-\varphi\right)\right]_{T_{c}-\frac{j_{1}}{F}}
\end{equation}
\begin{equation}
\tilde{\mathcal{T}}_{2}=\frac{9}{4}\left[3+2\omega\tan(\omega T-\varphi)\right]^{2}\left[\frac{27}{4}+c_{-}+\omega\tan(\omega T-\varphi)\left(6+\tan(\omega T-\varphi)\right)\right]_{T_{c}-\frac{j_{1}}{F}}.
\end{equation}
In the resonant case, we can expand as before about $T=T_{c}$: 
\begin{equation}
\mathcal{E}_{1}(T_{c})\approx\frac{\Upsilon^{3}e^{-\frac{9j_{1}}{2F}}\left(\frac{\mathcal{A}_{1}F}{2}+\frac{3}{2}\mathcal{A}_{2}+\frac{\text{1}}{4F}\mathcal{A}_{3}+\frac{81\left(4c_{-}+27\right)j_{1}^{2}}{32F^{3}}+\frac{9j_{1}\left(3\left(4c_{-}+45\right)j_{1}\tau(j_{1})+4c_{-}-4\omega^{2}+27\right)}{8F^{2}}\right)}{r_{1}^{3}\left(j_{1}+\Upsilon\right){}^{3}}\label{eq:E1}
\end{equation}
\begin{equation}
\mathcal{A}_{1}(j_{1})\equiv\tau^{2}(j_{1})\left(2+9\text{\ensuremath{j_{1}}}^{2}\tau^{2}(j_{1})\right)+\frac{7\text{\ensuremath{j_{1}}}^{2}}{(\text{\ensuremath{j_{1}}}+\Upsilon)^{4}}+\frac{2+4\text{\ensuremath{j_{1}}}\text{\ensuremath{\tau}(\ensuremath{j_{1}})}(5+11\text{\ensuremath{j_{1}}}\text{\ensuremath{\tau}(\ensuremath{j_{1}})})}{(\text{\ensuremath{j_{1}}}+\Upsilon)^{2}}\label{eq:a1}
\end{equation}
\begin{equation}
\mathcal{A}_{2}(j_{1})\equiv\text{\ensuremath{\tau}(\ensuremath{j_{1}})}(4+3\text{\ensuremath{j_{1}}}\text{\ensuremath{\tau}(\ensuremath{j_{1}})}(4+9\text{\ensuremath{j_{1}}}\text{\ensuremath{\tau}(\ensuremath{j_{1}})}))+\frac{\text{\ensuremath{j_{1}}}(10+43\text{\ensuremath{j_{1}}}\text{\ensuremath{\tau}(\ensuremath{j_{1}})})}{(\text{\ensuremath{j_{1}}}+\Upsilon)^{2}}\label{eq:a2}
\end{equation}
\begin{align}
\mathcal{A}_{3}(j_{1}) & \equiv27(1+7\text{\ensuremath{j_{1}}}\text{\ensuremath{\tau}(\ensuremath{j_{1}})})+4c_{-}+18\text{\ensuremath{j_{1}}}^{2}\tau^{2}(j_{1})\left(27+c_{-}\right)+12\text{\ensuremath{j_{1}}}\text{\ensuremath{\tau}(\ensuremath{j_{1}})}\left(c_{-}-\omega^{2}\right)\nonumber \\
 & +\frac{3\text{\ensuremath{j_{1}}}^{2}\left(63+2c_{-}\right)}{(\text{\ensuremath{j_{1}}}+\Upsilon)^{2}}\label{eq:a3}
\end{align}
\begin{equation}
\tau(j_{1})\equiv\frac{\sqrt{1-\frac{\omega^{2}\left(j_{1}+\Upsilon\right){}^{2}}{F^{2}}}}{j_{1}+\Upsilon}\approx\frac{1}{j_{1}+\Upsilon}
\end{equation}
where one should keep in mind that $\Upsilon$ and $r_{1}$ are of
order unity in the resonant scenarios where the expansion involving
$\Upsilon$ about the zero crossing of $\cos(\omega T-\varphi)$ has
been made. Since we are evaluating at $T_{c}-j_{1}/F$ which is farther
away from the zero crossing of $\cos(\omega T-\varphi)$ this expression
is only about 25\% accurate. Typically, the $\mathcal{A}_{1}F/2$
term dominates and this large coefficient pushes the error towards
larger than unity at $T_{c}$. Therefore $j_{1}$ needs to be made
as large as possible to induce the $\left(j_{1}+\Upsilon\right){}^{-3}$
in Eq.~(\ref{eq:E1}) to reduce $\mathcal{E}_{1}$. This is a motivation
for having a quadratic Taylor expansion compared to a linear Taylor
expansion since generically a Taylor expansion has a larger degree
of accuracy for higher order polynomials.\footnote{One might ask, why not then go to even higher orders in Taylor expansion?
That is because of Fig.~\ref{fig:Uncertainty-in-the} which tells
us that the higher order polynomial coefficients are not approximated
well for a given expansion point $T=T_{c}-j_{1}/F$. To rigorously
optimize, one would have to minimize the error in the $(n,j_{1})$
plane where $n$ is the degree of polynomial with which one is expanding.
However, we will be content with setting $n=2$ and maximizing $j_{1}$
to approximately minimize the error.}

Let's discuss the competing error $\mathcal{E}_{2}$ incurred from
Taylor expanding the exact solution to quadratic order (which is always
possible for any analytic solution in a sufficiently small neighborhood):
\begin{equation}
\mathcal{E}_{2}(T)=\left|\frac{1}{6}\underset{u\in[T_{c}-\frac{1}{F},T_{c}]}{\max}\partial_{T}^{3}\phi_{-}\left(u\right)\left(T-T_{c}+\frac{j_{1}}{F}\right)^{3}\right|.
\end{equation}
According to the equation of motion for $\phi_{-}$ (Eq.~(\ref{eq:backgroundeom})):
\begin{align}
\partial_{T}^{3}\phi_{-}\left(u\right) & =-\left[-3\left(3\dot{\phi}_{-}(u)+c_{-}\phi_{-}+(\phi_{+}\phi_{-}-F^{2})\phi_{+}\right)\right.\nonumber \\
 & \left.+c_{-}\dot{\phi}_{-}(u)+(2\phi_{+}(u)\phi_{-}(u)-F^{2})\dot{\phi}_{+}(u)+\phi_{+}^{2}(u)\dot{\phi}_{-}(u)\right].\label{eq:thirdderiveq}
\end{align}
Let's first see why it is a bit delicate to estimate the RHS. We know
that the largest contribution to the RHS of Eq.~(\ref{eq:thirdderiveq})
are from the the potential terms near its maximum since $F^{2}\dot{\phi}_{+}\gtrsim F^{4}$
: 
\begin{equation}
\partial_{T}^{3}\phi_{-}(u)\sim-\left[(2\phi_{+}(u)\phi_{-}(u)-F^{2})\dot{\phi}_{+}(u)+\phi_{+}^{2}(u)\dot{\phi}_{-}(u)\right].\label{eq:potonly}
\end{equation}
There is a partial cancellation in this expression since at least
at $T_{c}-j_{1}/F$, we have by construction 
\begin{equation}
\phi_{+}\phi_{-}\sim F^{2}\rightarrow\dot{\phi}_{+}\phi_{-}\sim-\dot{\phi}_{-}\phi_{+}
\end{equation}
making 
\begin{align}
\partial_{u}^{3}\phi_{-}(u) & \approx-\left[(\phi_{+}(u)\phi_{-}(u)-F^{2})\dot{\phi}_{+}(u)\right].
\end{align}
This cancellation fails more and more as $\phi_{+}\phi_{-}$ becomes
smaller and smaller compared to $F^{2}$ as $u$ approaches $T_{c}$.
This means we expect $\partial_{u}^{3}\phi_{-}(u)$ to be maximized
near $T_{c}$. On the other hand, since $(2\phi_{+}(u)\phi_{-}(u)-F^{2})\dot{\phi}_{+}(u)<0$
near $T_{c}-j_{1}/F$ (since that is where $\phi_{+}\phi_{-}\sim F^{2})$
whereas $\phi_{+}^{2}(u)\dot{\phi}_{-}(u)>0$ in this region, there
is a cancellation which could increase near $T_{c}$, making the exact
location of the maximum of $\partial_{u}^{3}\phi_{-}$ uncertain.
Nonetheless, as long as $T_{c}$ does not exactly represent the zero
of $\partial_{u}^{3}\phi_{-}(u)$, we expect from these arguments
that 
\begin{equation}
|\partial_{u}^{3}\phi_{-}(T_{c})|\sim O\left(\underset{u\in[T_{c}-\frac{1}{F},T_{c}]}{\max}\partial_{T}^{3}\phi_{-}\left(u\right)\right)
\end{equation}
which is what we will evaluate now.

Since the RHS of Eq.~(\ref{eq:potonly}) only involves first derivatives
and lower, we can use the assumed solution to evaluate these derivatives:
\begin{equation}
\phi_{-}(T)\approx\phi_{-}^{(0)}\left(T_{c}-\frac{j_{1}}{F}\right)+\dot{\phi}_{-}^{(0)}\left(T_{c}-\frac{j_{1}}{F}\right)\left(T-T_{c}+\frac{j_{1}}{F}\right)+\frac{1}{2}\ddot{\phi}_{-}^{(0)}\left(T_{c}-\frac{j_{1}}{F}\right)\left(T-T_{c}+\frac{j_{1}}{F}\right)^{2}
\end{equation}
which when evaluated at $T_{c}$ is 
\begin{equation}
\phi_{-}(T_{c})\approx\phi_{-}^{(0)}\left(T_{c}-\frac{j_{1}}{F}\right)+\dot{\phi}_{-}^{(0)}\left(T_{c}-\frac{j_{1}}{F}\right)\left(\frac{j_{1}}{F}\right)+\frac{1}{2}\ddot{\phi}_{-}^{(0)}\left(T_{c}-\frac{j_{1}}{F}\right)\left(\frac{j_{1}}{F}\right)^{2}.
\end{equation}
The derivative can also be evaluated at $T_{c}$: 
\begin{equation}
\dot{\phi}_{-}(T_{c})\approx\dot{\phi}_{-}^{(0)}\left(T_{c}-\frac{j_{1}}{F}\right)+\frac{1}{2}\ddot{\phi}_{-}^{(0)}\left(T_{c}-\frac{j_{1}}{F}\right)\left(\frac{j_{1}}{F}\right)^{2}.
\end{equation}
After using Eq.~(\ref{eq:r1def}), the largest number in $\partial_{T}^{3}\phi_{-}$
is $F$. Hence, we expand in powers of $F$ to obtain
\begin{equation}
\partial_{T}^{3}\phi_{-}(T_{c})\approx\frac{4r_{1}(3\Upsilon-2)F^{4}}{\Upsilon(\Upsilon+2)^{3}}+\frac{18r_{1}(\Upsilon-2)F^{3}}{(\Upsilon+2)^{3}}
\end{equation}
where $r_{1}$ is defined in Eq.~(\ref{eq:r1def}) and $\Upsilon$
is defined as 
\begin{equation}
\Upsilon\equiv\left(T_{z}-T_{c}\right)F\sim O(1)\label{eq:upsilondef}
\end{equation}
where $T_{z}$ is given by Eq.~(\ref{eq:Tz}). Note the mass matrix
becomes strongly off-diagonal at $T_{z}$: i.e.~ 
\begin{equation}
\frac{F^{2}}{\phi_{+}(0)e^{-\frac{3}{2}T_{z}}\sec(\varphi)}\approx O(F).\label{eq:envelopeeq}
\end{equation}
Because $(\Upsilon+2)^{3}$ can easily be of order $F$ (because of
the cubic power) and because $18$ can be of order $F$, these two
terms can compete. Also, $3\Upsilon-2$ can easily be negative. Hence
we arrive at the Taylor expansion error estimate 
\begin{equation}
\mathcal{E}_{2}(T_{c})=\left|\frac{1}{6}\left(\frac{4r_{1}(3\Upsilon-2)F}{\Upsilon(\Upsilon+2)^{3}}+\frac{18r_{1}(\Upsilon-2)}{(\Upsilon+2)^{3}}\right)j_{1}^{3}\right|\label{eq:E2}
\end{equation}
at $T_{c}$. The equivalent fractional error of Taylor expansion is
\begin{equation}
\left|\frac{\mathcal{E}_{2}(T_{c})}{\phi_{-}(T_{c})}\right|\approx\left|\frac{r_{1}}{r}\left(\frac{(2\Upsilon-\frac{4}{3})}{\Upsilon(\Upsilon+2)^{3}}+\frac{3(\Upsilon-2)}{(\Upsilon+2)^{3}F}\right)j_{1}^{3}\right|
\end{equation}
where 
\begin{equation}
\frac{r_{1}}{r}=\frac{\phi_{+}^{(0)}(T_{c})}{\phi_{+}(T_{c})}\approx1,
\end{equation}
$r$ and $r_{1}$ were defined in Eqs.~(\ref{eq:rrangedef}) and
(\ref{eq:r1def}). This error pushes the choice of $j_{1}$ towards
smaller values (i.e.~in the opposite direction of the push by $\mathcal{E}_{1}$).
To evaluate this error, we need a value of $\Upsilon$ (or equivalently
$T_{c}$) which is calculated with $j_{1}$ fixed. Hence, $\Upsilon$
itself is a function of $j_{1}$.

Ideally, we want to compute 
\begin{equation}
\frac{d}{dj_{1}}\left[\mathcal{E}_{1}(T_{c}(j_{1}))+\mathcal{E}_{2}(T_{c}(j_{1}))\right]=0\label{eq:derivativevanish}
\end{equation}
to minimize the errors. Because $\Upsilon$ itself depends on $j_{1}$,
the derivative is tedious to obtain.

\begin{figure}
\begin{centering}
\includegraphics[scale=0.4]{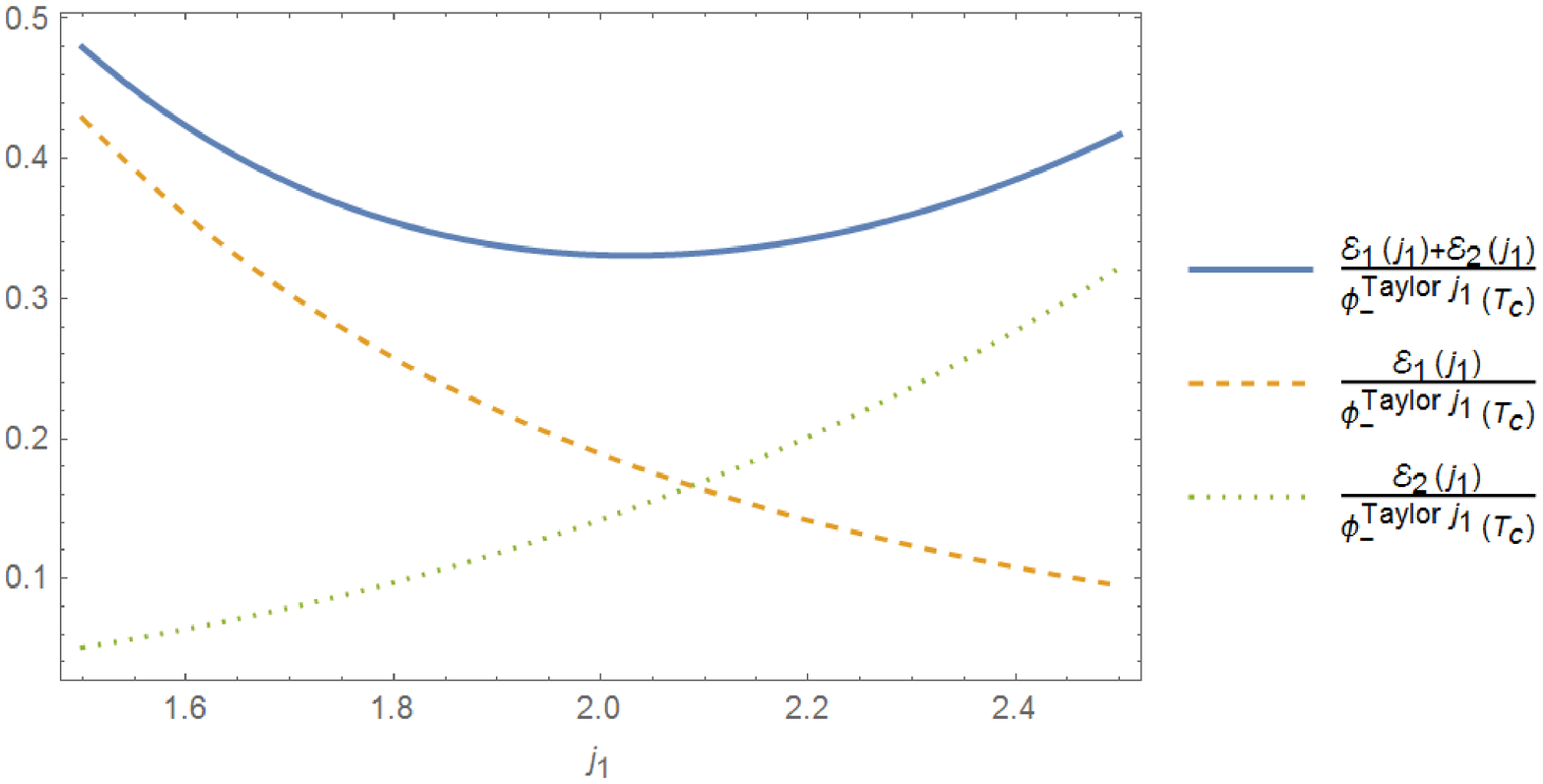}\includegraphics[scale=0.4]{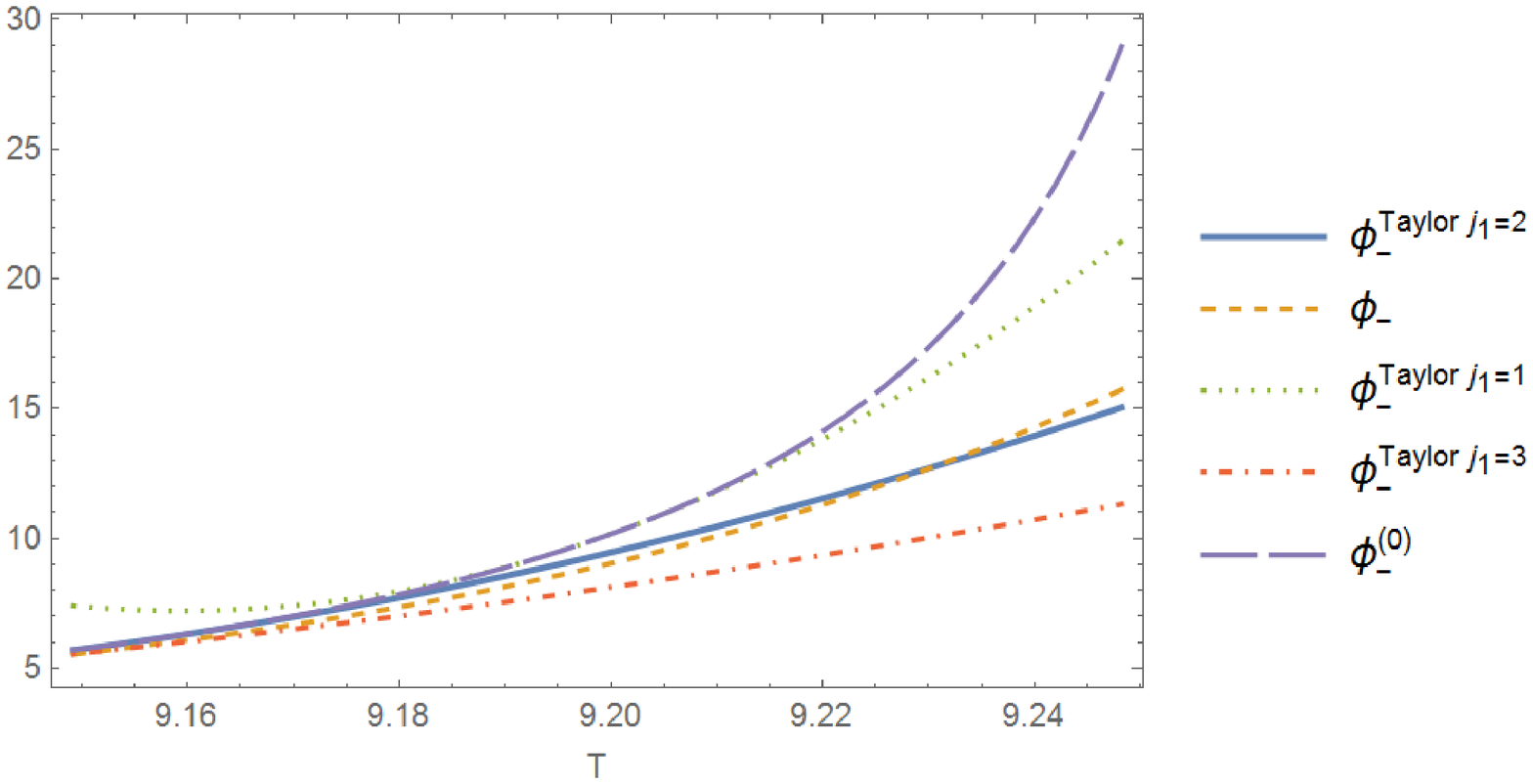} 
\par\end{centering}
\caption{\label{fig:Combined-error-for}(left) Combined error for the case
of the fiducial parameters of Fig.~\ref{fig:Numerical-background-solution}
(e.g.~$T_{c}=9.248$). The value of $\Upsilon\approx0.6$ corresponding
to the approximate location of $T_{c}$ has been fixed by hand and
not varied with $j_{1}$. The minimum error for the value of $\phi_{-}(T_{c})$
using Eq.~(\ref{eq:linear}) occurs at around $j_{1}=2$ corresponding
to a Taylor expansion at $T=T_{c}-2/F$. The denominator $\phi_{-}^{{\rm Taylor}\,j_{1}}(T_{c})$
used for this comparison plot has been made using Eq.~(\ref{eq:linear})
instead. (right) The right plot shows the approximate $\phi_{-}^{{\rm Taylor}\,j_{1}=2}$
of Eq.~(\ref{eq:linear}) matches the numerical solution $\phi_{-}$
very well compared to the approximation obtained with $j_{1}=1$.
It is clear that the solution where the Taylor expansion is done about
$j_{1}=2$ works much better than the approximation made through expanding
about $j_{1}=1$ or $j_{1}=3$ as predicted by the left plot. Also,
as expected and noted before, all the Taylor expansion approximations
in the $j_{1}\in[1,3]$ are much better than the zeroth $\lambda$
order solution $\phi_{-}^{(0)}$.}
\end{figure}

\begin{figure}
\begin{centering}
\includegraphics[scale=0.4]{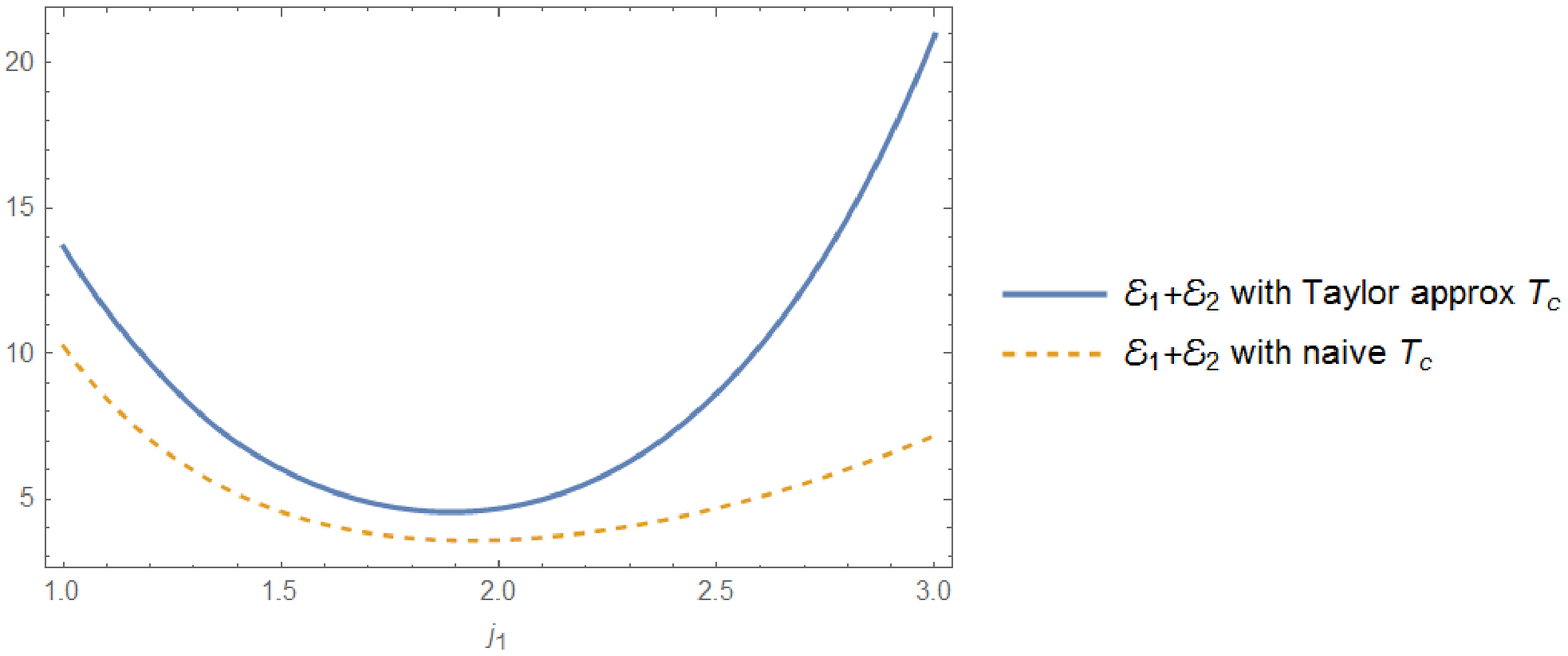}\includegraphics[scale=0.4]{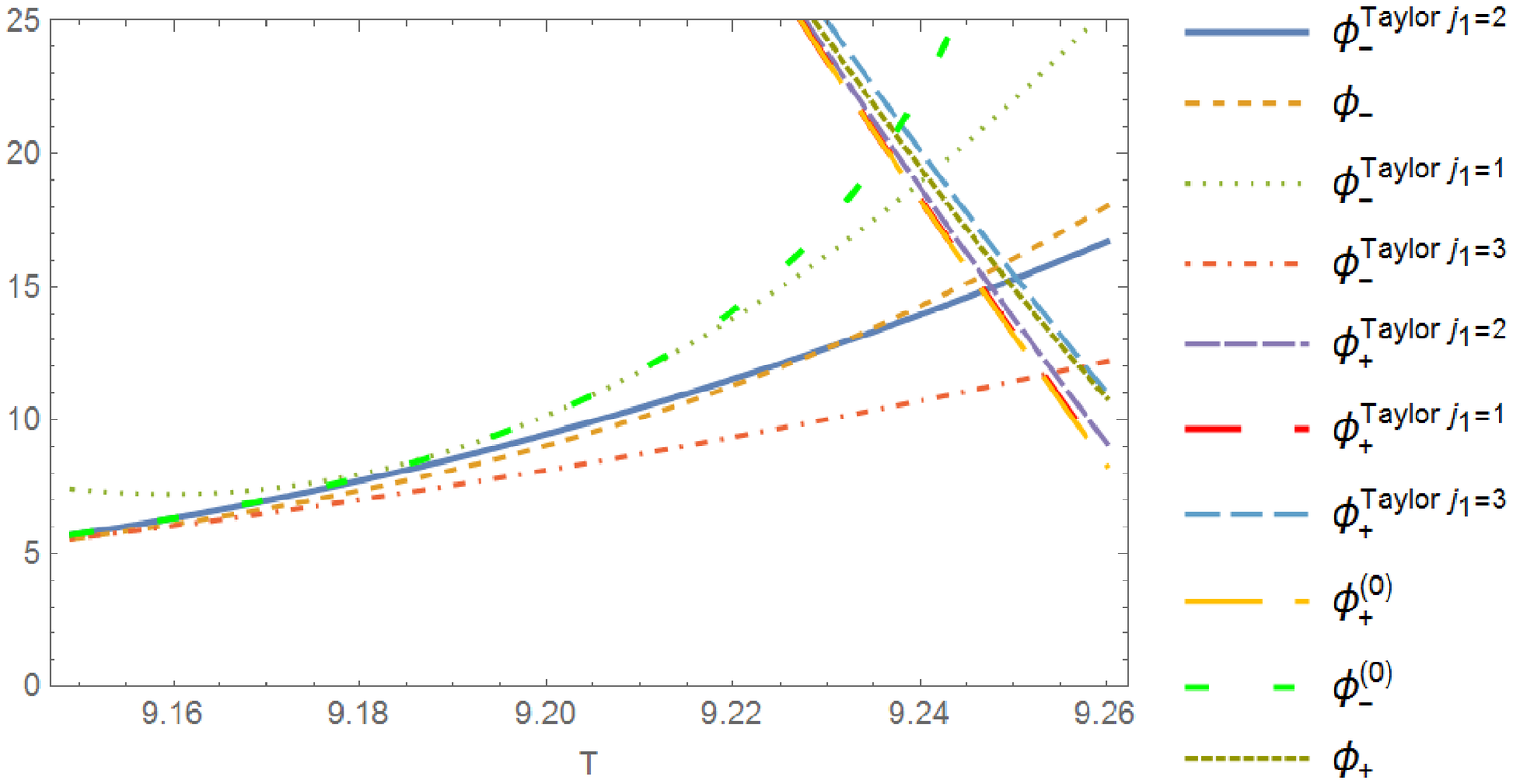} 
\par\end{centering}
\caption{\label{fig:error-with-naive} Here, we again consider the parametric
point of Fig.~\ref{fig:Numerical-background-solution}. Errors with
less accurate $T_{c}$ (the dashed curve) gives about the same $j_{1}$
location of the minimum error as the errors evaluated with the more
accurate $T_{c}$ (the solid curve). The exact definition of the more
accurate versus less accurate value of $T_{c}$ is explained in the
text. This insensitivity is expected to occur because of the steepness
of the slope of $\phi_{+}$ in the resonant case as seen the second
figure, where we put on top of Fig.~\ref{fig:Combined-error-for}
various $\phi_{+}$ approximations. Note unlike in the case of $\phi_{-}^{(0)}$,
the leading perturbative $\phi_{+}^{(0)}$ solution accurately describes
the numerical solution at $T_{c}\approx9.248$.}
\end{figure}

The combined error (in absolute values added instead of quadrature)
is plotted in Fig.~\ref{fig:Combined-error-for}.\footnote{Although we could add in quadrature to tighten the error estimates,
we here stay conservative both because we do not really know the distribution
shape of the errors and we want to keep the algebra simpler.} We see that the minimum error occurs when the Taylor expansion point
of Eq.~(\ref{eq:linear}) is $T\approx T_{c}-2/F$. The expected
error for the value of the function is at most around 35\%. However,
to get the plots (or equivalently, to use Eqs.~(\ref{eq:E1}) and
(\ref{eq:E2})), we need to compute $\Upsilon$ for a given model,
i.e. $T_{c}=T_{z}-\Upsilon F^{-1}$ using Eqs.~(\ref{eq:tccond}),
(\ref{eq:linear}), and the analogous equation for $\phi_{+}$ approximation:
\begin{equation}
\phi_{+}(T)\approx\phi_{+}^{(0)}\left(T_{c}-\frac{j_{1}}{F}\right)+\dot{\phi}_{+}^{(0)}\left(T_{c}-\frac{j_{1}}{F}\right)\left(T-T_{c}+\frac{j_{1}}{F}\right)+\frac{1}{2}\ddot{\phi}_{+}^{(0)}\left(T_{c}-\frac{j_{1}}{F}\right)\left(T-T_{c}+\frac{j_{1}}{F}\right)^{2}.
\end{equation}
However, in practice this does not need to be done. That is because
the minimum error generating $j_{1}$ is not sensitive to the exact
value of $T_{c}$. This is illustrated in Fig.~\ref{fig:error-with-naive}
where the errors are evaluated with two different estimates of $T_{c}$.
The Taylor approximation curve is generated by solving for $T_{c}$
self-consistently by solving 
\begin{equation}
\phi_{+}^{{\rm Taylor}}(T_{c})|_{j_{1}}=\phi_{-}^{{\rm Taylor}}(T_{c})|_{j_{1}}
\end{equation}
for $T_{c}=T_{c}(j_{1})$ which is now a function of $j_{1}$ and
evaluating $\mathcal{E}_{1}(j_{1},T)+\mathcal{E}_{2}(j_{1},T)$ evaluated
at $T=T_{c}(j_{1})$. The ``naive $T_{c}$'' curve is generated
by solving 
\begin{equation}
|\phi_{-}^{(0)}(T_{c}^{({\rm naive})})|=F
\end{equation}
and Eq.~(\ref{eq:envelopeeq}) is satisfied. Hence, we see $j_{1}\approx2$
gives the minimum error in the parametric case of Fig.~\ref{fig:Numerical-background-solution}.

Let's analyze the sensitivity of $j_{1}$ to the parameters more generally.
Start by rewriting Eq.~(\ref{eq:E1}) as 
\begin{equation}
\mathcal{E}_{1}(T_{c})\approx\frac{\Upsilon^{3}e^{-\frac{9j_{1}}{2F}}R_{1}}{r_{1}^{3}\left(j_{1}+\Upsilon\right){}^{3}}\label{eq:eps12}
\end{equation}
\begin{equation}
R_{1}\equiv\frac{\mathcal{A}_{1}F}{2}+\frac{3}{2}\mathcal{A}_{2}+\frac{\text{1}}{4F}\mathcal{A}_{3}+\frac{81\left(4c_{-}+27\right)j_{1}^{2}}{32F^{3}}+\frac{9j_{1}\left(3\left(4c_{-}+45\right)j_{1}\tau(j_{1})+4c_{-}-4\omega^{2}+27\right)}{8F^{2}}.\label{eq:r1}
\end{equation}
The solution to Eq.~(\ref{eq:derivativevanish}) is thus 
\begin{equation}
\left(\frac{-9}{2F}+\frac{-3(1+\partial_{j_{1}}\Upsilon)}{(j_{1}+\Upsilon)}+\frac{3\partial_{j_{1}}\Upsilon}{\Upsilon}+\frac{\partial_{j_{1}}R_{1}}{R_{1}}+\frac{\partial_{j_{1}}\Upsilon\partial_{\Upsilon}R_{1}}{R_{1}}\right)\mathcal{E}_{1}+\frac{3}{j_{1}}\mathcal{E}_{2}=0.\label{eq:minimize}
\end{equation}
To get an approximation for the solution, we approximate $R_{1}$
as a smooth polynomial in $j_{1}$, note $\frac{-9}{2F}$ is unimportant,
and assume that we are close to the solution of Eq.~(\ref{eq:derivativevanish})
such that $|\partial_{j_{1}}\Upsilon/\Upsilon|\ll1$ : 
\begin{equation}
\frac{d}{dj_{1}}\left[\mathcal{E}_{1}(T_{c})+\mathcal{E}_{2}(T_{c})\right]\approx\left(\frac{-3}{(j_{1}+\Upsilon)}+\frac{O(1)}{j_{1}}\right)\mathcal{E}_{1}+\frac{3}{j_{1}}\mathcal{E}_{2}.
\end{equation}
Setting this to zero gives 
\begin{equation}
\mathcal{E}_{2}=\left(\frac{j_{1}}{(j_{1}+\Upsilon)}-\frac{O(1)}{3}\right)\mathcal{E}_{1}
\end{equation}
or equivalently 
\begin{equation}
\left|\frac{1}{6}\left(\frac{4r_{1}(3\Upsilon-2)F}{\Upsilon(\Upsilon+2)^{3}}+\frac{18r_{1}(\Upsilon-2)}{(\Upsilon+2)^{3}}\right)j_{1}^{3}\right|=\left(\frac{j_{1}}{(j_{1}+\Upsilon)}-\frac{O(1)}{3}\right)\frac{\Upsilon^{3}e^{-\frac{9j_{1}}{2F}}R_{1}}{r_{1}^{3}\left(j_{1}+\Upsilon\right){}^{3}}.
\end{equation}
Hence, for $j_{1}\gtrsim2\Upsilon$ we can approximate 
\begin{equation}
j_{1}\approx\left(\frac{4\Upsilon^{3}R_{1}}{\left|\frac{4(3\Upsilon-2)F}{\Upsilon(\Upsilon+2)^{3}}+\frac{18(\Upsilon-2)}{(\Upsilon+2)^{3}}\right|}\right)^{1/6}.\label{eq:parametric}
\end{equation}
Since $\Upsilon$ stays near unity for different $c_{+}$ and $R_{1}\sim O(F)$,
we see that this $j_{1}$ varies slowly with different parameters
of $c_{+}$ owing to the $1/6$ power dependence. Hence, we will use
the result of Fig.~\ref{fig:error-with-naive} and use 
\[
j_{1}=2
\]
for the resonant cases. This should be a good approximation to about
$30\%$ accuracy since $1/6$ power reduces $O(1)$ uncertainties
to about this level. The $j_{1}$ variable is the analog of the $\left(2n\right)^{1/4}/\sqrt{\alpha}$
in Eq.~(\ref{eq:Tssolution}), and $j_{1}=2$ for $c_{+}=2.35$ is
consistent with taking $n=10$. This appendix has thus provided a
consistency check as well as an error estimate of Sec.~\ref{sec:Behavior-of-}.

\subsection{Solving for $T_{c}$}

Although the value of $j_{1}$ that minimizes the error is not very
sensitive to the value of $T_{c}$, we see from the second figure
of Fig.~\ref{fig:error-with-naive} that $\phi_{+}(T_{c})$ is sensitive
to $T_{c}$ . We defined $T_{c}$ to be the solution Eq.~(\ref{eq:tccond}),
which is approximately 
\begin{align}
\phi_{+}^{(0)}\left(T_{c}-\frac{j_{1}}{F}\right)+\dot{\phi}_{+}^{(0)}\left(T_{c}-\frac{j_{1}}{F}\right)\left(\frac{j_{1}}{F}\right)+\frac{1}{2}\ddot{\phi}_{+}^{(0)}\left(T_{c}-\frac{j_{1}}{F}\right)\left(\frac{j_{1}}{F}\right)^{2} & =\nonumber \\
\phi_{-}^{(0)}\left(T_{c}-\frac{j_{1}}{F}\right)+\dot{\phi}_{-}^{(0)}\left(T_{c}-\frac{j_{1}}{F}\right)\left(\frac{j_{1}}{F}\right)+\frac{1}{2}\ddot{\phi}_{-}^{(0)}\left(T_{c}-\frac{j_{1}}{F}\right)\left(\frac{j_{1}}{F}\right)^{2}.\label{eq:eqtosolve}
\end{align}
This equation and Eq.~(\ref{eq:derivativevanish}) together determine
$j_{1}$ and $\Upsilon$ (where $\Upsilon$ is a parameterization
of $T_{c}$ through Eq.~(\ref{eq:upsilondef})).

Use resonant condition 
\begin{equation}
\cos\left(\omega\left(T_{c}-\frac{j_{1}}{F}\right)-\varphi\right)\approx\omega\frac{j_{1}+\Upsilon}{F}
\end{equation}
to turn the trigonometric functions in this expression into polynomials:
\begin{align}
0 & =\frac{\phi_{+}(0)\omega\sec(\varphi)e^{\frac{3(-FT_{z}+j_{1}+\Upsilon)}{F}}\left(8F^{2}\Upsilon+j_{1}\left(-12F\Upsilon-j_{1}\left(4\omega^{2}-9\right)(j_{1}+\Upsilon)\right)\right)}{8F^{3}}\nonumber \\
 & -\frac{F\cos(\varphi)\left(8F^{2}\left(3j_{1}^{2}+3j_{1}\Upsilon+\Upsilon^{2}\right)+12Fj_{1}(j_{1}+\Upsilon)(2j_{1}+\Upsilon)+9j_{1}^{2}(j_{1}+\Upsilon)^{2}\right)}{8\phi_{+}(0)\omega(j_{1}+\Upsilon)^{3}}.
\end{align}
Although this equation can be linearized successively to obtain an
accurate solution, the algebra becomes significantly simpler with
only about $O(3\Upsilon/(FT_{z}))$ loss of precision if we drop the
$\Upsilon$ dependence on arising from the exponent 
\begin{align}
e^{\frac{3(-FT_{z}+j_{1}+\Upsilon)}{F}} & =\exp\left(-3T_{z}+\frac{3j_{1}}{F}+3\Upsilon F^{-1}\right)\\
 & \approx\exp\left(-3T_{z}+\frac{3j_{1}}{F}\right).
\end{align}
Use successive linearization (effective Newton's method) to obtain
a solution to the simplified nonlinear equation:~ 
\begin{equation}
T_{c}=T_{z}-\Upsilon F^{-1}\label{eq:upstc}
\end{equation}
\begin{equation}
\Upsilon\approx\Upsilon_{1}+\Upsilon_{2}\label{eq:upsilon}
\end{equation}
\begin{equation}
\Upsilon_{1}\equiv-\frac{j_{1}\left(3\frac{F^{4}}{\phi_{+}(0)^{2}}e^{3\text{\ensuremath{T_{z}}}}\left(8F^{2}+8Fj_{1}+3j_{1}^{2}\right)\cos^{2}(\varphi)+j_{1}^{4}\omega^{2}\left(4\omega^{2}-9\right)e^{\frac{3j_{1}}{F}}\right)}{j_{1}^{2}\omega^{2}e^{\frac{3j_{1}}{F}}\left(-8F^{2}+12Fj_{1}+j_{1}^{2}\left(4\omega^{2}-9\right)\right)-3\frac{F^{4}}{\phi_{+}(0)^{2}}e^{3\text{\ensuremath{T_{z}}}}\left(16F^{2}+12Fj_{1}+3j_{1}^{2}\right)\cos^{2}(\varphi)}\label{eq:ups1}
\end{equation}
where the definition of the rest of the $\Upsilon_{X}$ objects are
given in Eqs.~(\ref{eq:ups2def})-(\ref{eq:ups24def}). These expressions
indicate that $\Upsilon\sim O(1)$ since for example in the $\Upsilon_{1}$
contribution we have contributions such as 
\begin{equation}
3\frac{F^{4}}{\phi_{+}(0)^{2}}e^{3\text{\ensuremath{T_{z}}}}\left(8F^{2}+8Fj_{1}+3j_{1}^{2}\right)\cos^{2}(\varphi)\sim\frac{F^{4}}{F^{4}}O(F^{2})
\end{equation}
in the numerator\footnote{Recall that $\phi_{+}(0)e^{-3T_{Z}/2}\sim O(F^{2})$ in the resonant
scenarios.} with an $O(F^{2})$ in the denominator. Also, note that even though
$\Upsilon$ looks like it is sensitive to $j_{1}$, one can check
that there are cancellations when $j_{1}$ derivative of $\Upsilon$
is computed. This cancellation occurs because if the $j_{1}$ is chosen
in the region where the error $\mathcal{E}_{1}+\mathcal{E}_{2}$ of
Eq.~(\ref{eq:linear}) is minimized, the sensitivity on $j_{1}$
by construction is minimized.

For example, consider 
\begin{equation}
\frac{\partial\Upsilon}{\partial j_{1}}\sim\frac{\partial\Upsilon_{1}}{\partial j_{1}}=\frac{\partial}{\partial j_{1}}\left(\frac{\mathcal{N}_{\Upsilon1}}{\mathcal{D}_{\Upsilon2}}\right)
\end{equation}
where $\mathcal{N}_{\Upsilon1}$ is the numerator and $\mathcal{D}_{\Upsilon1}$
is the denominator of Eq.~(\ref{eq:ups1}). The derivative receives
contributions from the numerator and denominator (after combining
over a common denominator) 
\begin{equation}
\left(\frac{\partial\mathcal{N}_{\Upsilon1}}{\partial j_{1}}\right)\mathcal{D}_{\Upsilon2}=K(\frac{6F^{4}}{\phi_{+}^{2}(0)e^{-3T_{z}}\sec^{2}\varphi}+e^{\frac{3j_{1}}{F}}j_{1}^{2}\omega^{2}+...)\label{eq:firstnumer}
\end{equation}
\begin{equation}
-\mathcal{N}_{\Upsilon1}\left(\frac{\partial\mathcal{D}_{\Upsilon2}}{\partial j_{1}}\right)=K(-2e^{\frac{3j_{1}}{F}}j_{1}^{2}\omega^{2}+...)\label{eq:second}
\end{equation}
where $K$ is a common factor and we have displayed the leading terms\footnote{Leading terms at least for $c_{+}$ near $2.35$ parametric region.}
To see the cancellation between these two terms, Eq.~(\ref{eq:firstnumer})
can be rewritten using Eq.~(\ref{eq:upsilondef}): 
\begin{align}
K(\frac{6F^{4}}{\phi_{+}^{2}(0)e^{-3T_{z}}\sec^{2}\varphi}+e^{\frac{3j_{1}}{F}}j_{1}^{2}\omega^{2}+...) & =K\left(\frac{6F^{4}}{\phi_{+}^{2}(0)e^{-3(T_{z}-\Upsilon/F)}e^{-3\Upsilon/F}\sec^{2}\varphi}+e^{\frac{3j_{1}}{F}}j_{1}^{2}\omega^{2}+...\right)\\
 & =K\left(\left[\frac{\Upsilon^{2}6}{e^{-3\Upsilon/F}}e^{\frac{-3j_{1}}{F}}+j_{1}^{2}\right]e^{\frac{3j_{1}}{F}}\omega^{2}+...\right).
\end{align}
Comparing with Eq.~(\ref{eq:second}), we see that the cancellation
occurs because 
\begin{equation}
\left|\left(\frac{\Upsilon^{2}6}{e^{-3\Upsilon/F}}e^{-\frac{3j_{1}}{F}}+j_{1}^{2}\right)-2j_{1}^{2}\right|\ll2j_{1}^{2}.\label{eq:cancellation}
\end{equation}
One of the merits of this exercise is to see that this cancellation
is independent of $c_{+}$ in the resonant region considered here.
This also allows one to see $\Upsilon$ has to be in the approximate
region of the zero of the left hand side of Eq.~(\ref{eq:cancellation}).

In the case of $c_{+}=2.35$ considered in the plots such as \ref{fig:Numerical-background-solution},
$\Upsilon_{1}$ dominates over $\Upsilon_{2}$ by about a factor of
5. However, since the entire point of this messy exercise was to obtain
a good numerical estimate of $T_{c}$, we keep $\Upsilon_{2}$. In
this parametric point example, we find Eq.~(\ref{eq:upsilon}) to
evaluates to 
\begin{equation}
\Upsilon\approx0.64
\end{equation}
with $T_{z}=9.278$ while the numerical solution for this case is
\begin{equation}
\Upsilon^{{\rm numerical}}\approx0.61
\end{equation}
attesting to a good approximation (about 5\% error). Note this also
allows us to compute for example 
\begin{align}
\phi_{+}(T_{c}) & \approx\phi_{+}(0)\exp\left(\frac{-3}{2}\left[T_{z}-\frac{\Upsilon}{F}\right]\right)\sec(\varphi)\omega\Upsilon/F\\
 & \approx15.294
\end{align}
giving 
\begin{equation}
r_{1}\approx0.76
\end{equation}
(defined in Eq.~(\ref{eq:r1def})) very close to the numerical value
of $r_{1}^{{\rm numerical}}=0.78$. In evaluating this, we made usage
of Eqs.~(\ref{eq:Tz}) and (\ref{eq:omega}) as well.

The error in the more general case can be estimated as follows. To
account for the $u=$35\% type of error $\Delta\phi_{-}$ in the $\phi_{-}(T_{c})$
field value, note 
\begin{align}
\Delta T_{c} & \approx\frac{\Delta\phi_{-}}{\partial_{T}\phi_{+}(T_{c})-\partial_{T}\phi_{-}(T_{c})}\\
 & \approx\frac{uF}{\partial_{T}\phi_{+}(T_{c})}\\
 & \sim\frac{u}{F}
\end{align}
which means that the error in $\phi_{-}$ shifts $\Upsilon$ by $u$.
This is why $T_{c}$ has to be very accurately determined to obtain
$\Delta\phi_{-}$.

\section{\label{sec:Small}Small $\alpha$}

For field configuration where $\alpha<\alpha_{{\rm L}}$ (where $\alpha_{{\rm L}}$
is defined in Eq.~(\ref{eq:alpha1})) at transition, the resonant
conditions in Sec.~\ref{subsec:Resonant-scenarios} are not satisfied.
In such cases, the mass eigenvalue and the rotated eigenvector gradient
effects are less than $O(F)$ post-transition. As a result, the axion
mass transitions smoothly from $c_{+}$ to a massless state. In these
cases, the separation between the transition $T_{c}$ and the zero-crossing
$T_{z}$ is usually $O(1)$. However, by evaluating Eq.~(\ref{eq:perturbative})
at $T_{c}$, one finds that the leading order correction $\phi_{-}^{(1)}\sim2\alpha^{2}F$.
Hence, we see that the perturbative solution is still valid for $\alpha<\alpha_{{\rm L}}$
and the cumbersome nonperturbative computation is no longer necessary
for computing the value of the fields at $T_{c}.$ However, to compute
the spectrum, $-V_{1}$ still needs to be computed, and this requires
an accurate computation of $T_{c}$ using the nonperturbative computation
that we have presented in Sec.\ref{sec:Behavior-of-}. Using this
$-V_{1}$ and a slowly time-dependent mass squared function $m_{B}^{2}$,
the final mode amplitude can be computed. The absence of resonance
and a weak $V_{1}$ dip less than $O(F)$ results in a power spectrum
with the long wavelength region plateauing after the first bump without
any further noticeable bumps similar to an overdamped scenario.

\section{\label{sec:Adiabatic-approximation-for}Adiabatic approximation for
an oscillating time space potential}

Consider the following second order ODE with an oscillating time space
potential.
\[
\ddot{y}+A\beta(t)\,\cos(ft)y=0
\]
where $\beta(t)$ is a slow-varying envelope function with amplitude
$A$ while the harmonic oscillations are rapidly varying (large frequency
$f$). The solution $y(t)$ of the aforementioned ODE can be approximated
by separating into the IR and UV components. As long as this hierarchy
can be maintained we can approximate $y$ as
\begin{equation}
y=y_{s}+y_{f}
\end{equation}
whereby $y_{s}$ represents the slow-varying (IR) adiabatic behavior
superimposed with a fast high frequency (UV) noise $y_{f}$. Next,
we substitute this into our original equation to get,
\[
\ddot{y}_{s}+\ddot{y}_{f}+A\beta(t)\,\cos(ft)(y_{s}+y_{f})=0.
\]

\begin{figure}[H]
\begin{centering}
\includegraphics[scale=0.35]{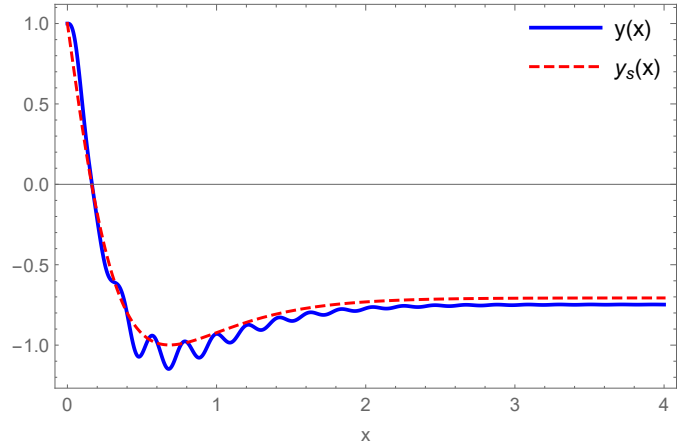} 
\par\end{centering}
\caption{\label{fig:adiabaticys}Comparison of exact and adiabatic solution
for the equation $\ddot{y}+3\dot{y}+200\,\sin(30\,x)e^{-\frac{3}{2}x}y=0,y(0)=1,\dot{y}(0)=0$.
Using the adiabatic or IR approximation, we solved the reduced equation
$\ddot{y}+3\dot{y}+\left(200/30\right)^{2}e^{-3x}y/2=0,y(0)=1,\dot{y}(0)=-200/30$.
Note that the initial conditions are modified.}
\end{figure}

We now apply the initial conditions. Assuming that the incoming function
is $y=y_{0}$ at some $t=t_{0}$ and has no UV behavior, the slow-varying
component $y_{s}$ will match appropriately with the $y_{0}$. The
fast-varying $y_{f}$ will then be matched with $0$ or be negligible.
Accordingly, over a small time-scale $\Delta t$, the UV component
will be initially sourced by the incoming IR component 
\[
\ddot{y}_{f}+A\beta(t)\,\cos(ft)y_{s}\sim0
\]
and by assuming a slow-varying envelope function $\beta(t)$ we obtain
up to a leading order
\[
y_{f}\sim\frac{A}{\Omega^{2}}\beta(t)\,\cos(ft)y_{s}.
\]
Note that $\left|y_{f}/y_{s}\right|\sim O\left(A/f^{2}\right)$ and
sets the scale of the UV component compared to IR. Next, we substitute
the UV solution into our original differential equation and integrate
out the UV scale over one time-period 
\begin{equation}
\ddot{y}_{s}+\frac{A^{2}}{2f^{2}}\beta^{2}(t)\,y_{s}\sim0.
\end{equation}
The above differential equation governs the dynamics of the IR component
subject to the initial conditions $y_{s}(t_{0})=y(t_{0})$ and $\dot{y}_{s}(t_{0})=\dot{y}(t_{0})-\dot{y}_{f}(t_{0})$.
By defining $\delta=A/f^{2}$, we note that if $\delta\ll1$, then
the above UV and IR treatment is also valid up to $O\left(\delta\right)$.
Fig.~\ref{fig:adiabaticys} gives plots of $y(x)$ and $y_{s}(x)$
obtained by solving the exact ode $\ddot{y}+3\dot{y}+200\,\sin(30\,x)e^{-\frac{3}{2}x}y=0,y(0)=1,\dot{y}(0)=0$
and its adiabatically reduced form respectively.

\section{\label{sec:Flat-deviation}Flat-deviation $\xi$}

\begin{figure}
\begin{centering}
\includegraphics[scale=0.5]{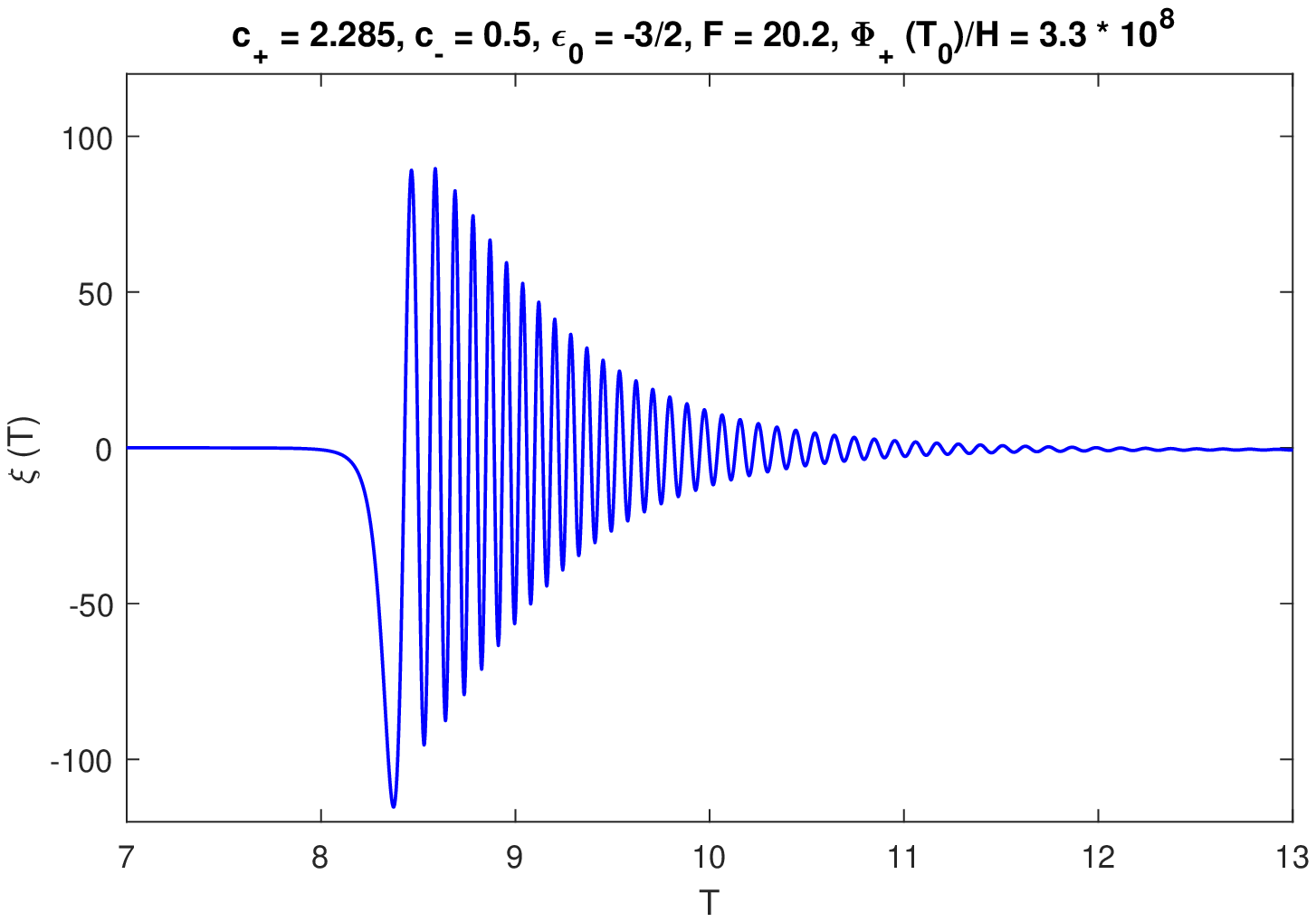}\includegraphics[scale=0.5]{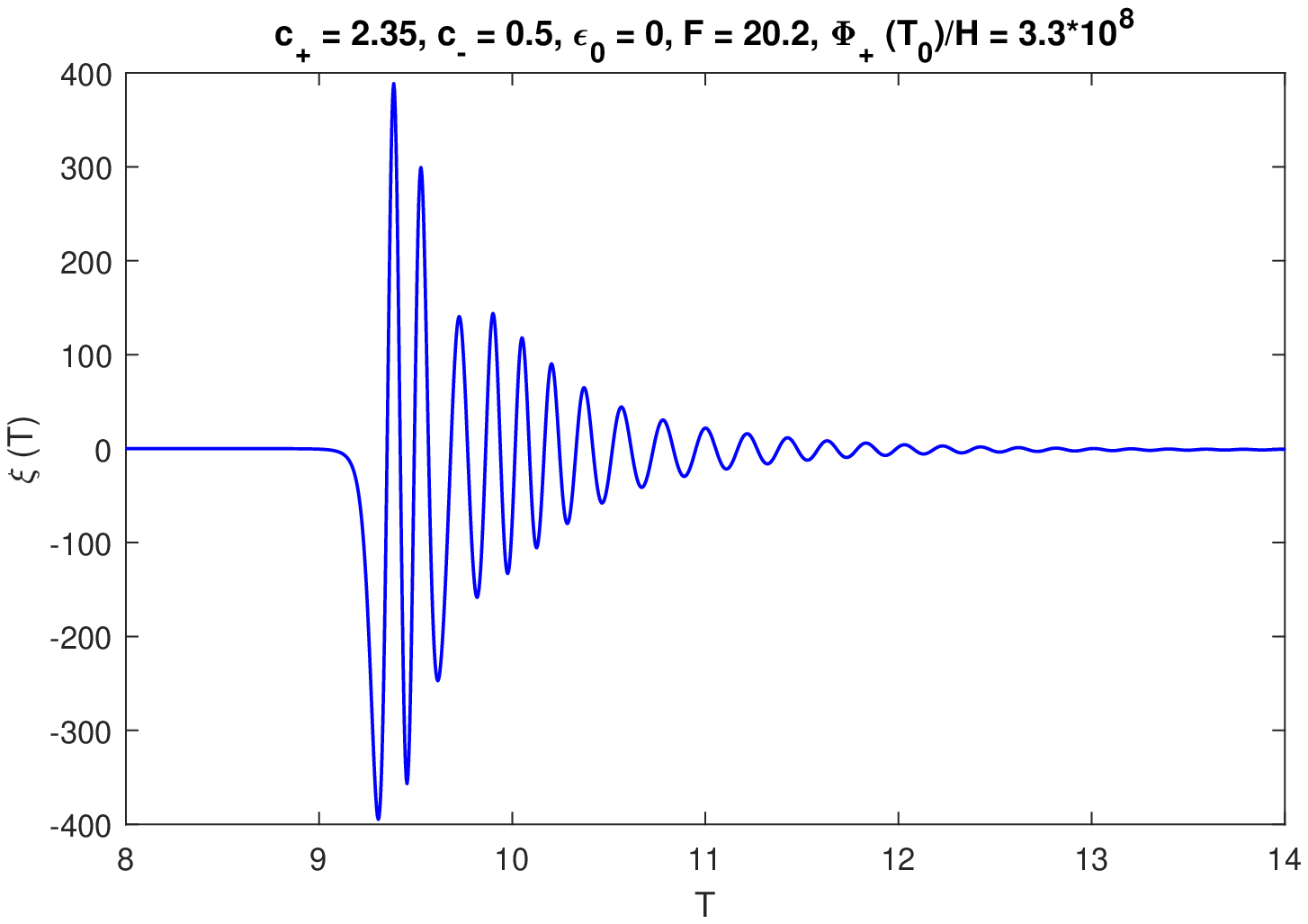} 
\par\end{centering}
\caption{\label{fig:flatdevcomparison}Comparison of numerical solutions to
flat-deviation $\xi(T)$ after transition for $c_{+}=2.285$, $\epsilon_{0}=-3/2$
(left) and $c_{+}=2.35$, $\epsilon_{0}=0$ (right) highlighting the
resonant cases. All other parameters were set at their fiducial values
$P_{A}$ of Fig.~\ref{fig:Comparison-of-Taylor}.}
\end{figure}

The quantity $\xi=\phi_{+}\phi_{-}-F^{2}$ defines a deviation from
the flat direction. This is a crucial measure as it controls the strongly
coupled dynamics of the two fields. To obtain a differential equation
for $\xi$ we start with the background field Eq.~(\ref{eq:backgroundeom}).
We multiply the two equations with $\phi_{-}$ and $\phi_{+}$ respectively
and add them together to obtain,
\[
-2\dot{\phi}{}_{+}\dot{\phi}{}_{-}+\partial_{T}^{2}(\phi_{+}\phi_{-})+3\partial_{T}(\phi{}_{+}\phi_{-})+(c_{+}+c_{-}+\phi_{-}^{2}+\phi_{+}^{2})\phi_{+}\phi_{-}-F^{2}(\phi_{-}^{2}+\phi_{+}^{2})=0
\]
which yields an effective equation for the flat-deviation
\begin{equation}
\ddot{\xi}+3\dot{\xi}+(\tilde{M}_{11}^{2}+\tilde{M}_{22}^{2})\xi=-(c_{+}+c_{-})F^{2}+2\dot{\phi}{}_{+}\dot{\phi}{}_{-}\label{eq:flatdeveqn}
\end{equation}
where $\tilde{M}_{ij}^{2}$ are the elements of the $\tilde{M}^{2}$
matrix. This is an interesting equation in that apart from the kinetic
mixing terms on the RHS, the equation for $\xi$ has been made to
``look'' linear, although it certainly is not because of $\tilde{M}_{11}^{2}+\tilde{M}_{22}^{2}$.
Consider an expansion about a neighborhood of $T_{m}$ defined to
be the zero of $\dot{\phi}_{+}$ : 
\begin{equation}
\dot{\phi}_{+}(T_{m})=0.
\end{equation}
In that neighborhood, the solution must behave as 
\begin{equation}
\phi_{+}(T)=\phi_{+}(T_{m})+\frac{1}{2}\ddot{\phi}_{+}(T_{m})(T-T_{m})^{2}+....
\end{equation}
Since $T_{m}$ comes $O(1/F)$ time after $T_{c}$ while $\phi_{-}$
has been increasing beyond $F$ while $\phi_{+}$ has been decreasing
towards $\phi_{+}(T_{m})$, we can approximate that $\phi_{-}$ has
a Taylor expansion in time: 
\begin{equation}
\phi_{-}(T)=\phi_{-}(T_{m})+\dot{\phi}_{-}(T_{m})(T-T_{m})+\frac{1}{2}\ddot{\phi}_{-}(T_{m})(T-T_{m})^{2}+...
\end{equation}
We then find 
\begin{equation}
\dot{\phi}_{+}(T)=\ddot{\phi}_{+}(T_{m})(T-T_{m})+\frac{1}{2}\dddot{\phi}_{+}(T_{m})(T-T_{m})^{2}+...
\end{equation}
\begin{equation}
\dot{\phi}_{-}(T)=\dot{\phi}_{-}(T_{m})+\ddot{\phi}_{-}(T_{m})(T-T_{m})+\frac{1}{2}\dddot{\phi}_{-}(T_{m})(T-T_{m})^{2}+...
\end{equation}
yielding 
\begin{equation}
\dot{\phi}_{+}\dot{\phi}_{-}=\ddot{\phi}_{+}(T_{m})\dot{\phi}_{-}(T_{m})(T-T_{m})+O[(T-T_{m})^{2}].\label{eq:phidotsq1}
\end{equation}
We also know 
\begin{align}
\tilde{M}_{11}^{2}+\tilde{M}_{22}^{2} & =c_{+}+c_{-}+[\phi_{-}(T_{m})+\dot{\phi}_{-}(T_{m})(T-T_{m})+...]^{2}+[\phi_{+}(T_{m})+\frac{1}{2}\ddot{\phi}_{+}(T_{m})(T-T_{m})^{2}+....]^{2}\\
 & =c_{+}+c_{-}+\phi_{-}^{2}(T_{m})+\phi_{+}^{2}(T_{m})+2\phi_{-}(T_{m})\dot{\phi}_{-}(T_{m})(T-T_{m})+O[(T-T_{m})^{2}].\label{eq:mass-sum}
\end{align}
Keeping to zeroth order in $T-T_{m}$, we put into Eq.~(\ref{eq:flatdeveqn})
the zeroth order terms in Eqs.~(\ref{eq:phidotsq1}) and (\ref{eq:mass-sum}).
The resulting Eq.~(\ref{eq:flatdeveqn}) has a solution in the vicinity
of $T_{m}$ 
\begin{equation}
\xi=-A\,e^{-\frac{3}{2}(T-T_{m})}\cos(\Omega(T_{m})(T-T_{m}))-\frac{(c_{+}+c_{-})F^{2}}{\Omega(T_{m})^{2}}+O\left((T-T_{m})^{3}\ddot{\phi}_{+}(T_{m})\dot{\phi}_{-}(T_{m})\right)\label{eq:}
\end{equation}
\begin{equation}
\Omega(T_{m})\equiv\sqrt{\tilde{M}_{11}^{2}(T_{m})+\tilde{M}_{22}^{2}(T_{m})-\frac{9}{4}}.
\end{equation}
For the remainder of our discussion, we will consider the following
approximate expression for the flat-deviation with a constant amplitude
$A$ and a slow-varying time-dependent frequency 
\begin{equation}
\xi=-A\,e^{-\frac{3}{2}(T-T_{m})}\cos\left(\int_{T_{m}}^{T}\Omega(t)dt\right)-\frac{(c_{+}+c_{-})F^{2}}{\Omega(T)^{2}}.\label{eq:flatdevfn}
\end{equation}
Flat-deviations of $O(F^{2})$ occur close to a zero-crossing of $\phi_{+}$
characterized by a strong nonlinear interaction between the two background
fields. After transition, when the background fields are settling
to their minima, the frequency of flat-deviation is $\sim O(F)$.
As the fields initially start out along the flat direction, $\xi\approx2\dot{\phi}{}_{+}\dot{\phi}{}_{-}/\phi_{+}^{2}\sim O\left(\phi{}_{-}/\phi_{+}\right)<1$
is negligible since $\phi_{+}\gg\phi{}_{-}$. When the fields reach
close to the transition, the $2\dot{\phi}{}_{+}\dot{\phi}{}_{-}$
term causes the fields to deviate away from flat direction. Later
when the fields have settled to their minima, the flat-deviation tends
to 
\begin{equation}
-\frac{(c_{+}+c_{-})F^{2}}{\tilde{M}_{11}^{2}+\tilde{M}_{22}^{2}}\approx-\sqrt{c_{+}c_{-}}.
\end{equation}
The kind of dynamic behavior described above can lead to resonance
which is characterized by a significant flat-deviation $\gtrsim O(0.1F^{2})$
as shown in Fig.~\ref{fig:flatdevcomparison}. 

\section{\label{sec:UV-and-IR-phi-fields}UV and IR decomposition of the background
fields}

Post transition, the background fields are strongly coupled through
the interaction term $\xi\phi_{\pm}$ as shown in the equations
\begin{align}
\ddot{\phi}_{+}+3\dot{\phi}{}_{+}+c_{+}\phi_{+}+\xi\phi_{-} & =0\\
\ddot{\phi}_{-}+3\dot{\phi}{}_{-}+c_{-}\phi_{-}+\xi\phi_{+} & =0
\end{align}
where $\xi$ is an oscillating function given in Eq.~(\ref{eq:flatdevfn}).
The dynamics of the coupled system post-transition can be understood
in terms of a UV and IR decomposition detailed in Appendix \ref{sec:Adiabatic-approximation-for}.
If the frequency of the oscillating function $\xi$ is much larger
than the root of its amplitude, the system exhibits a hierarchy between
the UV and IR states. We may then integrate out the UV degree of freedom
and retain the IR components to describe the adiabatic behavior of
the coupled system. In principle we can write
\begin{equation}
\phi_{\pm}\approx\phi_{\pm s}+\phi_{\pm f}\label{eq:sfdef}
\end{equation}
where the subscripts $s,f$ represent the slow (IR) and fast (UV)
component of the fields.

From Appendix \ref{sec:Adiabatic-approximation-for} and assuming
$A/\Omega_{s}^{2}\ll1$ (which we will justify below), we can write
down an approximate solution for the UV component of the $\phi_{\pm}$
fields as 
\begin{align}
\phi_{+f} & \approx-\frac{A}{\Omega^{2}}e^{-\frac{3}{2}\left(T-T'\right)}\sin(f\left(T\right))\phi_{-s}\nonumber \\
\phi_{-f} & \approx-\frac{A}{\Omega^{2}}e^{-\frac{3}{2}\left(T-T'\right)}\sin(f\left(T\right))\phi_{+s}\label{eq:phiUVcomponent}
\end{align}
where we have used Eq.~($\ref{eq:flatdevfn}$) in place of $\xi$
(with a time-dependent frequency term $f(T)=\int_{T'}^{T}\Omega(t)dt$)
and switched the cosine function in Eq.~($\ref{eq:flatdevfn}$) to
a sine for convenience. In terms of the UV and IR components we can
write flat-deviation as 
\begin{align}
\xi & =\phi_{+}\phi_{-}-F^{2}\\
 & \approx\phi_{+s}\phi_{-s}+\phi_{+f}\phi_{-s}+\phi_{+s}\phi_{-f}+\phi_{+f}\phi_{-f}-F^{2}.
\end{align}
Meanwhile from Eq.~(\ref{eq:phiUVcomponent}) we infer that
\begin{equation}
\xi\approx-Ae^{-\frac{3}{2}\left(T-T'\right)}\sin(f\left(T\right))\approx\phi_{+f}\phi_{-s}+\phi_{+s}\phi_{-f}
\end{equation}
subject to the approximation $\Omega^{2}\approx\phi_{+s}^{2}+\phi_{-s}^{2}\equiv\Omega_{s}^{2}$.
Thus, we will approximately take the frequency-squared of the $\xi$
function as the sum of the squares of the IR components of the background
fields. Since the UV components are smaller in amplitude compared
to IR, we obtain an approximate relationship between the IR components
of the background fields, 
\begin{equation}
\phi_{+s}\phi_{-s}\approx F^{2}-O\left(\phi_{+f}\phi_{-f}\right).
\end{equation}
This is an important result which indicates that if the background
fields can be factorized into UV and IR components, then the IR components
continue to follow the flat direction.

Post transition, $|\phi_{-}|$ begins to increase due to a positive
velocity of $O(F^{2})$ and becomes dominant compared to a decreasing
$|\phi_{+}|$. During this time, the UV integrated equation of motion
for the IR component of the dominant $\phi_{-}$ field is given below
\begin{equation}
\ddot{\phi}_{-s}+3\dot{\phi}{}_{-s}+c_{-}\phi_{-s}+\frac{A^{2}}{2\Omega_{s}^{2}}e^{-3\left(T-T_{2}\right)}\phi_{-s}+\sqrt{c_{+}c_{-}}\phi_{+s}\approx0\label{eq:phimseqn}
\end{equation}
while the smaller $\phi_{+s}$ field is obtained through the flat
direction condition $\phi_{+s}\phi_{-s}\approx F^{2}$. Note that
$\Omega_{s}^{2}\approx\phi_{+s}^{2}+\phi_{-s}^{2}$ is a function
of the background fields highlighting the non-linearity of the above
equation. However, to obtain an analytic solution we will consider
an average value for the parameter $\Omega_{s}$ over a half-oscillation
of $\phi_{-s}$ (since $\phi_{-s}$ increases up to a maximum and
then falls back towards $F$) such that a general solution to the
above damped oscillator equation for a constant average $\bar{\Omega}$
is given as (during the time when the term $\sqrt{c_{-}c_{+}}\phi_{+s}$
is not appreciable) 
\begin{equation}
\phi_{-s}=e^{-\frac{3}{2}\left(T-T_{2}\right)}\left(c_{1}J_{n_{1}}\left(\frac{A\sqrt{2}}{3\bar{\Omega}}e^{-\frac{3}{2}\left(T-T_{2}\right)}\right)+c_{2}J_{-n_{1}}\left(\frac{A\sqrt{2}}{3\bar{\Omega}}e^{-\frac{3}{2}\left(T-T_{2}\right)}\right)\right)\label{eq:phims_sol}
\end{equation}
where
\begin{align}
n_{1} & =\sqrt{1-4c_{-}/9}
\end{align}

\begin{align}
c_{1,2} & =\frac{\mp\pi}{2\sin\left(\pi n_{1}\right)}\left(\phi_{-s}(T_{2})\frac{A\sqrt{2}}{3\bar{\Omega}}J'_{\mp n_{1}}\left(\frac{A\sqrt{2}}{3\bar{\Omega}}\right)+\left(\phi_{-s}(T_{2})+2\dot{\phi}_{-s}(T_{2})/3\right)J_{\mp n_{1}}\left(\frac{A\sqrt{2}}{3\bar{\Omega}}\right)\right)
\end{align}
for 
\begin{align}
\phi_{-s}(T_{2}) & \approx\phi_{-}(T_{2})\\
\dot{\phi}_{-s}(T_{2}) & \approx\dot{\phi}_{-}(T_{2})-\left.\frac{\xi\phi_{+}}{\phi_{-}}\right|_{T_{2}}\label{eq:phims_atT2}
\end{align}
where $\phi_{-}(T)$ properties near $T_{2}$ can be obtained from
Eq.~(\ref{eq:posttransparam}). Meanwhile, the $\phi_{+s}$ field
is given by the flat direction 
\begin{equation}
\phi_{+s}\approx\frac{F^{2}}{\phi_{-s}}.\label{eq:flatdirconst}
\end{equation}
Since $\Omega$ is a time-dependent function of the background fields,
we apply a semi-numeric approach to estimate an average value of $\Omega$
between $T_{2}$ and the time $T_{*}$ when $\phi_{-s}\left(T_{*}\right)\approx4/3\phi_{-\min}$
for $c_{+}>c_{-}$. This choice of $T_{*}$ allows us to consider
both the situations where the two background fields may either cross
each other again after $T_{c}$ or not. The procedure involves matching
the analytical solution in Eq.~(\ref{eq:phims_sol}) to the numerical
results for the fit parameter $\bar{\Omega}$. Through this procedure
we obtain an empirical fit expression for $\bar{\Omega}$ as a function
of $\alpha$ and $F$: 
\begin{equation}
\bar{\Omega}\approx2.05F+\frac{0.1327F+0.0454F^{2}}{1+\exp\left(7.86\left(\alpha-\alpha_{0}\right)\right)}\label{eq:avgomega}
\end{equation}
where 
\begin{equation}
\alpha_{0}\equiv0.7442-0.0008F.
\end{equation}
When $\sqrt{2}A/(3\bar{\Omega})\,e^{-3/2\left(T-T_{2}\right)}\gg\left|3/4-4c_{-}/9\right|$,
the above solution has an oscillating behavior with a maximum frequency
$f_{{\rm IR}}$ 
\begin{equation}
f_{{\rm IR}}\approx\frac{\sqrt{2}}{3}\frac{A}{\bar{\Omega}}.\label{eq:IRscale}
\end{equation}
Within our parametric region of interest, the amplitude satisfies
$A<F^{2}$ from Eq.~(\ref{eq:Aapprox}) imposed by $\alpha\lesssim\alpha_{{\rm U}}$.
Since the transition occurs close to $F$, the two fields oscillate
over the equilibrium scale $F$ such that $\Omega_{s}\approx\sqrt{\phi_{+s}^{2}+\phi_{-s}^{2}}\geq\sqrt{2}F$
with an average value approximately $O(2F)$ at $T_{2}$. Therefore,
we see the self-consistency of the assumption that $A/\Omega_{s}^{2}\ll1$
for $T\geq T_{2}$. Between $T_{c}$ and $T_{2}$, the system of background
fields can be given by the cubic polynomial solution. Also, the IR
fields can oscillate momentarily with a frequency $f_{{\rm IR}}\sim O\left(0.1F\right)$
while the UV scales oscillate with frequency $\Omega_{s}\sim O\left(F\right)$.
Therefore, the hierarchy between the two scales is clearly established.

\begin{figure}
\begin{centering}
\includegraphics[scale=0.9]{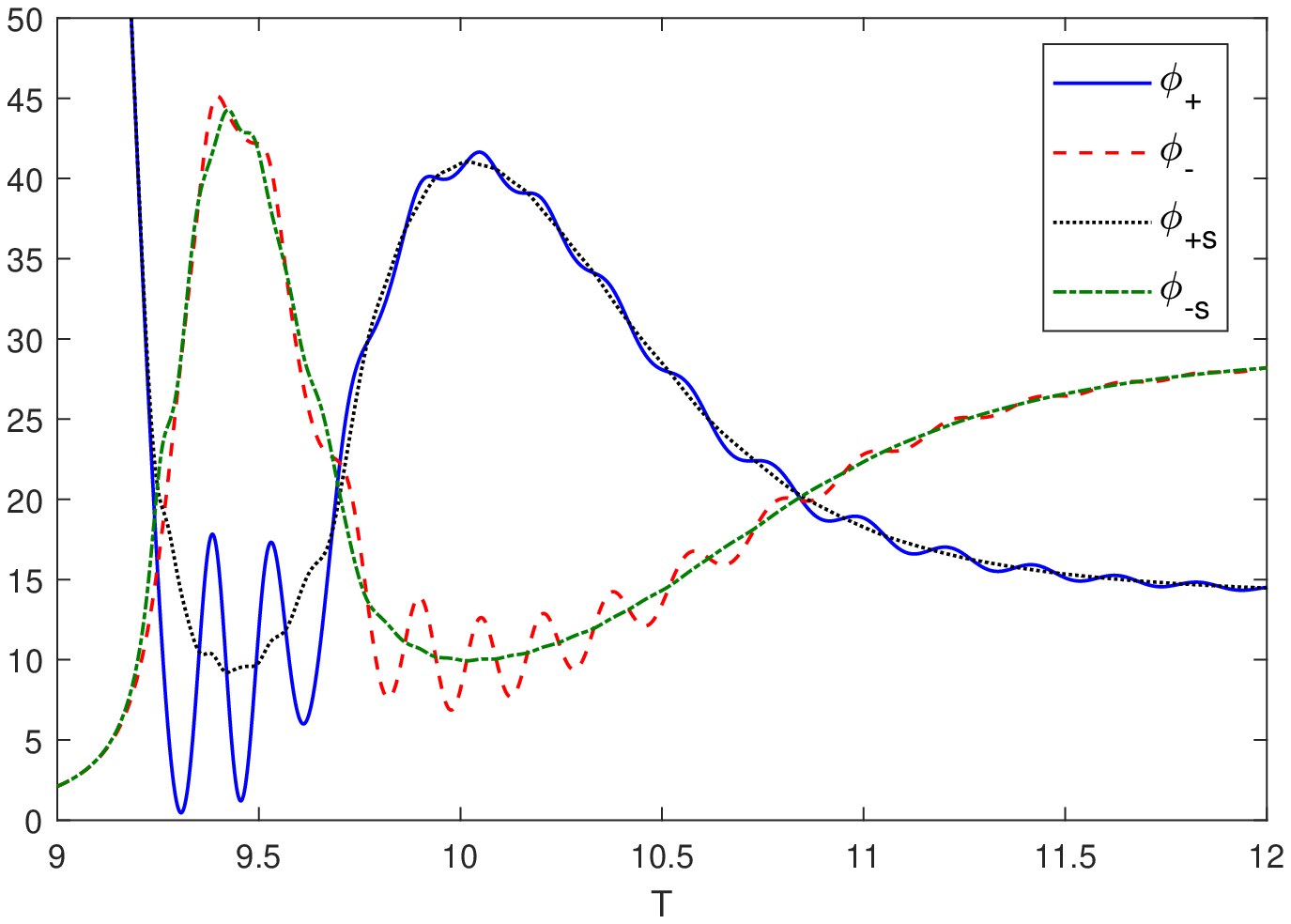} 
\par\end{centering}
\begin{centering}
\includegraphics[scale=0.9]{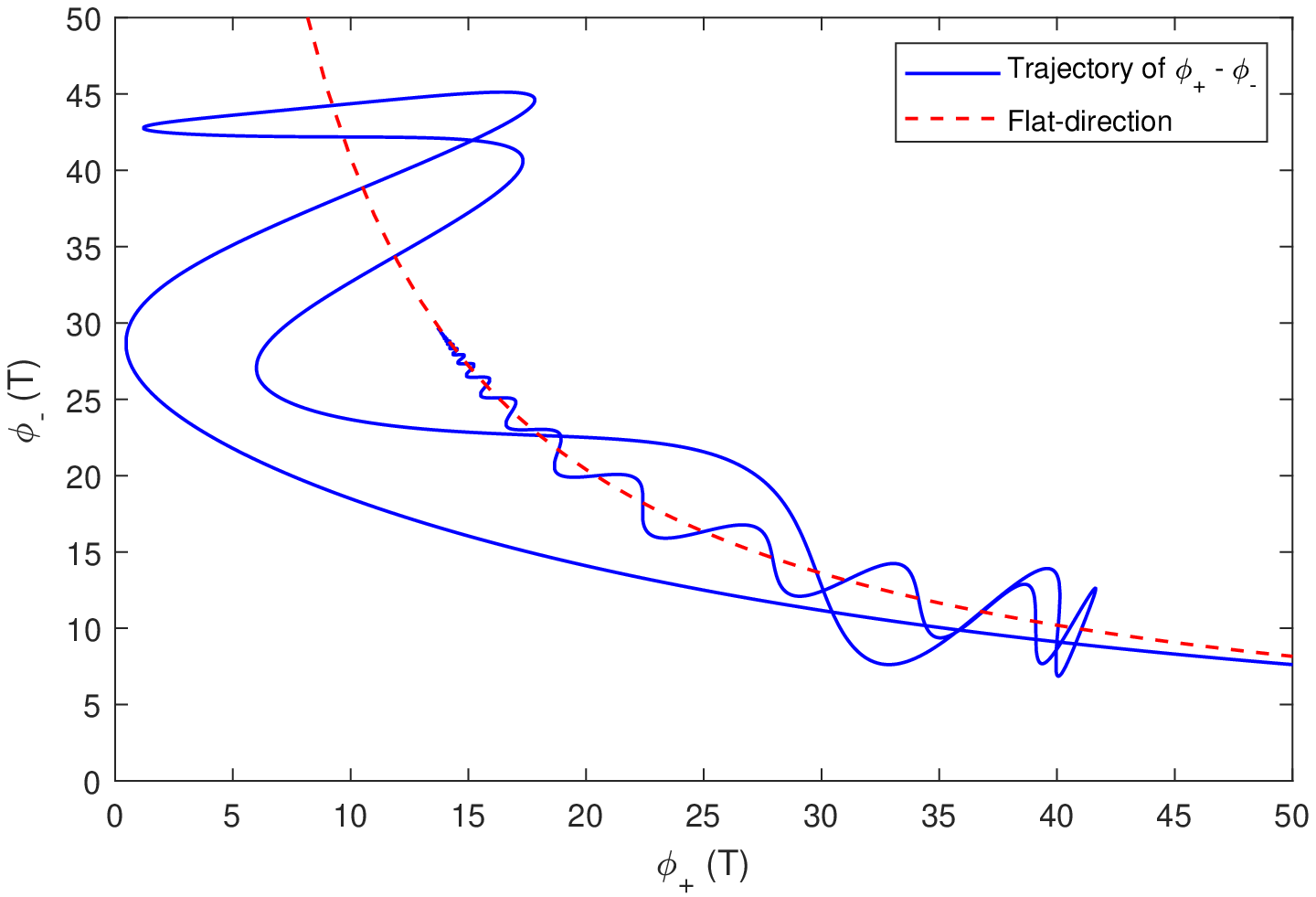} 
\par\end{centering}
\caption{\label{fig:UVandIRfreq}Plots showing UV and IR components of the
background fields $\phi_{\pm}$ for $c_{+}=2.35$ where the curves
for $\phi_{\pm}$ have been computed numerically by solving Eqs.~(\ref{eq:backgroundeom0})
and (\ref{eq:backgroundeom}). The IR components $\phi_{\pm s}$ have
been computed by subtracting $\phi_{\pm f}$ of Eq.~(\ref{eq:phiUVcomponent})
from the numerically computed $\phi_{\pm}$. On the right, the trajectory
of the background fields clearly highlights the deviations away from
the flat direction. All the parameters have been set to the $P_{A}$
set used in Fig.~\ref{fig:Comparison-of-Taylor}.}
\end{figure}

In summary, prior to transition the fields are best described via
primary frequency $\omega=\sqrt{c_{+}-9/4}$. After transition, we
can separate the fields into UV and IR components as long as $A/\Omega_{s}^{2}\ll1$.
As $T\rightarrow T_{\infty}$, the UV component decays away and the
IR components settle to the minima. Fig.~\ref{fig:UVandIRfreq} highlights
the above conclusions by showing UV and IR components of the background
fields $\phi_{\pm}$ for $c_{+}=2.35$ with all the other parameters
set to the $P_{A}$ set used in Fig.~\ref{fig:Comparison-of-Taylor}.

\section{\label{sec:Lightest-eigenvector}Lightest eigenvector}

We will now study the lightest mass eigenvector $e_{1}$ corresponding
to the lightest mass eigenvalue and derive analytical expression for
$\left(\partial_{T}e_{1}\right)^{2}$. We begin by defining the lightest
mass eigenvector,
\begin{equation}
e_{1}=\left[\begin{array}{c}
e_{11}\\
e_{21}
\end{array}\right]
\end{equation}
where the column matrix elements are 
\begin{align}
e_{11} & =\frac{e}{\sqrt{e^{2}+1}}\approx\frac{-\phi_{+}}{\sqrt{\phi_{+}^{2}+\phi_{-}^{2}}}\qquad e_{21}=\frac{1}{\sqrt{e^{2}+1}}\approx\frac{\phi_{-}}{\sqrt{\phi_{+}^{2}+\phi_{-}^{2}}}
\end{align}
with the definition 
\[
e=\frac{1}{2F^{2}}\left(\tilde{M}_{11}^{2}-\tilde{M}_{22}^{2}-\sqrt{(\tilde{M}_{22}^{2}-\tilde{M}_{11}^{2})^{2}+4F^{4}}\right).
\]
We would like to evaluate the derivative squared term $\left(\partial_{T}e_{1}\right)^{2}=\left(\partial_{T}e{}_{11}\right)^{2}+\left(\partial_{T}e{}_{21}\right)^{2}$.
Using the definitions from above, we note that $\partial_{T}e{}_{11}=\left(\partial_{T}e\right)/(e^{2}+1)^{3/2}$
and $\partial_{T}e{}_{21}=\partial_{T}e\left(-e/(e^{2}+1)^{3/2}\right)$
so that $\dot{e}{}_{1}\cdot\dot{e}{}_{1}=\dot{e}^{2}/\left(e^{2}+1\right)^{2}$.
By defining $g=\tilde{M}_{11}^{2}-\tilde{M}_{22}^{2}$ we expand $\dot{e}{}_{1}\cdot\dot{e}{}_{1}$,
\begin{align}
\dot{e}{}_{1}\cdot\dot{e}{}_{1} & =\dot{g}^{2}\left(\frac{e^{2}}{\left(e^{2}+1\right)^{2}}\,\frac{1}{g^{2}+4F^{4}}\right)\label{e1primesq}
\end{align}
with 
\begin{equation}
\dot{g}=2\dot{\phi}_{-}\phi_{-}-2\dot{\phi}_{+}\phi_{+}.
\end{equation}
Thus, $\dot{e}{}_{1}\cdot\dot{e}{}_{1}$ peak amplitude is related
to the relative velocity of the two fields as they cross each other.
The peak is maximized if the two fields approach from opposite directions
thus maximizing the relative velocity. Upon substituting the analytical
form of $\dot{g}$ into Eq.~\ref{e1primesq} and solving we obtain
that the first maxima close to $T_{c}$ occurs when the $|\phi_{+}|$
field is approximately $F$. The maximum amplitude is given as, 
\begin{align*}
(\dot{e}{}_{1}\cdot\dot{e}{}_{1})_{{\rm max}} & \approx\left.\dot{g}^{2}\left(\frac{e^{2}}{\left(e^{2}+1\right)^{2}}\,\frac{1}{g^{2}+4F^{4}}\right)\right|_{\phi_{+}\rightarrow F}.
\end{align*}
As $|\phi_{+}|\rightarrow F$, we have $-F^{2}\leq g<0$ so that we
can expand in terms of $\left(g+F^{2}\right)/F^{2}\ll1$ which simplifies
$(\dot{e}{}_{1}\cdot\dot{e}{}_{1})_{{\rm max}}$
\begin{align}
(\dot{e}{}_{1}\cdot\dot{e}{}_{1})_{{\rm max}} & \approx\left.\left(\frac{\dot{g}}{5F^{2}}\right)^{2}\left(1+\frac{4}{5}\left(\frac{g+F^{2}}{F^{2}}\right)+\frac{2}{25}\left(\frac{g+F^{2}}{F^{2}}\right)^{2}-\frac{28}{125}\left(\frac{g+F^{2}}{F^{2}}\right)^{3}\right)\right|_{\phi_{+}\rightarrow F}\label{eq:e1primemax-1}
\end{align}
which has a limiting case
\begin{equation}
\lim_{\alpha\gg1}(\dot{e}{}_{1}\cdot\dot{e}{}_{1})_{{\rm max}}\rightarrow\frac{4}{25}F^{2}\alpha^{2}.
\end{equation}
Using the above analytical expressions, we present a second order
polynomial fit in terms of parameter $\alpha$ for the Eq.~$\ref{eq:e1primemax-1}$
in the range $[0.25,1.5]$ 
\begin{equation}
(\dot{e}{}_{1}\cdot\dot{e}{}_{1})_{{\rm max}}\approx F^{2}\left(-0.0030+0.2156\alpha+0.1779\alpha^{2}\right)\qquad\alpha\in\left[0.25,1.5\right].\label{eq:e1primesq-polyfit}
\end{equation}
As the $\phi_{+}$ field rapidly rolls down from the Plank scale,
the first peak $(\dot{e}{}_{1}\cdot\dot{e}{}_{1})$ peak occurs slightly
before transition and is characterized by the the dominant $\omega=\sqrt{c_{+}-9/4}$
frequency. After transition, the $\phi_{\pm}$ fields can be divided
in to the UV and IR components. When the jump ETSP in Eq.~(\ref{eq:phimseqn})
is significant at $T_{2}$, it leads to an $O(0.1F)$ frequency oscillations
of the IR fields (see appendix \ref{sec:UV-and-IR-phi-fields}) such
that the fields cross again after transition (also characterized by
the zeros of $g\approx\phi_{-}^{2}-\phi_{+}^{2}$). Quantitatively,
this is equivalent to $\phi_{-s}\rightarrow F$ since $\phi_{+s}\approx F^{2}/\phi_{-s}$.
Thus, we can obtain approximate location of the second crossing of
the background fields by solving Eq.~(\ref{eq:phims_sol}) for the
time $T_{3}$ when $\phi_{-s}(T_{3})=F$.

Since additional crossings require a significant jump ETSP, we begin
with the Eq.~(\ref{eq:phims_sol}) for the $\phi_{-s}$ background
field and evaluate an approximate condition for the background fields
to cross after transition. As we are interested in cases where the
crossings are caused by the jump ETSP, we will neglect the $c_{-}$
term in Eq.~(\ref{eq:phimseqn}). Hence, we consider the following
equation
\begin{equation}
\ddot{\phi}_{-s}+3\dot{\phi}{}_{-s}+\frac{A^{2}}{2\bar{\Omega}^{2}}e^{-3\left(T-T_{2}\right)}\phi_{-s}\approx0
\end{equation}
which has the general solution 
\begin{equation}
\phi_{-s}(T\geq T_{2})\approx e^{-\frac{3}{2}(T-T_{2})}\left[c'_{1}J_{1}\left(me^{-\frac{3}{2}(T-T_{2})}\right)+c'_{2}Y_{1}\left(me^{-\frac{3}{2}(T-T_{2})}\right)\right]
\end{equation}
where the primed coefficients $c'_{1,2}$ are obtained similar to
the $c_{1,2}$ below Eq.(\ref{eq:phims_sol}) and $m=\sqrt{2}A/(3\bar{\Omega})$
is a function of $\alpha$.

As $T\rightarrow T_{\infty}$, we look for the minimum value of $m$
such that $\phi_{-s}\left(T>T_{2}\right)=F$. Hence, we equate
\begin{equation}
\lim_{T\rightarrow T_{\infty}}\phi_{-s}(T\geq T_{2})\approx\phi_{-s}(T_{2})J'_{1}\left(m\right)+\frac{\left(\phi_{-s}(T_{2})+2\dot{\phi}_{-s}(T_{2})/3\right)}{m}J_{1}\left(m\right)=F
\end{equation}

Since $\dot{\phi}_{-s}(T_{2})\sim O\left(.5F^{2}\right)\gg\phi_{-s}(T_{2})\sim O\left(F\right)$,
we reduce the above expression to
\begin{equation}
\frac{3}{F}\approx\frac{J_{1}\left(m\right)}{m}.\label{eq:alpha2conditionalexpr}
\end{equation}

Eq.~(\ref{eq:alpha2conditionalexpr}) gives us an approximate minimum
value of $m$ such that the two background fields cross each other
after $T_{c}$. For $c_{-}\ll1$ and $F\gg1$, we find that the minimum
value of $m$ saturates to about
\begin{equation}
m=z_{1}\approx3.8
\end{equation}
where $J_{1}(z_{1})=0.$ The term $A$ in Eq.~(\ref{eq:alpha2conditionalexpr})
can be evaluated using the nonperturbative cubic polynomial expansion
for the background fields from Eqs.~(\ref{eq:phippoly}) and (\ref{eq:phimpoly})
around $T_{c}$. The minimum value of $\alpha$ that satisfies the
conditional equality in Eq.~(\ref{eq:alpha2conditionalexpr}) is
defined as $\alpha_{2}$. It corresponds to a parameteric cutoff such
that for $\alpha\gtrsim\alpha_{2}$, the background fields cross each
other again after $T_{c}$. For $F=20.2$ (corresponding to the fiducial
parameter set $P_{A}$), we obtain $\alpha_{2}\approx0.87$. Similarly,
for a much larger value of $F=100$, we obtain $\alpha_{2}\approx0.6$
highlighting that $\alpha_{2}$ reduces with $F$. If the resonance
amplitude $A$ is large enough, the background fields can cross each
other more than once after $T_{c}$. This corresponds to the situation
where
\begin{equation}
m\gtrsim z_{2}\label{eq:alpha3conditionalexpr}
\end{equation}
where $z_{2}$ corresponds to the second zero of $J_{1}(z)$. Further,
we remind that each crossing of the background fields corresponds
to a $\left(\dot{e}_{1}\right)^{2}$ peak which is modeled as a $-V_{i}$
dip within our numerical mass model in Eq.~(\ref{eq:model}). Since
we limit ourselves to just two dips in this paper, we will consider
only those cases where $\alpha\lesssim\alpha_{3}$. 

\begin{figure}[H]
\begin{centering}
\includegraphics[scale=0.5]{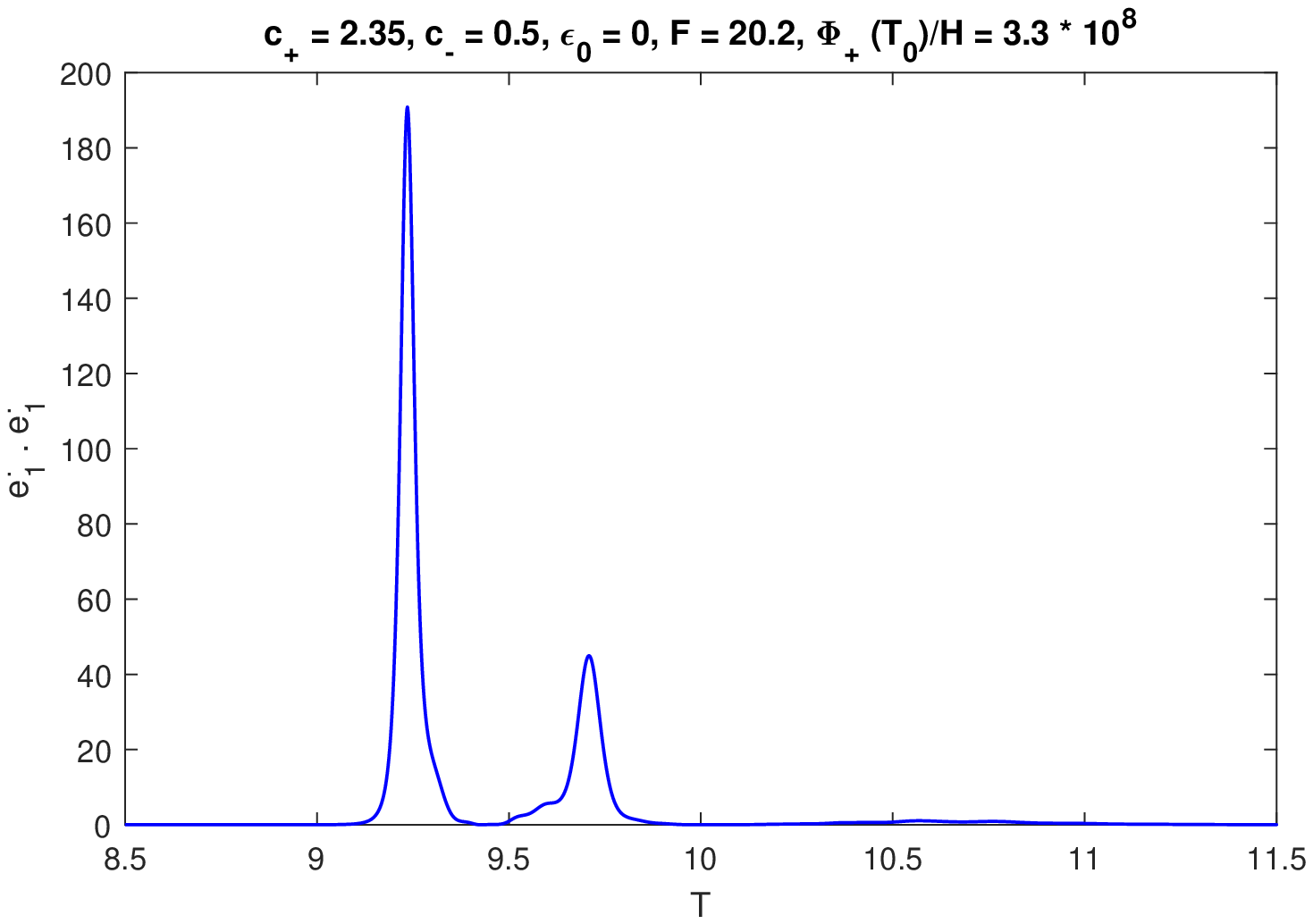}\includegraphics[scale=0.5]{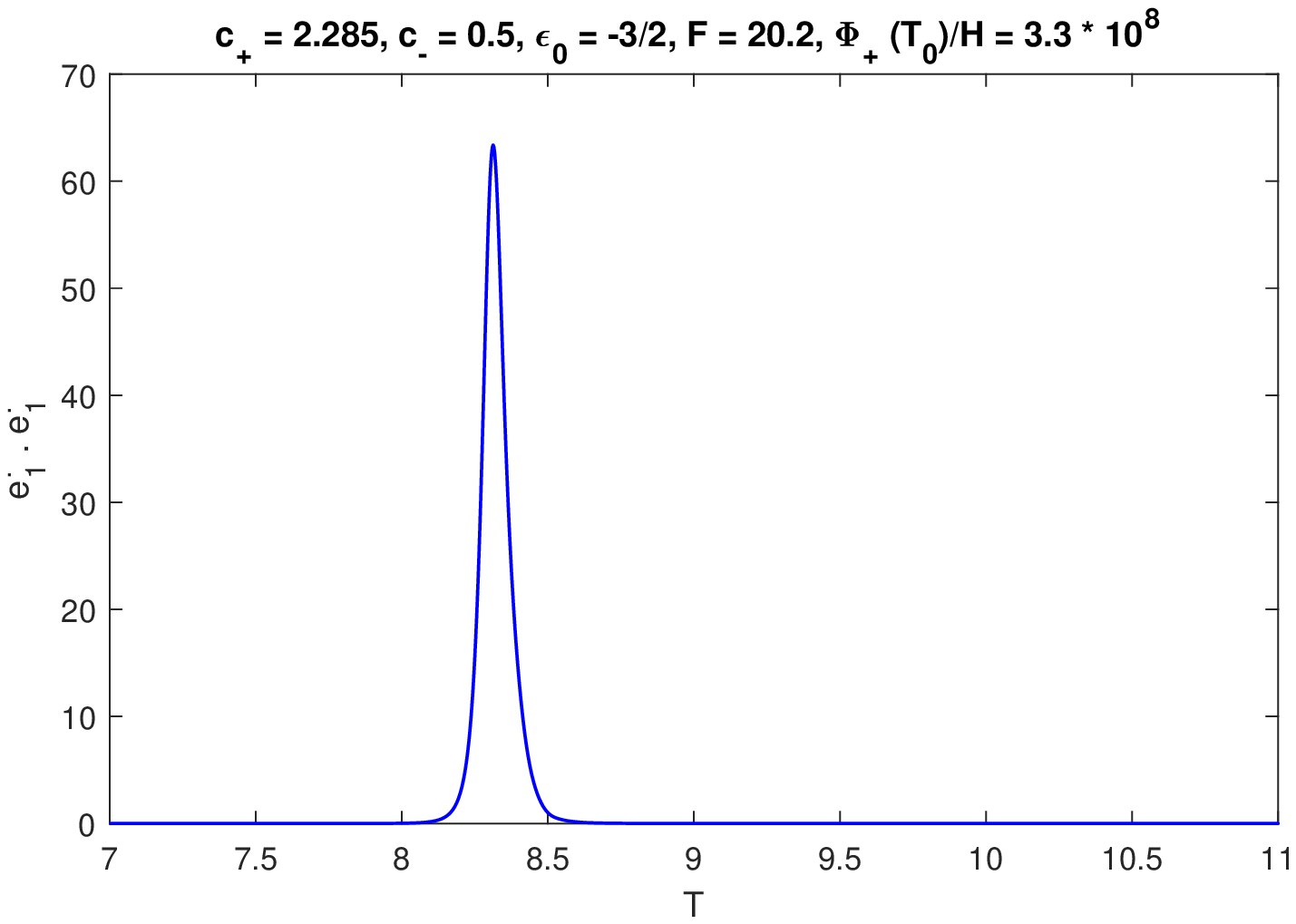} 
\par\end{centering}
\caption{\label{fig:e1primeplots}Plots depicting peaks of $\dot{e}{}_{1}\cdot\dot{e}{}_{1}$
(Eq.~(\ref{e1primesq})) for $c_{+}=2.35,\epsilon_{0}=0$ (left)
and $c_{+}=2.285,\epsilon_{0}=-3/2$ (right). All the parameters have
been set to the $P_{A}$ set used in Fig.~\ref{fig:Comparison-of-Taylor}.}
\end{figure}

For $\alpha\gtrsim\alpha_{2}$ cases, the $-V_{3}$ dip within our mass-model can
be evaluated using Eq.~(\ref{e1primesq}) wherein the peak amplitude
of the $\dot{e}{}_{1}\cdot\dot{e}{}_{1}$ function around $T_{3}$
is evaluated by substituting $\phi_{\pm s}$ into $g\approx\phi_{-}^{2}-\phi_{+}^{2}$
using solution provided in Eq\@.~(\ref{eq:phims_sol}) and Eq.~(\ref{eq:flatdirconst}).
Meanwhile, the maximum amplitude of these peaks located close to the
zeros of $g$ requires evaluation of $\dot{g}$ as observed in Eq.~(\ref{e1primesq}).
In terms of the IR and UV components, we rewrite $\dot{g}$ as
\begin{align}
\dot{g} & \approx\dot{g}_{s}+\dot{g}_{sf}+\dot{g}_{fs}+\dot{g}_{f}
\end{align}
where we identify $\dot{g}_{s}\approx2\left(\phi_{-s}\dot{\phi}_{-s}-\phi_{+s}\dot{\phi}_{+s}\right)$,
$\dot{g}_{f}\approx2\left(\phi_{-f}\dot{\phi}_{-f}-\phi_{+f}\dot{\phi}_{+f}\right)$,
$\dot{g}_{fs}\approx2\left(\phi_{-f}\dot{\phi}_{-s}-\phi_{+f}\dot{\phi}_{+s}\right)$
and $\dot{g}_{sf}\approx2\left(\phi_{-s}\dot{\phi}_{-f}-\phi_{+s}\dot{\phi}_{+f}\right)$.
Using Eqs.~(\ref{eq:phiUVcomponent}), one can show that the mixed
terms $\dot{g}_{fs}$ and $\dot{g}_{sf}$ cancel out due to an accidental
symmetry $\phi_{\pm}\rightarrow-\phi_{\pm}$ that exists in the potential
governing $\phi_{\pm}$, while the amplitude of the first derivative
of the UV component is given as 
\begin{equation}
\left.\dot{g}_{f}\right|_{T\sim T_{3}}\approx\left.\partial_{T}\left(\phi_{-f}^{2}-\phi_{+f}^{2}\right)\right|_{T\sim T_{3}}\approx\left(\frac{A^{2}e^{-3(T_{3}-T_{c})}}{\Omega_{s}^{3}}\right)g_{s}+\left(\frac{Ae^{-3/2(T_{3}-T_{c})}}{\Omega_{s}^{2}}\right)^{2}\dot{g}_{s}.
\end{equation}
Up to a linear order in Taylor expansion, we can approximate $g_{s}$
in the vicinity of $T_{3}$ as 
\begin{equation}
g_{s}(T)\approx\dot{g}_{s}\left(T_{3}\right)\left(T-T_{3}\right).
\end{equation}
Therefore, including the additional contributions from the UV term
$g_{f}$, we can approximate $\left(\dot{e}_{1}^{2}\right)_{T\sim T_{3}}$
peak in the vicinity of $T_{3}$ using 
\begin{equation}
\dot{g}(T\sim T_{3})\approx\dot{g}_{s}(T_{3})\left(1+\frac{A^{2}e^{-3(T_{3}-T_{c})}}{\Omega_{s}^{4}}\left(1+\Omega_{s}\left(T-T_{3}\right)\right)\right)
\end{equation}
within Eq. (\ref{e1primesq}). Through fitting, we find that up to
a $20\%$ error, the above evaluation procedure can be approximated
by the following simplified expression: 
\begin{align}
\left(\dot{e}_{1}^{2}\right)_{T=T_{j}} & \approx\begin{cases}
\left(\dot{e}_{1}^{2}\right)_{\max}e^{-3(T_{j}-T_{c})} & \frac{Ae^{-3/2(T_{j}-T_{c})}}{2F^{2}}>0.15\\
\left(\dot{e}_{1}^{2}\right)_{\max}\left(\frac{\dot{g}_{s}\left(T_{j}\right)}{\dot{g}_{s}\left(T_{c}\right)}\right)^{2} & \frac{Ae^{-3/2(T_{j}-T_{c})}}{2F^{2}}<0.15
\end{cases}
\end{align}
where $T_{j}$ refers to the time when the two background field cross
each other again after $T_{c}$ and $\dot{g}_{s}\left(T_{c}\right)$
is evaluated from the cubic polynomial solution for the background
fields in Eqs.~(\ref{eq:phippoly}) and (\ref{eq:phimpoly}). These
peaks are lower in magnitude due to the Hubble friction as shown in Fig.~\ref{fig:e1primeplots}.

Next we note that Eq.~\ref{eq:e1primemax-1} can be qualitatively
understood by the following factorization 
\begin{equation}
\dot{e}_{1}\cdot\dot{e}_{1}\sim\sum_{nm}d_{nm}(\phi_{+},\phi_{-})\frac{\dot{\phi}_{n}\dot{\phi}_{m}}{F^{2}}
\end{equation}
where the coefficients $d_{nm}$ are dimensionless and order $O(1)$.
The width $\Delta$ of a $\dot{e}_{1}\cdot\dot{e}_{1}$ peak is characterized
by 
\begin{equation}
\Delta\propto\frac{\Delta\phi_{{\rm max}}}{\dot{\phi}_{{\rm max}}}
\end{equation}
where $\phi_{\max}$ is the field at the maximum of $\dot{e}_{1}\cdot\dot{e}_{1}$
which gives us 
\begin{equation}
\frac{\Delta\phi_{{\rm max}}}{\dot{\phi}_{{\rm max}}}\sim\frac{\Delta\phi_{{\rm max}}\sqrt{{\rm larger\,\,eigenvalue}d_{nm}(\phi_{+},\phi_{-})}}{\sqrt{F^{2}\dot{e}_{1}\cdot\dot{e}_{1}}}.
\end{equation}
We shall therefore consider the following repatriation for the width
$\Delta$ 
\begin{equation}
\Delta\approx\frac{r_{\Delta}}{\sqrt{\dot{e}_{1}\cdot\dot{e}_{1}}}\label{eq:e1primesq-width}
\end{equation}
where $r_{\Delta}$ takes the following form: 
\begin{equation}
r_{\Delta}\approx\frac{\Delta\phi_{{\rm max}}\sqrt{{\rm larger\,\,eigenvalue}d_{nm}(\phi_{+},\phi_{-})}}{F}.
\end{equation}
Therefore, $r_{\Delta}$ has parametric dependence on $\Delta\phi_{{\rm max}}$
and $\sqrt{{\rm larger\,\,eigenvalue}d_{nm}(\phi_{+},\phi_{-})}$.
As $\alpha$ increases, $\Delta\phi_{{\rm max}}$ must also increase
due to increasing resonance amplitude. On the other hand, the corresponding
value of $\phi_{-}$ at the location of the peak reduces monotonically
with an increasing $\alpha$ which results in a smaller $\sqrt{{\rm larger\,\,eigenvalue}d_{nm}(\phi_{+},\phi_{-})}$.
Within the parametric region ($0.25\lesssim\alpha\lesssim1$), the
two competing behaviors are equally important and tend to cancel each
other out such that within the region of our interest we can approximate
$r_{\Delta}\approx0.72$ so that the width of the $\dot{e}_{1}\cdot\dot{e}_{1}$
peaks are approximately given as $\Delta\approx0.72/\sqrt{\dot{e}_{1}\cdot\dot{e}_{1}}$
within $10\%$ accuracy. Using Eq.~(\ref{eq:e1primesq-polyfit})
for the $(\dot{e}{}_{1}.\dot{e}{}_{1})_{{\rm max}}$ into Eq.~(\ref{eq:e1primesq-width}),
we obtain the following linear expression in $\alpha$ for the width
of the first $(\dot{e}_{1}\cdot\dot{e}_{1})$ dip 
\begin{equation}
\Delta_{(\dot{e}_{1})_{\max}^{2}}\approx\frac{\left(2.93-1.86\alpha\right)}{F}.\label{eq:e1prime-width-linearfit}
\end{equation}
In general, the width of these Gaussian-like peaks can also be evaluated
by taking the ratio of the total area under the peak to its maximum
amplitude.

\section{\label{sec:Lighter-mass-eigenvalue}Lighter mass eigenvalue $m_{1}^{2}$}

In this section, we will study the variation of lightest mass eigenvalue
over time. The lightest eigenvalues of the mass matrix $\tilde{M}^{2}$
is given by the expression
\begin{align}
m_{1}^{2} & =\frac{\tilde{M}_{11}^{2}+\tilde{M}_{22}^{2}}{2}\left(1-\sqrt{1+4\frac{F^{4}-\tilde{M}_{11}^{2}\tilde{M}_{22}^{2}}{(\tilde{M}_{22}^{2}+\tilde{M}_{11}^{2})^{2}}}\right).\label{eq:lightesteig}
\end{align}
In terms of $\Omega^{2}=\tilde{M}_{11}^{2}+\tilde{M}_{22}^{2}$ and
$\xi=\phi_{+}\phi_{-}-F^{2}$ we rewrite Eq.~(\ref{eq:lightesteig})
as
\begin{equation}
m_{1}^{2}=\frac{\Omega^{2}}{2}\left(1-\sqrt{1-4\left(\frac{\xi}{\Omega^{2}}\right)^{2}-8\frac{F^{2}}{\Omega^{2}}\left(\frac{\xi}{\Omega^{2}}\right)-4\frac{c_{+}\phi_{+}^{2}+c_{-}\phi_{-}^{2}+c_{-}c_{+}}{\Omega^{4}}}\right).\label{eq:m1eqn1}
\end{equation}
During the early-phase when $|\phi_{+}|\gg F$ and any transient oscillations
$\phi_{-}^{\mathrm{transient}}$ of $\phi_{-}$ are negligible i.e.
$\phi_{-}^{\mathrm{transient}}\ll\phi_{-}^{(0)}$, $m_{1}^{2}$ reduces
to,
\begin{align*}
m_{1}^{2} & \sim\frac{1}{2}\left(\tilde{M}_{11}^{2}+\tilde{M}_{22}^{2}-(\tilde{M}_{22}^{2}-\tilde{M}_{11}^{2})-\frac{2F^{4}}{\tilde{M}_{22}^{2}-\tilde{M}_{11}^{2}}\right),\\
 & \sim c_{+}+O\left(F^{2}\frac{\phi_{-}^{\mathrm{transient}}}{\phi_{+}}\right).
\end{align*}

As the fields then approach transition, the zero-order approximate
perturbative solution $\phi_{\pm}^{(0)}$ is no more valid. The lighter
mass eigenvalue transitions from $c_{+}$ to a negative dip. Post
this dip, $m_{1}^{2}$ oscillates due to the resonance before settling
down to zero. As explained in Appendix \ref{sec:UV-and-IR-phi-fields},
the background fields post $-V_{1}$ dip can be factorized in terms
of the UV and IR components. Using these components we can write $\Omega^{2}$
as 
\begin{align}
\Omega^{2} & \approx\phi_{+}^{2}+\phi_{-}^{2}\nonumber \\
 & \approx\phi_{+s}^{2}+\phi_{-s}^{2}+\phi_{+f}^{2}+\phi_{-f}^{2}+2\phi_{+s}\phi_{+f}+2\phi_{-s}\phi_{-f}\nonumber \\
 & \approx\Omega_{s}^{2}+\frac{4F^{2}}{\Omega_{s}^{2}}\xi+O\left(\Omega_{f}^{2}\right)
\end{align}
such that we can factor out the resonant UV oscillations over the
slow-varying $\Omega_{s}^{2}$ background.

Next we substitute the above factorization for $\Omega^{2}$ into
Eq.~(\ref{eq:m1eqn1}) and expand $m_{1}^{2}$ in powers of $\delta=\xi/\Omega_{s}^{2}$,
\begin{align*}
m_{1}^{2} & \approx\frac{c_{-}\phi_{-}^{2}+c_{+}\phi_{+}^{2}}{\Omega_{s}^{2}}+2F^{2}\left[\delta+\left(\frac{\Omega_{s}^{2}}{2F^{2}}-\frac{2F^{2}}{\Omega_{s}^{2}}\right)\delta^{2}+\frac{1}{2}\left(\frac{\Omega_{s}}{F}-\frac{4F^{3}}{\Omega_{s}^{3}}\right)^{2}\delta^{4}+O\left(\delta^{6}\right)\right].
\end{align*}
Within the parametric region of interest where $\left|\delta\right|\ll1$,
we drop all terms of $O\left(\delta^{3}\right)$ and higher such that
\begin{align}
m_{1}^{2}(T) & \approx2F^{2}\left[\delta+\left(\frac{\Omega_{s}^{2}}{2F^{2}}-\frac{2F^{2}}{\Omega_{s}^{2}}\right)\delta^{2}\right]+\frac{c_{-}\phi_{-}^{2}+c_{+}\phi_{+}^{2}}{\Omega_{s}^{2}}.\label{eq:mass_mL2}
\end{align}
The above expression highlights that the mass eigenvalue thus oscillates
in-tandem with the flat-deviation oscillations before settling down
to zero. The expansion leading to Eq.~($\ref{eq:mass_mL2}$) is appropriate
as long as the term $\xi/(\phi_{-s}^{2}+\phi_{+s}^{2})\lesssim O(0.2)$.
From Eq.~(\ref{eq:Aapprox}) we infer that $\left|\xi\right|\lesssim F^{2}$
for $\alpha\leq1$, meanwhile $\phi_{-}(T_{2})\approx2F$ and hence
the expansion term $\xi/(\phi_{-s}^{2}+\phi_{+s}^{2})$ tends to $\sim0.2$
as $\alpha\rightarrow1\equiv\alpha_{{\rm U}}$. To the contrary, if
the two fields tend to $O(1)$ momentarily, the above expansion breaks
down. However, the lighter mass eigenvalue in those conditions tends
to $-F^{2}$. This situation arises when the trajectory of the two
fields tends to be chaotic and unstable which is outside the scope
of our parametric region.

\section{\label{sec:Justification-for-the}Building the numerical model}

In Appendix \ref{sec:Lighter-mass-eigenvalue} we obtained an approximate
expression for the lighter mass eigenvalue in the limit $\xi/\Omega_{s}^{2}\ll1$.
Next we substitute for the flat-deviation $\xi$ from Eq.~\ref{eq:flatdevfn}
into the Eq.~\ref{eq:mass_mL2} 
\begin{equation}
m_{1}^{2}(T)\approx2F^{2}\left[\frac{\xi}{\Omega_{s}^{2}}+\left(\frac{\Omega_{s}^{2}}{2F^{2}}-\frac{2F^{2}}{\Omega_{s}^{2}}\right)\left(\frac{\xi}{\Omega_{s}^{2}}\right)^{2}\right]+m_{B}^{2}(T)\qquad\forall\frac{\xi}{\Omega^{2}}\ll1\label{eq:m1eqn2}
\end{equation}
where the slow-varying background term $-(c_{+}+c_{-})F^{2}/\Omega_{s}^{2}$
of $\xi$ from Eq.~(\ref{eq:flatdevfn}) has been absorbed within
$m_{B}^{2}(T)$ such that
\begin{equation}
m_{B}^{2}(T)\approx\frac{c_{-}\phi_{-s}^{2}+c_{+}\phi_{+s}^{2}}{\phi_{-s}^{2}+\phi_{+s}^{2}}-\frac{2F^{4}(c_{+}+c_{-})}{\left(\phi_{-s}^{2}+\phi_{+s}^{2}\right)^{2}}\label{eq:mB2}
\end{equation}
where $\phi_{\pm s}$ are the IR components of the $\phi_{\pm}$ fields
as given in Appendix \ref{sec:UV-and-IR-phi-fields}. We identify
$m_{B}^{2}(T)$ as a low-frequency axion mass of order $O\left(c_{-}\right)$
that eventually tends to zero when the background fields settle to
their respective minima. In this context, the effective axion mass
post transition comes from physics at two different frequency/energy
scales. In order to obtain an analytically solvable model, it is convenient
to integrate out the high frequency terms and obtain an effective
low frequency model in terms of the IR components. We begin by redefining
$\xi$ after the first dip as follows: 
\begin{equation}
\xi\approx Ae^{-\frac{3}{2}(T-T')}\sin\left(\int_{T'}^{T}\Omega_{s}(t)dt\right)
\end{equation}
where $T'\approx T_{c}+O(1/F)$ is the approximate time when the flat-deviation
$\xi$ first crosses zero after the initial $O\left(-F^{2}\right)$
dip. Thus, up to a quadratic expansion in $\xi/\Omega_{s}^{2}$, $m_{1}^{2}$
in Eq.~(\ref{eq:m1eqn2}) has the following harmonic expansion 
\begin{align}
m_{1}^{2} & \sim O\left(\sin\left(f\right)\right)+O\left(\sin^{2}\left(f\right)\right)\\
 & \sim O\left(\sin\left(f\right)\right)+O\left(1-\cos\left(2f\right)\right)
\end{align}
where $f=\int_{T'}^{T}\Omega_{s}(t)dt$. Applying the adiabatic approximation
method elucidated in Appendix \ref{sec:Adiabatic-approximation-for}
we integrate out the UV degree of freedom in $m_{1}^{2}$ to yield
\begin{align}
m_{1}^{2} & \approx\left(\frac{2F^{2}}{\Omega_{s}^{2}}\right)^{2}\frac{1}{2\Omega_{s}^{2}}A^{2}e^{-3(T-T')}+\frac{1}{\Omega_{s}^{2}}\left(1-\frac{4F^{4}}{\Omega_{s}^{4}}\right)\frac{A^{2}}{2}e^{-3(T-T')}+m_{B}^{2}(T)\qquad\forall\alpha\lesssim1\label{eq:m1eqn3}
\end{align}
where the first term in the above expression comes from the IR reduction
of the term linear in $\xi$ in Eq.~(\ref{eq:m1eqn2}) while the
second term is a positive offset due to the $O\left(\sin^{2}\left(f\right)\right)$
term from the $\xi^{2}$ quadratic term in Eq.~(\ref{eq:m1eqn2}).
Moreover, while the first term in Eq.~(\ref{eq:m1eqn3}) results
in a modification of the initial conditions at $T'$, the second term
does not. An important consequence of the IR reduction in Eq.~(\ref{eq:m1eqn3})
is that the contribution from the linear term cancels out. This cancellation
of the first order UV contribution is similar to the one we observed
and explained in Appendix \ref{sec:Lightest-eigenvector}. Thus, the
adiabatic reduction approximates the oscillating time space potential
in $m_{1}^{2}$ to a nonoscillating exponentially decaying positive
time space potential
\begin{align}
m_{1}^{2} & \approx e^{-3(T-T')}\beta^{2}\frac{A^{2}}{2}+m_{B}^{2}(T)\qquad\forall\alpha\lesssim1\label{eq:reducedpot}
\end{align}
where the prefactor 
\begin{equation}
\beta^{2}\approx\frac{1}{\Omega_{s}^{2}}.
\end{equation}
Since $\Omega_{s}^{2}$ is a time-varying function, we evaluate an
average value for $\beta^{2}$ within a half-oscillation of the background
fields using $e^{-3(T-T')}$ as the weighing function.\footnote{Note that this averaging of $\beta^{2}$ is different from the procedure
we carried out in Appendix \ref{sec:UV-and-IR-phi-fields} since there
we were concerned with a non-linear differential equation of the background
fields.} Using the analytic solutions for the background fields in Appendix
\ref{sec:UV-and-IR-phi-fields}, we obtain the following approximate
empirical fit expression for average $\left\langle \beta^{2}\right\rangle $:
\begin{align}
\left\langle \beta^{2}\right\rangle  & \approx F^{-2}\left(0.139+\frac{.14}{1.08+\exp\left(11\left(\alpha-0.72\right)\right)}\right).
\end{align}
Note that the above estimation is approximate and hence one of the
sources of uncertainty for all modes super-horizon at $T_{2}$. With
this approximation, the system is now analytically solvable where
the reduced low frequency mass-model has a jump ETSP $V_{2}$ defined
as 
\begin{equation}
e^{-3(T-T')}\frac{V_{2}}{2}\equiv e^{-3(T-T')}\left\langle \beta^{2}\right\rangle \frac{A^{2}}{2}
\end{equation}
at $T=T'\approx T_{2}$. Within the framework of this adiabatic reduction,
the $y_{1}$ axion mode function is expressed in terms of its slow
(IR) and fast (UV) components, 
\begin{equation}
y_{1}=y_{1s}+y_{1f}
\end{equation}
where we will neglect the $y_{1f}$ component since $\xi/\Omega_{s}^{2}\ll1$
(within the scope of the adiabatic reduction). The effective mass
squared ($m_{1}^{2}-\dot{e}_{1}\cdot\dot{e}_{1})$ for $y_{1s}$ mode
function is now generalized in terms of a low-frequency mass model
$m^{2}$ with a reduced mass eigenvalue in Eq.~(\ref{eq:reducedpot})
and the $-\dot{e}_{1}\cdot\dot{e}_{1}$ dips (explained in Appendix
\ref{sec:Lightest-eigenvector}) modeled as negative square wells/dips.

The first dip at the transition $T_{c}$ is obtained through a superposition
of the first $\dot{e}_{1}\cdot\dot{e}_{1}$ dip and a corresponding
dip due to the evolution of the mass eigenvalue from $c_{+}$ to an
oscillating function (due to a strong resonance between the $\phi_{\pm}$
fields). This explains the first $V_{1}$ dip of our model given in
Eqs.~(\ref{eq:model}) and (\ref{eq:e1prsqdip}). After this first
negative dip of $O\left(F^{2}\right)$, the effective mass squared
is governed by the exponentially decaying positive function $e^{-3(T-T_{2})}V_{2}/2$
of $O(F^{2}/20)$. Once the $V_{2}$ mass squared function decays
away, the parameter $V_{B}$ of $O\left(c_{-}\right)$ evaluated as
an average of the $m_{B}^{2}(T)$ defines the asymptotic behavior
of the $y_{1s}$ mode amplitude. The dynamics of $m_{B}^{2}$ function
and its effect on the mode amplitude is covered in Appendix \ref{sec:mB2}.

\section{\label{sec:mB2}Effective mass squared function $m_{B}^{2}$}

\begin{figure}
\begin{centering}
\includegraphics[scale=0.8]{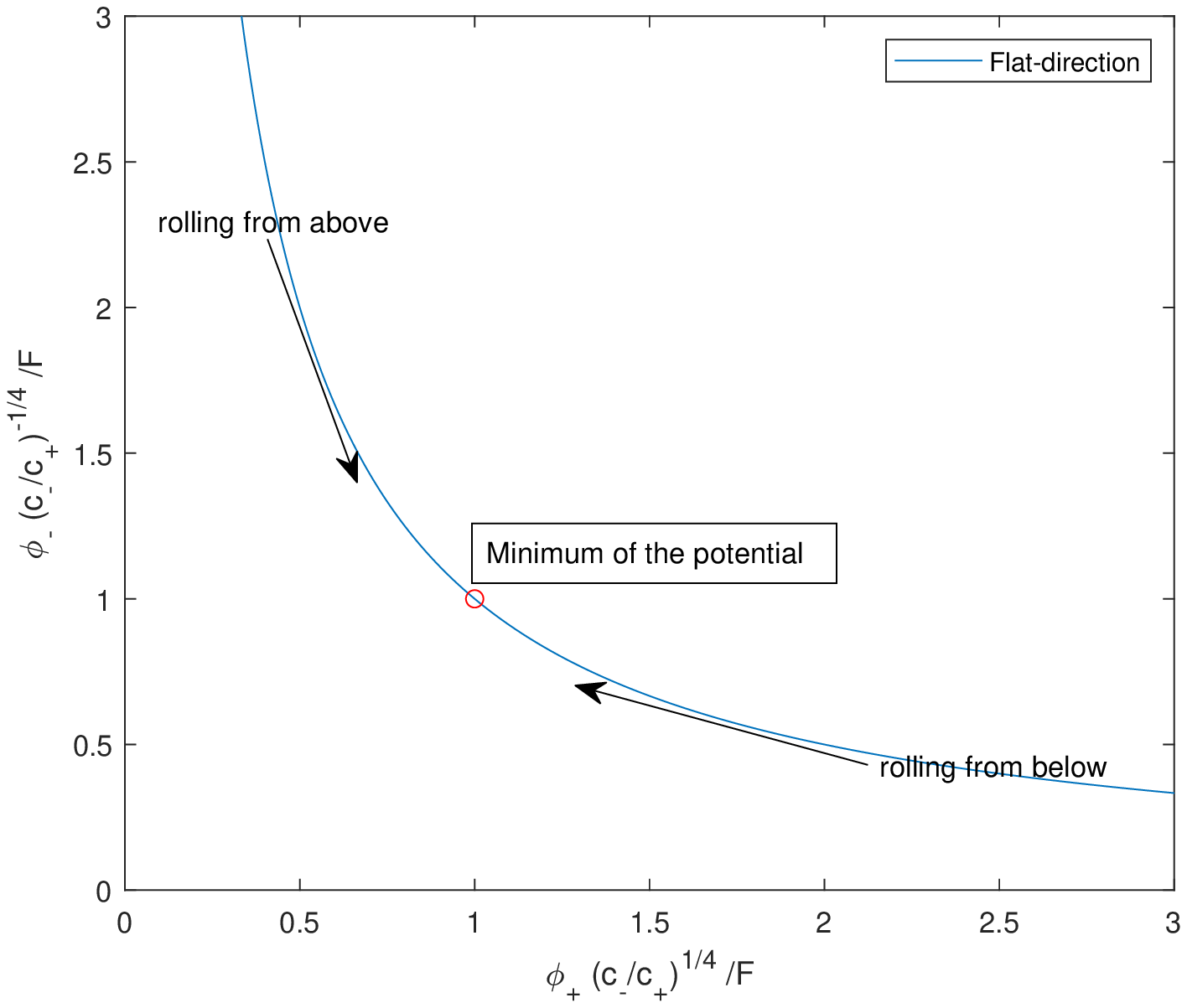} 
\par\end{centering}
\begin{centering}
\includegraphics[scale=0.49]{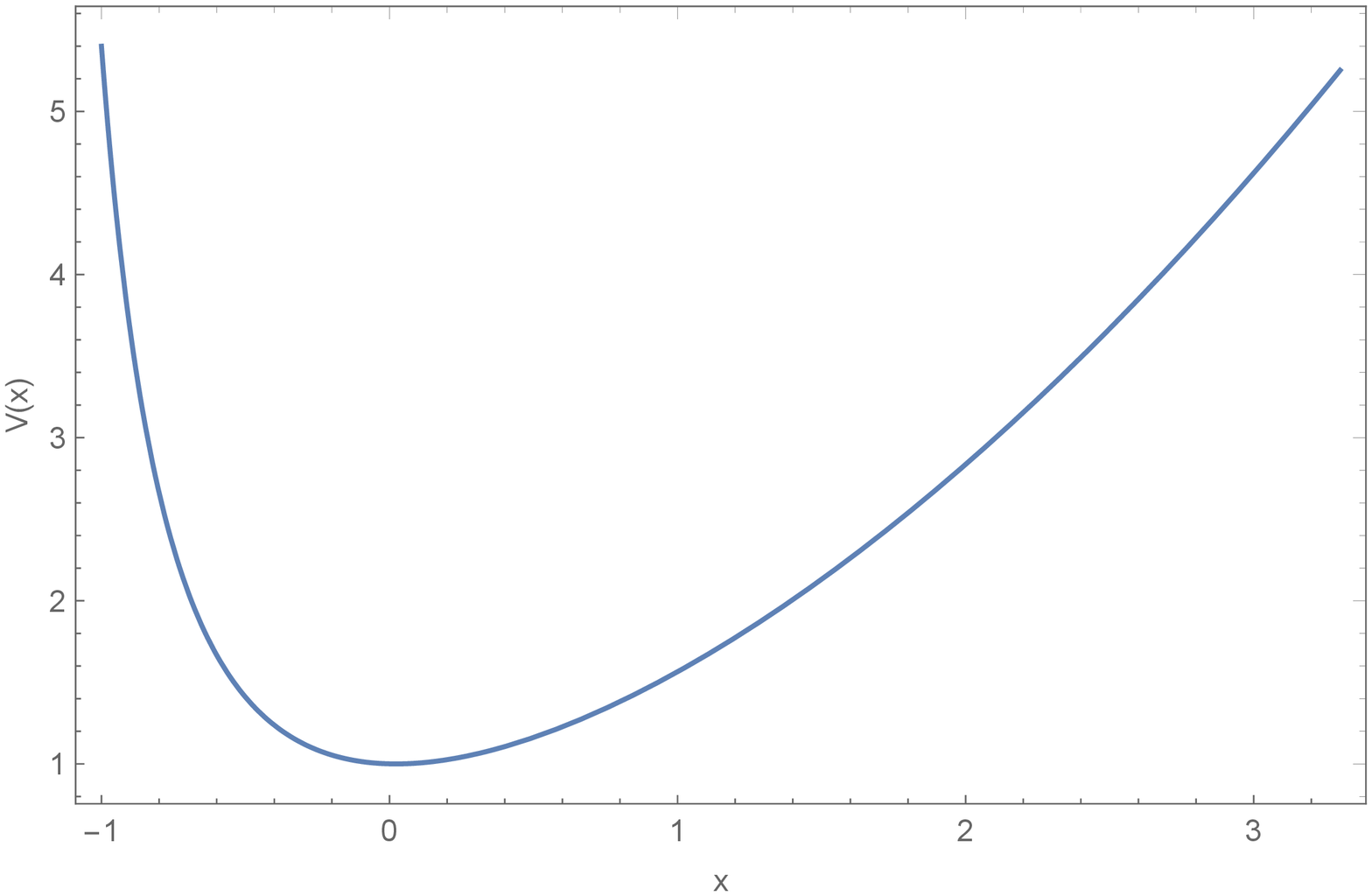} 
\par\end{centering}
\caption{\label{fig:flat-direction} In the first plot we show the two directions
from which the fields can settle to the minimum of the potential while
moving along the flat direction. In the second plot, the normalized
potential $V\left(\phi_{\pm}\right)$ is parameterized along the flat
direction via parameter $x$ for some fiducial $c_{\pm}$ such that
$c_{-}\ll9/4<c_{+}$. The minimum of the potential occurs at $x=0$
where $x>0$ corresponds to movement of the fields from above. We
note that the potential has a large gradient when moving from below
and falls to the minimum rapidly.}
\end{figure}

The effective axion mass after IR averaged $m_{y_{1}}^{2}$ makes
a positive jump transition at time $T_{2}$ (see Fig.~\ref{fig:Schematic-massmodel})
is derived from physics at two different frequency/energy scales.
These are specified by under-damped $O(F)$ oscillation of the lightest
mass eigenvalue and slowly varying part of the lightest mass squared
eigenvalue function $m_{B}^{2}$ given by Eq.~(\ref{eq:mB2}). In
this appendix, we will discuss the function $m_{B}^{2}$ in detail
and evaluate its effect on $y_{1}$ mode amplitude. From Eq.~(\ref{eq:mB2})
we note that dynamics of $m_{B}^{2}$ is characterized by the motion
of IR components of the fields $\phi_{\pm}$ along the flat direction
towards the minimum of the potential. Along this direction, the fields
can move towards the minimum either from above ($\phi_{-s}>\phi_{-\min}$)
or below ($\phi_{-s}<\phi_{-\min}$) as shown in Fig.~\ref{fig:flat-direction}.
When the fields have settled to their respective minimum, $m_{B}^{2}$
goes to zero. Interestingly then, if the $\phi_{-}$ field settles
from above the minimum then the condition $m_{B}^{2}>0$ is satisfied,
while if it moves from below, then $m_{B}^{2}<0$ subject to a few
conditions.

\begin{figure}
\begin{centering}
\includegraphics[scale=0.65]{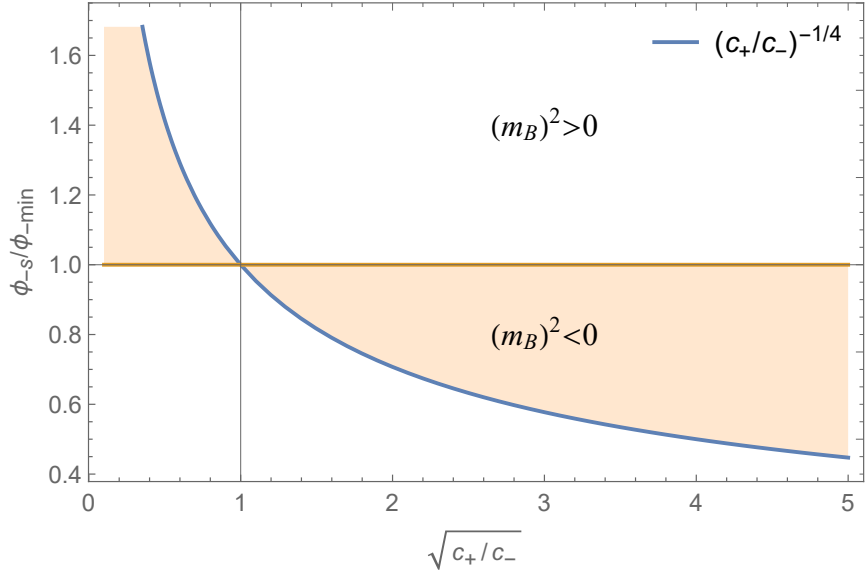} 
\par\end{centering}
\caption{\label{fig:mB2plot}Plot highlighting regions in $\left\{ \sqrt{c_{+}/c_{-}},\phi_{-s}/\phi_{-\min}\right\} $
parametric space where $m_{B}^{2}>0$ or $m_{B}^{2}<0$ corresponding
to the expression in Eq.~(\ref{eq:mB2-plotexpression}).}
\end{figure}

We shall now study this behavior starting from the expression for
$m_{B}^{2}$ of Eq.~(\ref{eq:mB2}) which can be rearranged to get
\begin{align}
m_{B}^{2} & \approx\frac{\left(\sqrt{c_{-}}\phi_{-s}^{2}+\sqrt{c_{+}}\phi_{+s}^{2}\right)^{2}-F^{4}\left(\sqrt{c_{+}}+\sqrt{c_{-}}\right)^{2}}{\left(\phi_{-s}^{2}+\phi_{+s}^{2}\right)^{2}}.\label{eq:mbsqintermed}
\end{align}
Next we parameterize 
\begin{equation}
\phi_{-s}(T)=n(T)\,\phi_{-\min}
\end{equation}
where $n(T)>1$ for $\phi_{-}$ moving from above and $n(T)<1$ when
$\phi_{-}$ moves from below. In terms of the function $n(T)$, Eq.~(\ref{eq:mbsqintermed})
becomes
\begin{align}
m_{B}^{2} & \propto\left(\sqrt{c_{+}/c_{-}}n^{2}+n^{-2}\right)^{2}-\left(\sqrt{c_{+}/c_{-}}+1\right)^{2}.
\end{align}
Hence, we find the following conditional expression for $m_{B}^{2}$
sign for $c_{+}>9/4\geq c_{-}$ 
\begin{equation}
\mathrm{sgn}\left(m_{B}^{2}\right)=\begin{cases}
-1 & \left(\frac{c_{-}}{c_{+}}\right)^{1/4}<\frac{\phi_{-s}}{\phi_{-\min}}<1\\
1 & {\rm otherwise}
\end{cases}\label{eq:mB2-plotexpression}
\end{equation}
as mapped in Fig.~\ref{fig:mB2plot}.

Therefore, dependent on the trajectory of $\phi_{-s},$ a prolonged
exponential decay or amplification of the super-horizon axion modes
behaves as 
\begin{equation}
y_{1}\propto\exp\left[\int dT\left(\frac{-3}{2}+\frac{3}{2}\sqrt{1-\frac{4}{9}m_{B}^{2}}\right)\right]\approx\exp\left[-\frac{1}{3}\int dTm_{B}^{2}\right]
\end{equation}
where the integral in the exponent is cut off when $m_{B}^{2}$ decays
faster than $1/T$. The lower limit of this integral corresponds to
when the exponentially decaying term proportional to $V_{2}$ in Eq.~(\ref{eq:model})
can be neglected.

In situations where $c_{-}\ll9/4<c_{+}$, the magnitude of the gradient
of the potential is much smaller when $\phi_{-s}>\phi_{-\min}$ than
when $\phi_{-s}<\phi_{-\min}$. Therefore, the fields evolve slowly
($O(1)$ time scale in $T$-dependent evolution) in the former scenario
and fall towards the minimum rapidly in the latter (see Fig.~(\ref{fig:flat-direction})).
Accordingly, the axion background mass is significant in the latter
case only when the fields are close to the minimum.\footnote{If $c_{-}>9/4$, then there is no appreciable $m_{B}^{2}$ since the
mass squared function time dependence due to the fast rolling fields
rapidly diminishes the magnitude of this function.}

When the jump ETSP is significant at $T_{2}$, it leads to an $O(0.1F)$
frequency oscillations of the IR fields (see appendix \ref{sec:UV-and-IR-phi-fields}).
These oscillations cause the fields to cross each other at least twice
after the transition. To see this, first note that the resonant conditions
force $\phi_{+s}$ to cross $\phi_{-s}$ (first crossing) to make
$\phi_{+s}>\phi_{-s}$. Next, the asymptotic values of these fields
are $\phi_{\pm{\rm min}}$ where $\phi_{-\min}>\phi_{+\min}$ for
$c_{+}>c_{-}$, and this fact requires a second crossing. These crossings
give rise to a second and third $\dot{e}_{1}\cdot\dot{e}_{1}$ peaks
(the first one being at $T_{c}$). Since the crossings occur at approximately
$F$ (due to the initial conditions making $\phi_{\pm s}(T_{c})\sim O(F)$),
we know $\phi_{-s}<\phi_{-{\rm min}}\sim\left(c_{+}/c_{-}\right)^{1/4}F>F$
at the second crossing. For $\alpha<1$, the interaction energy now
decays away due to the Hubble friction while $\phi_{-s}<\phi_{-\min}$
before $\phi_{-s}$ crosses $\phi_{-{\rm min}}$ again. Without the
interactions mediated by the flat deviation $\xi$, the $\phi_{-s}$
settles towards the minimum from below ($\phi_{-s}<\phi_{-\min}$)
and $m_{B}^{2}$ can become significant.

The parametric boundary of when the IR components of the $\phi_{\pm}$
fields cross each other at least once after $T_{2}$ is set by the
following condition provided in Appendix \ref{sec:Lightest-eigenvector}
\[
\frac{3}{F}\approx\frac{J_{1}\left(\sqrt{2}A/(3\bar{\Omega})\right)}{\sqrt{2}A/(3\bar{\Omega})}.
\]
The above provides an approximate minimum value of $\sqrt{2}A/(3\bar{\Omega})$
for the background fields to cross again after $T_{c}$. Substituting
for $A$ using Eq.~(\ref{eq:Aapprox}) and $\bar{\Omega}$ using
Eq.~(\ref{eq:avgomega}), we obtain a cutoff (boundary) in terms
of $\alpha$ defined to be $\alpha_{2}$. Therefore, we will consider
$\alpha<\alpha_{2}$ and $\alpha>\alpha_{2}$ cases separately.

\subsection{$\alpha<\alpha_{2}$}

For the parametric region $\alpha<\alpha_{2}$, the fields will not
cross again for $T>T_{2}$. To solve for this system, it is convenient
to divide the time regions based on whether or not $\sqrt{c_{-}c_{+}}\phi_{+s}$
term is appreciable compared to $c_{-}\phi_{-s}$ term in Eq.~(\ref{eq:phimseqn}).
We will call the region $[T_{2},T_{L}]$ the period when $\sqrt{c_{-}c_{+}}\phi_{+s}$
is negligible and $T>T_{L}$ the period when $\sqrt{c_{-}c_{+}}\phi_{+s}$
is important.

Consider the equation of the $\phi_{-s}$ field in the region $[T_{2},T_{L}]:$
\begin{equation}
\ddot{\phi}_{-s}+3\dot{\phi}_{-s}+c_{-}\phi_{-s}+\frac{A^{2}}{2\bar{\Omega}^{2}}e^{-3\left(T-T_{2}\right)}\phi_{-s}\approx0
\end{equation}
where we can safely neglect the effect from the asymptotic term $\sqrt{c_{-}c_{+}}\phi_{+}$.
The solution to the above equation is given in Eq.~(\ref{eq:phims_sol})
in Appendix \ref{sec:UV-and-IR-phi-fields} while $\phi_{+s}$ is
given by the flat direction condition in Eq.~(\ref{eq:flatdirconst}).

Due to a large positive velocity at $T_{2}$, the $\phi_{-s}$ reaches
a maximum and then moves slowly towards the minimum with an initial
exponential decay rate equal to 
\begin{equation}
\frac{3}{2}\left(1-\sqrt{1-4c_{-}/9}\right)\approx\frac{c_{-}}{3}+O\left(c_{-}^{2}\right).\label{eq:decayfactor-1}
\end{equation}
During this period when $\phi_{+s}/\phi_{-s}\ll1$ we can expand $m_{B}^{2}$
using Eq.~(\ref{eq:mB2}):
\begin{align}
m_{B}^{2} & \approx c_{-}-\frac{2F^{4}\left(c_{+}+c_{-}\right)}{\phi_{-s}^{4}}+\left(c_{+}-c_{-}+\frac{4F^{4}\left(c_{+}+c_{-}\right)}{\phi_{-s}^{4}}\right)\left(\frac{\phi_{+s}}{\phi_{-s}}\right)^{2}\\
 & \approx c_{-}+\delta m_{B}^{2}.
\end{align}
Note that as $T\rightarrow T_{\infty}$, the ratio $\phi_{-s}/\phi_{+s}$
gradually decreases. As the two fields then approach closer to their
respective minima, the interaction term $\xi\phi_{+s}\sim-\sqrt{c_{+}c_{-}}\phi_{+s}$
becomes important starting at time $T_{L}$, and the decay rate changes.
Therefore, the integral corresponding to the exponential decay of
the super-horizon mode amplitude during the first temporal phase where
the $\phi_{-s}$ field has an exponential decay factor given by Eq.~(\ref{eq:decayfactor-1})
is 
\begin{equation}
\int_{\tilde{T}}^{T_{L}}m_{B}^{2}dT\approx c_{-}\left(T_{L}-\tilde{T}\right)+\int_{\tilde{T}}^{T_{L}}\delta m_{B}^{2}dT\label{eq:exponentintegral}
\end{equation}
where 
\begin{equation}
\tilde{T}=\max\{T_{2},T_{V},\mathcal{T}_{K}\}\label{eq:Ttilde}
\end{equation}
and $\mathcal{T}_{K}$ is the time when the $K$-mode becomes super-horizon
i.e. $Ka(\mathcal{T}_{K})=3/2$. Thus, modes that exit the horizon
before transition ($\mathcal{T}_{K}<T_{2}$) have a K-independent
decay factor. The time $T_{V}$ is when the $V_{2}$ jump ETSP has
decayed and becomes negligible compared to $m_{B}^{2}(T)$.

\begin{figure}
\begin{centering}
\includegraphics[scale=0.9]{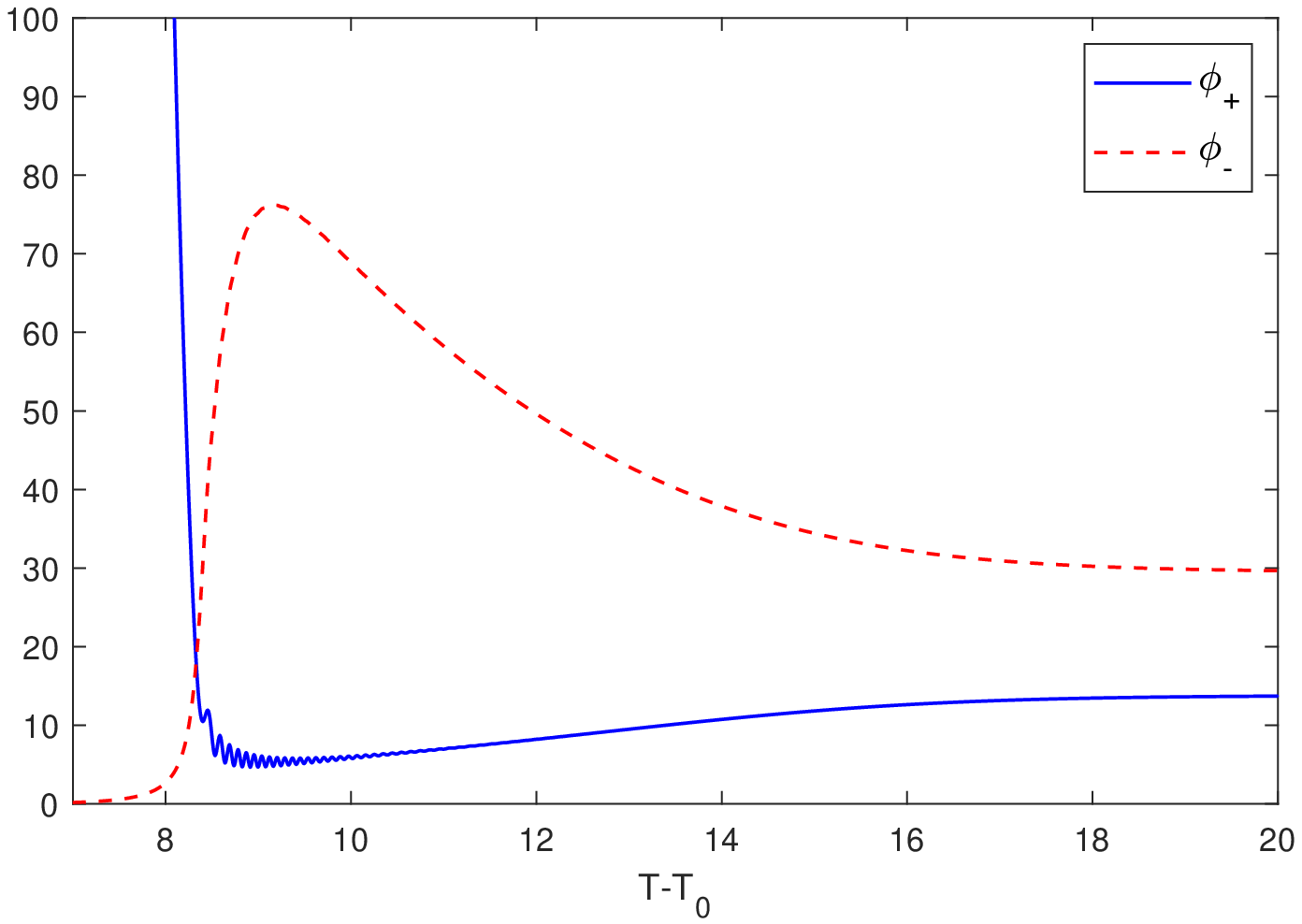} 
\par\end{centering}
\begin{centering}
\includegraphics[scale=0.9]{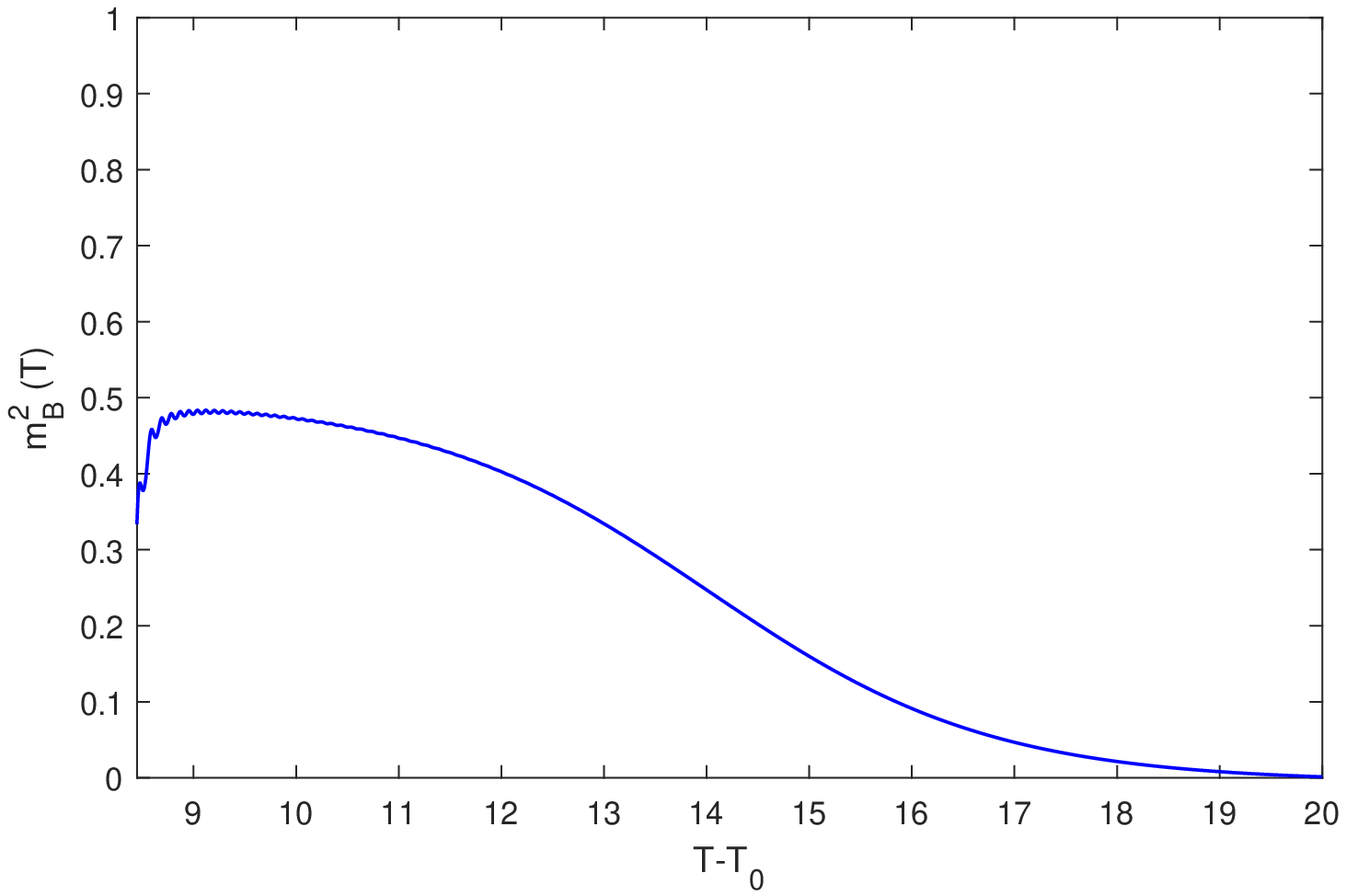} 
\par\end{centering}
\caption{\label{fig:slowroll}Plots of $\phi_{\pm s}$ and $m_{B}^{2}$ for
$\alpha\sim0.51$ where the curves have been computed numerically
by solving Eqs.~(\ref{eq:backgroundeom0}) and (\ref{eq:backgroundeom}).
Notice the slow-roll of the fields past $\phi_{-s}$ maximum. This
slow-roll results in an effective decay of the mode amplitude through
the $O(c_{-})$ axion mass squared function $m_{B}^{2}$ as explained
in the text. In this instance, the approximate mode amplitude decay
through the $\exp[-\frac{1}{3}\int m_{B}^{2}dT]$ factor is $1/2$.}
\end{figure}

Next, we consider the time period $T\in\left[T_{L},T_{\infty}\right]$
where the $\sqrt{c_{+}c_{-}}\phi_{+s}$ term is non-negligible compared
to $c_{-}\phi_{-s}$. By comparing the two terms and using the flat
direction $\phi_{+s}\approx F^{2}/\phi_{-s}$, we make an approximate
choice of $T_{L}$ as when
\begin{equation}
\phi_{-s}(T_{L})\approx\frac{4}{3}\phi_{-\min}
\end{equation}
such that
\begin{align}
\sqrt{c_{+}c_{-}}\phi_{+s}(T_{L}) & \approx0.5c_{-}\phi_{-s}(T_{L})
\end{align}
where additionally we note that at $T_{\infty}$
\begin{align}
\sqrt{c_{+}c_{-}}\phi_{+s}(T_{\infty}) & =c_{-}\phi_{-s}(T_{\infty}).
\end{align}
To derive field equations, we consider the field displacements $\delta\phi_{\pm}$
as
\begin{equation}
\delta\phi_{\pm s}=\phi_{\pm s}-\phi_{\pm\min}\label{eq:tot}
\end{equation}
that implies
\begin{equation}
\delta\phi_{-s}(T_{L})\approx\frac{\phi_{-\min}}{3}\label{eq:phimov3}
\end{equation}
in which case the terms quadratic in $\delta\phi_{\pm s}$ can be
neglected compared to $\phi_{\pm\min}$ with the minima of the fields
located at
\begin{equation}
\phi_{\pm\min}\approx\sqrt{-c_{\mp}+F^{2}\sqrt{\frac{c_{\mp}}{c_{\pm}}}}\,.
\end{equation}
Expand the expressions $c_{\pm}\phi_{\pm s}+\xi\phi_{\mp s}$ in equations
of motion Eqs.~(\ref{eq:backgroundeom0}) and (\ref{eq:backgroundeom})
in terms of $\delta\phi_{\pm s}$ to yield
\begin{align*}
c_{+}\phi_{+s}+\xi\phi_{-s} & \approx\left(c_{+}+\phi_{-min}^{2}\right)\delta\phi_{+s}+\left(2\phi_{+min}\phi_{-min}-F^{2}\right)\delta\phi_{-s}
\end{align*}
\begin{align*}
c_{-}\phi_{-s}+\xi\phi_{+s} & \approx\left(c_{-}+\phi_{+min}^{2}\right)\delta\phi_{-s}+\left(2\phi_{+min}\phi_{-min}-F^{2}\right)\delta\phi_{+s}
\end{align*}
where all terms quadratic in $\delta\phi_{\pm s}$ within $\left[T_{L},T_{\infty}\right]$
have been neglected. Hence, the effective mass matrix in Eqs.~(\ref{eq:backgroundeom0})
and (\ref{eq:backgroundeom}) has the following $T\rightarrow T_{\infty}$
asymptotic form
\begin{equation}
\lim_{T\rightarrow T_{\infty}}\tilde{M}^{2}\rightarrow\left[\begin{array}{cc}
c_{+}+\phi_{-min}^{2} & \,\,\,2\phi_{+min}\phi_{-min}-F^{2}\\
2\phi_{+min}\phi_{-min}-F^{2} & \,\,\,c_{-}+\phi_{+min}^{2}
\end{array}\right]
\end{equation}
with the smallest eigenvalue 
\begin{equation}
\lambda_{\text{min}}=\frac{4c_{-}c_{+}}{c_{-}+c_{+}}+O\left(\frac{1}{F^{2}}\right).
\end{equation}
The field motion is overdamped if $\lambda_{\min}<9/4$ which provides
an upper bound on $c_{-}$, 
\begin{equation}
c_{-}<\frac{9}{16}\left(1-\frac{9}{16c_{+}}\right)^{-1}.
\end{equation}
We will assume that the $c_{-}$ lies within this bound because that
ensures that there will be no second crossing of $\phi_{\pm}$ for
$T>T_{L}$. In terms of the smallest eigenvalue, the $\delta\phi_{\pm}$
field displacements along the approximate flat direction ($\xi\approx-\sqrt{c_{-}c_{+}}$)
can be expressed in the following general asymptotic form 
\begin{align}
\delta\phi_{\pm s} & \approx C_{\pm}e^{-\Lambda T}\label{eq:cmdef}
\end{align}
where 
\begin{align}
\Lambda & \approx\frac{3}{2}-\sqrt{9/4-\lambda_{\min}}
\end{align}
and the constants $C_{+}$ and $C_{-}$ have opposite signs such that
the fields follow the flat direction. Using Eq.~(\ref{eq:phims_sol}),
we can solve Eq.~(\ref{eq:phimov3}) to obtain
\begin{equation}
T_{L}\approx T_{2}-\frac{2/3}{\left(1-n_{1}\right)}\ln\left(\frac{2^{2-n_{1}}\Gamma\left(1-n_{1}\right)\phi_{-\min}}{3c_{2}}\left(\frac{A\sqrt{2}}{3\bar{\Omega}}\right)^{n_{1}}\right)\label{eq:TLeqn}
\end{equation}
and $C_{-}$ of Eq.~(\ref{eq:cmdef}). In the limit $c_{-}\ll1$,
$T_{L}$ in Eq. (\ref{eq:TLeqn}) reduces to 
\begin{equation}
T_{L}\approx T_{2}-\left(\frac{3}{c_{-}}\right)\ln\left(\frac{2\sin\left(\pi n_{1}\right)2^{2-n_{1}}\Gamma\left(1-n_{1}\right)\phi_{-\min}\,x^{n_{1}}}{\pi\left(3\phi_{-s}(T_{2})x\,\partial_{x}J{}_{n_{1}}\left(x\right)+\left(3\phi_{-s}(T_{2})+2\dot{\phi}_{-s}(T_{2})\right)J_{n_{1}}\left(x\right)\right)}\right)_{x=\frac{A\sqrt{2}}{3\bar{\Omega}}}.
\end{equation}
In situations where $\frac{A\sqrt{2}}{3\bar{\Omega}}\ll1$,
\begin{equation}
T_{L}\approx T_{2}-\left(\frac{3}{c_{-}}\right)\ln\left(\frac{4\phi_{-\min}n_{1}}{3\phi_{-s}\left(T_{2}\right)+\dot{\phi}_{-s}\left(T_{2}\right)}\right)
\end{equation}
and thus is independent of resonance term $A$. Combining with Eq.~(\ref{eq:tot}),
the $\phi_{-s}$ field solution is given as 
\begin{align}
\phi_{-s}(T) & \approx\phi_{-\min}\left(1+\frac{1}{3}e^{-\Lambda\left(T-T_{L}\right)}\right)\quad T_{L}\leq T<T_{\infty}
\end{align}
which together with Eq.~(\ref{eq:flatdirconst}) can be used to compute
the second term of Eq.~(\ref{eq:exponentintegral}): 
\begin{align}
\int_{\tilde{T}}^{T_{L}}\delta m_{B}^{2}dT & \approx-\frac{243}{1024}-\frac{73629c_{-}}{131072c_{+}}\nonumber \\
 & +\left(\frac{243-72c_{+}+81c_{-}/c_{+}+8c_{-}\left(2c_{+}-27\right)}{4/3}\right)\frac{F^{4}}{\dot{\phi}_{-s}^{4}(T_{2})}e^{\frac{4c_{-}}{3}\left(\tilde{T}-T_{2}\right)}+O\left(c_{-}^{2}\right)\,.
\end{align}
Next, we extend Eq.~(\ref{eq:exponentintegral}) integral to $T_{\infty}$:
\begin{equation}
\int_{\tilde{T}}^{T_{\infty}}m_{B}^{2}dT\approx c_{-}\left(T_{L}-\tilde{T}\right)+\int_{\tilde{T}}^{T_{L}}\delta m_{B}^{2}dT+\int_{T_{L}}^{T_{\infty}}m_{B}^{2}dT\,.\label{eq:extended}
\end{equation}
In terms of the $\phi_{-s}$ field equations derived above, the last
term is
\begin{equation}
\int_{T_{L}}^{T_{\infty}}\delta m_{B}^{2}dT\approx\frac{7}{12}+\frac{\left(45-14c_{+}\right)c_{-}}{54c_{+}}+O(c_{-}^{2})\hspace{1em},
\end{equation}
and substituting these, Eq.~(\ref{eq:extended}) becomes 
\begin{align}
\int_{\tilde{T}}^{T_{\infty}}m_{B}^{2}dT & \approx\frac{1063}{3072}+\frac{106793c_{-}}{393216c_{+}}+c_{-}\left(T_{L}-\tilde{T}\right)\nonumber \\
 & +\left(\frac{243-72c_{+}+81c_{-}/c_{+}+8c_{-}\left(2c_{+}-27\right)}{4/3}\right)\frac{F^{4}}{\dot{\phi}_{-s}^{4}(T_{2})}e^{\frac{4c_{-}}{3}\left(\tilde{T}-T_{2}\right)}\label{eq:mb2integral}
\end{align}
where $\tilde{T}$ is defined in Eq.~(\ref{eq:Ttilde}). In Fig.~\ref{fig:slowroll},
we give sample plots of $\phi_{\pm s}$ and $m_{B}^{2}$ for $\alpha\sim0.51$
where the curves have been computed numerically by solving Eqs.~(\ref{eq:backgroundeom0})
and (\ref{eq:backgroundeom}).

The above expression in Eq.~(\ref{eq:mb2integral}) is utilized in
Sec.~\ref{sec:features-of-isocurvspectrum} to give an approximate
decay of the super-horizon mode amplitude for cases where $\alpha<\alpha_{2}$.
Alternatively, we define a constant mass-model parameter $V_{B}$
in Eq.~(\ref{eq:model}) as an average value for the time-varying
$m_{B}^{2}$ function. Since $m_{B}^{2}\sim O\left(c_{-}\right)$
is much larger during the first time period $T\in\left[T_{2},T_{L}\right]$
than the second period $T\in\left[T_{L},T_{\infty}\right]$ and also
since $T_{L}-T_{2}\gg T_{\infty}-T_{L}$, we can approximate the $m_{B}^{2}(T)$
function by a mode-independent constant parameter $V_{B}$ during
the entire time interval from $T\in\left[T_{2},T_{L}\right]$ as follows,
\begin{align}
V_{B} & \approx\frac{1}{\left(T_{L}-T_{2}\right)}\int_{T_{2}}^{T_{\infty}}m_{B}^{2}dT\label{eq:constantmB2_eq1}\\
 & \approx c_{-}+\frac{1}{\left(T_{L}-T_{2}\right)}\left(\frac{1063}{3072}+\frac{106793c_{-}}{393216c_{+}}\right).
\end{align}
Therefore, during the time interval from $T_{2}$ to $T_{L}$, the
$y_{1}$ mode equation has the following form for single dip cases
\begin{equation}
\ddot{y_{1}}+3\dot{y}_{1}+\left(K^{2}e^{-2T}+V_{B}+\frac{V_{2}}{2}e^{-3\left(T-T_{2}\right)}\right)y_{1}\approx0
\end{equation}
with the general solution given in Eqs\@.~(\ref{eq:expopot1}) and
(\ref{eq:expopot2}) of Sec.~\ref{subsec:Independent-analytic-functions}.

\subsection{$\alpha>\alpha_{2}$}

For fields with $\alpha>\alpha_{2}$, the $m_{B}^{2}$ function during
the first temporal region is $O\left(c_{-}\right)$ and mostly insignificant
due to the oscillating IR fields. Within the second region, the IR
fields are overdamped and are moving asymptotically towards their
respective minima. Using the solution derived in previous subsection,
the $\phi_{-s}$ field can be expressed as 
\begin{align}
\phi_{-s}(T) & \approx\phi_{-\min}\left(1-\frac{1}{3}e^{-\Lambda\left(T-T_{L}\right)}\right)\quad T_{L}\leq T<T_{\infty}
\end{align}
where the negative $1/3$ factor indicates that the $\phi_{-s}$ field
is settling from below as explained previously. Hence, the $m_{B}^{2}$
function is positive and can lead to mode amplification. In terms
of 
\begin{equation}
n(T)=\left(1-\frac{1}{3}e^{-\Lambda\left(T-T_{L}\right)}\right),
\end{equation}
the $m_{B}^{2}$ function can be expressed as 
\begin{equation}
m_{B}^{2}\left(T\right)\approx\frac{c_{-}c_{+}\left(c_{+}n^{4}-c_{-}\right)\left(-1+n^{4}\right)}{\left(c_{+}n^{4}+c_{-}\right)^{2}}
\end{equation}
which can be used to evaluate the $m_{B}^{2}$ integral from $T_{L}$
to $T_{\infty}$. Note that the integral is independent of the location
of $T_{L}$. Similar to Eq.~(\ref{eq:constantmB2_eq1}), we average
out the $m_{B}^{2}$ integral during the second temporal region within
a time interval from $T_{4}\approx T_{3}+O(1/F)$ to $2/\Lambda$
where $T_{3}$ is defined within the model Eq.~(\ref{eq:model})
and is the time when the background fields cross each other again
after $T_{c}$ since $\alpha>\alpha_{2}$. Meanwhile, $\frac{2}{\Lambda}$
is an approximate time at which the $m_{B}^{2}$ integral is naturally
cutoff where $\Lambda$ is the smallest eigenvalue of the asymptotic
$\tilde{M}^{2}$ effective mass squared matrix.

\section{\label{sec:Decoupling-of-heavy}Decoupling of heavy modes}

In this section, we will estimate the effect of heavy mode mixing
and show that the scalar modes $y_{1,2}$ are effectively decoupled
at transition within the parametric region $\alpha\lesssim1$ such
that the heavier mode $y_{2}$ can be completely neglected within
a 20\% error margin. We begin with the $y_{1,2}$ mode mixing equations
from Eqs.~(\ref{eq:lightmode}) and (\ref{eq:heavymode}) with the
mode-mixing term $S_{ns}$ in the RHS defined in Eq.~(\ref{eq:Sns}),
\begin{align}
\ddot{y}_{1}+\gamma_{1}^{2}y_{1} & =S_{12}y_{2}\label{eq:HMeq1}\\
\ddot{y}_{2}+\gamma_{2}^{2}y_{2} & =S_{21}y_{1}\label{eq:HMEq2}
\end{align}
with 
\begin{equation}
S_{ns}(T)=-e_{n}\cdot\ddot{e}_{s}-2e_{n}\cdot\dot{e}_{s}\partial_{T}
\end{equation}
and 
\begin{equation}
\gamma_{i}^{2}=m_{i}^{2}-\dot{e}_{i}\cdot\dot{e}_{i}+k^{2}/a^{2}
\end{equation}
where we define $\gamma_{n}^{2}$ as the effective frequency squared.
Note the Hubble friction term has been removed by rescaling the mode
functions.

Let us assume that the eigenvector gradient term $\dot{e}_{1}\cdot\dot{e}_{1}$
peaks at a time $T_{*}$ when the background fields tend to cross
each other such that the kinetic energy corresponding to the relative
velocity of the two fields is maximized and the eigenvector gradient
$\dot{e}_{1,2}$ is larger than $O(F)$. Thus, the mode-mixing operator
$S_{ns}$ is significant in a small neighborhood $O\left(1/F\right)$
around $T_{*}$. We begin with the normalized eigenstates and rewrite
the gradient terms in $S_{ns}$ through an approximate Lorentzian
function $L_{ns}(T-T_{*})$ with peak at $T_{*}$ such that 
\begin{align}
e_{n}\cdot\partial_{T}e_{s} & \approx L_{ns}(T-T_{*})\\
e_{n}\cdot\partial_{T}^{2}e_{s} & \approx\partial_{T}L_{ns}(T-T_{*}).
\end{align}
Note that $e_{n}\cdot\dot{e}_{s}$ term is symmetric around $T_{*}$
while $e_{n}\cdot\ddot{e_{s}}$ is anti-symmetric. The second term
$-3\dot{e_{s}}\cdot e_{n}$ in $S_{ns}$ is due to Hubble friction
and can be removed by scaling the mode functions without affecting
our discussion.

During the early phase when $|\phi_{+}|\gg|\phi_{-}|$, the heavier
mode $y_{2}$ is forced driven by the lighter mode $y_{1}$. This
is similar to the perturbed solution for the $\phi_{\pm}$ background
fields where the $\phi_{-}$ field is effectively forced driven by
$\phi_{+}$. Accordingly, the $y_{2}$ mode has the following solution
\begin{equation}
y_{2}\approx\frac{S_{21}y_{1}}{\gamma_{2}^{2}}\label{eq:y2earlyphase}
\end{equation}
where $y_{1,2}$ satisfying the condition 
\begin{equation}
\left|\frac{y_{2}}{y_{1}}\right|\ll1.
\end{equation}
Meanwhile, the right hand side term $S_{12}y_{2}$ in the $y_{1}$mode
equation is negligible and thus 
\begin{equation}
\mathcal{O}_{1}y_{1}\approx0.
\end{equation}
Therefore, the lighter mode is decoupled from the heavier during the
early phase since $\gamma_{2}^{2}\sim O(\phi_{+}^{2})\gg O(F^{2})$.

Later at $T\approx T_{*}$ the eigenvector gradient terms in $S_{ns}$
become significant $O\left(F^{2}\right)$. At the same time, the effective
frequency squared $\gamma_{1,2}^{2}$ approach a local minima at $T\approx T_{*}$.
Consequently, $y_{2}$ amplitude reaches a local maximum close to
$T_{*}$ such that the heavy mode mixing effect due to the term $S_{12}y_{2}$
in the RHS of Eq.~(\ref{eq:lightmode}) cannot be neglected. Post
$T_{*}$, the heavier mode $y_{2}$ behaves like an underdamped harmonic
oscillator and undergoes rapid oscillations with a large frequency
$\sqrt{\gamma_{2}^{2}}$ due to the heavier mass eigenvalue.

To evaluate the heavy mode mixing effect, we will approximate the
function $L_{ns}(T-T_{*})$ as a rectangular ETSP of amplitude $E$
and width $\Delta T\approx1/E$ where $E\sim O(F)$\footnote{The following assumption can be verified using the analytical form
of the background fields from Sec.~\ref{subsec:Beyond-perturbation-theory}}
\begin{align}
\left|e_{n}\cdot\partial_{T}e_{s}\right| & \approx\begin{cases}
E & T_{1}<T<T_{2}\\
0 & {\rm otherwise}
\end{cases}.
\end{align}
Using this approximation, the $e_{n}\cdot\ddot{e_{s}}$ term peaks
at the boundaries $T_{1,2}$ and remains $0$ within the interval
$\left[T_{1},T_{2}\right]$.

\subsection{$y_{1}$ solution}

The general solution for the lighter mode $y_{1}$ from equation $\ref{eq:HMeq1}$
can be expressed as 
\begin{equation}
y_{1}=\left(c_{1}-\int f_{2}\frac{S_{12}y_{2}}{W}dT\right)f_{1}(T)+\left(c_{2}+\int f_{1}\frac{S_{12}y_{2}}{W}dT\right)f_{2}(T)\label{eq:y1generalsoln}
\end{equation}
where $f_{1,2}(T)$ are the linearly independent functions that solve
the homogeneous equation $\ddot{y}_{1}+\gamma_{1}^{2}y_{1}=0$. The
coefficients $c_{1,2}$ are obtained from the initial conditions at
$T=T_{1}$ and the integral terms correspond to the inhomogeneous
part on the RHS of $\ref{eq:HMeq1}$. $W$ is the associated Wronskian
of $f_{1,2}$. Within the interval $\left[T_{1},T_{2}\right]$, $0<\dot{e}_{1}\cdot\dot{e}_{1}\sim O(F^{2})$,
while $m_{1}^{2}<0$ due to the deviation of the background fields
away from the flat direction. Therefore, the effective lighter frequency
squared $\gamma_{1}^{2}=m_{1}^{2}-\dot{e}_{1}\cdot\dot{e}_{1}+k^{2}/a(T_{*})^{2}<0$
for $k^{2}/a(T_{*})^{2}\ll m_{1}^{2}-\dot{e}_{1}\cdot\dot{e}_{1}$.
Therefore, the lighter mode $y_{1}$ has the homogeneous solution
\begin{equation}
y_{1}^{h}(T)\approx y_{1}(T_{1})\cosh\left[\sqrt{9/4-\gamma_{1}^{2}}\left(T-T_{1}\right)\right]
\end{equation}
where 
\begin{align}
f_{1}(T) & =e^{\sqrt{-\gamma_{1}^{2}}T},\qquad f_{2}(T)=e^{-\sqrt{-\gamma_{1}^{2}}T}
\end{align}
and $\partial_{T}y_{1}(T_{1})/y_{1}(T_{1})\ll\sqrt{9/4-\gamma_{1}^{2}}\sim O(F)$
so that $c_{1,2}\approx y_{1}(T_{1})/2$. Note that a positive value
of $\int dT\,f_{2}S_{12}y_{2}/W$ integral accounts for a decrease
in the power of the lighter mode (dominated by the $f_{1}$ mode if
$k^{2}/a(T_{*})^{2}\ll m_{1}^{2}-\dot{e}_{1}\cdot\dot{e}_{1}$ such
that $\gamma_{1}^{2}<0$) due to the heavy mode coupling. As we will
show later, the integral term is indeed positive such that a finite
fraction of the power is removed by the heavier mode.

\subsection{$y_{2}$ solution}

To solve for the heavier mode $y_{2}$ within the interval $\left[T_{1},T_{2}\right]$,
we rewrite Eq.~(\ref{eq:HMEq2}) as
\begin{equation}
\ddot{y}_{2}+\gamma_{2}^{2}y_{2}=-2L_{21}(T_{*})\partial_{T}y_{1}.
\end{equation}
Assuming decoupling of the modes, we will substitute $y_{1}$ with
$y_{1}^{h}$.\footnote{Although this is a cyclic argument, we prove this by self-consistency
at the end.} The $y_{2}$ is then given by
\begin{equation}
y_{2}\approx\frac{-2L_{21}(T_{*})y_{1}(T_{1})}{\gamma_{2}^{2}+(9/4-\gamma_{1}^{2})}\sqrt{9/4-\gamma_{1}^{2}}\sinh\left[\sqrt{9/4-\gamma_{1}^{2}}\left(T-T_{1}\right)\right]+\frac{-\partial_{T}L_{21}(T_{1})y_{1}(T_{1})}{\gamma_{2}^{2}(T_{1})}\cos\left[\gamma_{2}\left(T-T_{1}\right)\right]
\end{equation}
where the first term is via the forced component in the RHS while
the second term is the homogeneous component that oscillates with
frequency $\gamma_{2}$ with initial conditions set at $T_{1}$. Using
the above solution for $y_{2}$, the RHS term $S_{12}y_{2}=-2L_{12}(T_{*})\partial_{T}y_{2}$
is evaluated as
\begin{align}
S_{12}y_{2} & \approx-2\left(\frac{-2L_{12}(T_{*})L_{21}(T_{*})y_{1}(T_{1})}{\gamma_{2}^{2}-\gamma_{1}^{2}}\left(-\gamma_{1}^{2}\right)\cosh\left[\sqrt{-\gamma_{1}^{2}}\left(T-T_{1}\right)\right]\right.\nonumber \\
 & \left.-\frac{-L_{12}(T_{*})\partial_{T}L_{21}(T_{1})y_{1}(T_{1})}{\gamma_{2}(T_{1})}\sin\left[\gamma_{2}\left(T-T_{1}\right)\right]\right).
\end{align}
Using the equations for the background fields in Sec.~\ref{subsec:Beyond-perturbation-theory},
$L_{12}(T_{*})L_{21}(T_{*})\approx-E^{2}$ and $L_{12}(T_{*})\partial_{T}L_{21}\approx-nE^{3}$
for $n\approx4/3$. Hence we have,
\begin{align}
S_{12}y_{2} & \approx4y_{1}(T_{1})E^{2}\left(\frac{\gamma_{1}^{2}}{\gamma_{2}^{2}-\gamma_{1}^{2}}\cosh\left[\sqrt{-\gamma_{1}^{2}}\left(T-T_{1}\right)\right]+\frac{nE/2}{\gamma_{2}(T_{1})}\sin\left[\gamma_{2}\left(T-T_{1}\right)\right]\right).
\end{align}

\subsection{Heavy mixing coefficient $\chi_{{\rm HM}}$}

We are now in a position to complete the $y_{1}$ solution in Eq.~(\ref{eq:y1generalsoln})
by solving the integral terms. Since, the homogeneous function $f_{1}$
dominates over $f_{2}$ we will only solve $\int dT\,f_{2}S_{12}y_{2}/W$
within the interval $\left[T_{1},T_{2}\right]$ when the $S_{ns}$
operator is significant. Using the Wronskian $W=-2\sqrt{-\gamma_{1}^{2}}$,
\begin{align}
\int f_{2}\frac{S_{12}y_{2}}{W}dT & =4y_{1}(T_{1})\frac{\gamma_{1}^{2}E^{2}}{\gamma_{2}^{2}-\gamma_{1}^{2}}\int_{T_{1}}^{T_{2}}e^{-\sqrt{-\gamma_{1}^{2}}T}\frac{\cosh\left[\sqrt{-\gamma_{1}^{2}}\left(T-T_{1}\right)\right]}{-2\sqrt{-\gamma_{1}^{2}}}dT\nonumber \\
+ & 4y_{1}(T_{1})\frac{nE^{3}/2}{\gamma_{2}(T_{1})}\int_{T_{1}}^{T_{2}}\frac{e^{-\sqrt{-\gamma_{1}^{2}}T}}{-2\sqrt{-\gamma_{1}^{2}}}\sin\left[\gamma_{2}\left(T-T_{1}\right)\right]dT
\end{align}
\begin{align}
 & \approx4y_{1}(T_{1})\frac{\gamma_{1}^{2}E^{2}}{\gamma_{2}^{2}-\gamma_{1}^{2}}\left(\frac{1+2\sqrt{-\gamma_{1}^{2}}\Delta T-e^{-2\sqrt{-\gamma_{1}^{2}}\Delta T}}{-8\left(-\gamma_{1}^{2}\right)}\right)\nonumber \\
 & +4y_{1}(T_{1})\frac{nE^{3}/2}{\gamma_{2}(T_{1})}\left(\frac{-\gamma_{2}+e^{-\sqrt{-\gamma_{1}^{2}}\Delta T}\left(\sqrt{-\gamma_{1}^{2}}\sin\left[\gamma_{2}\Delta T\right]+\gamma_{2}\cos\left[\gamma_{2}\Delta T\right]\right)}{2\sqrt{-\gamma_{1}^{2}}\left(\gamma_{2}^{2}-\gamma_{1}^{2}\right)}\right).
\end{align}The above can be further simplified as, 
\begin{equation}
\int f_{2}\frac{S_{12}y_{2}}{W}dT\approx4y_{1}(T_{1})\frac{E^{2}}{\gamma_{2}^{2}-\gamma_{1}^{2}}\left[\frac{1+2\sqrt{-\gamma_{1}^{2}}\Delta T}{8}+\frac{nE/4}{\sqrt{-\gamma_{1}^{2}}}\left(-1+e^{-\sqrt{-\gamma_{1}^{2}}\Delta T}\cos\left[\gamma_{2}\Delta T\right]\right)\right].
\end{equation}
We now define heavy mixing coefficient $\chi_{{\rm HM}}$ as 
\begin{equation}
\chi_{{\rm HM}}=\frac{\int f_{2}\frac{S_{12}y_{2}}{W}dT}{c_{1}}.\label{eq:heavymodemixingeffect}
\end{equation}
Using $c_{1}\approx y_{1}(T_{1})/2$ and $E\Delta T\approx1$, we
obtain
\begin{align}
\left.\chi_{{\rm HM}}\left(l_{1}^{2},l_{2}^{2}\right)\right|_{T=T_{*}} & \approx\frac{1}{l_{2}^{2}-l_{1}^{2}}\left(1+2\sqrt{-l_{1}^{2}}+\frac{2n}{\sqrt{-l_{1}^{2}}}\left(-1+e^{-\sqrt{-l_{1}^{2}}}\cos\left[l_{2}\right]\right)\right)\label{eq:chiHM}
\end{align}
where $l_{i}^{2}=\gamma_{i}^{2}/E^{2}$ for $E^{2}=\max\left(\dot{e}_{1}^{2}\right)$
and we approximate $n\approx4/3$. Since $m_{i}^{2}$ and $E^{2}$
are $O\left(F^{2}\right)$, the parameters $l_{i}^{2}$ are almost
$F$ independent for $F\gg1$.

\begin{figure}[H]
\begin{centering}
\includegraphics[scale=0.45]{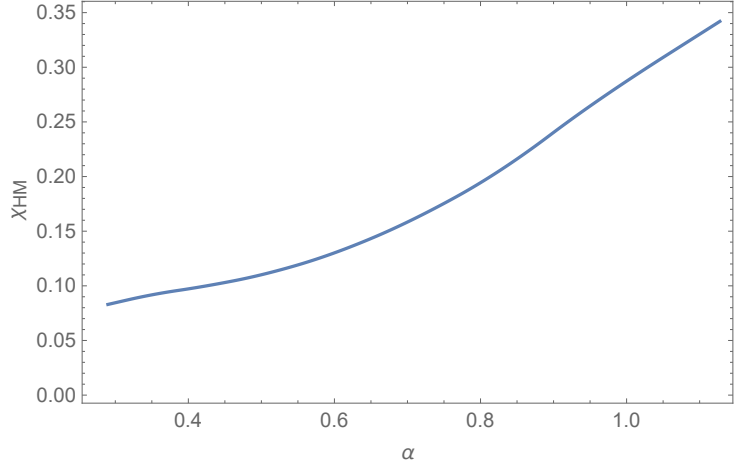} 
\par\end{centering}
\caption{\label{fig:chiHMplot}Analytical plot of the fractional reduction
in the amplitude of the lighter mode $y_{1}$ due to heavy mode mixing
evaluated at $T_{*}\approx T_{c}$ using Eq.~(\ref{eq:chiHM}) in
the limit $k^{2}/a(T_{*})^{2}\rightarrow0$ plotted with respect to
the parameter $\alpha$ defined in Eq.~(\ref{eq:alphacplus}). Numerical
results suggest that the estimation given in Eq.~(\ref{eq:chiHM})
is an approximate upper bound as explained in the text.}
\end{figure}

From Eq.~(\ref{eq:y1generalsoln}), we infer that the two modes shall
remain decoupled as long as $\max\left(\chi_{{\rm HM}}\right)\ll1$
where we define $\max\left(\chi_{{\rm HM}}\right)$ as a local maxima
in the vicinity of $T_{*}$ within a neighbourhood of $O(1/F)$. Note
that a positive value of $\chi_{{\rm HM}}$ accounts for a decrease
in the power of the lighter mode due to the heavy mode coupling. The
mixing between the two modes thus results in a significant proportion
of power transfer from the lighter mode to the heavier and as a result
the isocurvature power spectrum reduces. Fig.~\ref{fig:chiHMplot}
gives analytical plot of $\chi_{{\rm HM}}$ evaluated at $T_{*}\approx T_{c}$
in the limit $k^{2}/a(T_{*})^{2}\rightarrow0$ using Eq.~(\ref{eq:chiHM})
plotted with respect to the parameter $\alpha$ defined in Eq.~(\ref{eq:alphacplus}).
By considering a reasonable decoupling between the two modes for $\chi_{{\rm HM}}\lesssim0.2$,
we obtain an upper bound of $\alpha_{{\rm U}}\sim1$ for fields crossing
each other close to $T_{*}$. If we consider $F\gg1$ cases, then
the upper bound $\alpha_{{\rm U}}$ is almost $F$-independent. For
$\alpha>\alpha_{2}$, every subsequent crossing of the two background
fields post transition will give rise to similar $\dot{e}_{1}\cdot\dot{e}_{1}$
peaks. The effective heavy mixing is then a sum of the contributions
from each of these peaks. Since the peaks are exponentially suppressed
by Hubble friction, their contribution is significantly low. However,
for large $F$, the subsequent peaks can get closer to each other
and hence the net heavy mixing contribution can become significant. 

From Eq.~(\ref{eq:chiHM}), we infer that as $k$ increases from
zero, $l_{1}^{2}$ begins to reduce in magnitude such that $\chi_{{\rm HM}}$
initially reduces until $k/a(T_{*})\sim\sqrt{-m_{1}^{2}}$. Thereafter,
$\chi_{{\rm HM}}$ turns imaginary and begins to increase in magnitude.
The above analytical estimate is primarily valid as long as $\chi_{{\rm HM}}$
remains much less than unity since in order to estimate $\chi_{{\rm HM}}$
we have approximated $S_{21}y_{1}$ as $S_{21}y_{1}^{h}$ by substituting
with the homogeneous $y_{1}^{h}$ solution. As $\left|y_{1}\right|\leq\left|y_{1}^{h}\right|$,
one expects that the $\chi_{{\rm HM}}$ evaluated using the exact
$y_{1}$ solution should be lower than the above estimate as long
as Eq.~(\ref{eq:chiHM}) is valid. If $k$ eventually becomes large
enough such that the $k^{2}/a(T_{*})^{2}$ term dominates over the
remaining mass-squared terms, we obtain $\gamma_{1}^{2}=\gamma_{2}^{2}$
at $T_{*}$ and the modes are strongly coupled such that $\left|y_{2}\right|\rightarrow\left|y_{1}\right|$.
Similar strong coupling is possible if $E\gg m_{2}^{2}\sim F^{2}\sqrt{c_{+}/c_{-}}$.
However, in such cases, the corresponding value of $\Delta T\ll1/F$.
Since the coupled system of $y_{1,2}$ has only one dominant time
scale of $O\left(1/F\right)$, the two modes momentarily tend to $\left|y_{2}\right|\sim\left|y_{1}\right|$
at $T_{*}$ before $y_{1}$ returns back to the attractor solution
$y_{1}^{h}$. Hence, whenever $E\gg m_{2}^{2}$, the coupling between
the two modes can be neglected.

\bibliographystyle{JHEP2}
\bibliography{blueiso_constraints,Inflation_general,blue_isocurvature,axion_isocurvature_papers2,file,misc,kasuya_citations_and_other_misc_papers2}

\end{document}